\documentclass[12pt,twoside]{book}

\pdfoutput=1
\usepackage[english]{babel}
\usepackage[a4paper,tmargin=3truecm,bmargin=3truecm,rmargin=2.5truecm,
lmargin=2.5truecm,twoside,verbose=true]{geometry}
\usepackage{cancel,graphicx,subfig}
\usepackage[all]{xy}
\usepackage[pdftex]{hyperref}
\usepackage{amsmath,amssymb}
\usepackage[amsmath, hyperref, thmmarks]{ntheorem}
\usepackage{pdfpages}
\usepackage[sf,bf,compact,topmarks,calcwidth,pagestyles]{titlesec}
\usepackage{titletoc}

\titleformat{\chapter}[display]
{\filleft\normalfont\fontsize{30}{32}\sffamily\bfseries\mathversion{bold}}%
{\fontsize{36}{38}\selectfont\thechapter}{20pt}{\Huge}
\titlespacing*{\chapter} {0pt}{50pt}{40pt}
\titleformat{\section}
{\normalfont\sffamily\Large\bfseries\mathversion{bold}}{\thesection}{1em}{}
\titlespacing*{\section} {0pt}{3ex plus 1ex minus .2ex}{2ex plus .2ex}
\titleformat{\subsection}
{\normalfont\sffamily\large\bfseries\selectfont\mathversion{bold}}{\thesubsection}{1em}{}
\titlespacing*{\subsection} {0pt}{2.25ex plus .6ex minus .2ex}{1ex plus .2ex}
\titleformat{\subsubsection}
{\normalfont\sffamily\normalsize\bfseries\mathversion{bold}}{\thesubsubsection}{1em}{}
\titlespacing*{\subsubsection}{0pt}{2ex plus .7ex minus .2ex}{.7ex plus .2ex}
\titleformat{\paragraph}[runin]
{\normalfont\normalsize\bfseries\mathversion{bold}}{}{0pt}{}
\titlespacing*{\paragraph} {\oldParIndent}{2.25ex plus 1ex minus .2ex}{0.5em}
\titleformat{\subparagraph}[runin]
{\normalfont\normalsize\bfseries\mathversion{bold}}{\thesubparagraph}{1em}{}
\titlespacing*{\subparagraph} {\parindent}{3.25ex plus 1ex minus .2ex}{1em}

\renewpagestyle{plain}[\normalfont\sffamily\bfseries\mathversion{bold}]{
  \setfoot[][][]{}{}{}
  \sethead[][][]
  {}{}{}}
\renewpagestyle{myheadings}[\normalfont\sffamily\bfseries\mathversion{bold}]{
  \headrule
  \setfoot[][][]{}{}{}
  \sethead[\usepage][][%
  \ifthechapter{\chaptertitlename\ \thechapter\ -- \chaptertitle}{\chaptertitle}]
  {\ifthesection{\thesection\hspace{1.2em}\sectiontitle}{\chaptertitle}}%
  {}{\usepage}}
\renewpagestyle{empty}[]{
  \setfoot{}{}{}
  \sethead[][][]
  {}{}{}  }

\newcommand\caA{{\mathcal A}}
\newcommand\caB{{\mathcal B}}
\newcommand\caC{{\mathcal C}}
\newcommand\caF{{\mathcal F}}
\newcommand\caG{{\mathcal G}}
\newcommand\caL{{\mathcal L}}
\newcommand\caM{{\mathcal M}}
\newcommand\caS{{\mathcal S}}
\newcommand\caU{{\mathcal U}}
\newcommand\caZ{{\mathcal Z}}
\newcommand\wx{{\widetilde x}}
\newcommand\gone{{ \mathchoice {1\mskip-4mu\mathrm{l} } {1\mskip-4mu\mathrm{l} }{1\mskip-4.5mu\mathrm{l} } {1\mskip-5mu\mathrm{l}} }}
\newcommand\gR{{\mathbb R}}
\newcommand\gS{{\mathbb S}}
\newcommand\gK{{\mathbb K}}
\newcommand\gC{{\mathbb C}}
\newcommand\gH{{\mathbb H}}
\newcommand\gN{{\mathbb N}}
\newcommand\gZ{{\mathbb Z}}
\newcommand\Omr{\underline\Omega_\varepsilon}
\newcommand\algzero{{\mathsf 0}}
\newcommand\algA{\mathbf A}
\newcommand\algrA{\mathbf A^\bullet}
\newcommand\algB{\mathbf B}
\newcommand\algrB{\mathbf B^\bullet}
\newcommand\modM{{\boldsymbol M}}
\newcommand\modrM{{\boldsymbol M^\bullet}}
\newcommand\kg{{\mathfrak g}}

\newcommand\ksl{{\mathfrak{sl}}}
\newcommand\kX{{\mathfrak X}}
\newcommand\kY{{\mathfrak Y}}
\newcommand\kS{{\mathfrak S}}
\newcommand\eps{{\varepsilon}}
\newcommand\ad{{\text{\textup{ad}}}}

\newcommand{\grast}{\bullet}
\newcommand\fois{\mathord{\cdot}}
\DeclareMathOperator{\tr}{Tr} 
\DeclareMathOperator{\Hom}{\mathsf{Hom}}
\DeclareMathOperator{\Aut}{\mathsf{Aut}}
\newcommand\Der{{\text{\textup{Der}}}}
\newcommand\Int{{\text{\textup{Int}}}}
\newcommand\Lie{{\text{\textup{Lie}}}}
\newcommand\Out{{\text{\textup{Out}}}}
\newcommand\dd{{\text{\textup{d}}}}
\newcommand\invar{{\text{\textup{inv}}}}
\newcommand{\omi}[1]{\buildrel { \buildrel{#1}\over{\vee} } \over .}
\newcommand\norm{\mathord{\parallel}}
\newcommand\Omder{\underline\Omega_{\text{Der}}}
\newcommand\Matr{{M}}
\newcommand\Unit{{U}}
\newcommand\ksu{{\mathfrak{su}}}
\newcommand\symes{{\mathchoice{\textstyle\mathsf{S}}{\textstyle\mathsf{S}}%
{\scriptstyle\mathsf{S}}{\scriptscriptstyle\mathsf{S}}}} 
\newcommand\exter{{\textstyle\bigwedge}} 
\newcommand\Supp{{\text{\textup{Supp}}}}
\newcommand\ehH{{\mathcal H}}
\DeclareMathOperator{\Sp}{Sp}
\newcommand\otimeshat{\mathrel{\widehat\otimes}}
\newcommand\Dom{{\text{\textup{Dom}}}}
\newcommand\cutint{{\int \!\!\!\!\!\! -}}

\theoremstyle{break}
\theoremsymbol{}
\theorembodyfont{\slshape}
\theoremheaderfont{\normalfont\bfseries}
\theoremseparator{}
\newtheorem{Theorem}{Theorem}[section]
\newtheorem{theorem}[Theorem]{Theorem}
\newtheorem{Proposition}[Theorem]{Proposition}
\newtheorem{proposition}[Theorem]{Proposition}

\newtheorem{lemma}[Theorem]{Lemma}
\newtheorem{Corollary}[Theorem]{Corollary}
\newtheorem{corollary}[Theorem]{Corollary}

\theorembodyfont{\upshape}
\theoremsymbol{\ensuremath{\blacklozenge}}

\newtheorem{example}[Theorem]{Example}

\newtheorem{remark}[Theorem]{Remark}

\newtheorem{definition}[Theorem]{Definition}

\theoremstyle{nonumberplain}
\theoremheaderfont{\scshape}
\theorembodyfont{\normalfont}
\theoremsymbol{\ensuremath{\blacksquare}}

\newtheorem{proof}{Proof}
\qedsymbol{\ensuremath{_\blacksquare}}
\theoremclass{LaTeX}

\begin{document}
\includepdf[pages={1}]{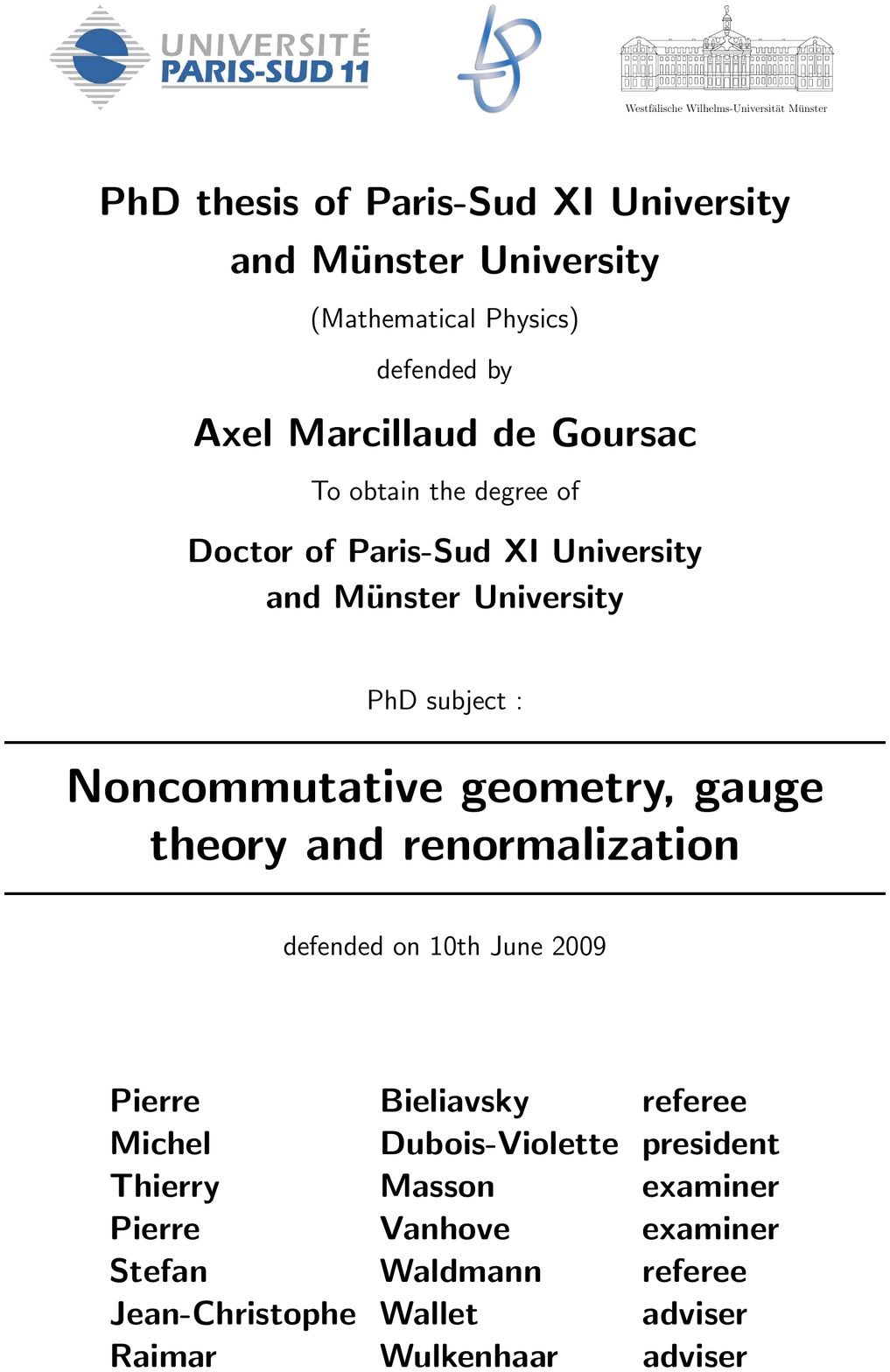}

\pagestyle{empty}

\tableofcontents

\vskip.5cm
\section*{R\'esum\'e}

De nos jours, la g\'eom\'etrie non-commutative est un domaine grandissant des math\'ema-
tiques, qui peut appara\^itre comme un cadre prometteur pour la physique moderne. Les th\'eories quantiques des champs sur des ``espaces non-commutatifs'' ont en effet \'et\'e tr\`es \'etudi\'ees, et sont sujettes \`a un nouveau type de divergence, le m\'elange ultraviolet-infrarouge. Cependant, une solution a r\'ecemment \'et\'e apport\'ee \`a ce probl\`eme par H. Grosse et R. Wulkenhaar en ajoutant \`a l'action d'un mod\`ele scalaire sur l'espace non-commutatif de Moyal, un terme harmonique qui la rend renormalisable.

Un des buts de cette th\`ese est l'extension de cette proc\'edure aux th\'eories de jauge sur l'espace de Moyal. En effet, nous avons introduit une nouvelle th\'eorie de jauge non-commutative, fortement reli\'ee au mod\`ele de Grosse-Wulkenhaar, et candidate \`a la renormalisabilit\'e. Nous avons  ensuite \'etudi\'e ses propri\'et\'es les plus importantes, notamment ses configurations du vide. Finalement, nous donnons une interpr\'etation math\'ematique de cette nouvelle action en terme de calcul diff\'erentiel bas\'e sur les d\'erivations, associ\'e \`a une superalg\`ebre.

Ce travail contient, outre les r\'esultats mentionn\'es ci-dessus, une introduction \`a la g\'eom\'etrie non-commutative, une introduction aux alg\`ebres $\eps$-gradu\'ees, d\'efinies dans cette th\`ese, et une introduction \`a la renormalisation des th\'eories quantiques de champs scalaires (point de vue wilsonien et BPHZ) et de jauge.
\vskip2cm

\section*{Zusammenfassung}

Die nichtkommutative Geometrie bildet als wachsendes Gebiet der
Mathematik einen vielversprechenden Rahmen f\"ur die moderne
Physik. Quantenfeldtheorien \"uber nichtkommutativen R\"aumen werden
zur Zeit intensiv studiert. Sie f\"uhren zu einer neuen Art von
Divergenzen, die ultraviolett-infrarot Mischung. Eine 
L\"osung dieses Problems wurde von H.~Grosse und R.~Wulkenhaar
durch Hinzuf\"ugen eines harmonischen Terms zur
Wirkung des $\phi^4$-Modells gefunden. Dadurch wird diese
Quantenfeldtheorie \"uber dem Moyal-Raum renormierbar.

Ein Ziel dieser Doktorarbeit ist die Verallgemeinerung dieses
harmonischen Terms auf Eichtheorien \"uber dem Moyal-Raum. 
Basierend auf dem Grosse-Wulkenhaar-Modell wird eine neue 
nichtkommutative Eichtheorie eingef\"uhrt, die begr\"undete Chancen hat,
renormierbar zu sein. Die wichtigsten Eigenschaften dieser
Eichtheorie, insbesondere die Vakuumskonfigurationen, werden
studiert. Schlie\ss{}lich wird mittels eines zu einer
Superalgebra assoziierten Derivationskalk\"uls eine mathematische
Interpretation dieser neuen Wirkung gegeben. 

Die Arbeit enth\"alt neben diesen Ergebnissen eine
Einf\"uhrung zur nichtkommutativen Geometrie und zu 
$\eps$-graduierten Algebren sowie eine Einf\"uhrung in die 
Renormierung von Quantenfeldtheorien f\"ur Skalarfelder 
(nach Wilson und BPHZ) und Eichfelder.
\vskip2cm

\section*{Abstract}

Nowadays, noncommutative geometry is a growing domain of mathematics, which can appear as a promising framework for modern physics. Quantum field theories on ``noncommutative spaces'' are indeed much investigated, and suffer from a new type of divergence called the ultraviolet-infrared mixing. However, this problem has recently been solved by H. Grosse and R. Wulkenhaar by adding to the action of a noncommutative scalar model a harmonic term, which renders it renormalizable.

One aim of this thesis is the extension of this procedure to gauge theories on the Moyal space. Indeed, we have introduced a new noncommutative gauge theory, strongly related to the Grosse-Wulkenhaar model, and candidate to renormalizability. We have then studied the most important properties of this action, and in particular its vacuum configurations. Finally, we give a mathematical interpretation of this new action in terms of a derivation-based differential calculus associated to a superalgebra.

This work contains among the results of this PhD, an introduction to noncommutative geometry, an introduction to $\eps$-graded algebras, defined in this thesis, and an introduction to renormalization of scalar (wilsonian and BPHZ point of view) and gauge quantum field theories.

\newpage

\section*{Acknowledgments}
\vskip1cm

First of all, I want to thank my PhD advisers Jean-Christophe Wallet and Raimar Wulkenhaar for having accepted to be in charge of this thesis. I am very grateful to them for having initated me into noncommutative geometry and quantum field theory both from the point of view of mathematics and physics. I have very much appreciated the scientific enthusiasm of Jean-Christophe and the teaching skills of Raimar, their patience and attention, which have permitted me to benefit from excellent working conditions.
\medskip

I also thank Henk Hilhorst, director of the Laboratoire de Physique Th\'eorique of Paris-Sud XI University, and Joachim Cuntz, Dekan of the Mathematisches Institut of the Westf\"alische Wilhelms University of M\"unster, for their welcome in this double Franco-German and mathematical-physical framework, which have been very enriching.
\medskip

Many thanks to Pierre Bieliavsky and Stefan Waldmann, who accepted to read and to judge this thesis, and to Michel Dubois-Violette, Thierry Masson, and Pierre Vanhove, who accepted to be examiners of the PhD defense. A special thank to Thierry whose lecture notes attracted me towards noncommutative geometry, and who taught me much in this area. I enjoyed to collaborate with him and his help on \LaTeX  has been precious.
\medskip

I want also to thank Adrian Tanasa for his collaboration and his precious advices, and all people I have discussed fruitfully with: Johannes Aastrup, Eric Cagnache, Yvon Georgelin, Jesper Grimstrup, Razvan Gurau, John Madore, Jacques Magnen, Pierre Martinetti, Mario Paschke, Vincent Rivasseau, Fabien Vignes-Tourneret, Wend Werner, and all people I cannot name here.

Many thanks to Philippe Boucaud, Olivier Brand-Froissac and Jean-Pierre Leroy for their help concerning my computer problems, Patricia Dubois-Violette for her bibliographic help, and Mireille Calvet and Gabriele Dierkes for their kindness.
\medskip

Finally, I want to thank my parents, my family and my friends for having supported during my studies and these three years of PhD thesis.

\pagestyle{myheadings}

\numberwithin{equation}{section}
\chapter*{Introduction}
\chaptermark{Introduction}

Noncommutative geometry is a new domain of mathematics, which joins several different areas of mathematics and physics. Its origin lies in quantum mechanics, describing at a microscopic level the laws of nature. Quantum mechanics motivated also in the first half of the twentieth century an important development in the theory of operator algebras, like the study of $C^\ast$-algebras and von Neumann algebras.

The essential principle of noncommutative geometry lies in the duality between spaces and (associative) algebras, so that properties of spaces can be algebraically characterized. Indeed, at the level of topology, there is an anti-equivalence between the category of locally compact Hausdorff spaces and the category of commutative $C^\ast$-algebras. In the case of a compact space $X$, the associated commutative $C^\ast$-algebra is the algebra $C(X)$ of complex-valued continuous functions on $X$. In the same way, to measurable spaces correspond commutative von Neumann algebras. This duality also exists in the framework of algebraic geometry, between affine algebraic varieties and finitely generated commutative unital reduced algebras for example.

\medskip

From classical mechanics to quantum mechanics, one changes the commutative algebra of functions on the phase space to a noncommutative operator algebra on a Hilbert space. One of the founder of noncommutative geometry \cite{Connes:1994}, A. Connes, realized that this change could also be done in the various mathematical domains mentionned above, as soon as there is a duality between spaces and commutative algebras. Indeed, one can speak about ``noncommutative topology'' by considering noncommutative $C^\ast$-algebras as functions algebras on ``noncommutative spaces'', which do not have concrete existence. It can be abstractly formalized by defining the category of (locally compact Hausdorff) {\it noncommutative} topological spaces as the dual category of the one of the (noncommutative) $C^\ast$-algebras. The noncommutative measure theory is then defined by noncommutative von Neumann algebras. This very rich way of thinking permits to generalize some important constructions and properties from spaces to corresponding noncommutative algebras.

\medskip

What do classical notions become, as smooth manifolds, fiber bundles, de Rham forms complex and its cohomology, group actions on spaces? What are their noncommutative algebraic analogs? Firstly, the algebraic version of vector bundles on a locally compact Hausdorff space $X$ is given by finitely generated projective modules on the commutative algebra $C(X)$, as described in the Serre-Swan Theorem. Since the K-theory groups of a topological space $X$ are defined from isomorphy classes of vector bundles on $X$, the K-theory can be extended to ``noncommutative topological spaces'', namely to $C^\ast$-algebras.

According to the results of Atiyah, Brown-Douglas-Fillmore, and Kasparov, the dual theory of K-theory, K-homology of the topological space $X$, is represented by the abstract elliptic operators on $X$. The coupling between the two dual theories is then given by: $\langle[D],[E]\rangle=$ the Fredholm index of the elliptic operator $D$ with coefficients in the vector bundle $E$. K-homology can also be extended to noncommutative $C^\ast$-algebras, and K-theory and K-homology of $C^\ast$-algebras are unified in the KK-theory of Kasparov.

In ordinary differential geometry, the Chern Character is a particular characteristic class, $ch(E)\in H^\bullet_{\text{dR}}(M)$ for all complex vector bundle $E\to M$, and it induces a isomorphism $ch:K^\bullet(M)\otimes\gR\to H^\bullet_{\text{dR}}(M)$ between K-theory and de Rham cohomology. There exists a noncommutative generalization of the Chern character, from the K-theory of an algebra to the periodic cyclic homology of this algebra. This cyclic homology introduced by A. Connes, appears then as a noncommutative version of the de Rham cohomology, which is an important tool of the differential geometry.

Another way to generalize the de Rham complex to the noncommutative case is to consider the differential calculus based on the derivations of a noncommutative algebra, introduced by M. Dubois-Violette \cite{DuboisViolette:1988cr}, and studied by Connes, Kerner, Madore, Masson,... Indeed, this differential calculus mimics what happens for a manifold $M$: the vector fields $\Gamma(M)$ are the derivations of the algebra $C^\infty(M)$, and the one-forms are the duals of vector fields. However, this differential calculus exists for all associative algebras, involves only algebraic informations, without constraints of functional analysis, so that the exact noncommutative analog of algebras $C^\infty(M)$ is not defined in this framework.

Recently, A. Connes has shown that a noncommutative analog of a smooth (compact oriented) manifold is a spectral triple $(\algA,\ehH,D)$, composed of an algebra $\algA$, a Hilbert space $\ehH$, on which $\algA$ is represented, and an (unbounded) selfadjoint operator $D$, which are respectively analogs of the algebra $C^\infty(M)$, the bundle of spinors, and the Dirac operator of a riemannian (compact orientable) manifold $M$. The spectral triples are related to K-homology and permit to introduce the local index formula (Connes, Moscovici) generalizing the index theorem of Atiyah-Singer.

Finally, an action of a (compact) topological group on a topological space can be generalized in the noncommutative framework by a quantum group, namely an action of a (noncommutative) Hopf algebra. To each right action of a topological group $G$ on a topological space $X$ is associated a coaction $\Gamma:C(X)\to C(X)\otimes C(G)\subset C(X\times G)$, defined by
\begin{equation*}
\forall\chi\in C(X),\quad \forall x\in X,\quad\forall g\in G,\quad \Gamma(\chi)(x,g)=\chi(x\fois g),
\end{equation*}
satisfying some constraints, which can be generalized to noncommutative algebras replacing $C(X)$ and $C(G)$. The group $G$ acts on itself by multiplication, so that the coaction defined in this case satisfies the axioms of a coproduct, and endows $C(G)$ with a structure of Hopf algebra (for an introduction to quantum groups, see \cite{Podles:1998}). Moreover, the Chern-Weil theory on the Lie group actions can be extended to the actions of Hopf algebras, by using the local index formula and the cyclic cohomology of Hopf algebras. This is still a lively domain. Note that noncommutative geometry is also related to other domains of mathematics like number theory, but this will not be exposed here.

\medskip

Let us also mention that a particular type of noncommutative algebras is given by deformation quantization of commutative algebras. Indeed, one can deform the commutative product of the algebra of smooth functions $C^\infty(M)$ on a Poisson manifold $M$ into a noncommutative product by using the Poisson structure. The general theorem of existence of such deformations has been settled by M. Kontsevich in 1997. Note that some systems described by quantum mechanics can be reformulated by a deformation quantization, where the commutative algebra of functions of the phase space is deformed in the direction of the canonical Poisson bracket into a noncommutative algebra of observables. This is also a lively domain of research.

\medskip

On another side, graded algebra has been studied for a long time \cite{Nastasescu:1982}, and its most well-known applications are the theories of supermanifolds, of graded associative and Lie algebras, and the supersymmetry in Physics. A natural generalization of $\gZ$-grading (or $\gZ_2$-grading) is the $\Gamma$-grading, where $\Gamma$ is an abelian group. Nevertheless, in order to recover similar properties as for $\gZ$-graded Lie algebras, one has to define an additional structure called the commutation factor $\eps$. One can now ask for a noncommutative geometry adapted to this setting of graded algebras. It would then correspond to some ``graded noncommutative spaces''. In this work, we will make a step in the direction of graded noncommutative geometry by studying the notion of $\eps$-graded associative algebra and defining an appropriate differential calculus.

\bigskip

Noncommutative geometry is then a growing domain of mathematics with numerous applications, whose construction and development are bound with physics. Indeed, certain effective models describing fractional quantum Hall effect have been interpreted by using noncommutative geometry. Moreover, it has been used to reproduce the standard model of particle physics, including a riemannian version of gravity. Let us analyze some heuristic arguments for the use of noncommutative geometry in fundamental physics.

Actual physical theories describing the four fundamental known interactions, namely gravitation, electromagnetism, strong and weak nuclear interactions, are well verified experimentally in their own domain of validity: general relativity describes the gravitation at large scales, even if the question of dark matter and energy is posed by cosmology and astrophysics; quantum physics, renormalizable quantum field theory and the standard model of particle physics at small scales. However, modern physics is investigating a unification of both theories because there exist concrete situations for which both theories are needed: neutrons stars, behavior of elementary particles in the neighborhood of a black hole, primordial universe, etc... But this unification encounters a big problem: the non-renormalizability of the Einstein-Hilbert action describing general relativity, so that this theory is not consistant in the quantum field theory formalism used in the standard model. To solve this problem, several ways have been proposed. String theory and quantum loop gravity are two of the most well-known examples, even if they will not be considered here. The use of noncommutative geometry in physics is a third possible way.

\medskip

With the progress of science, one can observe a relative mathematization of physics, since it needs more and more mathematics to define its notions, its measurable quantities. Indeed, in the last century, a part of the physics, kinematics, enters into geometry in the special relativity, by introducing the concept of a minkowskian space-time. Then, with the general relativity, gravitation is viewed as a part of the geometry thanks to its lorentzian metric. In the standard model at the classical level, interactions are expressed in geometrical terms with principal fiber bundles on the space-time. We see therefore that relativistic physics uses a lot of notions and tools of differential and spinorial geometry, such as Lie groups, connections, gauge transformations, invariants,...

On another side, quantum physics is expressed in terms of operator algebras, as we have seen it above. This permits to understand the quantification of observed measurements for the photoelectric effect for instance, by interpreting possible measures as eigenvalues of some selfadjoint operator; but also Heisenberg uncertainty's relations which are a direct consequence of the non-commutativity of two operators, in this formalism. Modern physics asks today for a unification of relativity and quantum physics, and we may deduce from our above analysis that this unification will be done by a supplementary step in the mathematization. Thus, a conceptual framework is required such that it can contain differential geometry and operator algebras. Such conditions are satisfied by noncommutative geometry.

The standard model of particle physics has been reproduced at the classical level, with a riemannian formulation of gravity, by a spectral triple whose algebra is given by the tensor product of $C^\infty(M)$, the smooth functions on a manifold $M$, by a matrix algebra of finite dimension \cite{Chamseddine:2006ep}. It would be an additional step in the direction described above. Indeed, it is a (noncommutative) generalization of Kaluza-Klein theory, which wanted to include directly some (electromagnetic) interactions in the space-time geometry, without fiber bundles but by additional dimensions, which had eventually to be compactified. However, by denoting $G=SU(3)\times SU(2)\times U(1)$ the structure group of the trivial fiber bundle used in the standard model on the space-time manifold $M$, the symmetry group of the standard model including gravity is then
\begin{equation*}
\caG=C^\infty(M,G)\rtimes \text{Diff}(M),
\end{equation*}
the semi-direct product group of diffeomorphisms on $M$ with gauge transformations, which admits $C^\infty(M,G)$ as a normal subgroup. Then, a result due to W.~Thurston, D.~Epstein and J.~Mather proved that it does not exist any ``commutative'' manifold $M'$ such that $\text{Diff}(M')\approx \caG$. On the contrary, the automorphisms group $\Aut(\algA)$ of a noncommutative algebra $\algA$ admits the inner group $\Int(\algA)$ as a normal subgroup, and by choosing
\begin{equation*}
\algA=C^\infty(M)\otimes(\gC\oplus\gH\oplus M_3(\gC)),
\end{equation*}
A. Connes shew that $\Aut(\algA)\approx\caG$. In this noncommutative model, interactions are then encoded by the matrix algebra, which corresponds to additional ``discrete dimensions''. Note however that this model stays at the classical level of field theory, and that the noncommutativity of matrices corresponds to the non-abelianity of the gauge theory, but not to all the quantum properties. Then, noncommutative geometry appears as a possible unifying framework for modern fundamental physics.

\bigskip

Another argument has been developed by Doplicher, Fredenhagen, Roberts \cite{Doplicher:1994zv}, in favor of the use of noncommutative geometry in fundamental physics. If one supposes that a system is described both by quantum field theory and general relativity, and that an observer can measure the position of any component of the system very precisely, until a Planck scale precision, then the necessary energy will be very large by Heisenberg uncertainties relations, so that a black hole can be generated. Therefore, the observation of the geometry would lead to a change of the geometry. To avoid this problem, one can release the axiom of commutativity of the position operators in the different directions and keep an unavoidable uncertainty on the observed position.

Even if the paper \cite{Doplicher:1994zv} does not consider it as a possible solution, the Moyal space, a deformation quantization of $\gR^{D}$, is a prototype of noncommutative space whose coordinate functions satisfy the relation:
\begin{equation*}
[x_\mu,x_\nu]_\star=i\Theta_{\mu\nu},
\end{equation*}
where $\Theta$ is some noncommutativity matrix. One can then define a field theory on this noncommutative space, and the most natural way for a scalar theory is to replace the usual commutative product by the Moyal $\star$-product in the action. One thus obtains:
\begin{equation*}
S=\int \dd^Dx\Big(\frac 12(\partial_\mu\phi)^2+\frac{\mu^2}{2}\phi^2+\lambda\,\phi\star\phi\star\phi\star\phi\Big).
\end{equation*}
Then, the Moyal product occurs only in the quartic interaction of the action, so that this theory can also be viewed as a field theory on the commutative space $\gR^D$ but with a non-local interaction.

\medskip

One of the key ingredients of quantum field theory is renormalization. A theory is said to be renormalizable if the vertex functional (corresponding to the quantum action) takes the same form as the classical action at a certain scale, which means that the infinite values appearing in the perturbative computations of Feynman diagrams can be reabsorbed in a finite number of parameters of the action. Two conditions are usually required to show the renormalizability of a theory: power-counting and locality.

For a scalar theory on the Moyal space, we have seen that the interaction is no longer local, so that the question of renormalizability of such a theory can be asked. Minwalla, Van Raamsdonk and Seiberg have discovered a new type of divergences in this theory \cite{Minwalla:1999px}, called Ultraviolet-Infrared (UV/IR) mixing, which renders the theory non-renormalizable, due to the non-locality of the Moyal product.

\medskip

However, H. Grosse and R. Wulkenhaar have proposed a satisfying solution to this problem by adding a harmonic term in the action:
\begin{equation*}
S=\int \dd^Dx\Big(\frac 12(\partial_\mu\phi)^2+\frac{\Omega^2}{2}\wx^2\phi^2+\frac{\mu^2}{2}\phi^2+\lambda\,\phi\star\phi\star\phi\star\phi\Big).
\end{equation*}
This theory is then renormalizable up to all orders in perturbation \cite{Grosse:2004yu}, and it is the first scalar real theory on a noncommutative space (or with non-local interaction) to be renormalizable. Since gauge theories are used in the standard model, one can ask wether a renormalizable gauge theory exists on the Moyal space. The standard generalization of usual gauge theory to the Moyal space:
\begin{equation*}
S=\frac{1}{4}\int\dd^Dx\Big(F_{\mu\nu}F_{\mu\nu}\Big),\text{ with }F_{\mu\nu}=\partial_\mu A_\nu-\partial_\nu A_\mu-i[A_\mu,A_\nu]_\star,
\end{equation*}
suffers also from the UV/IR mixing, and one wants to find the analog of the harmonic term for adding it to the gauge action, so that it becomes renormalizable. This was one of the aims of this thesis.

\medskip

A new model of gauge theory \cite{deGoursac:2007gq} is indeed proposed in this PhD thesis:
\begin{equation*}
S=\int\dd^Dx\left(\frac 14 F_{\mu\nu}\star F_{\mu\nu}+\frac{\Omega^2}{4}\{\caA_\mu,\caA_\nu\}_\star^2+\kappa\caA_\mu\star\caA_\mu\right),
\end{equation*}
where $\caA_\mu=A_\mu+\frac 12\wx_\mu$, and its properties are studied. Moreover, we introduce a differential calculus adapted to $\eps$-graded algebras \cite{deGoursac:2008bd} in order to interpret mathematically this new gauge action.

\bigskip

The plan of this PhD is as follows. In the chapter \ref{cha-gnc}, we will briefly recall the important results of noncommutative geometry: to the topological level correspond the $C^\ast$-algebras, to the level of measure theory correspond the von Neumann algebras, and at the level of differential geometry, three noncommutative structures are interesting for instance, derivation-based differential calculus which will be useful in the rest of this thesis, cyclic cohomology and spectral triples.

Then, in chapter \ref{cha-eps}, one will introduce the structure of $\eps$-graded algebras, a generalization of associative $\gZ$-graded algebras, and the derivation-based differential calculus adapted to these algebras will be exposed. It will be illustrated in particular for the matrix algebra. This study can be seen as a step in the direction of noncommutative graded geometry.

Chapter \ref{cha-ren} is a brief introduction to renormalization. We will study the links between the wilsonian and the BPHZ approach to renormalization for scalar theories, which will be useful in the chapter \ref{cha-moy}. One will also present the algebraic version of renormalization of gauge theories, and this will permit us to see the major directions in the proof of the renormalizability of the new gauge theory proposed in this thesis.

We will then present in chapter \ref{cha-moy} a brief introduction to deformation quantization, the construction of the Moyal space, and we will study the questions of renormalizability of scalar theories on such a space. The Grosse-Wulkenhaar model will be exposed in details, and its vacuum configurations computed.

Finally, in chapter \ref{cha-gauge}, we shall present the new gauge theory on the Moyal space, candidate to renormalizability, obtained by an effective action. We will also study its properties, and in particular its vacuum configurations. A mathematical interpretation in terms of a graded curvature will be given within the formalism introduced in chapter \ref{cha-eps}.

\numberwithin{equation}{section}
\chapter[Noncommutative Geometry]{Introduction to noncommutative geometry}
\label{cha-gnc}

\section{\texorpdfstring{Topology and $C^\ast$-algebras}{Topology and C*-algebras}}
\label{sec-gnc-cast}

In this section, we present the theorem of Gelfand-Na\"imark, which states that a locally compact Hausdorff space $X$ is entirely characterized by the commutative $C^\ast$-algebra of its continuous functions (vanishing at infinity) $C_0(X)$, and that each commutative $C^\ast$-algebra is of this type. This major result permits to introduce the continuous functional calculus. Moreover, complex vector bundles on such spaces $X$ correspond then to finitely generated projective modules on $C_0(X)$. See \cite{Blackadar:2006,Landsman:1998,Kadison:1983,Bratteli:1987} for more about $C^\ast$-algebras.

\subsection{Definitions}
\label{subsec-gnc-castdef}

\begin{definition}
\begin{itemize}
\item A {\it Fr\'echet algebra} $\algA$ is a complete topological algebra, for the topology defined by a countable family of algebra seminorms $(p_i)_i$ such that $\forall a\in\algA$, $a=0\Leftrightarrow (\forall i,\ p_i(a)=0)$.
\item A {\it Banach algebra} $\algA$ is a complete normed algebra, such that $\norm\gone\norm=1$ if $\algA$ is unital.
\item A {\it $C^\ast$-algebra} $\algA$ is a complete complex involutive normed algebra satisfying $\forall a\in \algA$, $\norm a^\ast a\norm=\norm a\norm^2$ (and $\norm \gone\norm=1$ if $\algA$ is unital).
\item A morphism between two $C^\ast$-algebras $\algA$ and $\algB$ is a morphism $\algA\to\algB$ of involutive associative algebras. It is then automatically continuous for the norm topology.
\end{itemize}
\end{definition}

\begin{proposition}[Properties]
\begin{itemize}
\item If $\algA$ is a $C^\ast$-algebra, then $\forall a\in \algA$, $\norm a^\ast\norm=\norm a\norm$.
\item If $\ehH$ is a complex Hilbert space, then $\caB(\ehH)$, the space of continuous endomorphisms (or bounded operators) of $\ehH$, endowed with the uniform norm, is a $C^\ast$-algebra.
\item A {\it representation} of the $C^\ast$-algebra $\algA$ is a morphism of $C^\ast$-algebras $\pi:\algA\to\caB(\ehH)$.
\item Any $C^\ast$-algebra is then isomorphic to a $C^\ast$-subalgebra of $\caB(\ehH)$, for a certain $\ehH$ (see subsection \ref{subsec-gnc-castgns}).
\item Conversely, any subalgebra of $\caB(\ehH)$, closed for the uniform topology and the involution, is a $C^\ast$-subalgebra of $\caB(\ehH)$.
\end{itemize}
\end{proposition}

In the following of this section, $\algA$ is a unital $C^\ast$-algebra and $\ehH$ a complex separable Hilbert space.

\subsection{Spectral theory}
\label{subsec-gnc-castspectr}

Here, one introduces the basics of the spectral theory for $C^\ast$-algebras.

\begin{definition}
For $a\in \algA$, one defines the {\it spectrum} of $a$ in $\algA$ by
\begin{equation*}
\Sp_\algA(a)=\{\lambda\in\gC,\ (a-\lambda\gone) \text{ not invertible in } \algA\}.
\end{equation*}
Then, $\Sp_\algA(a)$ is a compact subspace of the complex disk of radius $\norm a\norm$.

Moreover, $\Sp_\algA(a^\ast)=\{\overline\lambda,\ \lambda\in\Sp_\algA(a)\}$ and $\Sp_\algA(a^2)=\{\lambda^2,\ \lambda\in\Sp_\algA(a)\}$.
\end{definition}

\begin{definition}
\begin{itemize}
\item $a\in \algA$ is {\it normal} if $a^\ast a=aa^\ast$.
\item $a\in \algA$ is {\it selfadjoint} if $a^\ast=a$.
\item $a\in \algA$ is {\it positive} if $\exists \,b\in \algA$ such that $a=b^\ast b$. The set of positive elements is a convex cone denoted by $\algA_+$.
\item $a\in \algA$ is {\it unitary} if $a^\ast a=aa^\ast=\gone$.
\item $a\in \algA$ is a (orthogonal) {\it projector} si $a^2=a^\ast=a$.
\item $a\in \algA$ is a {\it partial isometry} if $a^\ast a$ is a projector.
\end{itemize}
\end{definition}

\begin{proposition}[Properties]
\begin{itemize}
\item Any selfadjoint $a\in\algA$ can be decomposed into $a=a_+-a_-$, with $a_+$ and $a_-$ positive elements.
\item If $a$ is selfadjoint (resp. positive, unitary, orthogonal projector), $\Sp_\algA(a)\subset\gR$ (resp. $\gR_+$, $\gS^1$, $\{0,1\}$). The converse is generally false.
\end{itemize}
\end{proposition}

\begin{definition}
The {\it spectral radius} of $a\in \algA$ is defined by: $r(a)=\sup\{|\lambda|,\ \lambda\in\Sp_\algA(a)\}=\lim_{n\to\infty}\norm a^n\norm^{\frac 1n}$. If $a$ is normal, $r(a)=\norm a\norm$.

For $a\in \algA$ and $\lambda\notin\Sp_\algA(a)$, one defines the resolvent of $a$ in $\lambda$ by: $R(a,\lambda)=(a-\lambda\gone)^{-1}$.
\end{definition}

\subsection{Duality in the commutative case}
\label{subsec-gnc-castdual}

\begin{definition}
\begin{itemize}
\item A {\it character} on the $C^\ast$-algebra $\algA$ is a morphism of $C^\ast$-algebras $\chi:\algA\to\gC$.
\item The {\it spectrum} of $\algA$ is then defined as the set of its characters and denoted by $\Sp(\algA)$. The weak${}^\ast$-topology on $\Sp(\algA)$ is defined by the convergence $\chi_n\to\chi$ if and only if $\forall a\in\algA$, $\chi_n(a)\to\chi(a)$.
\item The spectrum $\Sp(\algA)$ is a locally compact Hausdorff space for the weak${}^\ast$-topology.
\end{itemize}
\end{definition}

Let $X$ be a locally compact Hausdorff space. If $X$ is compact, then $C(X)$, the set of all continuous maps $X\to\gC$, endowed with the sup-norm $\norm\fois\norm_\infty$, is a unital $C^\ast$-algebra; else $C_0(X)$, the subalgebra of $C(X)$ of the functions vanishing at infinity, with the sup-norm, is also a $C^\ast$-algebra, but not unital.
\begin{definition}
Let $\algA$ be a $C^\ast$-algebra. Then, the {\it Gelfand transformation} is the following map:
\begin{equation*}
\algA\to C(\Sp(\algA))\qquad a\mapsto\widehat a,
\end{equation*}
with $\forall\chi\in\Sp(\algA)$, $\widehat a(\chi)=\chi(a)$. Then, $\forall a\in \algA$,
\begin{equation*}
\Sp_\algA(a)=\{\chi(a),\ \chi\in\Sp(\algA)\}=\widehat a(\Sp(\algA)).
\end{equation*}
\end{definition}

In the commutative case, the Gelfand transformation is bijective.

\begin{theorem}[Gelfand-Na\"imark]
\label{thm-gnc-gn}
If $\algA$ is a commutative $C^\ast$-algebra, the Gelfand transformation is an isomorphism of $C^\ast$-algebras:
\begin{itemize}
\item if $\algA$ is unital, there exists a unique (up to homeomorphism) compact Hausdorff space $X=\Sp(\algA)$, such that $\algA\approx C(X)$.
\item if $\algA$ is not unital, there exists a unique (up to homeomorphism) locally compact Hausdorff space, $X=\Sp(\algA)$, but non compact, such that $\algA\approx C_0(X)$.
\end{itemize}
\end{theorem}

This theorem shows the equivalence between considering a locally compact Hausdorff space and a commutative $C^\ast$-algebra. Then, thanks to this correspondence, a noncommutative $C^\ast$-algebra can be interpreted as the algebra of continuous functions of a ``noncommutative space'', which however does not exist. This is the essence of noncommutative geometry.

Furthermore, one can define the continuous functional calculus for a (noncommutative) $C^\ast$-algebra thanks to the Gelfand-Na\"imark theorem.

\begin{theorem}[Continuous functional calculus]
Let $a$ be a normal element of $\algA$, a unital $C^\ast$-algebra, and $C^\ast(a)$ the minimal commutative $C^\ast$-subalgebra of $\algA$ involving $\gone$, $a$ and $a^\ast$. Then, the spectrum of $a$ is independent from the algebra $\algA$:
\begin{equation*}
\Sp_\algA(a)=\Sp_{C^\ast(a)}(a)=\Sp(C^\ast(a)) \quad\text{and}\quad C^\ast(a)=C(\Sp(a)).
\end{equation*}

By considering the above identification, the Gelfand transformation $\widehat a:\Sp(a)\to\gC$ corresponds to the identity, and for any $f\in C(\Sp(a))$, one associates a unique element $f(a)\in C^\ast(a)\subset \algA$ such that:
\begin{equation*}
\overline f(a)=f(a)^\ast,\qquad\norm f(a)\norm=\norm f\norm_\infty,\qquad \Sp(f(a))=f(\Sp(a)).
\end{equation*}
It is called the {\it continuous functional calculus}.
\end{theorem}

\begin{definition}
For any element $a$ of $\algA$, one defines its {\it absolute value} by $|a|=\sqrt{a^\ast a}$, with the help of the continuous functional calculus.
\end{definition}

\subsection{GNS construction}
\label{subsec-gnc-castgns}

In this subsection, we recall the construction by Gelfand-Na\"imark-Segal of a Hilbert space and a representation from a $C^\ast$-algebra, thanks to its states.

\begin{definition}
A {\it state} on a unital $C^\ast$-algebra $\algA$ is a linear map $\omega:\algA\to\gC$, such that $\omega(\gone)=1$ et and $\forall a\in \algA_+$, $\omega(a)\geq0$. The set of all states is denoted by $\caS(\algA)$. Then $\caS(\algA)$ is a convex space and $\forall \omega\in\caS(\algA)$, $\omega$ is continuous.

A {\it pure state} of $\algA$ is an extremal point of the convex space $\caS(\algA)$.
\end{definition}
If $\algA$ is non-unital, we replace the condition $\omega(\gone)=1$ by $\norm\omega\norm=1$ for $\omega$ to be a state of $\algA$. Any character $\chi\in\Sp(\algA)$ is a state of $\algA$.

\begin{theorem}[GNS Representation]
\label{thm-gnc-gns}
To any state $\omega\in\caS(\algA)$, one associates uniquely (up to equivalence) a representation $\pi_\omega:\algA\to\caB(\ehH_\omega)$ and an element $\Omega_\omega\in\ehH_\omega$ such that:
\begin{align*}
&\overline{\pi_\omega(\algA)\Omega_\omega}=\ehH_\omega \quad(\Omega_\omega\text{ is {\it cyclic}}),\\
&\forall a\in \algA,\quad \omega(a)=\langle\Omega_\omega,\pi_\omega(a)\Omega_\omega\rangle.
\end{align*}
Moreover, $\pi_\omega$ is irreducible $\Leftrightarrow$ $\omega$ is a pure state of $\algA$.
\end{theorem}

Indeed, $(a,b)\mapsto\omega(b^\ast a)$ is a positive hermitian sesquilinear form on $\algA$. Let $\mathcal N_\omega=\{a\in \algA,\ \omega(a^\ast a)=0\}$. This form can be projected into a scalar product $\langle-,-\rangle$ on $\algA/\mathcal N_\omega$, and we denote $\ehH_\omega$ the Hilbert space obtained as the completion of $\algA/\mathcal N_\omega$ for this scalar product.

$\algA$ acts also by its product on $\ehH_\omega$, and this defines the GNS representation $\pi_\omega$ associated to $\omega$. One defines also the cyclic vector $\Omega_\omega$ as the projected of $\gone\in \algA$ on $\algA/\mathcal N_\omega$.

\begin{definition}
The {\it universal representation} of $\algA$ is defined by $\pi_U=\bigoplus_{\omega\in \caS(\algA)}\pi_\omega$. It is injective.
\end{definition}

\begin{definition}
If $\algA_1$ and $\algA_2$ are $C^\ast$-algebras, the tensor product $\algA_1\otimeshat \algA_2$ is the completion of $\algA_1\otimes \algA_2$ for the norm of $\caB(\ehH_1\otimeshat\ehH_2)$ thanks to the universal representations of $\algA_1$ and $\algA_2$. It is also a $C^\ast$-algebra.
\end{definition}

\begin{theorem}
Let $X$ a locally compact Hausdorff space.
\begin{itemize}
\item If $X$ is compact, the state space $\caS(\algA)$ of $\algA=C(X)$ consists of all probability measures on $X$.
\item If $X$ is compact, the pure state space of $\algA=C(X)$ is homeomorphic to $X$.
\item If $X$ is non compact, the pure state space of $\algA=C_0(X)$ is homeomorphic to $X$.
\end{itemize}
\end{theorem}

As a consequence, the pure state space of a commutative $C^\ast$-algebra $\algA$ is exactly its spectrum $\Sp(\algA)$.

\subsection{Vector bundles and projective modules}
\label{subsec-gnc-bundle}

We have seen what can be the noncommutative analog of a (locally compact Hausdorff) topological space, it is given by a noncommutative $C^\ast$-algebra. However, one can construct on topological spaces the notion of fiber bundles. Let us see how to interpret in a noncommutative way vector bundles on topological spaces.

\begin{definition}[bundles]
Let $X$ and $V$ be topological spaces.
\begin{itemize}
\item A {\it bundle} on base $X$ with typical fiber $V$ is a topological space $E$, endowed with a continuous surjection $\pi:E\to X$, such that $\forall x\in X$, there exists a neighborhood $U_\alpha$ of $x$ in $X$, and a homeomorphism $\phi_\alpha:\pi^{-1}(U_\alpha)\to U_\alpha\times V$ for which $\pi=p_X\circ\phi_\alpha$, with $p_X:X\times V\to X$ and $p_V:X\times V\to V$ the projections resp. on $X$ and $V$.
\item Moreover, $E$ is a {\it vector bundle} if $V$ is a finite dimensional vector space, $\forall x\in X$, $\pi^{-1}(x)$ is also a finite dimensional vector space whose relative topology (coming from the one of $E$) coincides with its vector space topology, and if $p_V\circ\phi_\alpha:\pi^{-1}(x)\to V$ is linear.
\end{itemize}
\end{definition}
The maps $\phi_\alpha$ are called local trivializations, and $\pi^{-1}(x)$ the fibers of $E$. A simple example of bundle is the trivial bundle $E=X\times V$, with $\pi(x,v)=x$, for $x\in X$ and $v\in V$. Let $n\in\gN$; a {\it complex vector bundle} on $X$ is a vector bundle $E$ on $X$ with typical fiber $V=\gC^n$.

\begin{definition}[sections]
Let $E$ be a vector bundle on $X$ with typical fiber $V$. A {\it section} on $E$ is a continuous map $s:X\to E$ such that $\pi\circ s=\text{id}_X$. The set of all sections of $E$ is a vector space denoted by $\Gamma(E)$.

If $X$ is a connected compact Hausdorff space, $\Gamma(E)$ is a right module on the commutative $C^\ast$-algebra $C(X)$ for the action: $\forall f\in C(X)$, $\forall s\in\Gamma(E)$, $\forall x\in X$,
\begin{equation*}
(s\,f)(x)=s(x)\,f(x).
\end{equation*}
\end{definition}

\begin{theorem}[Serre-Swan]
\label{thm-gnc-ss}
Let $X$ be a connected compact Hausdorff space. There is a bijective correspondence between complex vector bundles on $X$ and finitely generated projective modules on the commutative $C^\ast$-algebra $C(X)$. This correspondence is made by considering the module of sections of the vector bundle.
\end{theorem}

This theorem is used in $K$-theory in order to define groups associated to such spaces $X$, which contain informations on $X$. Of course, this theorem permits also to define $K$-theory for (noncommutative) $C^\ast$-algebras, since it is formulated in algebraic terms. This is the essence of noncommutative geometry to obtain classical theorems and definitions in algebraic terms so that they can be generalized into a noncommutative version.

\section{Measure theory and von Neumann algebras}
\label{sec-gnc-vn}

More details on the theory of von Neumann algebras can be found in \cite{Sakai:1971,Kadison:1983,Takesaki:2000}.

\subsection{Definition of von Neumann algebras}
\label{subsec-gnc-vndef}

Let $\ehH$ be a complex Hilbert space. One can construct two locally convex topologies on $\caB(\ehH)$, different from the uniform topology (associated to its norm).

\begin{definition}
\begin{itemize}
\item The {\it weak} topology on $\caB(\ehH)$ is induced by seminorms $T\to\langle x,T(y)\rangle$, $x,y\in\ehH$. Then, $T_n\to T$ weakly $\Leftrightarrow$ $\forall x,y\in\ehH$, $\langle x,T_n(y)\rangle\to\langle x,T(y)\rangle$.
\item The {\it strong} topology on $\caB(\ehH)$ is induced by seminorms $T\to\norm T(x)\norm$, $x\in\ehH$. Then, $T_n\to T$ strongly $\Leftrightarrow$ $\forall x\in\ehH$, $T_n(x)\to T(x)$.
\end{itemize}
The weak topology is weaker than the strong topology, which is weaker than the uniform topology.
\end{definition}

\begin{definition}
Let $\algA$ be a subset of $\caB(\ehH)$. On defines its {\it commutant} by:
\begin{equation*}
\algA'=\{T\in\caB(\ehH),\ \forall S\in\algA,\ [S,T]=0\}.
\end{equation*}
$\algA'$ is a unital subalgebra of $\caB(\ehH)$, even if $\algA$ is not an algebra.

Moreover, $\algA\subset \algA''=\algA^{(4)}=\dots$ and $\algA'=\algA^{(3)}=\dots$.
\end{definition}

The following theorem shows the importance of the notion of a von Neumann algebra, which can be viewed on several ways.

\begin{theorem}[Bicommutant]
Let $\algA$ be a unital $C^\ast$-subalgebra of $\caB(\ehH)$. Then, there is an equivalence between:
\begin{enumerate}
\item $\algA=\algA''$.
\item $\algA$ is weakly closed in $\caB(\ehH)$.
\item $\algA$ is strongly closed in $\caB(\ehH)$.
\item $\algA$ is the dual space of a Banach space (called its predual, unique up to isomorphism).
\end{enumerate}
If $\algA$ satisfies the previous conditions, $(\ehH,\algA)$ is called a {\it von Neumann algebra}, or a {\it $W^\ast$-algebra} if there is no reference to $\ehH$.
\end{theorem}
Every $W^\ast$-algebra is a $C^\ast$-algebra.

\begin{proposition}[Properties]
\begin{itemize}
\item The minimal von Neumann algebra involving a $C^\ast$-algebra $\algA$ is its bicommutant $\algA''$, also equal to its bidual $\algA^{\ast\ast}$.
\item Any subalgebra of $\caB(\ehH)$, closed for the weak topology and the involution, is a $W^\ast$-subalgebra of $\caB(\ehH)$.
\end{itemize}
\end{proposition}

In the following, $\ehH$ is a complex separable Hilbert space, and $(\ehH,\algA)$ a von Neumann algebra.

\begin{definition}
A von Neumann algebra $\algA$ is said to be {\it maximal abelian} if $\algA'=\algA$.

If an abelian von Neumann algebra admits a cyclic vector (see Theorem \ref{thm-gnc-gns}), then it is maximal abelian.
\end{definition}

\subsection{Duality in the commutative case}
\label{subsec-gnc-vndual}

Like in the case of commutative $C^\ast$-algebras and locally compact Hausdorff spaces, there is a duality between commutative von Neumann algebras and (localizable) measurable spaces.

\begin{example}
Let $(X,\mu)$ be a measurable space, with $\mu$ a finite measure on $X$. Then, the set $L^\infty(X,\mu)$ of all ($\mu$-essentially) bounded functions on $X$ is a commutative von Neumann algebra, and its predual is given by the set $L^1(X,\mu)$ of all $\mu$-integrable functions on $X$.
\end{example}

\begin{theorem}
\label{thm-gnc-vnc}
Let $\algA$ be a commutative von Neumann algebra. Then, there exists a localizable\footnote{direct sum of finite measurable spaces} measurable space $(X,\mu)$ such that $\algA\approx L^\infty(X,\mu)$.
\end{theorem}

In the commutative case, the links between von Neumann algebras and the previous constructions for $C^\ast$-algebras are summarized in the following theorem.
\begin{theorem}
Let $X$ be a locally compact Hausdorff space, and $\mu$ a finite measure on $X$. Then, $\mu$ defines a positive linear form on the commutative $C^\ast$-algebra $C_0(X)$:
\begin{equation*}
f\mapsto\int_X f(x)\dd\mu(x).
\end{equation*}
\begin{itemize}
\item Moreover, the cyclic GNS representation $\pi_\mu$ associated to this linear form is realized on the Hilbert space $\ehH_\mu=L^2(X,\mu)$ by multiplication by functions.
\item The von Neumann algebra generated by $\pi_\mu(C_0(X))$ is exactly $L^\infty(X,\mu)$, and it is maximal abelian.
\end{itemize}
\end{theorem}

\section{Noncommutative differential geometry}
\label{sec-gnc-geodiff}

\subsection{Algebraic geometry}
\label{subsec-gnc-alggeom}

The duality between spaces and commutative algebras is not a new story in algebraic geometry. For example, Hilbert established a duality between affine algebraic varieties and finitely generated commutative reduced algebras, in its {\it Nullstellensatz} (see \cite{Hartshorne:1977}). We recall here the definitions and the construction of this duality.
\begin{definition}
\begin{itemize}
\item An {\it affine algebraic variety} over a field $\gK$ is a subset of $\gK^n$, of zeros of a set of polynomials in $\gK[X_1,\dots,X_n]$.
\item A morphism of affine algebraic varieties between $V\subset\gK^n$ and $W\subset\gK^m$ is the restriction of a polynomial map $\gK^n\to\gK^m$.
\item A {\it reduced algebra} is an algebra without nilpotent element.
\item A morphism of finitely generated commutative reduced algebras is a morphism of algebras between finitely generated commutative reduced algebras.
\end{itemize}
\end{definition}

\begin{proposition}
\label{prop-gnc-alggeom}
\begin{itemize}
\item Let $V\subset\gK^n$ be an affine algebraic variety. Then, its coordinate ring $\gK[V]$, defined by
\begin{equation*}
\gK[V]=\gK[X_1,\dots,X_n]/\{P\in\gK[X_1,\dots,X_n],\ P_{|V}\equiv 0\}
\end{equation*}
is a finitely generated commutative unital reduced algebra.
\item Reciprocally, if $\algA$ is a finitely generated commutative unital reduced algebra with $n$ generators, then one can write 
\begin{equation*}
\algA\simeq\gK[X_1,\dots,X_n]/I,
\end{equation*}
where $I$ is a radical ideal. We associate to $\algA$ the affine algebraic variety
\begin{equation*}
V=\{x\in\gK^n,\quad \forall P\in I,\ P(x)=0\}.
\end{equation*}
Then, $\gK[V]\simeq\algA$, and this procedure defines an anti-equivalence of categories.
\end{itemize}
\end{proposition}

In the same way, there is a duality between compact Riemann surfaces and algebraic function fields \cite{Farkas:1980}. Noncommutative algebraic geometry, generalizing these notions to noncommutative algebras, is a lively domain \cite{Artin:1994}. These constructions and the dualities in the case of commutative $C^\ast$-algebras (see section \ref{sec-gnc-cast}) and commutative von Neumann algebras (see section \ref{sec-gnc-vn}) motivated the study of what could be such a duality in the framework of differential geometry \cite{Connes:1994}.

Throughout this section, we restrain our study to compact spaces, or equivalently to unital commutative algebras.

\subsection{Differential calculi}
\label{subsec-gnc-diffcalc}

The algebraic properties of the de Rham complex of a differentiable manifold can be extracted from their geometrical framework. Indeed, the concepts of forms and differential have been generalized for an arbitrary associative algebra.
\begin{definition}[Differential calculus]
\label{def-gnc-diffcalc}
Let $\algA$ be an associative unital algebra. A {\it differential calculus} on $\algA$ is a $\gN$-graded differentiable algebra $(\Omega^\bullet,\dd)$ such that $\Omega^0=\algA$ and $\dd(\gone)=0$.
\end{definition}
Note that $\Omega^\bullet$ is then a graded unital algebra, but $\algA$ and $\Omega^\bullet$ are not supposed to be respectively commutative and graded commutative.

For a smooth compact manifold $M$, the algebra $\algA=C^\infty(M)$ is unital and commutative, and the de Rham complex $(\Omega^\bullet_{\text{dR}}(M),\dd)$ is a graded commutative differential calculus in the sense of Definition \ref{def-gnc-diffcalc}.
\medskip

Let us now define a particular differential calculus for any unital algebra.
\begin{definition}[Universal differential calculus]
\label{def-gnc-univdiffcalc}
The {\it universal differential calculus} of an associative algebra $\algA$ is the free unital $\gN$-graded differentiable algebra $(\Omega_U^\bullet(\algA),\dd_U)$ generated by $\algA$ in degree $0$. It is a differential calculus in the sense of Definition \ref{def-gnc-diffcalc}.
\end{definition}
It is generated by elements of the form
\begin{equation*}
a_0\,\dd_U a_1\,\dd_U a_2\dots \dd_U a_n,
\end{equation*}
where $a_i\in\algA$. A concrete realization of this differential calculus is given by the identification $\Omega_U^n(\algA)\subset\algA^{\otimes(n+1)}$, in which the differential takes the form $\forall a_i\in\algA$,
\begin{multline*}
\dd_U(a_0\otimes\dots\otimes a_n)=\gone\otimes a_0\otimes\dots\otimes a_n\\
+\sum_{p=1}^n(-1)^p\, a_0\otimes\dots\otimes a_{p-1}\otimes\gone\otimes a_p\otimes\dots\otimes a_n\\
+(-1)^{n+1}a_0\otimes\dots\otimes a_n\otimes \gone,
\end{multline*}
and the product:
\begin{equation*}
(a_0\otimes\dots\otimes a_n)\fois (b_0\otimes\dots\otimes b_m)=a_0\otimes\dots\otimes a_{n-1}\otimes(a_n\fois b_0)\otimes b_1\otimes\dots\otimes b_m.
\end{equation*}
The cohomology of this differential calculus is trivial.
\medskip

We can define another differential calculus, based on the derivations of the algebra, which is the noncommutative analog of the de Rham complex. It has been introduced in \cite{DuboisViolette:1988cr,DuboisViolette:1994cr,Connes:1980,DuboisViolette:1995tz} (see \cite{DuboisViolette:1999cj,Masson:2007jb,Masson:2008uq} for reviews).

\begin{definition}[Derivation-based differential calculus]
Let $\algA$ be an associative unital algebra.\\
The {\it derivation-based differential calculus} $(\underline\Omega_\Der^\bullet(\algA),\dd)$ of $\algA$ is defined by:
\begin{equation*}
\underline\Omega^n_\Der(\algA)=\{\omega:(\Der(\algA))^n\to\algA,\quad \caZ(\algA)\text{-multilinear antisymmetric}\},
\end{equation*}
and $\underline\Omega^0_\Der(\algA)=\algA$. The space
\begin{equation*}
\underline\Omega_\Der^\bullet(\algA)=\bigoplus_{n=0}^\infty\underline\Omega^n_\Der(\algA)
\end{equation*}
is a $\gN$-graded differential algebra with the product: $\forall\omega\in\underline\Omega^p_\Der(\algA)$, $\forall\eta\in\underline\Omega^q_\Der(\algA)$, $\forall\kX_i\in\Der(\algA)$,
\begin{equation*}
(\omega\fois\eta)(\kX_1,\dots,\kX_{p+q}) =\frac{1}{p!q!}\sum_{\sigma\in\kS_{p+q}}(-1)^{|\sigma|}\omega(\kX_{\sigma(1)},\dots,\kX_{\sigma(p)})\fois\eta(\kX_{\sigma(p+1)},\dots,\kX_{\sigma(p+q)}),
\end{equation*}
and the differential:
\begin{multline*}
\dd\omega(\kX_1,\dots,\kX_{p+1})=\sum_{m=1}^{p+1} (-1)^{m+1}\kX_m\omega(\kX_1,\dots \omi{m} \dots,\kX_{p+1})\\
+\sum_{1\leq m<n\leq p+1}(-1)^{m+n}\omega([\kX_m,\kX_n],\dots \omi{m} \dots \omi{n} \dots,\kX_{p+1}).
\end{multline*}
It is a differential calculus in the sense of Definition \ref{def-gnc-diffcalc}.
\end{definition}

Let us denote by $\Omega^\bullet_\Der(\algA)$ the graded differential subalgebra of $\underline\Omega^\bullet_\Der(\algA)$ generated in degree $0$ by $\algA$. It is a differential calculus, and a quotient of the universal differential calculus.
\begin{example}
\label{ex-gnc-manifold}
In the case of a smooth compact manifold $M$, consider the associative unital algebra $\algA=C^\infty(M)$.
\begin{itemize}
\item Since $\algA$ is commutative, $\caZ(\algA)=\algA=C^\infty(M)$. But the right generalization of usual forms to the noncommutative framework involves a condition of $\caZ(\algA)$-multilinearity of the forms, and not $\algA$-multilinearity, because $\Der(\algA)$ is a $\caZ(\algA)$-bimodule and not a $\algA$-bimodule in general.
\item Moreover, a fondamental theorem of differential geometry permits to show that the derivations of $\algA$ are the vector fields of $M$: $\Der(\algA)=\Gamma(M)$.
\item The derivation-based differential calculus of $\algA$ is then the de Rham complex $\Omega_{\text{dR}}^\bullet(M)$ of $M$.
\end{itemize}
\end{example}

It is then possible to construct a theory of connections and gauge transformations associated to any differential calculus. In the particular case of derivation-based differential calculus, it leads to the following definition \cite{DuboisViolette:1998su}.
\begin{definition}[Connections, curvature]
Let $\algA$ be an associative unital algebra and $\modM$ be a right $\algA$-module. A linear map $\nabla_\kX:\modM\to\modM$, defined for any $\kX\in\Der(\algA)$, is said to be a {\it connection} on $\modM$ if $\forall a\in\algA$, $\forall z\in\caZ(\algA)$, $\forall\kX,\kY\in\Der(\algA)$, $\forall m\in\modM$,
\begin{align*}
&\nabla_\kX(ma)=(\nabla_\kX m) a+m\kX(a)\\
&\nabla_{(z\kX)}m=(\nabla_\kX m)z\\
&\nabla_{(\kX+\kY)}m=\nabla_\kX m+\nabla_\kY m.
\end{align*}
The {\it curvature} of a connection $\nabla$ is the linear map $R(\kX,\kY):\modM\to\modM$, defined for any $\kX,\kY\in\Der(\algA)$, by: $\forall m\in\modM$,
\begin{equation*}
R(\kX,\kY)m=[\nabla_\kX,\nabla_\kY]m-\nabla_{[\kX,\kY]}m.
\end{equation*}
\end{definition}
These definitions generalize the ones of the ordinary differential geometry, where for $\algA=C^\infty(M)$, the module $\modM$ represents a vector bundle on $M$, the center $\caZ(\algA)$ the functions $C^\infty(M)$, and the derivations $\Der(\algA)$ the vector fields $\Gamma(M)$. In the general case, it can be shown that the space of connections is an affine space, and that the curvature is a right $\algA$-module homomorphism.

\begin{definition}[Gauge group]
The {\it gauge group} of $\modM$ is the group of automorphisms of $\modM$ as a right $\algA$-module. Its elements are called gauge transformations.
\end{definition}

\begin{proposition}
The gauge group of $\modM$ acts on the space of connections by: for $\Phi$ a gauge transformation and $\nabla$ a connection, $\forall\kX\in\Der(\algA)$,
\begin{equation*}
\nabla_\kX^\Phi=\Phi\circ\nabla_\kX\circ\Phi^{-1}
\end{equation*}
is still a connection. The action on the corresponding curvature is given by $R^\Phi(\kX,\kY)=\Phi\circ R(\kX,\kY)\circ \Phi^{-1}$.
\end{proposition}
These notions coincide with the usual ones in the case of ordinary differential geometry $\algA=C^\infty(M)$.

Note that this formalism will be extended to the framework of $\eps$-graded algebras in chapter \ref{cha-eps}.

\begin{example}
\label{ex-gnc-matrix}
The matrix algebra \cite{DuboisViolette:1988ir} (see also \cite{Masson:2008}): $\algA=\Matr_n(\gC)=\Matr_n$.
\begin{itemize}
\item The center is trivial: $\caZ(\Matr_n)=\gC\gone$.
\item The derivations are inner: $\Der(\Matr_n)=\Int(\Matr_n)\approx\ksl_n$ (traceless elements). This isomorphism is made explicit by $\gamma\in\ksl_n\mapsto\ad_\gamma\in\Int(\Matr_n)$.
\item The differential calculus: $\underline\Omega_\Der^\bullet(\Matr_n)=\Omega^\bullet_\Der(\Matr_n)\approx \Matr_n\otimes\bigwedge^\bullet\ksl_n^\ast$, where the differential is the one of the complex of the Lie algebra $\ksl_n$ represented on $\Matr_n$ by the adjoint representation.
\item The canonical 1-form $i\theta\in\underline\Omega_\Der^1(\Matr_n)$, defined by $i\theta(\ad_\gamma)=\gamma-\frac 1n\tr(\gamma)\gone$, for $\gamma\in\Matr_n$, is the inverse isomorphism of $\ad:\ksl_n\to\Int(\Matr_n)$.
\item It satisfies a Maurer-Cartan type relation: $\dd(i\theta)-(i\theta)^2=0$.
\item The cohomology of the differential algebra $(\Omega_\Der^\bullet(\Matr_n),\dd)$ is
\begin{equation*}
H^\bullet(\Omega_\Der(\Matr_n),\dd)=H^\bullet(\ksl_n)=\mathcal{I}(\exter^\bullet\ksl_n^\ast),
\end{equation*}
the algebra of invariant elements for the natural Lie derivative.
\item For any connection $\nabla$ on $\Matr_n$, considered as a right module on itself, one defines $\omega(\kX)=\nabla_\kX\gone$, $\forall\kX\in\Der(\Matr_n)$. Then, $\nabla$ is completely determined by the 1-form $\omega$, and is denoted by $\nabla^\omega$: $\forall\kX\in\Der(\Matr_n)$, $\forall a\in\Matr_n$,
\begin{equation*}
\nabla^\omega_\kX a=\dd a(\kX)+\omega(\kX)\fois a.
\end{equation*}
\item The curvature of $\nabla^\omega$ writes: $R(\kX,\kY)(a)=\Omega(\kX,\kY)\fois a$, with
\begin{equation*}
\Omega=\dd\omega+\frac 12[\omega,\omega].
\end{equation*}
\item The gauge transformations $\Phi$ are determined by $\Phi(\gone)=g\in GL_n(\gC)$, and act as:
\begin{equation*}
a\mapsto g\fois a,\qquad \omega\mapsto g\fois\omega\fois g^{-1}+g\fois \dd g^{-1},\qquad \Omega\mapsto g\fois\Omega \fois g^{-1}.
\end{equation*}
\item For all $a\in\Matr_n$, $\dd a=[i\theta,a]$, the connection $\nabla^{-i\theta}$ is gauge invariant and its curvature is zero.
\end{itemize}
\end{example}

\begin{example}
\label{ex-gnc-matrfun}
The matrix valued functions algebra \cite{DuboisViolette:1989vq}: $\algA=C^\infty(M)\otimes \Matr_n$, where $M$ is a smooth paracompact manifold.
\begin{itemize}
\item The center: $\caZ(\algA)=C^\infty(M)$.
\item The derivations:
\begin{equation*}
\Der(\algA)=(\Der(C^\infty(M))\otimes\gC)\oplus(C^\infty(M)\otimes\Der(\Matr_n))= \Gamma(M)\oplus(C^\infty(M)\otimes\ksl_n),
\end{equation*}
so that each derivation $\kX\in\Der(\algA)$ can be written: $\kX=X+\ad_\gamma$, with $X\in\Gamma(M)$ and $\gamma\in\algA_0=C^\infty(M)\otimes\ksl_n$ (traceless elements in $\algA$).\\
$\Int(\algA)=\algA_0$ and $\Out(\algA)=\Gamma(M)$.
\item The differential calculus: $\underline\Omega_\Der^\bullet(\algA)=\Omega_\Der^\bullet(\algA)= \Omega_{\text{dR}}^\bullet(M)\otimes\Omega_\Der^\bullet(\Matr_n)$.
\item Like in Example \ref{ex-gnc-matrix}, a canonical noncommutative 1-form $i\theta$ can be defined by $i\theta(X+\ad_\gamma)=\gamma$. It splits the short exact sequence of Lie algebras and $C^\infty(M)$-modules:
\begin{equation*}
\label{eq-gnc-splittingmatrfun}
\xymatrix@1@C=25pt{{\algzero} \ar[r] & {\algA_0} \ar[r] & {\Der(\algA)} \ar[r] \ar@/_0.7pc/[l]_-{i\theta}& {\Gamma(M)} \ar[r] & {\algzero}}
\end{equation*}
\end{itemize}
\end{example}

In the above Example, if $M$ is compact, the noncommutative connections are also described by noncommutative 1-forms
$\omega(\kX)=\nabla_\kX(\gone)$, where
\begin{equation*}
\omega\in\Omega_\Der^1(\algA)=(\Omega_{\text{dR}}^1(M)\otimes\Matr_n)\oplus (C^\infty(M)\otimes\Omega_\Der^1(\Matr_n)).
\end{equation*}
The first part of $\omega$ (in $\Omega_{\text{dR}}^1(M)\otimes\Matr_n$) is an ordinary Yang-Mills potential, while the second part (in $C^\infty(M)\otimes\Omega_\Der^1(\Matr_n)$) is interpreted as a Higgs field in \cite{DuboisViolette:1989vq}. Indeed, the curvature can be decomposed in three components: the $\Omega_{\text{dR}}^2(M)\otimes\Matr_n$ part is the curvature of the ordinary Yang-Mills connection, the $\Omega_{\text{dR}}^1(M)\otimes\Omega_\Der^1(\Matr_n)$ part involves the covariant derivative of the Higgs field, and the $C^\infty(M)\otimes\Omega_\Der^2(\Matr_n)$ part a second order polynomial in the Higgs field. By considering the action constructed as a trace of the square of the curvature, one obtains an ordinary action, function of a Yang-Mills potential gauge-invariantly coupled to a Higgs field. Therefore, in this picture, the Higgs field can be seen as the noncommutative part of the noncommutative connection $\nabla$.

Note that Example \ref{ex-gnc-matrfun} has been generalized by considering the algebra of sections of the endomorphism bundle associated to a $SU(n)$-vector bundle \cite{Masson:1999ea}. It has permitted to interpret the ordinary connections on a $SU(n)$-vector bundle as a part of the noncommutative connections of its endomorphism algebra.

\subsection{Hochschild and cyclic homologies}
\label{subsec-gnc-homology}

The Hochschild homology of an associative algebra permits also to generalize the de Rham complex in the noncommutative framework. Moreover, the cyclic homology can be interpreted as the generalization of the de Rham cohomology, and is deeply related to Hochschild homology, universal differential calculi (see Definition \ref{def-gnc-univdiffcalc}), $K$-theory and characteristic classes. Note that we restrict to the special case of unital algebras. More details about Hochschild and cyclic homologies can be found in \cite{Masson:2008,Loday:1998,Cuntz:2004}.

\begin{definition}[Hochschild complex]
Let $\algA$ be an associative algebra with unit, and $\modM$ a bimodule on $\algA$. Then, the Hochschild differential complex is defined as the sum on the vector spaces:
\begin{equation*}
C_n(\algA,\modM)=\modM\otimes\algA^{\otimes n},
\end{equation*}
with $n\in\gN$, and with the boundary operator $b:C_n(\algA,\modM)\to C_{n-1}(\algA,\modM)$, given by:
\begin{multline}
b(m\otimes a_1\otimes\dots\otimes a_n)=(ma_1)\otimes a_2\otimes\dots\otimes a_n\\
+\sum_{i=1}^{n-1}(-1)^im\otimes a_1\otimes\dots\otimes (a_i\fois a_{i+1})\otimes\dots\otimes a_n +(-1)^n(a_n m)\otimes a_1\otimes\dots\otimes a_{n-1},\label{eq-gnc-hochschilddiff}
\end{multline}
for $m\in\modM$ and $a_i\in\algA$. Then, $b^2=0$ and the Hochschild complex is denoted by $(C_\bullet(\algA,\modM),b)$.
\end{definition}

In the case $\modM=\algA$, the complex is simply denoted by $C_\bullet(\algA)$, and its homology, called the {\it Hochschild homology} of $\algA$, $HH_\bullet(\algA)$.
\begin{proposition}
This homology respects direct sums of algebras: if $\algA$ and $\algB$ are two associative algebras, then
\begin{equation*}
HH_\bullet(\algA\oplus\algB)=HH_\bullet(\algA)\oplus HH_\bullet(\algB).
\end{equation*}
Moreover, if $\varphi:\algA\to\algB$ is a morphism of associative (unital) algebras, then it can be extended into a morphism of differentiable complexes $\varphi:C_\bullet(\algA)\to C_\bullet(\algB)$, and induces a morphism in homologies $\varphi_\sharp:HH_\bullet(\algA)\to HH_\bullet(\algB)$.
\end{proposition}

The Hochschild homology satisfies several properties which will not be exposed here. For example, it is Morita invariant. For further considerations on this subject, see \cite{Masson:2008} for instance. However, one also introduces a ``shuffle product'' on the Hochschild complex.

\begin{definition}[shuffle product]
Let $\algA$ and $\algB$ be two unital algebras. One defines a shuffle of type $(p,q)$ by a permutation $\sigma\in\kS_{p+q}$ such that $\sigma(1)<\sigma(2)<\dots<\sigma(p)$ and $\sigma(p+1)<\sigma(p+2)<\dots<\sigma(p+q)$. The set of all shuffles of type $(p,q)$ is denoted by $\mathfrak B(p,q)$. It acts (like $\kS_{p+q}$) on $C_{p+q}(\algA)$ by:
\begin{equation*}
\sigma(a_0\otimes a_1\otimes\dots\otimes a_{p+q})=a_0\otimes a_{\sigma^{-1}(1)}\otimes\dots\otimes a_{\sigma^{-1}(p+q)}.
\end{equation*}
Then, the {\it shuffle product}
\begin{equation*}
\text{Sh}_{p,q}:C_p(\algA)\otimes C_q(\algB)\to C_{p+q}(\algA\otimes\algB)
\end{equation*}
is defined by
\begin{multline*}
\text{Sh}_{p,q}((a_0\otimes\dots\otimes a_p)\otimes(b_0\otimes\dots\otimes b_q))=\\
\sum_{\sigma\in\mathfrak B(p,q)} (-1)^{|\sigma|}\sigma((a_0\otimes b_0)\otimes(a_1\otimes\gone)\otimes\dots \otimes(a_p\otimes\gone)\otimes(\gone\otimes b_1)\otimes\dots\otimes(\gone\otimes b_q)).
\end{multline*}
\end{definition}

We will denote $x\times y=\text{Sh}_{p,q}(x\otimes y)$, for $x\in C_p(\algA)$ and $y\in C_q(\algB)$.

\begin{proposition}
\begin{itemize}
\item The boundary application $b$ is a graded derivation of $C_\bullet(\algA\otimes\algB)$ of degree $-1$, namely: $\forall x\in C_p(\algA)$ and $\forall y\in C_q(\algB)$,
\begin{equation*}
b(x\times y)=(bx)\times y+(-1)^px\times(by).
\end{equation*}
\item The shuffle product induces an isomorphism in homologies (theorem of Eilenberg-Zilber):
\begin{equation*}
\text{Sh}_\sharp:HH_\bullet(\algA)\otimes HH_\bullet(\algB)\to HH_\bullet(\algA\otimes\algB).
\end{equation*}
\item If $\algA$ is commutative, the shuffle product (with $\algB=\algA$) can be composed with the product of $\algA$, and it provides an inner product $C_p(\algA)\otimes C_q(\algA)\to C_{p+q}(\algA)$, so that the Hochschild complex $C_\bullet(\algA)$ is a graded differential algebra and the Hochschild homology $HH_\bullet(\algA)$ is a graded commutative algebra.
\end{itemize}
\end{proposition}

\begin{example}
\begin{itemize}
\item If $\algA=\gC$, the Hochschild homology is trivial:
\begin{equation*}
HH_0(\gC)=\gC,\qquad HH_n(\gC)=\algzero,
\end{equation*}
for $n\geq 1$.
\item The definition of the Hochschild homology can be generalized for topological algebras $\algA$, by replacing in the Hochschild complex the tensor products by topological tensor products adapted to the topology of the algebra. The resulting homology is called the continuous Hochschild homology and denoted by $HH_\bullet^{\text{Cont}}(\algA)$.
\item If $M$ is a smooth (finite dimensional) locally compact manifold, consider $\algA=C^\infty(M)$ the algebra of its smooth functions. Then A. Connes computed the continuous Hochschild homology and found:
\begin{equation*}
HH_\bullet^{\text{Cont}}(C^\infty(M))=\Omega^\bullet_{\text{dR},\gC}(M),
\end{equation*}
the complexified de Rham forms. This result generalizes the Hochschild-Kostant-Rosenberg theorem.
\end{itemize}
\end{example}

\begin{definition}[Cyclic homology]
\begin{itemize}
\item Let us define on $C_\bullet(\algA)$ the cyclic operator $t$ and the norm operator $N$ by:
\begin{align*}
& t(a_0\otimes\dots a_n)=(-1)^n\, a_n\otimes a_0\otimes\dots\otimes a_{n-1},\\
&N=\sum_{i=0}^nt^i,
\end{align*}
satisfying
\begin{equation*}
t^{n+1}=\text{id},\qquad (1-t)N=N(1-t)=0.
\end{equation*}
\item On $C_\bullet(\algA)$, another differential than \eqref{eq-gnc-hochschilddiff} can be defined:
\begin{equation*}
b'(a_0\otimes\dots \otimes a_n)=\sum_{i=0}^{n-1}(-1)^i \,a_0\otimes\dots\otimes(a_i a_{i+1})\otimes\dots\otimes a_n.
\end{equation*}
Then, one has:
\begin{equation*}
b'^2=0, \qquad(1-t)b'=b(1-t),\qquad b'N=Nb.
\end{equation*}
\item The differential cyclic bicomplex $CC_{\bullet,\bullet}(\algA)$ is then described by:
\begin{equation*}
\xymatrix{ 
{\vdots} \ar[d]_-{b} & {\vdots} \ar[d]_-{-b'} & {\vdots} \ar[d]_-{b} & {\vdots} \ar[d]_-{-b'} & \\
{\algA^{\otimes n+1}} \ar[d]_-{b} & {\algA^{\otimes n+1}} \ar[l]_-{1-t} \ar[d]_-{-b'} &{\algA^{\otimes n+1}} \ar[l]_-{N} \ar[d]_-{b} & {\algA^{\otimes n+1}} \ar[l]_-{1-t} \ar[d]_-{-b'} & {\cdots} \ar[l]_-{N} \\
{\algA^{\otimes n}} \ar[d]_-{b} & {\algA^{\otimes n}} \ar[l]_-{1-t} \ar[d]_-{-b'} &{\algA^{\otimes n}} \ar[l]_-{N} \ar[d]_-{b} & {\algA^{\otimes n}} \ar[l]_-{1-t} \ar[d]_-{-b'} & {\cdots} \ar[l]_-{N} \\
{\vdots} \ar[d]_-{b} & {\vdots} \ar[d]_-{-b'} &{\vdots} \ar[d]_-{b} & {\vdots} \ar[d]_-{-b'} & \\
{\algA} & {\algA} \ar[l]_-{1-t} &{\algA} \ar[l]_-{N} & {\algA} \ar[l]_-{1-t} & {\cdots} \ar[l]_-{N}}
\end{equation*}
where $CC_{p,q}(\algA)=C_q(\algA)=\algA^{\otimes(q+1)}$, for $p$ the index of the column and $q$ the index of the line.
\item The {\it cyclic homology} of the algebra $\algA$ is the homology of the total complex $TCC_\bullet(\algA)$ defined by
\begin{equation*}
TCC_n(\algA)=\bigoplus_{p+q=n}CC_{p,q}(\algA),
\end{equation*}
and denoted by $HC_\bullet(\algA)$.
\end{itemize}
\end{definition}

\begin{proposition}[Properties]
This homology respects direct sums of algebras: if $\algA$ and $\algB$ are two algebras, then
\begin{equation*}
HC_\bullet(\algA\oplus\algB)=HC_\bullet(\algA)\oplus HC_\bullet(\algB).
\end{equation*}
Moreover, if $\varphi:\algA\to\algB$ is a morphism of (unital) algebras, then it can be extended into a morphism of differentiable bicomplexes $\varphi:CC_{\bullet,\bullet}(\algA)\to CC_{\bullet,\bullet}(\algB)$, and induces a morphism in homologies $\varphi_\sharp:HC_\bullet(\algA)\to HC_\bullet(\algB)$.
\end{proposition}

\begin{Proposition}
The Hochschild and cyclic homologies enter into the following long exact sequence:
\begin{equation*}
\xymatrix@1{ {\cdots} \ar[r] & {HH_n(\algA)} \ar[r]^-{I} & {HC_n(\algA)} \ar[r]^-{S} & {HC_{n-2}(\algA)} \ar[r]^-{B} & {HH_{n-1}(\algA)} \ar[r]^-{I} & {\cdots} }
\end{equation*}
\end{Proposition}

\begin{definition}[Cyclic homology]
\begin{itemize}
\item The periodic cyclic bicomplex $CC^{\text{per}}_{\bullet,\bullet}(\algA)$ is an extension on the left of $CC_{\bullet,\bullet}(\algA)$ and is given by:
\begin{equation*}
\xymatrix{ 
{} & {\vdots} \ar[d]_-{b} & {\vdots} \ar[d]_-{-b'} & {\vdots} \ar[d]_-{b} & {\vdots} \ar[d]_-{-b'} & \\
{\cdots} & {\algA^{\otimes n+1}} \ar[d]_-{b} \ar[l]_-{N} & {\algA^{\otimes n+1}} \ar[l]_-{1-t} \ar[d]_-{-b'} &{\algA^{\otimes n+1}} \ar[l]_-{N} \ar[d]_-{b} & {\algA^{\otimes n+1}} \ar[l]_-{1-t} \ar[d]_-{-b'} & {\cdots} \ar[l]_-{N} \\
{\cdots} & {\algA^{\otimes n}} \ar[d]_-{b} \ar[l]_-{N} & {\algA^{\otimes n}} \ar[l]_-{1-t} \ar[d]_-{-b'} &{\algA^{\otimes n}} \ar[l]_-{N} \ar[d]_-{b} & {\algA^{\otimes n}} \ar[l]_-{1-t} \ar[d]_-{-b'} & {\cdots} \ar[l]_-{N} \\
& {\vdots} \ar[d]_-{b} & {\vdots} \ar[d]_-{-b'} &{\vdots} \ar[d]_-{b} & {\vdots} \ar[d]_-{-b'} & \\
{\cdots} & {\algA} \ar[l]_-{N} & {\algA} \ar[l]_-{1-t} &{\algA} \ar[l]_-{N} & {\algA} \ar[l]_-{1-t} & {\cdots} \ar[l]_-{N} \\
{p=} & {-2} & {-1} & {0} & {1} & {\cdots}}
\end{equation*}
\item Then, the {\it periodic cyclic homology} is the homology of the total complex of $CC^{\text{per}}_{\bullet,\bullet}(\algA)$:
\begin{equation*}
TCC^{\text{per}}_n(\algA)=\bigoplus_{p+q=n}CC^{\text{per}}_{p,q}(\algA),
\end{equation*}
and is denoted by $HP_\bullet(\algA)$.
\end{itemize}
\end{definition}

\begin{proposition}
The periodic cyclic homology is naturally two-periodic:
\begin{equation*}
HP_n(\algA)\simeq HP_{n-2}(\algA).
\end{equation*}
\end{proposition}

\begin{example}
\begin{itemize}
\item For $\algA=\gC$, the cyclic and periodic homologies are given by:
\begin{align*}
& HC_{2n}(\gC)=\gC,\qquad HC_{2n+1}(\gC)=\algzero,\\
& HP_0(\gC)=\gC,\qquad HP_1(\gC)=\algzero.
\end{align*}
\item Like in the Hochschild homology case, the cyclic and periodic homology definitions can be generalized for topological algebras $\algA$, by replacing the tensor products by topological tensor products adapted to the topology of the algebra. The resulting homologies are called continuous cyclic and periodic homologies and denoted by $HC_\bullet^{\text{Cont}}(\algA)$ and $HP_\bullet^{\text{Cont}}(\algA)$.
\item Let $M$ be a smooth (finite dimensional) locally compact manifold, and $\algA=C^\infty(M)$ the algebra of its smooth functions. Then the continuous periodic cohomology computed by A. Connes is deeply related to the de Rham cohomology of $M$:
\begin{align*}
HP_0^{\text{Cont}}(C^\infty(M))=H^{\text{even}}_{\text{dR}}(M),\\
HP_1^{\text{Cont}}(C^\infty(M))=H^{\text{odd}}_{\text{dR}}(M).
\end{align*}
\end{itemize}
\end{example}

\subsection{Spectral triples}
\label{subsec-gnc-spectriple}

We present here the notion of spectral triple, introduced by A. Connes \cite{Connes:1994}, which plays the role of a noncommutative analog of smooth manifolds (and riemannian manifolds). See also \cite{Landi:1997,Madore:1999,Gracia-Bondia:2000}.

\begin{definition}
\label{def-gnc-sptr}
A {\it spectral triple} $(\algA,\ehH,D)$ of (metric) dimension $p$ is composed of a Hilbert space $\ehH$, an involutive algebra represented on $\ehH$, and a selfadjoint operator $D$ on $\ehH$, satisfying:
\begin{enumerate}
\item the $n^{\text{th}}$ characteristic value\footnote{The characteristic values of an operator are the eigenvalues of its absolute value} of the resolvent of $D$ is $O(n^{-1/p})$.
\item $\forall a,b\in\algA$, $[[D,a],b]=0$.
\item We set $\delta(T)=[|D|,T]$ for bounded operators $T\in\caL(\ehH)$ such that $T(\Dom(|D|))\subset\Dom(|D|)$ and $\delta(T)$ bounded. Then, $\forall a\in\algA$, $\forall m\in\gN^\ast$, $a\in\Dom(\delta^m)$ and $[D,a]\in\Dom(\delta^m)$.
\item Let $\pi_D(a_0\otimes a_1\otimes\dots\otimes a_p)=a_0[D,a_1]\dots[D,a_p]$, for $a_j\in\algA$. There exists a Hochschild cycle $c$ such that $\pi_D(c)=1$ if $p$ is odd, and $\pi_D(c)=\gamma$ if $p$ is even, with
\begin{equation*}
\gamma^\ast=\gamma,\qquad \gamma^2=1,\qquad \gamma D=-D\gamma.
\end{equation*}
\item $\ehH_\infty=\bigcap_{m\in\gN^\ast}\Dom(D^m)$ is a finitely generated projective $\algA$-module. It can be endowed with a hermitean structure $(\fois|\fois)$ defined by: $\forall x,y\in\ehH_\infty$, $\forall a\in\algA$,
\begin{equation*}
\langle x,ay\rangle=\cutint a(x|y)|D|^{-p},
\end{equation*}
where $\cutint$ is the Dixmier trace.
\end{enumerate}
\end{definition}

\begin{example}
Let $M$ be a compact oriented manifold, and $S$ be a spinor bundle on $M$.  $\algA=C^\infty(M)$, the space of smooth functions on $M$ is an involutive algebra, and $\ehH=L^2(M,S)$, the space of square-integrable sections of $S$, is a Hilbert space. By taking $D$ a Dirac operator of $M$ acting on $\ehH$, it can be shown that $(\algA,\ehH,D)$ is a spectral triple.
\end{example}

Then, A. Connes shew the following converse theorem \cite{Connes:2008sc}.
\begin{theorem}
Let $(\algA,\ehH,D)$ be a spectral triple (in the sense of Definition \ref{def-gnc-sptr}), with $\algA$ commutative, and
\begin{itemize}
\item for all $\algA$-endomorphisms $T$ of $\ehH_\infty$, $\forall m\in\gN^\ast$, $T\in\Dom(\delta^m)$ and $[D,T]\in\Dom(\delta^m)$,
\item the Hochschild cycle $c$ is antisymmetric.
\end{itemize}
Then, there exists a compact oriented smooth manifold $M$ such that $\algA=C^\infty(M)$.
\end{theorem}

\begin{proposition}
The map $\pi_D:\Omega_U^\bullet(\algA)\to\caL(\ehH)$ defined by: $\forall a_i\in\algA$,
\begin{equation*}
\pi_D(a_0\dd_U a_1\dots\dd_Ua_n)=a_0[D,a_1]\dots[D,a_n]
\end{equation*}
is a representation of the algebra $\Omega_U^\bullet(\algA)$ (see Definition \ref{def-gnc-univdiffcalc}).
\end{proposition}
The differential calculus of a spectral triple can therefore be seen as a quotient of the universal differential calculus of $\algA$. In the case of a smooth compact manifold, it coincides with the de Rham complex. Note that the previous representation is not a representation of differential algebras.
\medskip

A notion of distance can then be defined on states $\omega,\eta\in\caS(\algA)$ thanks to the operator $D$:
\begin{equation*}
d(\omega,\eta)=\sup_{a\in\algA}\{|\omega(a)-\eta(a)|,\ \norm[D,a]\norm\leq1\}.
\end{equation*}
It is a called the {\it spectral distance} on $\algA$. In the commutative case, it coincides with the usual geodesic distance on the points of the manifold, for the metric associated to the Dirac operator $D$, so that the spectral triple approach can be seen as a noncommutative generalization of riemannian geometry.

\numberwithin{equation}{section}
\chapter[Epsilon-graded algebras]{Epsilon-graded algebras and noncommutative geometry}
\label{cha-eps}

The noncommutative geometry based on the derivations of associative algebras, namely the differential calculus and the theory of connections (see subsection \ref{subsec-gnc-diffcalc}), can be extended to the framework of $\eps$-graded algebras. Lie superalgebras have been generalized by Rittenberg and Wyler \cite{Rittenberg:1978mr}, and by Scheunert \cite{Scheunert:1979}, by using the notion of commutation factor.

During this PhD, we have adapted this generalization to the case of associative algebras by defining $\eps$-graded algebras \cite{deGoursac:2008bd}. In this setting, derivations of the associative algebra are replaced by $\eps$-derivations, and one can extend the framework of differential calculus based on derivations to the $\eps$-graded algebras. This will be presented in this chapter, and illustrated on examples of $\eps$-graded commutative algebras, and matrix algebras.

\section{\texorpdfstring{General theory of the $\eps$-graded algebras}{General theory of the epsilon-graded algebras}}
\label{sec-eps-alg}

\subsection{Commutation factors and multipliers}
\label{subsec-eps-commfact}

We recall here the principal features of the commutation factors and the multipliers, that have been respectively introduced by Scheunert \cite{Scheunert:1979} and Schur (see also \cite{deGoursac:2008bd}). Let $\gK$ be a field, $\gK^\ast$ its multiplicative group, and $\Gamma$ an abelian group.

\subsubsection{Commutation factors}

\begin{definition}
\label{def-eps-commfactor}
A {\it commutation factor} is a map $\eps:\Gamma\times\Gamma\to\gK^\ast$ satisfying: $\forall i,j,k\in\Gamma$,
\begin{subequations}
\label{eq-eps-defepsilon}
\begin{align}
&\eps(i,j)\eps(j,i)=1_\gK\\
&\eps(i,j+k)=\eps(i,j)\eps(i,k)\\
&\eps(i+j,k)=\eps(i,k)\eps(j,k)
\end{align}
\end{subequations}
\end{definition}

One can define an equivalence relation on the commutation factors of an abelian group in the following way:
\begin{definition}
Two commutation factors $\eps$ and $\eps'$ are called equivalent if there exists $f\in\Aut(\Gamma)$, the group of automorphisms of $\Gamma$, such that: $\forall i,j\in\Gamma$,
\begin{equation*}
\eps'(i,j)=f^\ast\eps(i,j)=\eps(f(i),f(j)).
\end{equation*}
\end{definition}

Thanks to the axioms \eqref{eq-eps-defepsilon}, one defines the signature function $\psi_\eps:\Gamma\to\{-1_\gK,1_\gK\}$ of $\eps$ by $\forall i\in\Gamma$, $\psi_\eps(i)=\eps(i,i)$, which satisfies: $\forall f\in\Aut(\Gamma)$, $\psi_{f^\ast\eps}=f^\ast\psi_\eps$.
\begin{proposition}
Let $\eps$ be a commutation factor. Let us define
\begin{equation*}
\Gamma_\eps^0=\{i\in\Gamma,\quad\eps(i,i)=1_\gK\},\qquad \Gamma_\eps^1=\{i\in\Gamma,\quad\eps(i,i)=-1_\gK\},
\end{equation*}
then $\eps$ is called proper if $\Gamma_\eps^0=\Gamma$. This property is compatible with the equivalence relation on the commutation factors. If $\eps$ is not proper, $\Gamma_\eps^0$ is a subgroup of $\Gamma$ of index 2, and $\Gamma_\eps^0$ and $\Gamma_\eps^1$ are its residues.
\end{proposition}

In these notations, we define the signature factor of the commutation factor $\eps$ by: $\forall i,j\in\Gamma$,
\begin{itemize}
\item $s(\eps)(i,j)=-1_\gK$ if $i\in\Gamma_\eps^1$ and $j\in\Gamma_\eps^1$.
\item $s(\eps)(i,j)=1_\gK$ if not.
\end{itemize}
$s(\eps)$ is also a commutation factor, such that $\forall f\in\Aut(\Gamma)$, $s(f^\ast\eps)=f^\ast s(\eps)$.

\begin{lemma}
\label{lem-eps-thierry}
Let $\eps_1$ and $\eps_2$ be two commutation factors respectively on the abelian groups $\Gamma_1$ and $\Gamma_2$, over the same field $\gK$. Then, the map $\eps$ defined by: $\forall i_1,j_1\in\Gamma_1$, $\forall i_2,j_2\in\Gamma_2$,
\begin{equation}
\eps((i_1,i_2),(j_1,j_2))=\eps_1(i_1,j_1)\eps_2(i_2,j_2),\label{eq-eps-product}
\end{equation}
is a commutation factor on the abelian group $\Gamma=\Gamma_1\times\Gamma_2$, over $\gK$.
\end{lemma}

\begin{proposition}
\label{prop-eps-factcomm}
Let $\Gamma$ be a finitely generated abelian group and $\gK$ a field. Then, $\Gamma$ is the direct product of a finite number of cyclic groups, whose generators are denoted by $\{e_r\}_{r\in I}$. Any commutation factor on $\Gamma$ over $\gK$ takes the form: $\forall i=\sum_{r\in I}\lambda_re_r,j=\sum_{s\in I}\mu_se_s\in\Gamma$ ($\lambda_r,\mu_s\in\gZ$),
\begin{equation*}
\eps(i,j)=\prod_{r,s\in I}\eps(e_r,e_s)^{\lambda_r\mu_s},
\end{equation*}
with the condition: if $m_{rs}$ is the greatest common divisor of the orders $m_r\geq0$ of $e_r$ and $m_s\geq0$ of $e_s$ in $\Gamma$,
\begin{align*}
&\forall r\in I,\text{ such that } m_r\text{ is odd, }\eps(e_r,e_r)=1_\gK,\\
&\forall r\in I,\text{ such that } m_r\text{ is even, }\eps(e_r,e_r)=\pm 1_\gK,\\
&\forall r,s\in I,\quad \eps(e_r,e_s)^{m_{rs}}=1_\gK
\end{align*}
\end{proposition}
The proof is straightforward and has been given in \cite{Scheunert:1979} or \cite{deGoursac:2008bd}. This Proposition gives the explicit form of a commutation factor on a finitely generated abelian group, but this is not a classification of such factors. In general, it is not easy to obtain this classification and it is related to the theory of multipliers as we will see below, but in the following example, coming from \cite{Scheunert:1979}, things are simplifying.

\begin{example}
\label{ex-eps-factcomm}
Let $\Gamma=\gZ_p^n$, with $p$ a prime number, $\gK$ a field of characteristic different from $p$, and $\alpha\neq 1_\gK$ a $p$th root of unity in $\gK$. Then, any commutation factor on $\Gamma$ over $\gK$ takes the unique following form: $\forall i,j\in\Gamma$,
\begin{equation*}
\eps(i,j)=\alpha^{\varphi(i,j)},
\end{equation*}
where $\varphi$ is a bilinear form on the vector space $\gZ_p^n$ over the field $\gZ_p$, which is:
\begin{itemize}
\item symmetric if $p=2$,
\item skew-symmetric if $p\geq3$. In this case, $\eps$ is proper.
\end{itemize}
To equivalent commutation factors correspond equivalent (in the sense of linear algebra) bilinear forms.
\end{example}
\begin{proof}
We recall here the proof of \cite{Scheunert:1979}. Let $\eps$ be a commutation factor on $\Gamma$ over $\gK$. Since $p$ is prime and different from the characteristic of $\gK$, there exists a $p$th root of unity in $\gK$, $\alpha\neq1_\gK$, and all the $p$th roots of unity are powers of $\alpha$. Using Proposition \ref{prop-eps-factcomm}, if $\{e_r\}_{r\in I}$ are the canonical generators of $\Gamma=(\gZ_p)^n$, $\forall r,s\in I$, $\eps(e_r,e_s)^p=1_\gK$, so that there exists $m_{rs}\in\gZ_p$ such that $\eps(e_r,e_s)=\alpha^{m_{rs}}$ and $m_{sr}=-m_{rs}$.

Then, $\forall i=\sum_{r\in I}\lambda_re_r,j=\sum_{s\in I}\mu_se_s\in\Gamma$ ($\lambda_r,\mu_s\in\gZ$), $\eps(i,j)=\alpha^{\varphi(i,j)}$, where $\varphi(i,j)=\sum_{r,s\in I}\lambda_r\mu_sm_{rs}$ is a bilinear form. If $p\geq3$, $\forall r\in I$, $m_{rr}=0$ and $\varphi$ is skew-symmetric. If $p=2$, $\forall r,s\in I$, $m_{sr}=-m_{rs}=m_{rs}$ and $\varphi$ is symmetric.
\end{proof}

\subsubsection{Schur multipliers}

We will now study the theory of multipliers of an abelian group, due to Schur (also for non-abelian groups), which is related to the classification of commutation factors. Let us now recall the standard definition of a factor set, closely related to a projective representation.
\begin{definition}
\label{def-eps-multiplier}
Let $\Gamma$ be an abelian group, and $\gK$ a field. A {\it factor set} is an application $\sigma:\Gamma\times\Gamma\to\gK^\ast$ such that: $\forall i,j,k\in\Gamma$,
\begin{equation}
\sigma(i,j+k)\sigma(j,k)=\sigma(i,j)\sigma(i+j,k).\label{eq-eps-deffactorset}
\end{equation}
Two factor sets $\sigma$ and $\sigma'$ are said to be equivalent if there exists an application $\rho:\Gamma\to\gK^\ast$ such that: $\forall i,j\in\Gamma$,
\begin{equation*}
\sigma'(i,j)=\sigma(i,j)\rho(i+j)\rho(i)^{-1}\rho(j)^{-1}.
\end{equation*}
The quotient of the set of factor sets by this equivalence relation is an abelian group, for the product of $\gK$, and is called the multiplier group $M_\Gamma$ of $\Gamma$. Each class $[\sigma]\in M_\Gamma$ is called a {\it multiplier}.

If $\sigma$ is a factor set and $f\in\Aut(\Gamma)$, the pullback $f^\ast\sigma$ defined by: $\forall i,j\in\Gamma$,
\begin{equation*}
f^\ast\sigma(i,j)=\sigma(f(i),f(j)),
\end{equation*}
is also a factor set. Moreover, this operation is compatible with the above equivalence relation, so that the pullback can now be defined on the multipliers: $f^\ast[\sigma]=[f^\ast\sigma]$. This defines an equivalence relation on the multipliers, which is not compatible with the product.

A more refined equivalence relation involves also subgroups of $\Gamma$: if $[\sigma]$ and $[\sigma']$ are multipliers and $\Gamma_0$ and $\Gamma_1$ are subgroups of $\Gamma$, $([\sigma],\Gamma_0)$ and $([\sigma'],\Gamma_1)$ are called equivalent if there exists $f\in\Aut(\Gamma)$ such that $f(\Gamma_1)=\Gamma_0$ and $[\sigma']=f^\ast[\sigma]$.
\end{definition}

Note that the equation \eqref{eq-eps-deffactorset} is related to the definition of cocycles of the group $\Gamma$, while the equivalence of factor sets can be reexpressed in terms of coboundaries, so that multipliers are in fact exponentials of the cohomology classes $H^{2}(\Gamma,\gK)$ of the group $\Gamma$.

\subsubsection{Relation between commutation factors and multipliers}

If $\eps$ is a commutation factor on $\Gamma$ over $\gK$, notice that, due to Definition \ref{def-eps-commfactor}, it is a factor set of $\Gamma$. But there is a deeper relation between commutation factors and factor sets, given by the following theorem \cite{Scheunert:1979,deGoursac:2008bd}:
\begin{theorem}
\label{thm-eps-multcomm}
Let $\Gamma$ be an abelian group, and $\gK$ a field.
\begin{itemize}
\item Any multiplier $[\sigma]$ defines a unique proper commutation factor $\eps_\sigma$ on $\Gamma$ by: $\forall i,j\in\Gamma$,
\begin{equation}
\eps_\sigma(i,j)=\sigma(i,j)\sigma(j,i)^{-1}.\label{eq-eps-multcomm}
\end{equation}
\item If $\Gamma$ is finitely generated, any proper commutation factor $\eps$ on $\Gamma$ can be constructed from a multiplier $[\sigma]$ by \eqref{eq-eps-multcomm}.
\item If, in addition, $\gK$ is algebraically closed, then $\eps$ is constructed from a unique multiplier $[\sigma]$.
\item For $\Gamma$ finitely generated and $\gK$ algebraically closed, if two proper commutation factors $\eps_\sigma$ and $\eps_{\sigma'}$ are equivalent, then $[\sigma']$ is a pullback of $[\sigma]$.
\end{itemize}
\end{theorem}
\begin{proof}
This theorem has been proved in \cite{Scheunert:1979,deGoursac:2008bd}.
\begin{itemize}
\item Let $\sigma$ be a factor set, and $\eps_\sigma(i,j)=\sigma(i,j)\sigma(j,i)^{-1}$, for $i,j\in\Gamma$.\\
Then, $\eps_\sigma(i,j)\eps_\sigma(j,i)=1_\gK$, and $\forall i,j,k\in\Gamma$,
\begin{multline*}
\eps_\sigma(i,j+k)\eps_\sigma(i,j)^{-1}\eps_\sigma(i,k)^{-1}=\\
\sigma(i,j+k)\sigma(i,j)^{-1}\sigma(i,k)^{-1} (\sigma(j+k,i)\sigma(j,i)^{-1}\sigma(k,i)^{-1})^{-1}.
\end{multline*}
By using three times the property \eqref{eq-eps-deffactorset}, one obtains
\begin{equation*}
\sigma(j+k,i)=\sigma(i,j+k)\sigma(j,i)\sigma(k,i)\sigma(i,j)^{-1}\sigma(i,k)^{-1},
\end{equation*}
which proves that $\eps_\sigma(i,j+k)=\eps_\sigma(i,j)\eps_\sigma(i,k)$. The third axiom of Definition \ref{def-eps-commfactor} can be treated in a similar way. If $\sigma'(i,j)=\sigma(i,j)\rho(i+j)\rho(i)^{-1}\rho(j)^{-1}$, then $\sigma'(i,j)\sigma'(j,i)^{-1}=\sigma(i,j)\sigma(j,i)^{-1}$.
\item Let us suppose that $\Gamma$ is finitely generated, and $\{e_r\}_{r\in I}$ a system of generators, where $I$ is an ordered (finite) set. Let $\eps$ be a proper commutation factor on $\Gamma$. Define $\sigma:\Gamma\times\Gamma\to\gK^\ast$ by $\forall i=\sum_{r\in I}\lambda_re_r,j=\sum_{s\in I}\mu_se_s\in\Gamma$ ($\lambda_r,\mu_s\in\gZ$),
\begin{equation*}
\sigma(i,j)=\prod_{r<s}\eps(e_r,e_s)^{\lambda_r\mu_s}.
\end{equation*}
Since $\forall r,s\in I$, $\eps(e_r,e_r)=1_\gK$ and $\eps(e_r,e_s)=\eps(e_s,e_r)^{-1}$, one has\\
$\eps(i,j)=\sigma(i,j)\sigma(j,i)^{-1}$. And $\forall k=\sum_{r\in I}\nu_re_r\in\Gamma$ ($\nu_r\in\gZ$),
\begin{equation*}
\sigma(i,j+k)\sigma(j,k)=\prod_{r<s}\eps(e_r,e_s)^{\lambda_r\mu_s+\lambda_r\nu_s+\mu_r\nu_s}=\sigma(i,j)\sigma(i+j,k).
\end{equation*}
\item Let $\Gamma$ be a finitely generated abelian group, $\gK$ an algebraically closed field, and $\eps$ a commutation factor on $\Gamma$ over $\gK$. Suppose that $\eps$ is constructed through \eqref{eq-eps-multcomm} from two factor sets $\sigma$ and $\sigma'$. Then $\forall i,j\in\Gamma$, $\sigma(i,j)\sigma'(i,j)^{-1}=\sigma(j,i)\sigma'(j,i)^{-1}$, which means that $\sigma\sigma'^{-1}$ is a symmetric factor set. Since $\gK$ is algebraically closed, $\sigma\sigma'^{-1}$ is equivalent to one, and $\sigma$ and $\sigma'$ are in the same multiplier.
\item If $\Gamma$ is finitely generated and $\gK$ algebraically closed, consider two equivalent commutation factors $\eps$ and $\eps'$: $\eps'=f^\ast\eps$, with $f\in\Aut(\Gamma)$. Then, there exists a unique multiplier $[\sigma]$ such that $\eps=\eps_\sigma$. $\forall i,j\in I$,
\begin{equation*}
\eps'(i,j)=\sigma(f(i),f(j))\sigma(f(j),f(i))^{-1}=\eps_{f^\ast\sigma}(i,j).
\end{equation*}
By unicity of the associated multiplier of $\eps'$, we obtain the result.
\end{itemize}
\end{proof}

\begin{Corollary}
Let $\Gamma$ be a finitely generated abelian group, and $\gK$ an algebraically closed field. Then,
\begin{itemize}
\item the proper commutation factors on $\Gamma$ are classified by the equivalence classes (by pullback) of the multipliers of $\Gamma$.
\item the non-proper commutation factors on $\Gamma$ are classified by the equivalence classes of multipliers and subgroups of index 2 of $\Gamma$.
\end{itemize}
\end{Corollary}
\begin{proof}
The first point is a direct consequence of Theorem~\ref{thm-eps-multcomm}.

For the second point, if $\eps_1$ and $\eps_2$ are non-proper equivalent commutation factors on $\Gamma$, then there exists $f\in\Aut(\Gamma)$ such that $\eps_2=f^\ast\eps_1$, and $f(\Gamma_{\eps_2}^0)=\Gamma_{\eps_1}^0$. $\forall \alpha=1,2$, we decompose $\eps_\alpha=s(\eps_\alpha)\widetilde\eps_\alpha$, with $\widetilde\eps_\alpha$ proper commutation factors. Using now Theorem~\ref{thm-eps-multcomm}, there exist unique multipliers $[\sigma_\alpha]$ such that $\widetilde\eps_\alpha=\eps_{\sigma_\alpha}$, and they satisfy $[\sigma_2]=f^\ast[\sigma_1]$. Then, $([\sigma_1],\Gamma_{\eps_1}^0)$ and $([\sigma_2],\Gamma_{\eps_2}^0)$ are equivalent.

Conversely, if $\forall\alpha=1,2$, $\eps_\alpha=s(\eps_\alpha)\eps_{\sigma_\alpha}$, with $[\sigma_\alpha]$ multipliers such that $([\sigma_1],\Gamma_{\eps_1}^0)$ and $([\sigma_2],\Gamma_{\eps_2}^0)$ are equivalent, then there exists $f\in\Aut(\Gamma)$ such that $f(\Gamma_{\eps_2}^0)=\Gamma_{\eps_1}^0$ and $[\sigma_2]=f^\ast[\sigma_1]$. It means that $s(\eps_2)=f^\ast s(\eps_1)$ and $\eps_{\sigma_2}=f^\ast\eps_{\sigma_1}$, so that $\eps_2=f^\ast\eps_1$.
\end{proof}

\subsection{\texorpdfstring{Definition of $\eps$-graded algebras and properties}{Definition of epsilon-graded algebras and properties}}
\label{subsec-eps-defalg}

The notion of commutation factors has been introduced in the context of graded Lie algebras \cite{Scheunert:1979} and gives rise to the following definition:
\begin{definition}[$\eps$-Lie algebra]
Let $\kg^\bullet$ be a $\Gamma$-graded vector space, $\eps$ a commutation factor on $\Gamma$, and $[-,-]_\eps:\kg^\bullet\times\kg^\bullet\to\kg^\bullet$ a bilinear product homogeneous of degree 0 satisfying
\begin{align}
[a,b]_\eps &=-\eps(|a|,|b|)[b,a]_\eps\nonumber\\
[a,[b,c]_\eps]_\eps &=[[a,b]_\eps,c]_\eps+\eps(|a|,|b|)[b,[a,c]_\eps]_\eps,\label{eq-eps-jacobi}
\end{align}
$\forall a,b,c\in\kg^\bullet$ homogeneous, where $|a|\in\Gamma$ is the degree of $a\in\kg^\bullet$. The couple $(\kg^\bullet, [-,-]_\eps)$ is called an {\it $\eps$-Lie algebra}.

An $\eps$-Lie algebra for which the product $[-,-]_\eps$ is always $0$ is called an abelian $\eps$-Lie algebra.
\end{definition}

\subsubsection{$\eps$-graded algebras}

We can now introduce the notion of $\eps$-graded algebra \cite{Bourbaki:2007,deGoursac:2008bd}:
\begin{definition}[$\eps$-graded associative algebra]
Let $\algrA$ be an associative unital $\Gamma$-graded $\gK$-algebra, endowed with a commutation factor $\eps$ on $\Gamma$, then $(\algrA,\eps)$ will be called an {\it $\eps$-graded (associative) algebra}.
\end{definition}

Notice that the $\eps$-structure is only related to the algebra $\algrA$ through the grading abelian group $\Gamma$. In particular, the product in the algebra is not connected to this structure. In the following, such an $\eps$-graded (associative) algebra will be denoted simply by $\algrA_\eps$ or even $\algrA$ if no confusion arises.

Any $\gZ$-graded associative algebra is an $\eps$-graded associative algebra for the natural commutation factor on $\gZ$, so that the theory described below can be applied to any $\gZ$-graded (associative) algebra. In the same way, Lie superalgebras (see \cite{Kac:1977,Fuks:1986}) are particular $\eps$-Lie algebras.

\begin{remark}
\label{rem-eps--bigrading}
Using Lemma~\ref{lem-eps-thierry}, if $\algA^{\bullet, \bullet}$ is an associative unital $\Gamma_1\times\Gamma_2$-bigraded $\gK$-algebra equipped with two commutation factors $\eps_1$ and $\eps_2$ for the two gradings separately, then it is also a $\eps$-graded algebra for the product grading $\Gamma_1\times\Gamma_2$ with $\eps$ defined by \eqref{eq-eps-product}.
\end{remark}

If $\algrA$ is an $\eps$-graded algebra, one can construct its underlying $\eps$-Lie algebra using the bracket defined by
\begin{equation*}
[a,b]_\eps=a\fois b-\eps(|a|,|b|)\ b\fois a.
\end{equation*}
$\forall a,b\in\algrA$ homogeneous. We will denote by $\algrA_{\Lie, \eps}$ this structure.

\begin{definition}[$\eps$-graded commutative algebra]
$\algrA_\eps$ is called an $\eps$-graded commutative algebra if $\algrA_{\Lie, \eps}$ is an abelian $\eps$-Lie algebra.
\end{definition}

For the case of $\gZ$-graded algebras, depending on the commutation factor, one gets as $\eps$-graded commutative algebras either commutative and graded algebras (for the trivial commutation factor) or graded commutative algebras (for the natural commutation factor).

\begin{definition}[$\eps$-center]
Let $\algrA_\eps$ be an $\eps$-graded algebra. The {\it $\eps$-center} of $\algrA_\eps$ is the $\eps$-graded commutative algebra
\begin{equation*}
\caZ^\bullet_\eps(\algA)=\{a\in\algrA,\ \forall b\in\algrA\ [a,b]_\eps=0\}.
\end{equation*}
\end{definition}

Depending on the choice of the $\eps$-structure on a $\Gamma$-graded algebra, this $\eps$-center can be very different.
\medskip

Let us now mention some elementary constructions using $\eps$-graded algebras. Let $\algrA$ and $\algrB$ be two $\eps$-graded algebras with the same commutation factor $\eps$.

A morphism of $\eps$-graded algebras is defined to be a morphism of associative unital $\Gamma$-graded algebras $\chi:\algrA\to\algrB$. As a consequence, $\chi$ is also a morphism of $\eps$-Lie algebras between $\algrA_{\Lie, \eps}$ and $\algrB_{\Lie, \eps}$. 

The tensor product of the two $\eps$-graded algebras $\algrA$ and $\algrB$ is the $\eps$-graded algebra defined  as the $\Gamma$-graded vector space $(\algA \otimes \algB)^\bullet$ for the total grading, equipped with the product given by 
\begin{equation*}
(a\otimes b)\fois(c\otimes d)=\eps(|b|,|c|)(a\fois c)\otimes(b\fois d).
\end{equation*}
$\forall a,c\in\algrA$ and $\forall b,d\in\algrB$ homogeneous.

An {\it $\eps$-trace} on $\algrA$ is a linear map $T:\algA\to\gK$, which satisfies
\begin{equation}
T(a\fois b)=\eps(|a|,|b|)T(b\fois a).\label{eq-eps-deftrace}
\end{equation}
$\forall a,b\in\algrA$ homogeneous.

\bigskip
The structure of module compatible with an $\eps$-graded algebra $\algrA$ is simply the structure of $\Gamma$-graded module. $\modrM$ is a $\Gamma$-graded module on $\algrA$ if it is a $\Gamma$-graded vector space and a module on $\algrA$ such that $\modM^i\algA^j\subset \modM^{i+j}$ (for right modules) $\forall i,j\in\Gamma$. The space of homomorphisms of $\modrM$ is an $\eps$-graded algebra and will be denoted by $\Hom^\bullet_\algA(\modM,\modM)$.

\subsubsection{$\eps$-derivations}

Let us now introduce the key object which permits to introduce a differential calculus adapted to this situation.

\begin{definition}[$\eps$-derivations]
An {\it $\eps$-derivation} on the $\eps$-graded algebra $\algrA$ is a linear combination of homogeneous linear maps $\kX: \algrA \to \algrA$ of degree $|\kX|\in\Gamma$, such that
\begin{equation}
\kX(a\fois b)=\kX(a)\fois b+\eps(|\kX|,|a|)\ a\fois\kX(b),\label{eq-eps-defderiv}
\end{equation}
$\forall a,b\in\algA$ with $a$ homogeneous.

We denote by $\Der^\bullet_\eps(\algA)$ the $\Gamma$-graded space of $\eps$-derivation on the $\eps$-graded algebra $\algrA$.
\end{definition}

Notice that this definition makes explicit reference to the $\eps$-structure, so that $\Der^\bullet_\eps(\algA)$ really depends on it.

\begin{proposition}[Structure of $\Der^\bullet_\eps(\algA)$]
The space $\Der^\bullet_\eps(\algA)$ is an $\eps$-Lie algebra for the bracket
\begin{equation*}
[\kX,\kY]_\eps = \kX\kY - \eps(|\kX|,|\kY|) \kY \kX.
\end{equation*}

It is also a $\caZ^\bullet_\eps(\algA)$-bimodule for the product structure
\begin{equation}
(z\fois \kX)(a)=\eps(|z|,|\kX|)(\kX\fois z)(a)=z\fois (\kX(a)),\label{eq-eps-defmodder}
\end{equation}
$\forall\kX\in\Der^\bullet_\eps(\algA)$, $\forall z\in\caZ^\bullet_\eps(\algA)$ and $\forall a\in\algrA$ homogeneous.
\end{proposition}

Notice that the left and right module structures are equivalent modulo the factor $\eps(|z|,|\kX|)$. So that we mention it as a module structure, not as a bimodule one. In order to take into account this extra factor, it would be convenient to introduce the notion of $\eps$-central bimodule, as a straightforward adaptation of the notion of central bimodule defined in \cite{DuboisViolette:1994cr,DuboisViolette:1995tz}. We will not go further in this direction.

As usual, an inner $\eps$-derivation on $\algrA$ is an $\eps$-derivation which can be written as
\begin{equation*}
b \mapsto \ad_a(b) = [a,b]_\eps,
\end{equation*}
for an $a \in \algrA$. We denote by
\begin{equation*}
\Int^\bullet_\eps(\algA)=\{\ad_a,\ a\in\algrA\}
\end{equation*}
the space of inner $\eps$-derivations on $\algrA$.

\begin{proposition}
$\Int^\bullet_\eps(\algA)$ is an $\eps$-Lie ideal and a $\caZ^\bullet_\eps(\algA)$-module.

This permits one to define the quotient
\begin{equation*}
\Out^\bullet_\eps(\algA)=\Der^\bullet_\eps(\algA)/\Int^\bullet_\eps(\algA)
\end{equation*}
as an $\eps$-Lie algebra and a $\caZ^\bullet_\eps(\algA)$-module. This is the space of outer $\eps$-derivations on $\algrA$.
\end{proposition}

From these considerations, one then gets the two short exact sequences of $\eps$-Lie algebras and $\caZ_\eps^\bullet(\algA)$-modules:
\begin{gather*}
\xymatrix@1@C=25pt{{\algzero} \ar[r] & {\caZ^\bullet_\eps(\algA)} \ar[r] & {\algrA} \ar[r]^-{\ad} & {\Int^\bullet_\eps(\algA)} \ar[r] & {\algzero}}\\[3mm]
\xymatrix@1@C=25pt{{\algzero} \ar[r] & {\Int^\bullet_\eps(\algA)} \ar[r] & {\Der^\bullet_\eps(\algA)} \ar[r] & {\Out^\bullet_\eps(\algA)} \ar[r] & {\algzero}}
\end{gather*}

Notice that the notion of $\eps$-Lie algebra has been generalized in \cite{Larsson:2005} where quasi-hom-Lie algebras are introduced, and the notion of $\eps$-derivations is a particular case of $(\sigma,\tau)$-derivations (see \cite{Hartwig:2006}). However, in the framework of $\eps$-graded algebras, these two structures are compatible together, since the same commutation factor is used in their definitions.

\subsubsection{$\eps$-Hochschild cohomology}

\begin{definition}
$(\algrA,\dd)$ is called an $\eps$-graded differential algebra if $\algrA$ is an $\eps$-graded algebra and $\dd$ is a homogeneous $\eps$-derivation of $\algrA$ such that $\dd^2=0$. Note that we do not assume any degree for $\dd$.
\end{definition}
For the case of $\gZ$-graded algebras, this definition is compatible with the usual one only if we assume that the differential $\dd$ is of degree $1$. But it is not possible for any arbitrary grading group $\Gamma$.
\medskip

Let us now define the graded $\eps$-Hochschild cohomology of an $\eps$-graded algebra $\algrA$ and a $\algrA$-bimodule $\modrM$, where $\eps$ is a commutation factor on the abelian group $\Gamma$ over the field $\gK$.
\begin{definition}
The $(n,k)$-cochains ($n\in\gN$, $k\in\Gamma$) for the $\eps$-Hochschild complex of $\algrA$ and $\modrM$ are given by the multilinear maps $\omega$ from $\algA^{\otimes n}$ to $\modM$, such that $\forall a_1,\dots,a_n\in\algrA$ homogeneous, $\omega(a_1\otimes\dots\otimes a_n)\in\modM^{k+|a_1|+\dots+|a_n|}$. The space of the $(n,k)$-cochains is denoted by $C^{n,k}(\algA,\modM)$. Setting $C^{0,k}(\algA,\modM)=\modM^k$, the complex of cochains writes
\begin{equation*}
C^{\bullet,\bullet}(\algA,\modM)=\bigoplus_{n\in\gN,k\in\Gamma}C^{n,k}(\algA,\modM),
\end{equation*}
and is a bimodule on $\algrA$. The differential associated to this complex is $\delta_\eps:C^{n,k}(\algA,\modM)\to C^{n+1,k}(\algA,\modM)$, defined by $\forall \omega\in C^{n,|\omega|}(\algA,\modM)$, $\forall a_0,\dots,a_n\in\algrA$ homogeneous,
\begin{multline}
(\delta_\eps\omega)(a_0\otimes\dots a_n)=\eps(|\omega|,|a_0|)a_0\omega(a_1\otimes\dots\otimes a_n)+(-1)^{n+1}\omega(a_0\otimes\dots\otimes a_{n-1})a_n\\
+\sum_{i=1}^n(-1)^i\omega(a_0\otimes\dots a_{i-1}\otimes(a_{i-1}\fois a_i)\otimes a_{i+1}\otimes\dots\otimes a_n).\label{eq-eps-defdiffch}
\end{multline}
Then, $(C^{\bullet,\bullet}(\algA,\modM),\delta_\eps)$ is called the $\eps$-Hochschild complex of $\algrA$ and $\modrM$. Its cohomology is denoted by $H_\eps^{\bullet,\bullet}(\algA,\modM)$.
\end{definition}
Note that we omit from now on the subscript $\gK$ for the unit of the field: $-1=-1_\gK$.

\begin{proposition}
For $\modrM=\algrA$, we obtain that $H_\eps^{0,\bullet}(\algA,\algA)=\caZ_\eps^\bullet(\algA)$ and $H_\eps^{1,\bullet}(\algA,\algA)=\Out_\eps^\bullet(\algA)$.
\end{proposition}
\begin{proof}
From \eqref{eq-eps-defdiffch}, we deduce that a $(0,k)$-cocycle $a\in\algA^k$ satisfies: $\forall b\in\algrA$ homogeneous, $(\delta_\eps a)(b)=\eps(|a|,|b|)b\fois a-a\fois b=0$. This means that $a\in\caZ_\eps^k(\algA)$.

A $(1,k)$-cocycle $\omega$ verifies: $\forall a,b\in\algrA$ homogeneous, $(\delta_\eps\omega)(a\otimes b)=\eps(k,|a|)a\fois\omega(b)+\omega(a)\fois b-\omega(a\fois b)=0$, or $\omega(a\fois b)=\omega(a)\fois b+\eps(k,|a|)a\fois\omega(b)$. Therefore, $\omega\in\Der_\eps^k(\algA)$. A $(1,k)$-coboundary $\eta$ is of the form $\eta(b)=\eps(|a|,|b|)b\fois a-a\fois b$, for a certain $a\in\algA^k$. As a consequence, $\eta\in\Int_\eps^k(\algA)$, and $H_\eps^{1,\bullet}(\algA,\algA)=\Out_\eps^\bullet(\algA)$.
\end{proof}

In the following, we use the following notations for the abelian group $\widetilde{\Gamma}=\gZ\times\Gamma$ and the commutation factor $\widetilde{\eps}((p,i),(q,j))=(-1)^{pq}\eps(i,j)$.

\begin{proposition}[Cup-product]
One can introduce a product on $C^{\bullet,\bullet}(\algA,\algA)$, called the cup-product, which associate to $\omega\in C^{m,|\omega|}(\algA,\algA)$ and $\eta\in C^{n,|\eta|}(\algA,\algA)$, $\omega\cup\eta\in C^{m+n,|\omega|+|\eta|}(\algA,\algA)$ defined by: $\forall a_1,\dots,a_{m+n}\in\algrA$ homogeneous,
\begin{equation}
(\omega\cup\eta)(a_1\otimes\dots\otimes a_{m+n})=\eps(|\eta|,|a_1|+\dots+|a_m|)\omega(a_1\otimes\dots\otimes a_m)\fois\eta(a_{m+1}\otimes\dots\otimes a_{m+n}).\label{eq-eps-cupproduct}
\end{equation}
With the cup-product, $(C^{\bullet,\bullet}(\algA,\algA),\delta_\eps)$ is an $\widetilde\eps$-graded differential algebra ($|\delta_\eps|=(1,0)$), and $H_\eps^{\bullet,\bullet}(\algA,\algA)$ is a $\widetilde{\eps}$-graded commutative algebra.
\end{proposition}

\begin{proposition}
For $\omega\in C^{m,|\omega|}(\algA,\algA)$ and $\eta\in C^{n,|\eta|}(\algA,\algA)$, one can define the composition of $\omega$ and $\eta$. If $i\in\{1,\dots,m\}$, and $a_1,\dots,a_{m+n-1}\in\algrA$ are homogeneous, we set
\begin{multline*}
(\omega\circ_i\eta)(a_1\otimes\dots\otimes a_{m+n-1})=\eps(|\eta|,|a_1|+\dots+|a_{i-1}|)\fois\\
\omega(a_1\otimes\dots\otimes a_{i-1}\otimes\eta(a_i\otimes\dots\otimes a_{i+n-1})\otimes a_{i+n}\otimes\dots\otimes a_{m+n-1}),
\end{multline*}
and
\begin{equation*}
\omega\circ\eta=\sum_{i=1}^m(-1)^{(n-1)(i-1)}\omega\circ_i\eta\ \in C^{m+n-1,|\omega|+|\eta|}(\algA,\algA).
\end{equation*}
Endowed with the bracket $[\omega,\eta]=\omega\circ\eta-(-1)^{(m-1)(n-1)}\eps(|\omega|,|\eta|)\eta\circ\omega$, and the cup-product \eqref{eq-eps-cupproduct}, the Hochschild cohomology $H_\eps^{\bullet,\bullet}(\algA,\algA)$ is a bigraded commutative Poisson algebra.
\end{proposition}

\subsection{Relationship with superalgebras}
\label{subsec-eps-relsuper}

In \cite{Scheunert:1979}, it has been noticed that there exists functorial relations between $\eps$-Lie algebras and Lie superalgebras. We extend this correspondence to the case of $\eps$-graded algebra.

\begin{proposition}
\label{prop-eps-algrsigma}
Let $\eps$ be a commutation factor on the abelian group $\Gamma$ over the field $\gK$, $\algrA$ an $\eps$-graded algebra, and $\sigma$ a factor set of $\Gamma$. One can endow the $\Gamma$-graded vector space $\algrA$ with the product:
\begin{equation*}
\forall a,b\in\algrA\text{ homogeneous, }a\fois_\sigma b=\sigma(|a|,|b|)a\fois b,
\end{equation*}
and denote it by $\algrA_\sigma$. Then, $\algrA_\sigma$ is an $\eps\eps_\sigma$-graded algebra, where $\eps_\sigma$ is given by Equation \eqref{eq-eps-multcomm}.

Moreover, the bracket of $\algrA_{\sigma,\Lie}$ can be expressed in terms of the one of $\algrA_\Lie$: $\forall a,b\in\algrA$ homogeneous,
\begin{equation*}
[a,b]^\sigma_{\eps\eps_\sigma}=\sigma(|a|,|b|)[a,b]_\eps.
\end{equation*}
\end{proposition}
\begin{proof}
We check the associativity of the product of $\algrA_\sigma$. For $a,b,c\in\algrA_\sigma$ homogeneous,
\begin{align*}
a\fois_\sigma (b\fois_\sigma c)&=\sigma(|a|,|b|+|c|)\sigma(|b|,|c|) a\fois b\fois c,\\
(a\fois_\sigma b)\fois_\sigma c&=\sigma(|a|,|b|)\sigma(|a|+|b|,|c|) a\fois b\fois c.
\end{align*}
The associativity of this product is therefore a direct consequence of the associativity of the product of $\algrA$ and of the defining relation \eqref{eq-eps-deffactorset} of a factor set.
\end{proof}

\begin{theorem}
Let $\eps$ be a commutation factor on the abelian group $\Gamma$ over the field $\gK$ and $\sigma$ a factor set of $\Gamma$. The correspondence
\begin{equation*}
\Phi_\sigma:\algrA\to\algrA_\sigma,
\end{equation*}
between the category of $\eps$-graded algebras and the category of $\eps\eps_\sigma$-graded algebras, has the following functorial properties:
\begin{enumerate}
\item $\Phi_\sigma$ is bijective and its inverse is given by $\Phi_{\sigma^{-1}}$.
\item $\Phi_\sigma$ induces a correspondence between the associated $\eps$ and $\eps\eps_\sigma$-Lie algebras, which can be extended to arbitrary $\eps$ and $\eps\eps_\sigma$-Lie algebras.
\item Let $\algrA$ be an $\eps$-graded algebra and $V^\bullet$ be a $\Gamma$-graded subspace of $\algrA$. Then, $V^\bullet$ is an $\eps$-graded subalgebra of $\algrA$ if and only if $V^\bullet$ is an $\eps\eps_\sigma$-graded subalgebra of $\algrA_\sigma$.
\item Let $\algrA$ and $\algrB$ be $\eps$-graded algebras. Then, $\phi:\algrA\to\algrB$ is a homomorphism of $\eps$-graded algebras if and only if $\phi:\algrA_\sigma\to\algrB_\sigma$ is a homomorphism of $\eps\eps_\sigma$-graded algebras.
\end{enumerate}
\end{theorem}
\begin{proof}
This is a consequence of Proposition \ref{prop-eps-algrsigma}.
\end{proof}

\begin{proposition}
The above correspondence is compatible with Hochschild cohomology.

For example,
\begin{align*}
\delta^\sigma_{\eps\eps_\sigma}\omega_\sigma(a\otimes b\otimes c)=&\eps\eps_\sigma(|\omega|,|a|)a\fois_\sigma(\omega_\sigma(b\otimes_\sigma c))-\omega_\sigma((a\fois_\sigma b)\otimes_\sigma c)\\
&+\omega_\sigma(a\otimes_\sigma (b\fois_\sigma c))-\omega_\sigma(a\otimes_\sigma b)\fois_\sigma c\\
=&\sigma(|\omega|,|a|+|b|+|c|)\sigma(|a|,|b|+|c|)\sigma(|b|,|c|)\delta_\eps\omega(a\otimes b\otimes c),
\end{align*}
where $\delta^\sigma$ means that the products and the tensor products in \eqref{eq-eps-defdiffch} have to be twisted by $\sigma$.
\end{proposition}

One can give another construction for this correspondence.
\begin{definition}
Let $\Gamma$ be an abelian group, $\gK$ a field, and $\sigma$ a factor set of $\Gamma$. We define here $S=\gK\rtimes_\sigma\Gamma$, the {\it crossed-product} of $\gK$ by $\Gamma$ relatively to $\sigma$. $S$ is the algebra of functions $\Gamma\to\gK$ which vanish outside of a finite number of elements of $\Gamma$, with product: $\forall f,g\in S$, $\forall k\in\Gamma$,
\begin{equation*}
(f\fois g)(k)=\sum_{i+j=k}\sigma(i,j)f(i)g(j).
\end{equation*}
Let $(e_k)_{k\in\Gamma}$ be its canonical basis, given by $\forall i,k\in\Gamma$, $e_k(i)=\delta_{ik}$. It satisfies $\forall i,j\in\Gamma$, $e_i\fois e_j=\sigma(i,j)e_{i+j}$. With the $\Gamma$-grading given by $S^k=\gK e_k$, $S^\bullet$ is an $\eps_\sigma$-graded commutative algebra.
\end{definition}

\begin{example}
\label{ex-eps-cliffordalg}
Let $\Gamma=(\gZ_2)^n$, for $n\in\gN^\ast$, and $\gK$ be a field whose characteristic is different from $2$. Let $\sigma$ be the factor set of $\Gamma$ defined by: $\forall i,j\in\Gamma$,
\begin{equation}
\sigma(i,j)=(-1_\gK)^{\sum_{1\leq p<q\leq n}i_p j_q}.\label{eq-eps-exsigma1}
\end{equation}
Then, $\gK\rtimes_\sigma\Gamma$ is isomorphic to a Clifford algebra associated to the vector space $\gK^n$.

For $\gK=\gC$ and the factor set $\sigma$ given by \eqref{eq-eps-exsigma1}, $\gC\rtimes_\sigma(\gZ_2)^n$ is isomorphic to the Clifford algebra $\caC l(n,\gC)$. This is still true for any factor set equivalent to $\sigma$.

For $\gK=\gR$, let $r,s\in\gN$ such that $r+s=n$, and $\eta\in\{\pm 1\}^n$ given by: $\eta_p=1$ if $p\leq r$, and $\eta_p=-1$ otherwise. Let also $\sigma$ be the factor set of $(\gZ_2)^n$ defined by: $\forall i,j\in(\gZ_2)^n$,
\begin{equation}
\sigma(i,j)=\left(\prod_{1\leq p<q\leq n}(-1)^{i_p j_q}\right)\prod_{p=1}^n\eta_p^{i_pj_p}.\label{eq-eps-exsigma2}
\end{equation}
Then, $\gR\rtimes_\sigma(\gZ_2)^n$ is isomorphic to the real Clifford algebra $\caC l(r,s)$. To any factor set equivalent to \eqref{eq-eps-exsigma2} corresponds a real Clifford algebra but with a potentially different signature $(r,s)$.
\end{example}

\begin{proposition}
Let $\eps$ be a commutation factor on the abelian group $\Gamma$ over the field $\gK$, $\algrA$ an $\eps$-graded algebra, $\sigma$ a factor set of $\Gamma$ and $S^\bullet=\gK\rtimes_\sigma\Gamma$. Then, $\hat\algA^\bullet$, defined by:
\begin{equation*}
\forall k\in\Gamma,\quad \hat\algA^k=S^k\otimes\algA^k
\end{equation*}
is an $\eps\eps_\sigma$-graded algebra for the product $(f\otimes a)\fois(g\otimes b)=(f\fois g)\otimes(a\fois b)$. Moreover, $\hat\algA^\bullet$ is isomorphic to $\algrA_\sigma$ by: $a\in\algrA_\sigma\mapsto e_{|a|}\otimes a$.
\end{proposition}

\begin{theorem}
Let $\Gamma$ be a finitely generated abelian group, $\gK$ an algebraically closed field, $\eps$ a commutation factor on $\Gamma$ over $\gK$, and $\algrA$ an $\eps$-graded algebra.

Let $[\sigma]$ be the multiplier of $\Gamma$ given by Theorem \ref{thm-eps-multcomm} such that $\eps=s(\eps)\eps_\sigma$.
Then, $\algrA_\sigma\approx\hat\algA^\bullet$ is a $s(\eps)$-graded algebra (therefore a superalgebra), and $\Phi_\sigma$ is a functorial correspondence between the category of $\eps$-graded algebras and the category of $\Gamma$-graded superalgebras.
\end{theorem}

\section{\texorpdfstring{Noncommutative geometry based on $\eps$-derivations}{Noncommutative geometry based on epsilon-derivations}}
\label{sec-eps-epsncg}

\subsection{Differential calculus}
\label{subsec-eps-diffcalc}

In this subsection, we generalize the derivation-based differential calculus introduced in \cite{DuboisViolette:1988cr,DuboisViolette:1994cr} (see subsection \ref{subsec-gnc-diffcalc}) to the setting of $\eps$-graded algebras \cite{deGoursac:2008bd}. Note that some propositions to extend this construction to certain class of graded algebras have been presented for instance in \cite{Lecomte:1992, Grosse:1999}.

In the notations of subsection \ref{subsec-eps-defalg}, let $\algrA$ be an $\eps$-graded algebra and $\modrM$ a right $\algrA$-module. Let us introduce the differential calculus based on the $\eps$-derivations of $\algrA$.

\begin{definition}[Differential calculus]
For $n\in\gN$ and $k\in\Gamma$, let $\Omr^{n,k}(\algA,\modM)$ be the space of $n$-linear maps $\omega$ from $(\Der^\bullet_\eps(\algA))^n$ to $\modrM$, such that $\forall\kX_1,\dots,\kX_n\in\Der^\bullet_\eps(\algA)$, $\forall z\in\caZ^\bullet_\eps(\algA)$ homogeneous,
\begin{align}
\omega(\kX_1,\dots,\kX_n) &\in\modM^{k+|\kX_1|+\dots+|\kX_n|},\nonumber\\
\omega(\kX_1,\dots,\kX_n\fois z)&=\omega(\kX_1,\dots,\kX_n)z,\nonumber\\
\omega(\kX_1,\dots,\kX_i,\kX_{i+1},\dots,\kX_n)&=-\eps(|\kX_i|,|\kX_{i+1}|)\omega(\kX_1,\dots,\kX_{i+1},\kX_i,\dots,\kX_n),\label{eq-eps-defcalcdiff}
\end{align}
and $\Omr^{0,k}(\algA,\modM)=\modM^k$, where $\kX\fois z$ was defined by \eqref{eq-eps-defmodder}. 
\end{definition}

From this definition, it follows that the vector space $\Omr^{\bullet,\bullet}(\algA,\modM)$ is $\gN\times\Gamma$-graded, and $\Omr^{n,\bullet}(\algA,\modM)$ is a right module on $\algrA$. 

In the case $\modrM=\algrA$, we use the notation $\Omr^{\bullet,\bullet}(\algA) = \Omr^{\bullet,\bullet}(\algA,\algA)$.

\begin{proposition}
\label{prop-eps-proddiff}
Endowed with the following product and differential, $\Omr^{\bullet,\bullet}(\algA)$ is a differential algebra: $\forall\omega\in\Omr^{p,|\omega|}(\algA)$, $\forall\eta\in\Omr^{q,|\eta|}(\algA)$, and $\forall\kX_1,\dots,\kX_{p+q}\in\Der^\bullet_\eps(\algA)$ homogeneous, the product is
\begin{equation}
(\omega\fois\eta)(\kX_1,\dots,\kX_{p+q}) =\frac{1}{p!q!}\sum_{\sigma\in\kS_{p+q}}(-1)^{|\sigma|}f_1\ \omega(\kX_{\sigma(1)},\dots,\kX_{\sigma(p)})\fois\eta(\kX_{\sigma(p+1)},\dots,\kX_{\sigma(p+q)}),\label{eq-eps-defprod}
\end{equation}
and the differential is
\begin{multline}
\dd\omega(\kX_1,\dots,\kX_{p+1})=\sum_{m=1}^{p+1} (-1)^{m+1}f_2\ \kX_m\omega(\kX_1,\dots \omi{m} \dots,\kX_{p+1})\\
+\sum_{1\leq m<n\leq p+1}(-1)^{m+n}f_3\ \omega([\kX_m,\kX_n]_\eps,\dots \omi{m} \dots \omi{n} \dots,\kX_{p+1}),\label{eq-eps-defdiff}
\end{multline}
where the factors $f_i$ are given by
\begin{align*}
f_1&=\prod_{m<n,\sigma(m)>\sigma(n)}\eps(|\kX_{\sigma(n)}|,|\kX_{\sigma(m)}|)\prod_{m\leq p} \eps(|\eta|,|\kX_{\sigma(m)}|)\\
f_2&=\eps(|\omega|,|\kX_m|)\prod_{a=1}^{m-1}\eps(|\kX_a|,|\kX_m|)\\
f_3&=\eps(|\kX_n|,|\kX_m|)\prod_{a=1}^{m-1}\eps(|\kX_a|,|\kX_m|)\prod_{a=1}^{n-1}\eps(|\kX_a|,|\kX_n|).
\end{align*}

$\Omr^{\bullet,\bullet}(\algA)$ is an $\widetilde{\eps}$-graded differential algebra for the abelian group $\widetilde{\Gamma}=\gZ\times\Gamma$ and the commutation factor $\widetilde{\eps}((p,i),(q,j))=(-1)^{pq}\eps(i,j)$. Note that the degree of the differential is $|\dd|=(1,0)$.

For any right $\algrA$-module $\modrM$, $\Omr^{\bullet,\bullet}(\algA,\modM)$ is a right module on $\Omr^{\bullet,\bullet}(\algA)$ for the following action:
$\forall\omega\in\Omr^{p,|\omega|}(\algA,\modM)$, $\forall\eta\in\Omr^{q,|\eta|}(\algA)$, and $\forall\kX_1,\dots,\kX_{p+q}\in\Der^\bullet_\eps(\algA)$ homogeneous,
\begin{equation}
(\omega\eta)(\kX_1,\dots,\kX_{p+q}) =\frac{1}{p!q!}\sum_{\sigma\in\kS_{p+q}}(-1)^{|\sigma|}f_1\ \omega(\kX_{\sigma(1)},\dots,\kX_{\sigma(p)})\eta(\kX_{\sigma(p+1)},\dots,\kX_{\sigma(p+q)}),\label{eq-eps-defmodomega}
\end{equation}
where $f_1$ is still the one given above.
\end{proposition}

\begin{proof}
It is a straightforward calculation to check that \eqref{eq-eps-defprod} and \eqref{eq-eps-defdiff} are compatible and turning $\Omr^{\bullet,\bullet}(\algA)$ into an $\widetilde{\eps}$-graded differential algebra, in the same way that \eqref{eq-eps-defmodomega} turns $\Omr^{\bullet,\bullet}(\algA,\modM)$ into a right $\Omr^{\bullet,\bullet}(\algA)$-module. Then, Lemma~\ref{lem-eps-thierry} and Remark~\ref{rem-eps--bigrading}  give the end of the proposition.
\end{proof}

In low degrees (in the $\gZ$ part), the differential and the product takes the following form, $\forall a\in\algrA$, $\forall\omega,\eta\in\Omr^{1,\bullet}(\algA)$, $\forall\kX,\kY\in\Der^\bullet_\eps(\algA)$ homogeneous,
\begin{align*}
\dd a(\kX)&=\eps(|a|,|\kX|)\kX(a),\\
\dd\omega(\kX,\kY)&=\eps(|\omega|,|\kX|)\kX(\omega(\kY))-\eps(|\omega|+|\kX|,|\kY|)\kY(\omega(\kX))-\omega([\kX,\kY]_\eps),\\
(\omega\fois\eta)(\kX,\kY)&=\eps(|\eta|,|\kX|)\omega(\kX)\fois\eta(\kY)-\eps(|\eta|+|\kX|,|\kY|)\omega(\kY)\fois\eta(\kX).
\end{align*}

One has to be aware of the fact that even if the $\eps$-Lie algebra of derivations is finite dimensional, the vector space $\Omr^{\bullet,\bullet}(\algA)$ can be infinite dimensional. This is indeed the case for Example~\ref{ex-eps-supermatrix}, where the symmetric part of this differential calculus amounts to an infinite dimensional part.

\begin{proposition}
If $\algrA$ is an $\eps$-graded commutative algebra then $\Omr^{\bullet,\bullet}(\algA)$ is an $\widetilde{\eps}$-graded commutative differential algebra.
\end{proposition}

\begin{proof}
This is a straightforward computation.
\end{proof}

\begin{definition}[Restricted differential calculus]
Let $\kg^\bullet$ be an $\eps$-Lie subalgebra of $\Der^\bullet_\eps(\algA)$ and a module on $\caZ^\bullet_\eps(\algA)$. The restricted differential calculus $\Omr^{\grast,\grast}(\algA|\kg)$ associated to $\kg^\bullet$ is defined as the space of $n$-linear maps $\omega$ from $(\kg^\bullet)^n$ to $\algrA$ satisfying the axioms \eqref{eq-eps-defcalcdiff}, with the above product \eqref{eq-eps-defprod} and differential \eqref{eq-eps-defdiff}. It is also an $\widetilde\eps$-graded differential algebra.
\end{definition}

\begin{proposition}[Cartan operation]
Let $\kg^\bullet$ be an $\eps$-Lie subalgebra of $\Der_\eps^\bullet(\algA)$. $\kg^\bullet$ defines canonically a Cartan operation on $(\Omr^{\bullet,\bullet}(\algA),\dd)$ in the following way. 

For each $\kX\in\Der_\eps^\bullet(\algA)$, the inner product with $\kX$ is the map $i_\kX:\Omr^{n,k}(\algA)\to\Omr^{n-1,k+|\kX|}(\algA)$ such that, $\forall\omega\in\Omr^{n,|\omega|}(\algA)$ and $\forall\kX,\kX_1,\dots,\kX_{n-1}\in\kg^\bullet$ homogeneous,
\begin{equation}
i_\kX\omega(\kX_1,\dots,\kX_{n-1})=\eps(|\kX|,|\omega|)\omega(\kX,\kX_1,\dots,\kX_{n-1}),\label{eq-eps-definnerprod}
\end{equation}
and $i_\kX\Omr^{0,\grast}(\algA)=\algzero$. $i_\kX$ is then an $\widetilde{\eps}$-derivation of the algebra $\Omr^{\bullet,\bullet}(\algA)$ of degree $(-1,|\kX|)$. 

The associated Lie derivative $L_\kX$ to $i_\kX$ is
\begin{equation}
L_\kX=[i_\kX,\dd]=i_\kX\dd+\dd i_\kX:\Omr^{n,k}(\algA)\to\Omr^{n,k+|\kX|}(\algA),\label{eq-eps-deflieder}
\end{equation}
which makes it into an $\widetilde{\eps}$-derivation of $\Omr^{\bullet,\bullet}(\algA)$ of degree $(0,|\kX|)$, where the bracket in \eqref{eq-eps-deflieder} comes from the commutation factor $\widetilde{\eps}$ of $\Omr^{\bullet,\bullet}(\algA)$.

Then, the following properties are satisfied, $\forall\kX,\kY\in\kg^\bullet$ homogeneous,
\begin{align*}
[i_\kX,i_\kY]&=i_\kX i_\kY+\eps(|\kX|,|\kY|)i_\kY i_\kX=0, &[L_\kX,i_\kY]&=L_\kX i_\kY-\eps(|\kX|,|\kY|)i_\kY L_\kX=i_{[\kX,\kY]_\eps},\\
[L_\kX,\dd]&=L_\kX\dd-\dd L_\kX= 0, &[L_\kX,L_\kY]&=L_\kX L_\kY-\eps(|\kX|,|\kY|)L_\kY L_\kX=L_{[\kX,\kY]_\eps}.
\end{align*}
\end{proposition}

\begin{proof}
$\forall\omega\in\Omr^{n,|\omega|}(\algA)$, $\forall\eta\in\Omr^{p,|\eta|}(\algA)$ and $\forall\kX,\kY,\kX_1,\dots,\kX_{n-2}\in\kg^\bullet$ homogeneous, one has:
\begin{equation*}
i_\kX(\omega\fois \eta)=(i_\kX \omega)\fois\eta+(-1)^n\eps(|\kX|,|\omega|)\omega(i_\kX\eta),
\end{equation*}
due to the definition of $i_\kX$ \eqref{eq-eps-definnerprod} and of the product \eqref{eq-eps-defprod}. Then, the third axiom of \eqref{eq-eps-defcalcdiff} implies:
\begin{align*}
i_\kX i_\kY\omega(\kX_1,\dots,\kX_{n-2})&=\eps(|\kX|+|\kY|,|\omega|)\eps(|\kX|,|\kY|)\omega(\kY,\kX,\kX_1,\dots,\kX_{n-2})\\
&=-\eps(|\kX|,|\kY|)i_\kY i_\kX\omega(\kX_1,\dots,\kX_{n-2}),
\end{align*}
so that $[i_\kX,i_\kY]=0$. Furthermore, $L_\kX\dd=\dd i_\kX\dd=\dd L_\kX$.
\begin{equation*}
[L_\kX,i_\kY]=i_\kX\dd i_\kY+\dd i_\kX i_\kY-\eps(|\kX|,|\kY|)(i_\kY i_\kX\dd+i_\kY\dd i_\kX).
\end{equation*}
By a long but straightforward calculation, using \eqref{eq-eps-defdiff} and \eqref{eq-eps-definnerprod}, one finds $[L_\kX,i_\kY]=i_{[\kX,\kY]_\eps}$. Finally, the above results permit one to show that
\begin{align*}
[L_\kX,L_\kY]&=L_\kX(i_\kY\dd+\dd i_\kY)-\eps(|\kX|,|\kY|)(i_\kY\dd+\dd i_\kY)L_\kX\\
&=i_{[\kX,\kY]_\eps}\dd+\dd i_{[\kX,\kY]_\eps}=L_{[\kX,\kY]_\eps}.
\end{align*}
\end{proof}

\subsection{\texorpdfstring{$\eps$-connections and gauge transformations}{epsilon-connections and gauge transformations}}
\label{subsec-eps-conn}

Let us now generalize the notion of noncommutative connections (see subsection \ref{subsec-gnc-diffcalc}) and gauge theory to the framework of $\eps$-graded algebras. Let $\modrM$ be a right module on the $\eps$-graded algebra $\algrA$.

\begin{definition}[$\eps$-connections]
A linear map,
\begin{equation*}
\nabla:\modrM\to\Omr^{1,\bullet}(\algA,\modM),
\end{equation*} 
is called an {\it $\eps$-connection} if $\forall a\in\algrA$, $\forall m\in\modrM$,
\begin{equation}
\nabla(ma)=\nabla(m)a+m\dd a.\label{eq-eps-defconn}
\end{equation}
\end{definition}

Note that an $\eps$-connection $\nabla$ can be decomposed in homogeneous elements:
\begin{equation*}
\nabla=\sum_{k\in\Gamma}\nabla_k,
\end{equation*}
where $|\nabla_k|=k\in\Gamma$. Then, $\nabla_0$ is a connection of degree $0$ and $\forall k\neq0$, $\nabla_k$ is an homomorphism of $\algrA$-modules homogeneous of degree $k$.

\begin{proposition}
Let $\nabla$ be an $\eps$-connection on $\modrM$. Then, it can be extended as a linear map 
\begin{equation*}
\nabla:\Omr^{p,\bullet}(\algA,\modM)\to\Omr^{p+1,\bullet}(\algA,\modM)
\end{equation*}
using the relation $\forall\omega\in\Omr^{p,|\omega|}(\algA,\modM)$ and $\forall\kX_1,\dots,\kX_{p+1}\in\Der^\bullet_\eps(\algA)$ homogeneous,
\begin{multline}
\nabla(\omega)(\kX_1,\dots,\kX_{p+1})=\sum_{m=1}^{p+1} (-1)^{m+1}f_4\ \nabla(\omega(\kX_1,\dots \omi{m} \dots,\kX_{p+1}))(\kX_m)\label{eq-eps-defnabla}\\
+\sum_{1\leq m<n\leq p+1}(-1)^{m+n}f_5\ \omega([\kX_m,\kX_n]_\eps,\dots \omi{m} \dots \omi{n} \dots,\kX_{p+1}),
\end{multline}
where the factors $f_i$ are given by:
\begin{align*}
f_4&=\prod_{a=m+1}^{p+1}\eps(|\kX_m|,|\kX_a|),\\
f_5&=\eps(|\kX_n|,|\kX_m|)\prod_{a=1}^{m-1}\eps(|\kX_a|,|\kX_m|)\prod_{a=1}^{n-1}\eps(|\kX_a|,|\kX_n|).
\end{align*}

Then, $\nabla$ satisfies the following relation, $\forall\omega\in\Omr^{p,|\omega|}(\algA,\modM)$, $\forall\eta\in\Omr^{q,|\eta|}(\algA)$ homogeneous,
\begin{equation}
\nabla(\omega\eta)=\nabla(\omega)\eta+(-1)^p\omega\dd\eta.\label{eq-eps-propconn}
\end{equation}

The obstruction for $\nabla$ to be an homomorphism of right $\algrA$-module is measured by its curvature $R=\nabla^2$, which takes the following form, $\forall m\in\modrM$, $\forall\kX,\kY\in\Der^\bullet_\eps(\algA)$ homogeneous,
\begin{equation}
R(m)(\kX,\kY)=\eps(|\kX|,|\kY|)\nabla(\nabla(m)(\kY))(\kX)-\nabla(\nabla(m)(\kX))(\kY)-\nabla(m)([\kX,\kY]_\eps).\label{eq-eps-defcurv}
\end{equation}
\end{proposition}

\begin{proof}
The formula \eqref{eq-eps-defnabla} is inspired by the formula \eqref{eq-eps-defdiff} in order that the definition of this extension is well defined. Proving \eqref{eq-eps-propconn} is therefore like proving that $\dd$ is an $\widetilde\eps$-derivation of degree $(1,0)$ in the proposition \ref{prop-eps-proddiff}, it is a long but straightforward computation.
\end{proof}

\begin{proposition}
The space of all $\eps$-connections on $\modrM$ is an affine space modeled on the vector space $\Hom^\bullet_\algA(\modrM,\Omr^{1,\bullet}(\algA,\modM))$. Furthermore, the curvature $R$ associated to an $\eps$-connection $\nabla$ is a homomorphism of right $\algrA$-modules.
\end{proposition}
\begin{proof}
Let $\nabla$ and $\nabla'$ be two $\eps$-connections, and $\Psi=\nabla-\nabla'$. Then, $\forall a\in\algrA$ and $\forall m\in\modrM$,
\begin{equation*}
\Psi(ma)=\nabla(m)a-\nabla'(m)a=\Psi(m)a.
\end{equation*}
Therefore, $\Psi\in\Hom^\bullet_\algA(\modrM,\Omr^{1,\bullet}(\algA,\modM))$.

In the same way, from $R=\nabla^2$, using \eqref{eq-eps-defconn} and \eqref{eq-eps-propconn}, one obtains:
\begin{equation*}
R(ma)=\nabla^2(m)a-\nabla(m)\dd a+\nabla(m)\dd a+m\dd^2 a=R(m)a.
\end{equation*}
\end{proof}

\begin{definition}[Gauge group]
The {\it gauge group} of $\modrM$ is defined as the group of automorphisms $\Aut_\algA(\modM,\modM)$ of the right $\algrA$-module $\modrM$. Its elements are called gauge transformations. 

Each gauge transformation $\Phi$ is extended to an automorphism of $\Omr^{\bullet,\bullet}(\algA,\modM)$, considered as a right $\Omr^{\bullet,\bullet}(\algA)$-module, in the following way, $\forall\omega\in\Omr^{p,|\omega|}(\algA,\modM)$ and\\ $\forall\kX_1,\dots,\kX_{p}\in\Der^\bullet_\eps(\algA)$,
\begin{equation*}
\Phi(\omega)(\kX_1,\dots,\kX_p)=\Phi(\omega(\kX_1,\dots,\kX_p)).
\end{equation*}
\end{definition}

\begin{proposition}
The gauge group of $\modrM$ acts on the space of its $\eps$-connections in the following way: for $\Phi\in\Aut_\algA^0(\modM,\modM)$ and $\nabla$ an $\eps$-connection,
\begin{equation*}
\nabla^\Phi=\Phi\circ\nabla\circ\Phi^{-1}
\end{equation*}
is again an $\eps$-connection. The induced action of $\Phi$ on the associated curvature is given by
\begin{equation*}
R^\Phi=\Phi\circ R\circ\Phi^{-1}.
\end{equation*}
\end{proposition}

\begin{proof}
The axioms \eqref{eq-eps-defconn} are satisfied for $\nabla^\Phi$: $\forall a\in\algrA$ and $\forall m\in\modrM$,
\begin{align*}
\nabla^\Phi(ma)&=\Phi\circ\nabla(\Phi^{-1}(m)a)\\
&=(\nabla^\Phi(m)a)+m\dd a.
\end{align*}
The proof is trivial for $R^\Phi=\nabla^\Phi\circ\nabla^\Phi$.
\end{proof}

\subsection{Involutions}
\label{subsec-eps-invol}

Let us now take a look at the notion of involution in this framework.
\begin{definition}
Let $\gK$ be a field, with an involution, that is  a ring-morphism of $\gK$, $\lambda\mapsto\overline{\lambda}$, such that $\forall \lambda\in\gK$, $\overline{\overline{\lambda}}=\lambda$. Then, a commutation factor $\eps$ on the abelian group $\Gamma$ over $\gK$ is called involutive if: $\forall i,j\in\Gamma$,
\begin{equation*}
\overline{\eps(i,j)}=\eps(j,i).
\end{equation*}
\end{definition}

\begin{definition}[Involution, unitarity and hermiticity]
\label{def-eps-invol}
Let $\Gamma$ be an abelian group, $\gK$ an involutive field, and $\eps$ an involutive commutation factor on $\Gamma$ over $\gK$. An {\it involution} on an $\eps$-graded algebra $\algrA$ is a map $\algA^i\to\algA^{-i}$, for any $i\in\Gamma$, denoted by $a\mapsto a^\ast$, such that $\forall a,b\in\algrA$, $\forall \lambda\in\gK$,
\begin{align}
(a+\lambda b)^\ast=a^\ast+\overline{\lambda}b^\ast,&\qquad (a^\ast)^\ast=a,\label{eq-eps-axinvol1}\\
(a\fois b)^\ast&=b^\ast\fois a^\ast.\label{eq-eps-axinvol2}
\end{align}

The unitary group of $\algrA$ associated to $\ast$ is defined and denoted by
\begin{equation*}
\caU(\algA)=\{g\in\algrA,\ g^\ast\fois g=\gone\}.
\end{equation*}

An $\eps$-trace $T$ on $\algrA$ is real if, $\forall a\in\algrA$, $T(a^\ast)=\overline{T(a)}$.

An $\eps$-Lie algebra $\kg^\bullet$ is called involutive if there exists a map $\kg^i\to\kg^{-i}$, for any $i\in\Gamma$, denoted by $a\mapsto a^\ast$, satisfying the axioms \eqref{eq-eps-axinvol1} and $\forall a,b\in\kg^\bullet$,
\begin{equation*}
[a,b]^\ast=[b^\ast,a^\ast].
\end{equation*}

Then, $\algrA_\Lie$ is an involutive $\eps$-Lie algebra.
\end{definition}

A {\it hermitean structure} on a right $\algrA$-module $\modrM$ is a sesquilinear form $\langle-,-\rangle:\modM^i\times\modM^j\to\algA^{j-i}$, for $i,j\in\Gamma$, such that $\forall m,n\in\modrM$, $\forall a,b\in\algrA$,
\begin{equation}
\label{eq-eps-hermiteanstructure}
\langle m,n\rangle^\ast=\langle n,m\rangle,\qquad \langle ma,nb\rangle=a^\ast\langle m,n\rangle b.
\end{equation}

\begin{remark}[Involution and $\eps$-structure]
Notice that the definition of the involution as it is given by \eqref{eq-eps-axinvol2} does not make direct reference to the $\eps$-structure, as well as the hermitean structure defined by \eqref{eq-eps-hermiteanstructure}.

Another natural definition of the involution can be considered, where an explicit reference to $\eps$ appears in the second relation in \eqref{eq-eps-axinvol2} and in the two relations in \eqref{eq-eps-hermiteanstructure}. It satisfies:
\begin{align*}
\overline{\eps(i,j)}&=\eps(i,j),\\
(a\fois b)^\ast&=\eps(|a|,|b|)b^\ast\fois a^\ast,\\
\langle m,n\rangle^\ast&=\eps(|m|,|n|)\langle n,m\rangle,\\
\langle ma,nb\rangle&=\eps(|m|,|a|)a^\ast\langle m,n\rangle b.
\end{align*}
\end{remark}

An involution (with the conventions of Definition \ref{def-eps-invol}) on an $\eps$-graded algebra $\algrA$ induces another one on the $\eps$-Lie algebra $\Der_\eps^\bullet(\algA)$ by: $\forall\kX\in\Der_\eps^\bullet(\algA)$, $\forall a\in\algrA$ homogeneous,
\begin{equation*}
\kX^\ast(a)=-\eps(|\kX^\ast|,|a|)(\kX(a^\ast))^\ast.
\end{equation*}
Indeed, $[\kX,\kY]^\ast=[\kY^\ast,\kX^\ast]$. Note that $(\ad_a)^\ast=\ad_{a^\ast}$, this has motivated the sign in the previous definition. A hermitean $\eps$-derivation is an $\eps$-derivation $\kX$ such that $\kX^\ast=\kX$. Decomposed in its homogeneous elements $\kX=\sum_{k\in\Gamma}\kX_k$ (with $|\kX_k|=k$), it gives: $\forall k\in\Gamma$, $\forall a\in\algrA$ homogeneous,
\begin{equation*}
\kX_k(a)^\ast=\eps(|a|,|\kX_k|)\kX_{-k}(a^\ast).
\end{equation*}
The space of hermitean $\eps$-derivations is not a graded vector subspace of $\Der_\eps^\bullet(\algA)$.

\begin{definition}
The involution on $\algrA$ defines a map $\Omr^{p,i}(\algA)\to\Omr^{p,-i}(\algA)$, for $p\in\gN$, and $i\in\Gamma$, by: $\forall\omega\in\Omr^{p,|\omega|}(\algA)$, $\forall\kX_1,\dots,\kX_p\in\Der_\eps^\bullet(\algA)$ homogeneous,
\begin{equation*}
\omega^\ast(\kX_1,\dots,\kX_p)=\prod_{i=1}^p\eps(|\omega^\ast|,|\kX_i|)\prod_{1\leq i<j\leq p}\eps(|\kX_i|,|\kX_j|)\ (\omega(\kX_1^\ast,\dots,\kX_p^\ast))^\ast.
\end{equation*}
This definition coincides with the previous one for $p=0$, and satisfies: $\forall\omega\in\Omr^{p,|\omega|}(\algA)$, $\forall\eta\in\Omr^{q,|\eta|}(\algA)$,
\begin{align*}
(\omega\fois\eta)^\ast&=(-1)^{pq}\ \eta^\ast\fois\omega^\ast,\\
[\omega,\eta]_{\widetilde\eps}^\ast&=(-1)^{pq}[\eta^\ast,\omega^\ast]_{\widetilde\eps},\\
(\dd\omega)^\ast&=-\dd(\omega^\ast).
\end{align*}
\end{definition}
Note that the map $\ast$ is not an involution for the $\widetilde\eps$-graded algebra $\Omr^{\bullet,\bullet}(\algA)$.

\begin{proposition}
Let $\algrA$ be an involutive $\eps$-graded algebra, and $\modrM$ a right $\algrA$-module. Then, any hermitean structure on $\modrM$ can be extended to the following structure\\
$\langle-,-\rangle:\Omr^{p_1,i_1}(\algA,\modM)\times \Omr^{p_2,i_2}(\algA,\modM)\to\Omr^{p_1+p_2,i_2-i_1}(\algA)$ by:
\begin{multline*}
\langle\omega_1,\omega_2\rangle(\kX_1,\dots,\kX_{p_1+p_2})=\\
\frac{1}{p_1!p_2!}\sum_{\sigma\in\kS_{p_1+p_2}} (-1)^{|\sigma|}f_6\langle\omega_1(\kX^\ast_{\sigma(1)},\dots,\kX^\ast_{\sigma(p_1)}), \omega_2(\kX_{\sigma(p_1+1)},\dots,\kX_{\sigma(p_1+p_2)})\rangle,
\end{multline*}
where $f_6$ is given by:
\begin{multline*}
f_6=\prod_{m\leq p_1}\eps(|\omega_1^\ast|+|\omega_2|,|\kX_{\sigma(m)}|)\prod_{m<n<p_1,\, \sigma(m)<\sigma(n)} \eps(|\kX_{\sigma(m)}|,|\kX_{\sigma(n)}|)\\
\prod_{m<n,\, n>p_1,\,\sigma(m)>\sigma(n)}\eps(|\kX_{\sigma(n)}|,|\kX_{\sigma(m)}|)
\end{multline*}
This structure satisfies: $\forall\omega_1\in\Omr^{p_1,|\omega_1|}(\algA,\modM)$, $\forall\omega_2\in\Omr^{p_2,|\omega_2|}(\algA,\modM)$, $\forall\eta_1\in\Omr^{q_1,|\eta_1|}(\algA)$ and $\forall\eta_2\in\Omr^{q_2,|\eta_2|}(\algA)$,
\begin{equation*}
\langle\omega_1,\omega_2\rangle^\ast=(-1)^{p_1p_2}\langle\omega_2,\omega_1\rangle,\qquad \langle\omega_1\eta_1,\omega_2\eta_2\rangle=(-1)^{p_1q_1}\eta_1^\ast\langle\omega_1,\omega_2\rangle\eta_2.
\end{equation*}
\end{proposition}

\begin{proposition}
An $\eps$-connection $\nabla$ is called hermitean if the following identity: $\forall m,n\in\modrM$,
\begin{equation*}
\langle\nabla(m),n\rangle+\langle m,\nabla(n)\rangle=\dd(\langle m,n\rangle),
\end{equation*}
holds on the hermitean $\eps$-derivations. Then, it satisfies: $\forall\omega_1\in\Omr^{p_1,|\omega_1|}(\algA,\modM)$ and $\forall\omega_2\in\Omr^{p_2,|\omega_2|}(\algA,\modM)$,
\begin{equation*}
\langle\nabla(\omega_1),\omega_2\rangle+(-1)^{p_1}\langle \omega_1,\nabla(\omega_2)\rangle=\dd(\langle \omega_1,\omega_2\rangle),
\end{equation*}
on the hermitean $\eps$-derivations.

A gauge transformation $\Phi$ is called unitary if $\forall m,n\in\modrM$,
\begin{equation*}
\langle\Phi(m),\Phi(n)\rangle=\langle m,n\rangle.
\end{equation*}
Then, it satisfies on all $\eps$-derivations:
\begin{equation*}
\langle\Phi(\omega_1),\Phi(\omega_2)\rangle=\langle \omega_1,\omega_2\rangle.
\end{equation*}
\end{proposition}
\begin{proof}
We decompose $\Phi=\sum_{k\in\Gamma}\Phi_k$ into its homogeneous elements. Since $\Phi$ is unitary, one has:
\begin{equation*}
\forall p\in\Gamma,\quad \sum_{-k+l=p}\langle\Phi_k(m),\Phi_l(n)\rangle=\delta_{p0}\langle m,n\rangle.
\end{equation*}
For example, if $\omega_1,\omega_2\in\Omr^{1,\bullet}(\algA,\modM)$ are homogeneous, then $\forall \kX_1,\kX_2\in\Der^\bullet_\eps(\algA)$ homogeneous,
\begin{align*}
\langle\Phi(\omega_1),\Phi(\omega_2)\rangle&(\kX_1,\kX_2)\\
=&\sum_{k,l\in\Gamma}\eps(-k+l-|\omega_1|+|\omega_2|,|\kX_1|)\Big(\langle\Phi_k(\omega_1)(\kX_1^\ast),\Phi_l(\omega_2)(\kX_2)\rangle\\
& -\eps(|\kX_1|,|\kX_2|)\langle\Phi_k(\omega_1)(\kX_2^\ast),\Phi_l(\omega_2)(\kX_1)\rangle\Big)\\
=&\sum_{p\in\Gamma}\sum_{-k+l=p}\eps(p-|\omega_1|+|\omega_2|,|\kX_1|)\Big(\delta_{p0}\langle\omega_1(\kX_1^\ast),\omega_2(\kX_2)\rangle\\
&-\eps(|\kX_1|,|\kX_2|)\delta_{p0}\langle\omega_1(\kX_2^\ast),\omega_2(\kX_1)\rangle\Big)\\
=&\langle\omega_1,\omega_2\rangle(\kX_1,\kX_2).
\end{align*}
\end{proof}

\begin{proposition}
The space of hermitean $\eps$-connections on $\modrM$ is stable under the group of unitary gauge transformations.
\end{proposition}
\begin{proof}
Let $\Phi$ be a unitary gauge transformation, $\nabla$ a hermitean connection, and $m,n\in\modrM$. Then, one has on the hermitean $\eps$-derivations:
\begin{align*}
\langle\nabla^\Phi(m),n\rangle+\langle m,\nabla^\Phi(n)\rangle &=\langle\nabla\circ\Phi^{-1}(m),\Phi^{-1}(n)\rangle+\langle\Phi^{-1}(m),\nabla\circ\Phi^{-1}(n)\rangle\\
&=\dd(\langle\Phi^{-1}(m),\Phi^{-1}(n)\rangle)\\
&=\dd(\langle m,n\rangle).
\end{align*}
\end{proof}

\section{\texorpdfstring{Application to some examples of $\eps$-graded algebras}{Application to some examples of epsilon-graded algebras}}
\label{sec-eps-examples}

\subsection{\texorpdfstring{$\eps$-graded commutative algebras}{epsilon-graded commutative algebras}}
\label{subsec-eps-excomm}

Firstly, a non-graded associative algebra can be seen as an $\eps$-graded algebra, so that the differential calculus based on the derivations of an associative algebra, presented in subsection \ref{subsec-gnc-diffcalc}, is a particular case of the formalism of this chapter. Consequently, for the commutative non-graded algebra $\algA=C^\infty(M)$ of the functions of a smooth compact manifold $M$, the differential calculus of subsection \ref{subsec-eps-diffcalc} is the de Rham complex for the manifold $M$, as shown in Example \ref{ex-gnc-manifold}.
\medskip

In the $\gZ_2$-graded case, one can consider for example the superalgebra
\begin{equation*}
\algrA=C^\infty(\gR^{p|q})\simeq C^\infty(\gR^p)\otimes \Lambda^\bullet(\theta_1,\dots,\theta_q),
\end{equation*}
where $\theta_i$ are anticommuting generators. This describes a particular case of supermanifolds \cite{DeWitt:1984,Leites:1980,Tuynman:2005,Rogers:2007}.

Then, each element $f\in\algrA$ can be decomposed into:
\begin{equation*}
f=\sum_{I\subset\{1,\dots,q\}}f_I(x_1,\dots,x_p)\theta^I,
\end{equation*}
where $(x_i)$ is the canonical coordinate system of $\gR^p$, $I$ are ordered subsets of $\{1,\dots,q\}$, and $\theta^I=\prod_{i\in I}\theta_i$. One defines $s(I,J)$ as the number of inversions needed to render $I\cup J$ as an ordered subset of $\{1,\dots,q\}$. Then, $\forall I,J$, $\theta^I\theta^J=0$ if $I\cap J\neq\emptyset$, while
\begin{equation*}
\theta^I\theta^J=(-1)^{s(I,J)}\theta^{I\cup J},
\end{equation*}
if $I\cap J=\emptyset$. One has: $s(J,I)=|I||J|-s(I,J)$, and $\forall f,g\in\algrA$,
\begin{equation*}
f\fois g=\sum_{I\cap J=\emptyset}\, f_I\fois g_J (-1)^{s(I,J)}\theta^{I\cup J}.
\end{equation*}
An involution can be defined by $(\theta_i)^\ast=\theta_i$ and $(f_I)^\ast=\overline{f_I}$, such that $(\theta^I)^\ast=(-1)^{\frac{|I|(|I|-1)}{2}}\theta^I$.

The superderivations are given by:
\begin{equation*}
\Der^\bullet(\algA)=\left[\Gamma(\gR^p)\otimes\Lambda^\bullet(\theta_1,\dots,\theta_q)\right] \oplus \left[C^\infty(\gR^p)\otimes\Der^\bullet(\Lambda^\bullet(\theta_1,\dots,\theta_q))\right]
\end{equation*}
and any derivation of $\Lambda^\bullet(\theta_1,\dots,\theta_q)$ can be written as $i_K$, where $K\in(\gC^q)^\ast\otimes\Lambda^\bullet(\theta_1,\dots,\theta_q)$ and $i$ is the inner product. Then, the differential calculus (based on superderivations) takes the form:
\begin{equation*}
\underline\Omega_\eps^{\bullet,\bullet}(\algA)=\Omega^\bullet_{\text{dR}}(\gR^p)\otimes\Omega^{\bullet,\bullet}(\Lambda^\bullet(\theta_1,\dots,\theta_q)).
\end{equation*}

\subsection{\texorpdfstring{$\eps$-graded matrix algebras with elementary grading}{epsilon-graded matrix algebras with elementary grading}}
\label{subsec-eps-exmatrixelem}

Let $\Gamma$ be an abelian group, $\gK$ an involutive field, and $\algA=\Matr_D(\gK)$ the algebra of $D\times D$ matrices on $\gK$, such that $\algA$ is a $\Gamma$-graded algebra:
\begin{equation}
\algrA=\bigoplus_{\alpha\in\Gamma}\algA^\alpha.\label{eq-eps-gradingmatrix}
\end{equation}
Let $(E_{ij})_{1 \leq i, j \leq D}$ be the canonical basis of $\algrA$, whose product is as usual
\begin{equation}
E_{ij}\fois E_{kl}=\delta_{jk}E_{il}\label{eq-eps-prodmat}.
\end{equation}
\begin{definition}[Elementary grading]
The grading \eqref{eq-eps-gradingmatrix} of $\algrA$ is called {\it elementary} if there exists a map $\varphi:\{1,\dots,D\}\to\Gamma$ such that $\forall i,j\in\{1,\dots,D\}$, $E_{ij}$ is homogeneous of degree $|E_{ij}|=\varphi(i)-\varphi(j)$.
\end{definition}

In this subsection, we suppose that the grading \eqref{eq-eps-gradingmatrix} is elementary. Then, one can check that the usual conjugation (coming from the involution on $\gK$) is an involution for $\algrA$. Furthermore, $\algrA$ can be characterized by the following result:
\begin{proposition}[\cite{Bahturin:2002}]
The matrix algebra $\algrA=\Matr_D(\gK)$, with an elementary $\Gamma$-grading, is isomorphic, as a $\Gamma$-graded algebra, to the endomorphism algebra of some $D$-dimensional $\Gamma$-graded vector space on $\gK$.
\end{proposition}
More general gradings on $\algA$ have been classified in \cite{Bahturin:2002}.
\medskip

Let $\eps:\Gamma\times\Gamma\to\gK^\ast$ be an involutive commutation factor, which turns the elementary $\Gamma$-graded involutive algebra $\algrA$ into an $\eps$-graded involutive algebra. See \cite{deGoursac:2008bd} for the results of this subsection.

\begin{proposition}
\label{prop-eps-centermatrix}
The $\eps$-center of $\algrA$ is trivial:
\begin{equation*}
\caZ_\eps^\bullet(\algA)=\caZ_\eps^0(\algA)=\gK\gone.
\end{equation*}
\end{proposition}

\begin{proof}
Let $A\in\caZ_\eps^\bullet(\algrA)$ written as $A=\sum_{i,j=1}^Da_{ij}E_{ij}$. Then, due to \eqref{eq-eps-prodmat}, we immediately get, $\forall k,l\in\{1,\dots,D\}$,
\begin{equation*}
0=[A,E_{kl}]_\eps=\sum_{i,j=1}^D(a_{ik}\delta_{jl}-\eps(\varphi(l)-\varphi(j),\varphi(k)-\varphi(l))a_{lj}\delta_{ik})E_{ij}.
\end{equation*}
Therefore, $\forall i,j,k,l$,
\begin{equation*}
a_{ik}\delta_{jl}=\eps(\varphi(l)-\varphi(j),\varphi(k)-\varphi(l))a_{lj}\delta_{ik}.
\end{equation*}
For $i\neq k$ and $j=l$, we get $a_{ik}=0$, and for $i=k$ and $j=l$, $a_{ii}=a_{jj}$. This means that $A\in\gK\gone$.
\end{proof}

\begin{proposition}[$\eps$-traces]
\label{prop-eps-tracematrixalgebra}
For any $A=(a_{ij})\in\algrA$, the expression
\begin{equation*}
\tr_\eps(A)=\sum_{i=1}^D\eps(\varphi(i),\varphi(i))a_{ii}
\end{equation*}
defines a real $\eps$-trace on $\algrA$. Moreover, the space of $\eps$-traces on $\algrA$ is one-dimensional.
\end{proposition}

\begin{proof}
Let $T$ be an $\eps$-trace on $\algrA$. From \eqref{eq-eps-deftrace}, we get, $\forall i,j,k,l$, $T(E_{ij}\fois E_{kl})=\eps(\varphi(i)-\varphi(j),\varphi(k)-\varphi(l))T(E_{kl}\fois E_{ij})$. Using \eqref{eq-eps-prodmat}, one can obtain
\begin{equation*}
\delta_{jk}T(E_{il})=\eps(\varphi(i)-\varphi(j),\varphi(k)-\varphi(l))\delta_{il}T(E_{kj}).
\end{equation*}
Then, with $i\neq l$ and $j=k$, one gets $T(E_{il})=0$. On the other hand, with $i=l$ and $j=k$, one has $T(E_{ii})=\eps(\varphi(i)-\varphi(j),\varphi(j)-\varphi(i))T(E_{jj}) =\eps(\varphi(i),\varphi(i))\eps(\varphi(j),\varphi(j))T(E_{jj})$. So $\tr_\eps$ is an $\eps$-trace on $\algrA$ and there exists $\lambda\in\gK$ such that $T=\lambda\tr_\eps$.
\end{proof}

In the following, we will denote by $\ksl_\eps^\bullet(D,\gK)$ the $\eps$-Lie subalgebra of $\algrA$ of $\eps$-traceless elements.

\begin{proposition}
\label{prop-eps-derivmatrix}
All the $\eps$-derivations of $\algrA$ are inner:
\begin{equation*}
\Out_\eps^\bullet(\algA)=\algzero.
\end{equation*}
Moreover, the space of hermitean $\eps$-derivations of $\algrA$ is isomorphic to the space of antihermitean matrices.
\end{proposition}

\begin{proof}
Let $\kX\in\Der_\eps^\bullet(\algA)$ homogeneous, decomposed on the basis as
\begin{equation*}
\kX(E_{ij})=\sum_{k,l=1}^D\kX_{ij}^{kl}E_{kl},
\end{equation*}
with $\kX_{ij}^{kl}=0$ if $\varphi(k)-\varphi(l)-\varphi(i)+\varphi(j)\neq|\kX|$ where $|\kX|$ is the degree of $\kX$. Then, due to \eqref{eq-eps-defderiv}, we have, $\forall i,j,k,l$,
\begin{equation}
\kX(E_{ij}\fois E_{kl})=\kX(E_{ij})\fois E_{kl}+\eps(|\kX|,\varphi(i)-\varphi(j))E_{ij}\fois \kX(E_{kl}).\label{eq-eps-deriv1}
\end{equation}
Using \eqref{eq-eps-prodmat}, this can be written as
\begin{equation*}
\sum_{a,b=1}^D\delta_{jk}\kX_{il}^{ab}E_{ab}=\sum_{a,b=1}^D(\delta_l^b\kX_{ij}^{ak}+\eps(|\kX|,\varphi(i)-\varphi(j))\delta_i^a\kX_{kl}^{jb})E_{ab}.
\end{equation*}
For $b=l$, and by changing some indices, we obtain, $\forall i,j,k,l,a$,
\begin{equation}
\kX_{ij}^{kl}=\delta_{jl}\kX_{ia}^{ka}-\eps(|\kX|,\varphi(i)-\varphi(j))\delta_i^k\kX_{la}^{ja}.\label{eq-eps-deriv2}
\end{equation}

On the other hand, let us define $M_\kX=\sum_{k=1}^D\kX(E_{ka})\fois E_{ak}=\sum_{k,l=1}^D\kX_{ka}^{la}E_{lk} \in \algrA$ for an arbitrary $a$. Then, using \eqref{eq-eps-prodmat}, we find
\begin{equation*}
[M_\kX,E_{ij}]_\eps=\sum_{k,l=1}^D(\delta_j^l\kX_{ia}^{ka}-\eps(\varphi(j)-\varphi(l),\varphi(i)-\varphi(j))\delta_i^k\kX_{la}^{ja})E_{kl}.
\end{equation*}
Since $\kX_{la}^{ja}=0$ for $\varphi(j)-\varphi(l)\neq|\kX|$, \eqref{eq-eps-deriv2} implies that $[M_\kX,E_{ij}]_\eps=\sum_{k,l=1}^D\kX_{ij}^{kl}E_{kl}=\kX(E_{ij})$, and $\kX$ is an inner derivation generated by $M_\kX$.

The statement about hermiticity can be checked easily.
\end{proof}

\subsubsection{Differential calculus}

We then describe the differential calculus based on $\eps$-derivations for the algebra $\algrA$. In order to do that, we introduce the following algebra, which is the analog of the exterior algebra in the present framework.

\begin{definition}[$\eps$-exterior algebra]
Let $V^\bullet$ be a $\Gamma$-graded vector space on $\gK$ and $\eps$ a commutation factor on $\Gamma$. One defines the $\eps$-exterior algebra on $V^\bullet$, denoted by $\exter_{\eps}^\bullet V^\bullet$, as the tensor algebra of $V^\bullet$ quotiented by the ideal generated by
\begin{equation*}
\{\kX\otimes\kY+\eps(|\kX|,|\kY|)\kY\otimes\kX,\,\,\kX,\kY\in V^\bullet\,\text{homogeneous}\}.
\end{equation*}
\end{definition}
$\exter_{\eps}^\bullet V^\bullet$ is a $(\gZ\times\Gamma)$-graded algebra, and it is easy to show that one has the following factor decomposition
\begin{equation*}
\exter_{{}_\eps}^\bullet V^\bullet= \Big(\bigotimes_{\substack{\alpha\in\Gamma\\ \eps(\alpha,\alpha)=1}}\exter^\bullet V^\alpha\Big)\otimes\Big(\bigotimes_{\substack{\alpha\in\Gamma\\ \eps(\alpha,\alpha)=-1}}\symes^\bullet V^\alpha\Big),
\end{equation*}
where $\exter^\bullet V^\alpha$ is the exterior algebra of $V^\alpha$ and $\symes^\bullet V^\alpha$ is its symmetric algebra.

\bigskip
Let $\eps$ be a commutation factor on an abelian group $\Gamma$, $\kg^\bullet$ an $\eps$-Lie algebra for this commutation factor and $V^\bullet$ a $\Gamma$-graded vector space of representation of $\kg^\bullet$. Let us introduce the $(\gZ\times\Gamma)$-graded vector space $\Omega_\eps^{\bullet,\bullet}(\kg,V)=V^\bullet\otimes \exter_{\eps}^\bullet(\kg^\ast)^\bullet$.

\begin{proposition}
\label{prop-eps-gv}
$\Omega_\eps^{\bullet,\bullet}(\kg,V)$ is a $(\gZ\times\Gamma)$-graded differential complex for the differential of degree $(1,0)$ defined by 
\begin{multline}
\label{eq-eps-defdiff2}
\dd\omega(\kX_1,\dots,\kX_{p+1})=\sum_{m=1}^{p+1} (-1)^{m+1}g_1\ \kX_m\omega(\kX_1,\dots \omi{m} \dots,\kX_{p+1})\\
+\sum_{1\leq m<n\leq p+1}(-1)^{m+n}g_2\ \omega([\kX_m,\kX_n],\dots \omi{m} \dots \omi{n} \dots,\kX_{p+1}),
\end{multline}
$\forall\omega\in\Omega_\eps^{p,|\omega|}(\kg,V)$ and $\forall\kX_1,\dots,\kX_{p+1}\in\kg^\bullet$ homogeneous, where
\begin{align*}
g_1&=\eps(|\omega|,|\kX_m|)\prod_{a=1}^{m-1}\eps(|\kX_a|,|\kX_m|)\\
g_2&=\eps(|\kX_n|,|\kX_m|)\prod_{a=1}^{m-1}\eps(|\kX_a|,|\kX_m|)\prod_{a=1}^{n-1}\eps(|\kX_a|,|\kX_n|).
\end{align*}

Moreover, in the case where $V^\bullet$ is an $\eps$-graded algebra and $\kg^\bullet = V^\bullet_{\Lie,\eps}$ is its associated $\eps$-Lie algebra, acting by the adjoint representation on $V^\bullet$, $\Omega_\eps^{\bullet,\bullet}(\kg,V)$ is a $(\gZ\times\Gamma)$-graded differential algebra for the product,
\begin{equation}
\label{eq-eps-defprod2}
(\omega\fois\eta)(\kX_1,\dots,\kX_{p+q}) =\frac{1}{p!q!}\sum_{\sigma\in\kS_{p+q}}(-1)^{|\sigma|}g_3\ \omega(\kX_{\sigma(1)},\dots,\kX_{\sigma(p)})\fois\eta(\kX_{\sigma(p+1)},\dots,\kX_{\sigma(p+q)}),
\end{equation}
$\forall\omega\in\Omega_\eps^{p,|\omega|}(\kg,V)$, $\forall\eta\in\Omega_\eps^{q,|\eta|}(\kg,V)$, and $\forall\kX_1,\dots,\kX_{p+q}\in\kg^\bullet$ homogeneous, where
\begin{equation*}
g_3 = \prod_{m<n,\sigma(m)>\sigma(n)}\eps(|\kX_{\sigma(n)}|,|\kX_{\sigma(m)}|)\prod_{m\leq p}\eps(|\eta|,|\kX_{\sigma(m)}|).
\end{equation*}
\end{proposition}

\begin{proof}
The product \eqref{eq-eps-defprod2} and the differential \eqref{eq-eps-defdiff2} are formally the same as the product \eqref{eq-eps-defprod} and the differential \eqref{eq-eps-defdiff}.
\end{proof}

This permits one to show the following Theorem, which gives the structure of the $\eps$-derivation-based differential calculus on the elementary $\eps$-graded matrix algebra $\algrA$.

\begin{theorem}[The $\eps$-derivation-based differential calculus]
\label{thm-eps-calcdiff}
Let $\algrA=\Matr_D(\gK)$ be an $\eps$-graded matrix algebra with elementary grading, and suppose that:
\begin{equation}
\tr_\eps(\gone)=\sum_{i=1}^D \eps(\varphi(i),\varphi(i)) \neq0,\label{eq-eps-condsimpl}
\end{equation}
Then:
\begin{itemize}
\item The adjoint representation $\ad:\ksl_\eps^\bullet(D,\gK)\to\Der_\eps^\bullet(\algA)$ is an isomorphism of $\eps$-Lie algebras.
\item Let $\ad_\ast:\Omega_\eps^{\bullet,\bullet}(\ksl_\eps(D,\gK),\algA)\to\Omr^{\bullet,\bullet}(\algA)$ be the push-forward of $\ad$ and let $(E_i)$ be a basis of $\algrA$. Then, for any $\omega\in\Omega_\eps^{k,\bullet}(\ksl_\eps(D,\gK),\algA)$, one has the relation:
\begin{equation}
\ad_\ast(\omega)(\ad_{E_{i_1}},\dots,\ad_{E_{i_k}})=\omega(E_{i_1},\dots,E_{i_k}).\label{eq-eps-exprbasead}
\end{equation}
\item $\ad_\ast$ is an isomorphism of $(\gZ\times\Gamma)$-graded differential algebras.
\end{itemize}
\end{theorem}

\begin{proof}
Since $\ksl_\eps(D,\gK)\cap\caZ_\eps(\algA)=\algzero$, by Proposition~\ref{prop-eps-centermatrix} and using the condition~\eqref{eq-eps-condsimpl}, the kernel of the surjective map $\ad:\ksl_\eps^\bullet(D,\gK)\to\Der_\eps^\bullet(\algA)=\Int_\eps^\bullet(\algA)$ is trivial, so that it is an isomorphism of vector spaces. Due to the $\eps$-Jacobi identity \eqref{eq-eps-jacobi}, one has $\forall a,b\in\algrA$, $\ad_{[a,b]_\eps}=[\ad_a,\ad_b]_\eps$, which proves that $\ad:\ksl_\eps^\bullet(D,\gK)\to\Der_\eps^\bullet(\algA)$ is an isomorphism of $\eps$-Lie algebras.

Equation~\eqref{eq-eps-exprbasead} is then a trivial consequence of this isomorphism.

Let us stress that $\ksl_\eps(D,\gK)$ is only a subalgebra of the associated $\eps$-Lie algebra of $\algrA$, but the result of Proposition~\ref{prop-eps-gv} generalizes to this case, so that $\Omega_\eps^{\bullet,\bullet}(\ksl_\eps(D,\gK),\algA)$ is a $(\gZ\times\Gamma)$-graded differential algebra. It is straightforward to see that $\ad_\ast$ is a morphism of $(\gZ\times\Gamma)$-graded differential algebras. Indeed, the products and the differentials have the same formal definitions, respectively \eqref{eq-eps-defprod} and \eqref{eq-eps-defprod2}, \eqref{eq-eps-defdiff} and \eqref{eq-eps-defdiff2}, for the two complexes $\Omr^{\bullet,\bullet}(\algA)$ and $\Omega_\eps^{\bullet,\bullet}(\ksl_\eps(D,\gK),\algA)$. Equation~\eqref{eq-eps-exprbasead} shows that $\ad_\ast$ is injective. For $(E_i)$ a basis of $\algA$, adapted to the decomposition $\algrA=\gC\gone\oplus\ksl_\eps^\bullet(D,\gK)$ (with $E_0=\gone$), $(\theta^i)$ its dual basis, and $\eta\in\Omr^{k,\bullet}(\algA)$, we set $\eta(\ad_{E_{i_1}},\dots,\ad_{E_{i_k}})=\eta^j_{i_1,\dots,i_k}E_j\in\algrA$. Then $\eta=\ad_\ast(\eta^j_{i_1,\dots,i_k}E_j\theta^{i_1}\wedge\dots\wedge\theta^{i_k})$, with $\eta^j_{i_1,\dots,i_k}=0$ if one of $i_1,\dots,i_k$ is zero. Therefore, $\ad_\ast$ is surjective, and it is an isomorphism of bigraded differential algebras.
\end{proof}

Notice that there exist some examples of $\eps$-graded matrix algebras for which the condition \eqref{eq-eps-condsimpl} is not satisfied (see Example~\ref{ex-eps-supermatrix} below with $m=n$).

\subsubsection{$\eps$-connections}

Let us now describe explicitely the space of $\eps$-connections.
\begin{proposition}
The space of $\eps$-connections on $\algrA$, where $\algrA$ is considered as a module over itself, is an affine space modeled on the vector space
\begin{equation*}
\Omr^{1,0}(\algA)=\sum_{\alpha\in\Gamma} \algA^\alpha\otimes\ksl_\eps^\alpha(D,\gK)^\ast,
\end{equation*}
and involving the (trivial) $\eps$-connection $\dd$. 

Moreover, $\ad^{-1}$ can be seen as a $1$-form, and $\dd+\ad^{-1}$ defines a gauge invariant $\eps$-connection.
\end{proposition}

\begin{proof}
Let $\nabla$ be an $\eps$-connection. Define $\omega(\kX)=\nabla(\gone)(\kX)$, $\forall\kX\in\Der_\eps^\bullet(\algA)$. Then, $\omega\in\Omr^{1,0}(\algA)$ and, $\forall a\in\algrA$, $\nabla(a)=\dd a+\omega\fois a$.

On the other hand, $\ad^{-1}:\Der_\eps^\bullet(\algA)\to\ksl_\eps^\bullet(D,\gK)\subset\algrA$ is a $1$-form of degree $0$. Since all $\eps$-derivations of $\algrA$ are inner, one has, $\forall \kX\in\Der_\eps^\bullet(\algA)$, $\kX(a)=[\ad^{-1}(\kX),a]_\eps$. Then, the gauge invariance of the connection $\dd+\ad^{-1}$ can be proved in a similar way as in Proposition \ref{prop-gauge-conninv} (see also Example \ref{ex-gnc-matrix}).
\end{proof}

The noncommutative $1$-form $\ad^{-1}$ defined here is the exact analog of the noncommutative $1$-form $i\theta$ defined in the context of the derivation-based differential calculus on the matrix algebra (see subsection \ref{subsec-gnc-diffcalc}), and it gives rise also to a gauge invariant connection. This shows that some of the results obtained in the non-graded case remain valid in this new context. But one has to be careful that the condition~\eqref{eq-eps-condsimpl} has to be fulfilled.

\subsubsection{Concrete examples of elementary $\eps$-graded matrix algebras}

We give here concrete example of $\eps$-graded complex matrix algebras in the case of an abelian group $\Gamma$, freely generated by a finite number of generators $\{e_r\}_{r\in I}$, and when $\eps$ is a commutation factor on $\Gamma$ over $\gK=\gC$.

We first recall the definitions given in \cite{Rittenberg:1978mr} of color algebras and superalgebras. If, $\forall r\in I$, $\eps(e_r,e_r)=1$, then the $\eps$-graded algebra $\algrA$ is called a {\it color algebra}. Otherwise, it is a {\it color superalgebra}.

In the following examples, we will consider three gradings: the trivial case $\Gamma=\algzero$, the usual case $\Gamma=\gZ_2$, and a third case $\Gamma=\gZ_2\times\gZ_2$. Proposition~\ref{prop-eps-factcomm} and Example~\ref{ex-eps-factcomm} determine the possible commutation factors for theses groups.
\begin{itemize}
\item For $\Gamma=\gZ_2$, the most general commutation factor is given by: $\forall i,j\in\gZ_2$,
\begin{equation*}
\eps(i,j)=\eps_1^{ij},\quad\text{where }\eps_1\in\{-1,1\}.
\end{equation*}
\item For $\Gamma=\gZ_2\times\gZ_2$, $\forall i=(i_1,i_2)\in\Gamma$ and $\forall j=(j_1,j_2)\in\Gamma$,
\begin{equation*}
\eps(i,j)=\eps_1^{i_1j_1}\eps_2^{i_2j_2}\eps_3^{i_1j_2-i_2j_1},\quad\text{where }\eps_1,\eps_2,\eps_3\in\{-1,1\}.
\end{equation*}
\end{itemize}

\begin{example}[$\Gamma=\algzero$]
\label{ex-eps-nogrmatrix}
We consider here the trivial grading $\Gamma=\algzero$ on the complex matrix algebra $\algA=\Matr(n)=\Matr_n$, so that the commutation factor is also trivial: $\eps(i,j)=1$. The $\eps$-commutator $[-,-]_\eps$ and the trace $\tr_\eps$ are the usual non-graded ones for matrices. The $\eps$-center and the $\eps$-derivations of this $\eps$-graded algebra are given by: $\caZ_\eps(\algA)=\gC\gone$ and $\Der_\eps(\algA)=\Int_\eps(\algA)=\ksl_n$, the usual Lie algebra of traceless matrices. The $\eps$-derivation-based differential calculus coincides with the (usual) derivation-based differential calculus studied in \cite{DuboisViolette:1988cr, DuboisViolette:1988ir} (see Example \ref{ex-gnc-matrix}):
\begin{equation*}
\underline{\Omega}^\bullet_{\eps}(\Matr(n))=\underline{\Omega}^\bullet_{\Der}(\Matr(n))\approx\Matr(n)\otimes\Big(\exter^\bullet{\ksl_n}^\ast\Big)
\end{equation*}
It is finite-dimensional and its cohomology is:
\begin{equation*}
H^\bullet(\Omr(\Matr(n)),\dd)=H^\bullet(\ksl_n)=\mathcal{I}(\exter^\bullet\ksl_n^\ast),
\end{equation*}
the algebra of invariant elements for the natural Lie derivative. 
\end{example}

\begin{example}[$\Gamma=\gZ_2$]
\label{ex-eps-supermatrix}
Consider now the case $\Gamma=\gZ_2$, with the usual commutation factor $\eps(i,j)=(-1)^{ij}$, and the matrix superalgebra $\algrA=\Matr(m,n)$. This superalgebra is represented by $(m+n)\times(m+n)$ matrices:
\begin{equation*}
M=\begin{pmatrix} M_{11}&M_{12} \\ M_{21}& M_{22} \end{pmatrix}\in\algrA,
\end{equation*}
where $M_{11}$, $M_{12}$, $M_{21}$ and $M_{22}$ are respectively $m\times m$, $m\times n$, $n\times m$ and $n\times n$ (complex) matrices. The $\gZ_2$-grading is defined such that $M_{11}$ and $M_{22}$ correspond to degree $0\in\gZ_2$, whereas $M_{12}$ and $M_{21}$ are in degree $1\in\gZ_2$, so that this grading is elementary. Using Proposition~\ref{prop-eps-tracematrixalgebra}, we find that
\begin{equation*}
\tr_\eps(M)=\tr(M_{11})-\tr(M_{22}),
\end{equation*}
is an $\eps$-trace, $\caZ_\eps^\bullet(\algA)=\gC\gone$ and $\Der_\eps^\bullet(\algA)=\Int_\eps^\bullet(\algA)$. 

Notice that when $m=n$, one has $\tr_\eps(\gone)=0$, so that condition \eqref{eq-eps-condsimpl} is not satisfied.

If $m \neq n$, one can suppose, by convention, that $m>n$. In that case, $\tr_\eps(\gone)\neq0$, and using Theorem~\ref{thm-eps-calcdiff}, one gets 
\begin{equation*}
\Der_\eps^\bullet(\algA)=\ksl^\bullet_\eps(m,n)=\ksl^0_\eps(m,n)\oplus\ksl^1_\eps(m,n),
\end{equation*}
and the associated differential calculus based on these superderivations is given by:
\begin{equation}
\Omr^{\bullet,\bullet}(\Matr(m,n))\approx\Matr^{\bullet}(m,n)\otimes\Big(\exter^\bullet\ksl^0_\eps(m,n)^\ast\Big) \otimes\Big(\symes^\bullet\ksl^1_\eps(m,n)^\ast\Big).\label{eq-eps-calcdiffsupermatrix}
\end{equation}
Note that it involves the symmetric algebra of the odd part of $\ksl^\bullet_\eps(m,n)^\ast$, which means that $\Omr^{\bullet,\bullet}(\Matr(m,n))$ is infinite dimensional as soon as $n>0$ (remember that $m>n$), even if $\ksl_\eps^\bullet(m,n)$ is finite dimensional. This is a key difference with the non-graded case (see Example~\ref{ex-eps-nogrmatrix}). The cohomology of \eqref{eq-eps-calcdiffsupermatrix} has been computed in \cite{Grosse:1999} and is given by:
\begin{equation*}
H^{\bullet,\bullet}(\Omr(\Matr(m,n)),\dd)=H^{\bullet,0}(\Omr(\Matr(m,n)),\dd)=H^\bullet(\ksl_m),
\end{equation*}
This is exactly the cohomology of the (non-graded) Lie algebra $\ksl_m$.
\end{example}

\begin{example}[$\Gamma=\gZ_2\times\gZ_2$]
\label{ex-eps-coloralgebra}
Let us now consider the case of $\Gamma=\gZ_2\times\gZ_2$ gradings, with the commutation factor $\eps(i,j)=(-1)^{i_1j_2+i_2j_1}$. Let $\algrA=\Matr(m,n,r,s)$ be the $\eps$-graded algebra of $(m+n+r+s)\times(m+n+r+s)$ matrices defined as follow: any element in $\algrA$ is written as:
\begin{equation}
M=\begin{pmatrix} M_{11}&M_{12}&M_{13}&M_{14} \\ M_{21}& M_{22}&M_{23}&M_{24} \\ M_{31}&M_{32}&M_{33}&M_{34} \\ M_{41}&M_{42}&M_{43}&M_{44} \end{pmatrix}\in\algrA,\label{eq-eps-colormatrix}
\end{equation}
where $M_{ij}$ are rectangular matrices. The grading is such that $M_{11}$, $M_{22}$, $M_{33}$ and $M_{44}$ correspond to degree $(0,0)\in\Gamma$; $M_{12}$, $M_{21}$, $M_{34}$ and $M_{43}$ correspond to degree $(1,0)$; $M_{13}$, $M_{24}$, $M_{31}$ and $M_{42}$ correspond to degree $(0,1)$; $M_{14}$, $M_{23}$, $M_{32}$ and $M_{41}$ correspond to degree $(1,1)$. This is an elementary grading, $\algrA$ is a color algebra in the sense of \cite{Rittenberg:1978mr}, and is therefore a less trivial example of $\eps$-graded algebra than the usual matrix algebra or the super matrix algebra described in Examples~\ref{ex-eps-nogrmatrix} and \ref{ex-eps-supermatrix}. 

\begin{equation*}
\tr_\eps(M)=\tr(M_{11})+\tr(M_{22})+\tr(M_{33})+\tr(M_{44})=\tr(M),
\end{equation*}
is an $\eps$-trace, and $\caZ_\eps^\bullet(\algA)=\gC\gone$ and $\Der_\eps^\bullet(\algA)=\Int_\eps^\bullet(\algA)=\ksl^\bullet_\eps(m,n,r,s)$.
\end{example}

\begin{example}[$\Gamma=\gZ_2\times\gZ_2$]
Consider the same grading group $\Gamma$ on the same algebra $\algrA$ as in Example~\ref{ex-eps-coloralgebra}, but with a different commutation factor: $\eps(i,j)=(-1)^{i_1j_1+i_2j_2}$. $\algrA$ is then a color superalgebra but not a color algebra (here $\eps((1,0),(1,0))=-1$). General results lead us to the $\eps$-trace,
\begin{equation*}
\tr_\eps(M)=\tr(M_{11})-\tr(M_{22})-\tr(M_{33})+\tr(M_{44}),
\end{equation*}
and one has $\caZ_\eps^\bullet(\algA)=\gC\gone$, $\Der_\eps^\bullet(\algA)=\Int_\eps^\bullet(\algA)$. If $m+s\neq n+r$, one gets $\Der_\eps^\bullet(\algA)=\ksl^\bullet_\eps(m,n,r,s)$.
\end{example}

The explicit computation of the commutators $[-,-]_\eps$ for the $\eps$-Lie algebras of $\eps$-derivations are not given here because they give rise to cumbersome expressions. Nevertheless, let us mention that they are different for the four above cases, and therefore the $\eps$-Lie algebras of $\eps$-derivations, $\Der_\eps^\bullet(\algA)$, are different for these four examples.

\subsection{\texorpdfstring{$\eps$-graded matrix algebras with fine grading}{epsilon-graded matrix algebras with fine grading}}
\label{subsec-eps-exmatrixfine}

In this subsection, we study the case of a fine-grading for the matrix algebra $\algrA$. The results can be found in \cite{deGoursac:2008bd}.

\begin{definition}[Fine grading]
Let $\algrA=\Matr_D(\gK)$ be the matrix algebra on $\gK$, graded by an abelian group $\Gamma$.

The grading \eqref{eq-eps-gradingmatrix} of $\algrA$ is called {\it fine} if $\forall\alpha\in\Gamma$, $\dim_\gK(\algA^\alpha)\leq 1$. Then, we define the support of the grading:
\begin{equation*}
 \Supp(\algrA)=\{\alpha\in\Gamma,\, \algA^\alpha\neq\algzero\}.
\end{equation*}
\end{definition}

Let $\Gamma$ be an abelian group, $\gK$ an involutive algebraically closed field, and $\algA=\Matr_D(\gK)$ the algebra of $D\times D$ matrices on the field $\gK$, such that $\algrA$ is a fine $\Gamma$-graded algebra. Let $(e_\alpha)_{\alpha\in\Supp(\algrA)}$ be a homogeneous basis of $\algrA$.

\begin{proposition}
With the above hypotheses on $\algrA$, $\algrA$ is a graded division algebra, namely all non-zero homogeneous elements of $\algrA$ are invertible. Moreover, $\Supp(\algrA)$ is a subgroup of $\Gamma$.

The fine grading of $\algrA$ is determined by the choice of the basis $(e_\alpha)$ and the factor set $\sigma:\Supp(\algrA)\times\Supp(\algrA)\to\gK^\ast$ defined by: $\forall\alpha,\beta\in\Supp(\algrA)$,
\begin{equation*}
e_\alpha\fois e_\beta=\sigma(\alpha,\beta)e_{\alpha+\beta}.
\end{equation*}
Furthermore, $\algrA$ is a fine $\eps_\sigma$-graded commutative algebra, where $\eps_\sigma$ is defined in \eqref{eq-eps-multcomm}.
\end{proposition}
\begin{proof}
The proof of the first property is in \cite{Bahturin:2002}. It uses the fact that $\algrA$, as an associative algebra, does not contain any proper ideal.

$\forall\alpha,\beta\in\Supp(\algrA)$, $e_\alpha\fois e_\beta$ is proportional to $e_{\alpha+\beta}$ and is different from $0$ because of the above property. This defines $\sigma(\alpha,\beta)$. Then, $\sigma$ is a factor set (see Definition \ref{def-eps-multiplier}) since $\algrA$ is associative. For $\alpha,\beta\in\Supp(\algrA)$, one has:
\begin{equation*}
e_\alpha\fois e_\beta=\sigma(\alpha,\beta)e_{\alpha+\beta}=\eps_\sigma(\alpha,\beta)e_\beta\fois e_\alpha,
\end{equation*}
with $\eps_\sigma(\alpha,\beta)=\sigma(\alpha,\beta)\sigma(\beta,\alpha)^{-1}$ (see \eqref{eq-eps-multcomm}).
\end{proof}
Note that $e_0=\sigma(0,0)\gone$. One can notice that the Clifford algebras (see Example \ref{ex-eps-cliffordalg}) are typical examples of fine-graded matrix algebras.

The algebra $\algrA=\Matr_D(\gK)$ has therefore a natural commutation factor $\eps_\sigma$, and one can ask about the properties of $\algrA$ if it is endowed with another general commutation factor $\eps$ on $\Gamma$. We have to ``compare'' $\eps$ with $\eps_\sigma$.

\begin{proposition}
Let $\eps_1$ and $\eps_2$ be two commutations factors on $\Gamma$ over $\gK$. We denote
\begin{equation*}
\Gamma_{\eps_1,\eps_2}=\{i\in\Gamma,\,\forall j\in\Gamma,\,\eps_1(i,j)=\eps_2(i,j)\}.
\end{equation*}
$\Gamma_{\eps_1,\eps_2}$ is a subgroup of $\Gamma$ compatible with the signature decomposition, that is: $\forall i\in\gZ_2$,
\begin{equation*}
\Gamma_{\eps_1,\eps_2}\cap\Gamma_{\eps_1}^i=\Gamma_{\eps_1,\eps_2}\cap\Gamma_{\eps_2}^i.
\end{equation*}
\end{proposition}

\begin{proposition}[$\eps$-center]
\label{prop-eps-centfine}
The $\eps$-center of $\algrA$, whose fine grading is associated to the factor set $\sigma$, is given by:
\begin{equation*}
\caZ_\eps^\bullet(\algA)=\bigoplus_{\alpha\in\Gamma_{\eps,\eps_\sigma}}\algA^\alpha.
\end{equation*}
\end{proposition}
\begin{proof}
Suppose that there exists $\alpha\in\Supp(\algrA)$ such that $e_\alpha\in\caZ^\bullet_\eps(\algA)$. Then, $\forall\beta\in\Supp(\algrA)$,
\begin{equation*}
[e_\alpha,e_\beta]_\eps=\sigma(\alpha,\beta)(1-(\eps\eps_\sigma^{-1})(\alpha,\beta))e_{\alpha+\beta}.
\end{equation*}
$\forall\beta$, $[e_\alpha,e_\beta]_\eps=0$ $\Leftrightarrow$ $\forall\beta$, $\eps(\alpha,\beta)=\eps_\sigma(\alpha,\beta)$, that is $\alpha\in\Gamma_{\eps,\eps_\sigma}$.
\end{proof}

Let us also define the set (potentially empty):
\begin{equation*}
R_{\eps_1,\eps_2}=\{i\in\Gamma,\,\forall j\in\Gamma,\, \eps_1(i-j,j)=\eps_2(i-j,j)\},
\end{equation*}
for two commutation factors $\eps_1$ and $\eps_2$ on $\Gamma$ over $\gK$.

\begin{proposition}
This set satisfies:
\begin{align*}
\forall i\in R_{\eps_1,\eps_2},&\quad -i\in R_{\eps_1,\eps_2}.\\
\forall i,j\in R_{\eps_1,\eps_2},&\quad i+j\in\Gamma_{\eps_1,\eps_2}.
\end{align*}
Moreover, if $\psi_{\eps_1}=\psi_{\eps_2}$, then $R_{\eps_1,\eps_2}=\Gamma_{\eps_1,\eps_2}$, else $R_{\eps_1,\eps_2}\cap \Gamma_{\eps_1,\eps_2}=\emptyset$, where $\psi_\eps$ is the signature function of $\eps$, defined in subsection \ref{subsec-eps-commfact}.
\end{proposition}

\begin{proposition}[$\eps$-traces]
The $\eps$-traces on $\algrA$ are the linear maps $\algA\to\gK$ vanishing outside of
\begin{equation*}
\bigoplus_{\alpha\in R_{\eps,\eps_\sigma}}\algA^\alpha.
\end{equation*}
\end{proposition}
\begin{proof}
Let $T:\algrA\to\gK$ be an $\eps$-trace. $\forall\alpha,\beta\in\Supp(\algrA)$, we have $T([e_\alpha,e_\beta]_\eps)=0$. Then, from the proof of Proposition \ref{prop-eps-centfine}, one obtains $(1-\eps\eps_\sigma^{-1}(\alpha,\beta))T(e_{\alpha+\beta})=0$. After a change of variables $\alpha\to\alpha-\beta$,
\begin{equation*}
\forall\alpha,\beta\in\Supp(\algrA,\quad (1-\eps\eps_\sigma^{-1}(\alpha-\beta,\beta))T(e_\alpha)=0.
\end{equation*}
And then, $\forall\alpha\in\Supp(\algrA)$, ($T(e_\alpha)=0$ or $\alpha\in R_{\eps,\eps_\sigma}$).
\end{proof}

For any homogeneous $\eps$-derivation $\kX$ of $\algrA$, we define its coordinates $(x_\alpha)_{\alpha\in\Supp(\algrA)}$ as: $\forall\alpha\in\Supp(\algrA)$,
\begin{equation*}
\kX(e_\alpha)=\sigma(|\kX|,\alpha)\ x_\alpha\ e_{\alpha+|\kX|}.
\end{equation*}

\begin{theorem}[$\eps$-derivations]
The coordinates of a homogeneous $\eps$-derivation $\kX$ of $\algrA$ satisfy: $\forall\alpha,\beta\in\Supp(\algrA)$,
\begin{equation*}
x_{\alpha+\beta}=x_\alpha+(\eps\eps_\sigma^{-1})(|\kX|,\alpha)x_\beta.
\end{equation*}
Moreover, $\kX$ is inner $\Leftrightarrow$ $x_\alpha$ is proportional (independently of $\alpha$) to $1-(\eps\eps_\sigma^{-1})(|\kX|,\alpha)$.

The exact sequence of $\eps$-Lie algebras and $\caZ_\eps^\bullet(\algA)$-modules is canonically split:
\begin{gather*}
\xymatrix@1@C=25pt{{\algzero} \ar[r] & {\Int^\bullet_\eps(\algA)} \ar[r] & {\Der^\bullet_\eps(\algA)} \ar[r] & {\Out^\bullet_\eps(\algA)} \ar[r] & {\algzero}}
\end{gather*}
 and there are only two possibilities for $\kX$:
\begin{enumerate}
\item If $|\kX|\in\Gamma_{\eps,\eps_\sigma}$, $\kX$ is outer and given by a group morphism of $\Supp(\algrA)$ into $\gK$.
\item Otherwise, $\kX$ is inner.
\end{enumerate}
\end{theorem}
\begin{proof}
\begin{itemize}
\item Let $\kX$ be an homogeneous $\eps$-derivation of $\algrA$. The defining relation \eqref{eq-eps-defderiv} of $\eps$-derivations: $\forall\alpha,\beta\in\Supp(\algrA)$,
\begin{equation*}
\kX(e_\alpha\fois e_\beta)=\kX(e_\alpha)\fois e_\beta+\eps(|\kX|,\alpha)e_\alpha\fois\kX(e_\beta),
\end{equation*}
can be reexpressed in terms of the coordinates of $\kX$:
\begin{equation*}
\sigma(|\kX|,\alpha+\beta)x_{\alpha+\beta}\sigma(\alpha,\beta)=\sigma(|\kX|,\alpha)x_\alpha\sigma(\alpha+|\kX|,\beta)+\eps(|\kX|,\alpha)\sigma(|\kX|,\beta)x_\beta\sigma(\alpha,\beta+|\kX|).
\end{equation*}
Thanks to \eqref{eq-eps-deffactorset} and \eqref{eq-eps-multcomm}, one obtains
\begin{equation*}
x_{\alpha+\beta}=x_\alpha+(\eps\eps_\sigma^{-1})(|\kX|,\alpha)x_\beta.
\end{equation*}
\item If $\kX$ is inner, then there exists $\lambda\in\gK$ such that $\forall\alpha\in\Supp(\algrA)$,
\begin{equation*}
\kX(e_\alpha)=\lambda[e_{|\kX|},e_\alpha]_\eps=\lambda\sigma(|\kX|,\alpha)(1-\eps\eps_\sigma^{-1}(|\kX|,\alpha))e_{|\kX|+\alpha}.
\end{equation*}
Due to the definition of the coordinates $(x_\alpha)$ of $\kX$, one has
\begin{equation*}
x_\alpha=\lambda(1-\eps\eps_\sigma^{-1}(|\kX|,\alpha)).
\end{equation*}
The converse is straightforward.
\item If $|\kX|\in\Gamma_{\eps,\eps_\sigma}$, $x_{\alpha+\beta}=x_\alpha+x_\beta$ and $x:\Supp(\algrA)\to\gK$ is a group morphism. Moreover, the only possible inner derivation is $0$. Consequently, $\kX$ can be seen as an outer $\eps$-derivation.
\item If $|\kX|\notin\Gamma_{\eps,\eps_\sigma}$, there exists $\alpha\in\Supp(\algrA)$ such that $\eps\eps_\sigma^{-1}(|\kX|,\alpha)\neq1$. Since $\forall\alpha,\beta\in\Supp(\algrA)$,
\begin{equation*}
x_\alpha+\eps\eps_\sigma^{-1}(|\kX|,\alpha)x_\beta=x_{\alpha+\beta}=x_\beta+\eps\eps_\sigma^{-1}(|\kX|,\beta)x_\alpha,
\end{equation*}
one concludes that $\forall\beta\in\Supp(\algrA)$,
\begin{equation*}
x_\beta=\frac{1-\eps\eps_\sigma^{-1}(|\kX|,\beta)}{1-\eps\eps_\sigma^{-1}(|\kX|,\alpha)}x_\alpha,
\end{equation*}
and, by the above property of inner $\eps$-derivations, $\kX$ is inner. The grading, and more precisely the belonging of the grading to $\Gamma_{\eps,\eps_\sigma}$ or not, provides a splitting of the $\eps$-derivations exact sequence.
\end{itemize}
\end{proof}

\begin{corollary}
\label{cor-eps-fineresults}
In the case $\eps=\eps_\sigma$, one has
\begin{align*}
\caZ_{\eps_\sigma}^\bullet(\algA)&=\algrA.\\
\Der_{\eps_\sigma}^\bullet(\algA)&=\Out_{\eps_\sigma}^\bullet(\algA).
\end{align*}
Moreover, the $\eps_\sigma$-traces of $\algrA$ are the linear maps $\algA\to\gK$.
\end{corollary}
\begin{proof}
Indeed, $\Gamma_{\eps_\sigma,\eps_\sigma}=R_{\eps_\sigma,\eps_\sigma}=\Gamma$. Note that $\forall\alpha,\beta\in\Supp(\algrA)$, $[e_\alpha,e_\beta]_{\eps_\sigma}=0$.
\end{proof}

We notice that the results of Corollary \ref{cor-eps-fineresults} are very different from the one obtained in subsection \ref{subsec-eps-exmatrixelem}, namely Proposition \ref{prop-eps-centermatrix} and Proposition \ref{prop-eps-derivmatrix}, which are close to the results of the non-graded case on the matrices (see subsection \ref{subsec-gnc-diffcalc}). This shows that the extension to the $\eps$-graded case of the framework of derivations is non-trivial and can provide interesting examples.

Note that the dimension has not been used for the moment in this subsection, so that the latter results remain true for any fine-graded division algebra. If we take into account that $\algrA$ is finite dimensional, one obtains:
\begin{corollary}
For $\algrA=\Matr_D(\gC)$ with fine grading, and $D<\infty$,
\begin{equation*}
\Out_\eps^\bullet(\algA)=\algzero.
\end{equation*}
\end{corollary}
\begin{proof}
If $\Supp(\algrA)$ is finite, the only group morphism $\Supp(\algrA)\to\gC$ is trivial.
\end{proof}

\begin{example}
\begin{itemize}
\item For $\Gamma=\algzero$, the only possibility of fine-graded complex matrix algebra is $\algrA=\gC$.
\item For $\Gamma=\gZ_2$, there is no fine-graded matrix algebra $\algrA$ such that $\Supp(\algrA)=\gZ_2$.
\end{itemize}
\end{example}

\begin{example}
For $\Gamma=\gZ_2\times\gZ_2$, the only possibility of fine-graded complex matrix algebra $\algrA$ such that $\Supp(\algrA)=\Gamma$, is $\algrA=\Matr_2$, the two by two complex matrix algebra.
\begin{itemize}
\item $\algA^{(0,0)}=\gC\gone$, and, up to a permutation, $\algA^{(1,0)}=\gC\tau_1$, $\algA^{(0,1)}=\gC\tau_2$ and $\algA^{(1,1)}=\gC\tau_3$, where the $\tau_i$ are the Pauli matrices:
\begin{equation*}
\tau_1=\begin{pmatrix} 0 & 1  \\ 1 & 0 \end{pmatrix},\qquad \tau_2=\begin{pmatrix} 0 & -i  \\ i & 0 \end{pmatrix},\qquad \tau_3=\begin{pmatrix} 1 & 0  \\ 0 & -1 \end{pmatrix}.
\end{equation*}
\item The factor set $\sigma$ associated to the algebra $\algrA$ is given by:
\begin{multline*}
\sigma((1,0),(0,1))=-\sigma((0,1),(1,0))=-\sigma((1,0),(1,1))\\
=\sigma((1,1),(1,0)) =\sigma((0,1),(1,1))=-\sigma((1,1),(0,1))=i,
\end{multline*}
for the non-trivial terms. The associated proper commutation factor is then: $\forall i,j\in\Gamma$,
\begin{equation*}
\eps_\sigma(i,j)=(-1)^{i_1j_2+i_2j_1}.
\end{equation*}
\item Since $\Gamma$ is a product of cyclic groups, and $\gC$ is a field of characteristic zero, there is no non-zero group morphism $\Gamma\to\gC$.
\item If $\algrA$ is endowed with its associated commutation factor $\eps_\sigma$, it is a color algebra. Then, by Corollary \ref{cor-eps-fineresults}, one obtains:
\begin{equation*}
\caZ^\bullet_{\eps_\sigma}(\Matr_2)=\Matr_2,\qquad \Der^\bullet_{\eps_\sigma}(\Matr_2)=\Out^\bullet_{\eps_\sigma}(\Matr_2)=\algzero,
\end{equation*}
and the $\eps_\sigma$-traces are the linear maps $\Matr_2\to\gC$. Moreover, the differential calculus is trivial:
\begin{equation*}
\underline\Omega^{\bullet,\bullet}_{\eps_\sigma}(\Matr_2)=\underline\Omega^{0,\bullet}_{\eps_\sigma}(\Matr_2)=\Matr_2.
\end{equation*}
\item If $\algrA$ is endowed with the commutation factor $\eps(i,j)=(-1)^{i_1j_1+i_2j_2}$, it is a color superalgebra. Since $\Gamma_{\eps,\eps_\sigma}=\{(0,0),(1,1)\}$ and $R_{\eps,\eps_\sigma}=\{(1,0),(0,1)\}$, one has:
\begin{equation*}
\caZ_{\eps}^\bullet(\Matr_2)=\gC\gone\oplus\gC\tau_3,\qquad \Der_{\eps}^\bullet(\Matr_2)=\Int_\eps^\bullet(\Matr_2) =\gC\,\ad_{\tau_1}\oplus\gC\,\ad_{\tau_2},
\end{equation*}
and the $\eps$-traces are the linear maps $\caZ_{\eps}(\Matr_2)\to\gC$. One can also notice that $\gC\tau_1\oplus\gC\tau_2=\ksl^1(1,1)$ in the notations of Example \ref{ex-eps-supermatrix}, so that the differential calculus writes:
\begin{equation*}
\underline\Omega^{\bullet,\bullet}_{\eps}(\Matr_2)=\Matr_2\otimes_{\ksl^0(1,1)}\Big(\symes^\bullet\ksl^1(1,1)^\ast\Big).
\end{equation*}
\end{itemize}
\end{example}

Note that it has been proved in \cite{Bahturin:2002} that every graded matrix algebra can be decomposed as the tensor product of an elementary graded matrix algebra and a fine graded matrix algebra. In this subsection, we have seen that fine graded matrix algebras are naturally related to commutation factors and that the theory of comparison of commutation factors permits to characterize properties of such algebras.

\numberwithin{equation}{section}
\chapter[Renormalization of QFT]{An introduction to renormalization of QFT}
\label{cha-ren}

We present here two approaches of renormalization for scalar theories, namely the wilsonian and the BPHZ approaches and links between them, which have been often used in the literature concerning the Moyal space (see Chapter \ref{cha-moy}). Then, the algebraic renormalization of gauge theories is exposed, related to BPHZ renormalization.

\section[Wilsonian renormalization of scalar QFT]{Renormalization of scalar theories in the wilsonian approach}
\label{sec-ren-wilson}

The perturbative renormalization has been introduced by Feynman, Dyson, Schwinger and others (see for example \cite{Dyson:1949}) in order to deal with infinite values generated by quantum field theory. Then, by using the ideas of Wilson \cite{Wilson:1974} on the renormalization in statistical physics, Polchinski \cite{Polchinski:1984} gave a physical interpretation to the renormalization in particle physics, with the key notion of scale of observation of a theory.

\subsection{Scalar field theory}
\label{subsec-ren-scalar}

In this chapter, we will use the functional form of the quantum field theory (QFT). Note that the first historical approach of QFT was the canonical quantization, but the functional form of QFT is deeply related to the wilsonian approach to renormalization that we want to present here. We will therefore have to deal with notions which need precautions to define like the functional measure and the functional integral, because the space of fields is infinite-dimensional in the continuous case. But here, we will avoid this whole discussion, which is outside of our point, and consider the definitions and properties in this chapter for a finite space of positions (like a finite lattice), and so a finite-dimensional space of fields, following \cite{Salmhofer:1999}, and then, take implicitely the continuum and thermodynamical limit in order to obtain $\gR^D$ as the usual space of positions.

However, one can sometimes see in the literature the theory directly exposed on the continuous space. This approach may be not equivalent with the one presented here since the proofs and computations involved in QFT may not commute with the continuum limit. Anyway, for the perturbative theory, physicists switch the functional integral and the perturbation expansion in the coupling constant ($\lambda$ in the scalar case, see Equation \eqref{eq-ren-defpotential}), although there is no theorem allowing such a switch both in the discrete and the continuous case, since the partition function \eqref{eq-ren-defpartition} is not a well defined quantity in the variable $\lambda$ near 0 \cite{Rivasseau:1991}. We do not go further in this direction which concerns the constructive field theory.

For simplicity reasons, our notations will be related to the continuous case, but one can keep in mind that the whole formalism is done and well-defined (except perhaps the point we mentionned above) in these notations for a discrete space, and after all the computations, the continuum limit is taken. This ambiguity which simplifies the notations will however not disturb us in the understanding of renormalization, because it uses other cut-offs in the impulsions space than the ones of the discrete space of positions. We recall that this section provides a non-exhaustive introduction to wilsonian renormalization. We refer the reader to \cite{Salmhofer:1999} for a more complete exposition on this subject with the proofs.

Moreover, the continuous space of positions $\gR^D$ will be euclidean in this chapter, because the field theories further considered on the Moyal space in this work are also euclidean.

\subsubsection{Functional integration}

Let $\Gamma=\gR^D$ be the (euclidean) space of positions (in fact, $\Gamma=\eps\gZ^D/L\gZ^D$, for $L/\eps\in 2\gN^\ast$, is a finite space on which one takes the limit $L\to\infty$ and $\eps\to0$), and $\Gamma^\ast=\gR^D$ be the dual of $\Gamma$, the space of impulsions ($\Gamma^\ast=\frac{2\pi}{L}\gZ^D/\frac{2\pi}{\eps}\gZ^D$). One defines the integral on $\Gamma$ and $\Gamma^\ast$ by: for $f:\Gamma\to\gR$ a function,
\begin{align*}
\int_\Gamma \dd x\ f(x)&=\int_{\gR^D}\dd^Dx\ f(x),\\
\int_{\Gamma^\ast} \dd p\ f(p)&=\int_{\gR^D}\frac{\dd^Dp}{(2\pi)^D}\ f(p),
\end{align*}
in the continuous case. Our Fourier conventions are:
\begin{align}
\hat\phi(p)&=\int_\Gamma \dd x\ e^{-ipx}\phi(x),\quad\forall p\in\Gamma^\ast,\nonumber\\
\phi(x)&=\int_{\Gamma^\ast}\dd p\ e^{ipx}\hat\phi(p),\quad\forall x\in\Gamma.\label{eq-ren-fourierconv}
\end{align}
The scalar products on $\Gamma$ and $\Gamma^\ast$:
\begin{align*}
(\phi,\psi)_\Gamma&=\int_\Gamma\dd x\ \overline{\phi(x)}\psi(x),\\
(\hat\phi,\hat\psi)_{\Gamma^\ast}&=\int_{\Gamma^\ast}\dd p\ \overline{\hat\phi(p)}\hat\psi(p),
\end{align*}
satisfy the Parseval identity: $\forall\phi,\psi$,
\begin{equation*}
(\phi,\psi)_\Gamma=(\hat\phi,\hat\psi)_{\Gamma^\ast},
\end{equation*}
so that we will omit the index $\Gamma$ or $\Gamma^\ast$ from now on. The Laplacian takes the form $\Delta=\partial_\mu\partial_\mu$, where an implicit summation is made on $\mu$. Note that all the indices of the space $\Gamma$ are down to recall that we are in the euclidean case. The Fourier transform of $\Delta$ is $\hat\Delta(p)=-p^2$.
\medskip

The measure $\dd\phi=\prod_{x\in\Gamma}\dd\phi(x)$, where $\phi:\Gamma\to\gR$ is a field variable, is not well defined in the continuous case. For $C:\Gamma\times\Gamma\to\gR$ an invertible positive operator, whose eigenvalues are strictly positive, and satisfying $\int_\Gamma\dd y\ C(x,y)C^{-1}(y,z)=\delta(x-z)$, and for $J:\Gamma\to\gC$, one has
\begin{equation*}
\int\dd\phi\ e^{-\frac 12(\phi,C^{-1}\phi)+(J,\phi)}=\sqrt{\det(2\pi C)}e^{\frac 12(J,CJ)},
\end{equation*}
so that the measure
\begin{equation*}
\dd\mu_C(\phi)=(\det(2\pi C))^{-\frac12}\ e^{-\frac12(\phi,C^{-1}\phi)}\dd\phi,
\end{equation*}
is well-defined in the continuous case, and is called the {\it gaussian measure on the fields} with mean $0$ and covariance $C$. Note that typical field configurations $\phi$ on the support of this measure are not regular. This gaussian measure satisfies the properties:
\begin{align*}
\int\dd\mu_C(\phi)\ e^{i(J,C)}&=e^{-\frac12(J,CJ)},\\
\int\dd\mu_C(\phi)\ \phi(x)\phi(y)&=C(x,y),
\end{align*}
which means that the covariance $C$ corresponds to the propagator of the free theory.
\medskip

One can also define derivatives $\frac{\delta}{\delta\phi(x)}$ on the space of fields $\phi$ (in the discrete case, and then take the continuum limit). Some useful identities are given by:
\begin{align*}
\frac{\delta}{\delta\phi(x)}\phi(y)&=\delta(x-y),\\
\frac{\delta}{\delta\phi(x)}e^{(J,\phi)}&=\overline{J(x)},\\
\frac{\delta}{\delta\phi(x)}\int_\Gamma\dd y\ F(y,\phi(y))&=\partial_2 F(x,\phi(x)),
\end{align*}
where $\partial_2F$ means the derivative of $F$ relative to its second variable.

\subsubsection{Action of the theory}

Let us now define the {\it action} functional of the theory considered in this subsection:
\begin{equation}
S(\phi)=\frac12(\phi,(-\Delta+m_\phi^2)\phi)-V(\phi),\label{eq-ren-defaction}
\end{equation}
where the parameter $m_\phi>0$ is the mass of the theory, and the potential is (for the moment) given by
\begin{equation}
V(\phi)=-\lambda\int_\Gamma\dd x\ \phi(x)^4.\label{eq-ren-defpotential}
\end{equation}
$\lambda\geq0$ is called the coupling constant of the theory. We can then introduce some objects which depend from this action. The {\it covariance} (or {\it propagator}) of the theory is
\begin{equation}
C=(-\Delta+m_\phi^2)^{-1}.\label{eq-ren-defpropag}
\end{equation}
For $J:\Gamma\to\gR$, we denote the {\it partition function} of the theory by:
\begin{equation}
Z_\lambda(J)=\frac{1}{\sqrt{\det(2\pi C)}}\int\dd\phi\ e^{-S(\phi)+(J,\phi)}=\int\dd\mu_C(\phi)\ e^{V(\phi)+(J,\phi)},\label{eq-ren-defpartition}
\end{equation}
where $\sqrt{\det(2\pi C)}$ is just a constant normalization factor. The correlation functions, or Green functions are:
\begin{align*}
\langle\phi(x_1)\dots\phi(x_m)\rangle&=\frac{1}{Z_\lambda(0)}\int\dd\mu_C(\phi)\ e^{V(\phi)}\phi(x_1)\dots\phi(x_m)\\
&=\frac{1}{Z_\lambda(0)}\left(\frac{\delta^m}{\delta J(x_1)\dots\delta J(x_m)}Z_\lambda(J)\right)_{J=0},
\end{align*}
and the connected correlation functions:
\begin{equation*}
\langle\phi(x_1)\dots\phi(x_m)\rangle_c=\left(\frac{\delta^m}{\delta J(x_1)\dots\delta J(x_m)}\log(Z_\lambda(J))\right)_{J=0}.
\end{equation*}
The explicit expression of the propagator \eqref{eq-ren-defpropag} is:
\begin{equation*}
C(x,y)=\int_{\Gamma^\ast}\dd p\ \frac{e^{ip(x-y)}}{p^2+m_\phi^2}=(2\pi)^{-\frac D2}\left(\frac{m_\phi}{|x-y|}\right)^{\frac D2-1}K_{\frac D2-1}(m_\phi|x-y|),
\end{equation*}
for $x\neq y$, and where $K$ is a modified Bessel function. Note that $C(x,x)$ does not exist for $D\geq2$, so that some Feynman diagrams are divergent. This problem will be removed by the Wick ordering, but other divergent quantities will also appear in Feynman graphs due to the behavior of the propagator. It will be solved by the renormalization procedure, and the Wick ordering is in fact a first step of the renormalization.

\subsubsection{Wick ordering}

We do not recall here the definition and combinatoric properties of Feynman graphs (see for example \cite{Salmhofer:1999}). The {\it Wick ordering}, denoted by $\Omega_C$ because it depends from the covariance $C$, is defined on the polynomials  of $\phi(x)$ ($x\in\Gamma$) by:
\begin{align*}
\Omega_C(1)&=1,\\
\Omega_C(\phi(x_1)\dots\phi(x_m))&=(-i)^m\left(\frac{\delta^m}{\delta J(x_1)\dots\delta J(x_m)}e^{\frac12(J,CJ)+i(J,\phi)}\right)_{J=0}.
\end{align*}
$\Omega_C$ is invariant under permutations of the $x_i$'s, and if one denotes
\begin{equation*}
\Delta_C=\frac12\left(\frac{\delta}{\delta\phi},C\frac{\delta}{\delta\phi}\right)=\frac12\int_{\Gamma^2}\dd x\dd y\ \frac{\delta}{\delta\phi(x)}C(x,y)\frac{\delta}{\delta\phi(y)},
\end{equation*}
to be the Laplacian on the fields, then:
\begin{equation*}
\Omega_C(\phi(x_1)\dots\phi(x_m))=e^{-\Delta_C}\phi(x_1)\dots\phi(x_m).
\end{equation*}
A change of the potential in \eqref{eq-ren-defaction} by
\begin{equation*}
V_C(\phi)=\Omega_C(-\lambda\int_\Gamma\dd x\ \phi(x)^4),
\end{equation*}
corresponds to suppress all Feynman graphs which involve the so-called tadpole diagram. Some examples of expressions of Wick ordered polynomials:
\begin{align*}
\Omega_C&(\phi(x_1))=\phi(x_1)\\
\Omega_C&(\phi(x_1)\phi(x_2))=\phi(x_1)\phi(x_2)-C(x_1,x_2)\\
\Omega_C&(\phi(x_1)\phi(x_2)\phi(x_3))=\phi(x_1)\phi(x_2)\phi(x_3)-C(x_1,x_2)\phi(x_3)\\
&-C(x_1,x_3)\phi(x_2)-C(x_2,x_3)\phi(x_1)\\
\Omega_C&(\phi(x_1)\phi(x_2)\phi(x_3)\phi(x_4))=\phi(x_1)\phi(x_2)\phi(x_3)\phi(x_4)-C(x_1,x_2)\phi(x_3)\phi(x_4)\\
&-C(x_1,x_3)\phi(x_2)\phi(x_4)-C(x_1,x_4)\phi(x_2)\phi(x_3)-C(x_2,x_3)\phi(x_1)\phi(x_4)\\
&-C(x_2,x_4)\phi(x_1)\phi(x_3)-C(x_3,x_4)\phi(x_1)\phi(x_2)+C(x_1,x_2)C(x_3,x_4)\\
&+C(x_1,x_3)C(x_2,x_4)+C(x_1,x_4)C(x_2,x_3)
\end{align*}

\subsection{Effective action and equation of the renormalization group}
\label{subsec-ren-erg}

For more details about this approach of renormalization, we refer the reader to the ideas of Wilson \cite{Wilson:1974}, the original work of Polchinski \cite{Polchinski:1984}, and to further works \cite{Keller:1992,Salmhofer:1999,Kopper:1998}.

\subsubsection{Regularization}

To regularize the theory \eqref{eq-ren-defaction} and remove the problem of divergent Feynman graphs, we use a smooth cut-off for the propagator. Let $K\in\caC^\infty(\gR_+,[0,1])$ such that
\begin{itemize}
\item $K(t)=0$ if $t\geq4$,
\item $K(t)=1$ if $t\leq1$,
\item $-1\leq K'(t)<0$ if $t\in]1,4[$.
\end{itemize}
Then, the regularized propagator takes the following form:
\begin{equation}
C_{0\Lambda_0}(x,y)=\int_{\Gamma^\ast}\dd p\ \frac{e^{ip(x-y)}}{p^2+m_\phi^2}K\left(\frac{p^2}{\Lambda_0^2}\right),\label{eq-ren-defpropreg}
\end{equation}
where $\Lambda_0>>1$ is called an ultraviolet (UV) cut-off. For $0<\Lambda<\Lambda_0$, we define
\begin{equation*}
C_{\Lambda\Lambda_0}(x,y)=C_{0\Lambda_0}(x,y)-C_{0\Lambda}(x,y)=\int_{\Gamma^\ast}\dd p\ e^{ip(x-y)}\hat C_{\Lambda\Lambda_0}(p),
\end{equation*}
with:
\begin{equation*}
\hat C_{\Lambda\Lambda_0}(p)=\frac{1}{p^2+m_\phi^2}\Big(K\left(\frac{p^2}{\Lambda_0^2}\right)-K\left(\frac{p^2}{\Lambda^2}\right)\Big)\geq0,
\end{equation*}
so that $C_{\Lambda\Lambda_0}$ is a covariance operator. $\hat C_{\Lambda\Lambda_0}$ is vanishing outside of $[\Lambda,2\Lambda_0]$. $\Lambda$ is called an infrared (IR) cut-off.

The potential $V$ of \eqref{eq-ren-defaction} has also to be regularized in the UV cut-off $\Lambda_0$,
\begin{equation}
V^{(\Lambda_0)}(\phi)=\Omega_{C_{0\Lambda_0}}(-\lambda\int_\Gamma\dd x\ \phi(x)^4),\label{eq-ren-potreg}
\end{equation}
but we do not need an IR cut-off $\Lambda$, in order to get a well-defined partition function \eqref{eq-ren-defpartition}.

\subsubsection{Effective action}

It is now time to introduce the {\it effective action} $\caG$ by:
\begin{equation}
e^{\caG(\phi,C_{\Lambda\Lambda_0},V^{(\Lambda_0)})}=\int\dd\mu_{C_{\Lambda\Lambda_0}}(\psi)\ e^{V^{(\Lambda_0)}(\psi+\phi)}.\label{eq-ren-defeffective}
\end{equation}
Then, one obtains the following identity
\begin{equation*}
\caG(\phi,C_{\Lambda\Lambda_0},V^{(\Lambda_0)})=-\frac12(\phi,C^{-1}_{\Lambda\Lambda_0}\phi)+\log(Z_\lambda^{\Lambda\Lambda_0}(C^{-1}_{\Lambda\Lambda_0}\phi)),
\end{equation*}
where $Z^{\Lambda\Lambda_0}_\lambda$ is the partition function for the covariance $C_{\Lambda\Lambda_0}$ and the interaction potential $V^{(\Lambda_0)}$.
\medskip

A major theorem of QFT (proved in \cite{Salmhofer:1999} for example) says that $\caG(\phi,C_{\Lambda\Lambda_0},V^{(\Lambda_0)})$ is the generating functional for the connected, amputated Green functions, so that if one has the decomposition
\begin{multline}
\caG(\phi,C_{\Lambda\Lambda_0},V^{(\Lambda_0)})=\\
\sum_{r=1}^\infty\lambda^r\sum_{m=0}^{\overline m(r)}\int_{(\Gamma^\ast)^m}\prod_{k=1}^m\dd p_k\ \delta(p_1+\dots+p_m)G_{mr}^{\Lambda\Lambda_0}(p_2,\dots,p_m)\Omega_{C_{0\Lambda}}(\hat\phi(p_1)\dots\hat\phi(p_m)),\label{eq-ren-decompgreen}
\end{multline}
where $\overline m(r)$ is a number fixed by the theory, then the $G_{mr}^{\Lambda\Lambda_0}$ are the sum of the amplitudes of all connected and amputated Feynman graphs, with $m$ external lines and $r$ internal vertices, for the theory with propagator $C_{\Lambda\Lambda_0}$ and interaction $V^{(\Lambda_0)}$.

For $0<\Lambda<\Lambda'<\Lambda_0$ and $p\in\Gamma^\ast$,
\begin{equation*}
\hat C_{\Lambda\Lambda_0}(p)=\hat C_{\Lambda\Lambda'}(p)+\hat C_{\Lambda'\Lambda_0}(p),
\end{equation*}
so that one can prove \cite{Salmhofer:1999} the semi-group property of the effective action:
\begin{equation}
\caG(\phi,C_{\Lambda\Lambda_0},V^{(\Lambda_0)})=\caG(\phi,C_{\Lambda\Lambda'},\caG(\fois,C_{\Lambda'\Lambda_0},V^{(\Lambda_0)})).\label{eq-ren-semigroup}
\end{equation}
In this picture, $G_{mr}^{\Lambda'\Lambda_0}$ can be seen as a connected and amputated Green function for the propagator $C_{\Lambda'\Lambda_0}$ and the interaction $V^{(\Lambda_0)}$, but also as the vertex function for the interaction with UV cut-off $\Lambda'$. Indeed, the semi-group property identifies the Green function $G_{mr}^{\Lambda\Lambda_0}$ for the propagator $C_{\Lambda\Lambda_0}$ and the interaction $V^{(\Lambda_0)}$ with the Green function for the propagator $C_{\Lambda\Lambda'}$ and the interaction given by the vertex function $G_{m'r'}^{\Lambda'\Lambda_0}$ at the UV cut-off $\Lambda'$. In \eqref{eq-ren-decompgreen}, it is therefore natural that the covariance of the Wick ordering be $C_{0\Lambda}$ because the fields have already been integrated between $\Lambda$ and $\Lambda_0$, and this contribution is in the Green functions $G_{mr}^{\Lambda\Lambda_0}$. Note also that \eqref{eq-ren-semigroup} generates only a semi-group, even if the historical denomination is ``renormalization group''.

The interaction potential can also be decomposed into components:
\begin{equation}
V^{(\Lambda_0)}(\phi)=\sum_{r=1}^\infty\lambda^r\sum_{m=0}^{\overline m(r)}\int_{(\Gamma^\ast)^m}\prod_{k=1}^m\dd p_k\ \delta(p_1+\dots+p_m)V_{mr}^{(\Lambda_0)}(p_2,\dots,p_m)\Omega_{C_{0\Lambda}}(\hat\phi(p_1)\dots\hat\phi(p_m)),\label{eq-ren-decomppot}
\end{equation}
and its expression \eqref{eq-ren-potreg} permits to obtain its components:
\begin{equation*}
V_{mr}^{(\Lambda_0)}=-\delta_{r1}\delta_{m4}.
\end{equation*}

\subsubsection{Renormalization Group Equation}

Let us now derive the Renormalization Group Equation (RGE). By taking the derivative with respect to $\Lambda$ in Equation \eqref{eq-ren-defeffective}, one obtains the RGE in the functional form:
\begin{equation}
\partial_\Lambda\caG=(\partial_\Lambda\Delta_{\Lambda\Lambda_0})\caG+\frac12\left(\frac{\delta\caG}{\delta\phi}, (\partial_\Lambda C_{\Lambda\Lambda_0})\frac{\delta\caG}{\delta\phi}\right),\label{eq-ren-rgefunc}
\end{equation}
where $\Delta_{\Lambda\Lambda_0}=\frac12\left(\frac{\delta}{\delta\phi}, C_{\Lambda\Lambda_0}\frac{\delta}{\delta\phi}\right)$ is the functional Laplacian associated to $C_{\Lambda\Lambda_0}$ in the field space. Note that Equation \eqref{eq-ren-rgefunc} is the infinitesimal form of the semi-group property \eqref{eq-ren-semigroup}.
\medskip

The RGE can be reexpressed as:
\begin{equation}
\partial_\Lambda G_{mr}^{\Lambda\Lambda_0}(k_2,\dots,k_m)=\frac12\gS_m Q_{mr}^{\Lambda\Lambda_0}(k_2,\dots,k_m),\label{eq-ren-rgecomp}
\end{equation}
with the following notations:
\begin{align*}
&(\gS_m F)(k_2,\dots,k_m)=\frac{1}{m!}\sum_{\sigma\in\kS_m}F(k_{\sigma(2)},\dots,k_{\sigma(m)}),\\
&Q_{mr}^{\Lambda\Lambda_0}(k_2,\dots,k_m)=\int\dd\kappa_{mr}\int_{(\Gamma^\ast)^l}\Big(\prod_{i=1}^l\dd p_i \hat C_{0\Lambda}(p_i)\Big)\partial_\Lambda\hat C_{\Lambda\Lambda_0}(p_0) G_{m'r'}^{\Lambda\Lambda_0}(\underline p,\underline k')G_{m''r''}^{\Lambda\Lambda_0}(-\underline p,\underline k''),\label{eq-ren-defQ}\\
&\int\dd\kappa_{mr}(m',m'',r',r'',l)=\sum_{\substack{r',r''\geq1\\ r'+r''=r}}\,\sum_{(m',m'',l)\in\caM_{r'r''m}} m'm'' l! {\begin{pmatrix} m'-1\\ l \end{pmatrix}}{\begin{pmatrix} m''-1\\ l \end{pmatrix}},\\
&\caM_{r'r''m}=\{(m',m'',l),\ l\in\gN,\ m'\in\{1,\dots,\overline m(r')\},\nonumber\\
&\qquad m''\in\{1,\dots,\overline m(r'')\},\ m'+m''=m+2l+2\},\\
&k_1=-\sum_{j=2}^m k_j,\quad \underline k'=(k_1,\dots,k_{m'-1-l}),\quad \underline k''=(k_{m'-l},\dots,k_m),\\
&\underline p=(p_1,\dots,p_l)\in\gR^{lD},\quad p_0=-\sum_{j=1}^l p_j-\sum_{j=1}^{m'-l-1}k_j.
\end{align*}
Note that the form of \eqref{eq-ren-rgecomp} will permit to show some properties inductively on $r\in\gN^\ast$, since $r=r'+r''$ in the decomposition of graphs. This will be crucial to find boundaries.

The RGE is a dynamical system whose initial condition is
\begin{equation}
\caG(\phi,C_{\Lambda\Lambda_0},V^{(\Lambda_0)})_{|\Lambda=\Lambda_0}=V^{(\Lambda_0)},\label{eq-ren-initcondfunc}
\end{equation}
or equivalently in component form, $\forall r,m$,
\begin{equation}
G_{mr}^{\Lambda_0\Lambda_0}(k_2,\dots,k_m)=V^{(\Lambda_0)}_{mr}=-\delta_{r1}\delta_{m4}.\label{eq-ren-initcondcomp}
\end{equation}
Note also that the RGE implies:
\begin{equation*}
G_{m1}^{\Lambda\Lambda_0}(k_2,\dots,k_m)=V^{(\Lambda_0)}_{m1}=-\delta_{m4}
\end{equation*}
It can also be proved inductively in $m$ that $G_{mr}^{\Lambda\Lambda_0}=0$ whenever $m$ is odd.

Let us briefly summarize the previous discussion in one theorem.
\begin{theorem}
Let $\caG(\phi,C_{\Lambda\Lambda_0},V^{(\Lambda_0)})$ be a functional of the field $\phi$, and $G_{mr}^{\Lambda\Lambda_0}$ its components given by Equation \eqref{eq-ren-decompgreen}.

Then, the following 4 definitions of $\caG$ to be the effective action are equivalent:
\begin{itemize}
\item $e^{\caG(\phi,C_{\Lambda\Lambda_0},V^{(\Lambda_0)})}=\int\dd\mu_{C_{\Lambda\Lambda_0}}(\psi)\ e^{V^{(\Lambda_0)}(\psi+\phi)}$.
\item Each $G_{mr}^{\Lambda\Lambda_0}$ is the sum of the amplitudes of all connected and amputated Feynman graphs, with $m$ external lines and $r$ internal vertices, for the theory with propagator $C_{\Lambda\Lambda_0}$ and interaction $V^{(\Lambda_0)}$.
\item $\caG$ satisfies the dynamical system: $\partial_\Lambda\caG=(\partial_\Lambda\Delta_{\Lambda\Lambda_0})\caG+\frac12\left(\frac{\delta\caG}{\delta\phi}, (\partial_\Lambda C_{\Lambda\Lambda_0})\frac{\delta\caG}{\delta\phi}\right)$, with the initial condition: $\caG(\phi,C_{\Lambda\Lambda_0},V^{(\Lambda_0)})_{|\Lambda=\Lambda_0}=V^{(\Lambda_0)}$.
\item The $G_{mr}^{\Lambda\Lambda_0}$ satisfy the dynamical system:
\begin{equation*}
\partial_\Lambda G_{mr}^{\Lambda\Lambda_0}(k_2,\dots,k_m)=\frac12\gS_m Q_{mr}^{\Lambda\Lambda_0}(k_2,\dots,k_m),
\end{equation*}
in the above notations, and with the initial condition: $G_{mr}^{\Lambda_0\Lambda_0}(k_2,\dots,k_m)=V^{(\Lambda_0)}_{mr}$.
\end{itemize}
\end{theorem}

We have seen that the theory can be ill-defined (if $D\geq3$). It has been regularized by cut-offs $\Lambda$ and $\Lambda_0$, but we want to remove them. The semi-group property of the effective action permits to show that it is possible to remove the IR cut-off $\Lambda$. Indeed, if for $\Lambda_1>0$ there exists a finite limit $\lim_{\Lambda_0\to\infty}G_{mr}^{\Lambda_1\Lambda_0}$, then it holds also for $G_{mr}^{0\Lambda_0}$ since
\begin{equation*}
\caG(\phi,C_{0\Lambda_0},V^{(\Lambda_0)})=\caG(\phi,C_{0\Lambda_1},\caG(\fois,C_{\Lambda_1\Lambda_0},V^{(\Lambda_0)})).
\end{equation*}
The strategy to remove the UV cut-off $\Lambda_0$ consists in changing the initial conditions \eqref{eq-ren-initcondfunc} or \eqref{eq-ren-initcondcomp} by including the divergences. A theory is said to be {\it renormalizable} if one can adjust a finite number of parameters in $V^{(\Lambda_0)}$ (the initial condition) such that the $G_{mr}^{\Lambda\Lambda_0}$ converge for $\Lambda_0\to\infty$. Note that it is important to impose a finite number of parameters to be adjusted (by experiments), otherwise the theory would not be predictable. In this procedure, the RGE is unchanged, so that it gives rise to an effective action, but this new one is different from the oldest (bare) one, and it is called the renormalized effective action. The euclidean $\varphi^4$ theory in $D=4$ dimensions is renormalizable, and we will sketch the proof in the next subsection (for more details, see \cite{Salmhofer:1999}).

\subsection{\texorpdfstring{Renormalization of the usual $\varphi^4$ theory in four dimensions}{Renormalization of the usual phi-4 theory in four dimensions}}
\label{subsec-ren-renphi4}

We first expose the general strategy for any dimension $D$. The RGE permits to find boundaries for the connected (and amputated) Green functions $G_{mr}^{\Lambda\Lambda_0}$, and more generally for $D^\alpha G_{mr}^{\Lambda\Lambda_0}$, where $D^\alpha$ is the derivative for the multi-index $\alpha\in\gN^{(m-1)D}$. The terms to be renormalized can be found among those for which the boundaries are divergent when $\Lambda_0\to\infty$. Then, one changes accordingly the initial condition of the renormalization group flow for these terms in order to obtain the renormalized theory, and it can be shown that the renormalized Green functions $G_{mr}^{\Lambda\Lambda_0}$ are bounded when $\Lambda_0\to\infty$. Moreover, these quantities converge because $\lim_{\Lambda_0\to\infty}\partial_{\Lambda_0}G_{mr}^{\Lambda\Lambda_0}=0$.

\subsubsection{Boundaries for Green functions}

Let us now enter into details. Define
\begin{equation*}
A_{2\Lambda,\eta}(F)=\sup\{|F(p_2,\dots,p_m)|,\,\forall k\in\{2,\dots,m\}\, \norm p_k\norm\leq\max(2\Lambda,\eta)\},
\end{equation*}
for a bounded function $F:\gR^{(m-1)D}\to\gC$.

Our goal is to obtain boundaries for $A_{2\Lambda,\eta}(D^\alpha G_{mr}^{\Lambda\Lambda_0})$. Notice that we have imposed $\norm p_k\norm\leq 2\Lambda$ or $\norm p_k\norm\leq\eta$ because the internal momenta in \eqref{eq-ren-defQ} are integrated over the support of $C_{0\Lambda}$, that is $\norm p_k\norm\leq 2\Lambda$. Moreover, we keep the freedom for an arbitrary $\eta>0$ in order to obtain boundaries for any values of the external momenta.

Thanks to the RGE, one has:
\begin{align*}
A_{2\Lambda,\eta}(\partial_\Lambda G_{mr}^{\Lambda\Lambda_0}) &=\frac12A_{2\Lambda,\eta}(\gS_mQ_{mr}^{\Lambda\Lambda_0})\leq \frac12 A_{2\Lambda,\eta}(Q_{mr}^{\Lambda\Lambda_0})\\
&\leq\frac12\int\dd\kappa_{mr}\ \norm\partial_{\Lambda}C_{\Lambda\Lambda_0}\norm_\infty(\norm C_{0\Lambda}\norm_1)^l A_{2\Lambda,\eta}(G_{m'r'}^{\Lambda\Lambda_0})A_{2\Lambda,\eta}(G_{m''r''}^{\Lambda\Lambda_0}).
\end{align*}
It can also be proved \cite{Salmhofer:1999} that there exist constants $K_1$ and $K_2$ such that $\forall\Lambda\in[3,\Lambda_0]$,
\begin{align*}
\norm\partial_\Lambda C_{\Lambda\Lambda_0}\norm_\infty&\leq\frac{K_1}{\Lambda^3},\\
\norm C_{0,\Lambda}\norm_1&\leq K_2\log(\Lambda)\quad\text{for }D=2,\\
\norm C_{0\Lambda}\norm_1&\leq K_2\Lambda^{D-2}\quad\text{for }D\geq 3.
\end{align*}
By using the Taylor expansion
\begin{equation}
G_{mr}^{\Lambda\Lambda_0}=G_{mr}^{\Lambda_0\Lambda_0}-\int_\Lambda^{\Lambda_0}\dd s\ \partial_s G_{mr}^{s\Lambda_0},\label{eq-ren-taylor}
\end{equation}
and the initial condition $G_{mr}^{\Lambda_0\Lambda_0}=V_{mr}^{(\Lambda_0)}$, one obtains:
\begin{equation*}
A_{2\Lambda,\eta}(G_{mr}^{\Lambda\Lambda_0})\leq A_{2\Lambda,\eta}(V_{mr}^{(\Lambda_0)})+\int_\Lambda^{\Lambda_0}\dd s\ A_{2s,\eta}(\partial_s G_{mr}^{s\Lambda_0}).
\end{equation*}
For the case $D=4$ and $V^{(\Lambda_0)}$ given by \eqref{eq-ren-potreg}, it gives:
\begin{equation*}
A_{2\Lambda,\eta}(G_{mr}^{\Lambda\Lambda_0})\leq 1+\int\dd\kappa_{mr}\frac{K_1 K_2^l}{2}\int_\Lambda^{\Lambda_0}\dd s\ s^{2l-3}A_{2s,\eta}(G_{m'r'}^{s\Lambda_0})A_{2s,\eta}(G_{m''r''}^{s\Lambda_0}).
\end{equation*}

In the same way, one obtains a boundary for $A_{2\Lambda,\eta}(D^\alpha G_{mr}^{\Lambda\Lambda_0})$. By Leibniz's rule,
\begin{align*}
D^\alpha Q_{mr}^{\Lambda\Lambda_0}(\underline k)=&\int\dd\kappa_{mr}\int_{(\Gamma^\ast)^l}(\prod_{i=1}^l\dd p_iC_{0,\Lambda}(p_i))\sum_{\alpha'+\alpha''+\alpha'''=\alpha}\frac{\alpha!}{\alpha'!\alpha''!\alpha'''!}D^{\alpha'}G_{m'r'}^{\Lambda\Lambda_0}(\underline p,\underline k')\\
&D^{\alpha''}G_{m''r''}^{\Lambda\Lambda_0}(-\underline p,\underline k'')D^{\alpha'''}\partial_\Lambda C_{\Lambda\Lambda_0}(p_0).
\end{align*}
There exists a constant $K_1(\alpha)$ such that
\begin{equation*}
\norm D^\alpha\partial_\Lambda C_{\Lambda\Lambda_0}\norm_\infty\leq\frac{K_1(\alpha)}{\Lambda^{3+|\alpha|}},
\end{equation*}
and
\begin{equation*}
A_{2\Lambda,\eta}(D^\alpha G_{mr}^{\Lambda\Lambda_0})\leq\int_\Lambda^{\Lambda_0}\dd s\ A_{2s,\eta}(D^\alpha\partial_s G_{mr}^{s\Lambda_0}),
\end{equation*}
so that:
\begin{align}
A_{2\Lambda,\eta}(D^\alpha G_{mr}^{\Lambda\Lambda_0})\leq \int\dd\kappa_{mr}\int_\Lambda^{\Lambda_0}\dd s\ &\sum_{\alpha'+\alpha''+\alpha'''=\alpha}\frac{\alpha!}{\alpha'!\alpha''!\alpha'''!}\frac{K_1(\alpha''')K_2^l}{2}\ s^{2l-3-|\alpha'''|}\nonumber\\
&A_{2s,\eta}(D^{\alpha'}G_{m'r'}^{s\Lambda_0})A_{2s,\eta}(D^{\alpha''}G_{m''r''}^{s\Lambda_0}),\label{eq-ren-majorgen}
\end{align}
for the case $D=4$ and $V^{(\Lambda_0)}$ given by \eqref{eq-ren-potreg}.

Applying this formula for $(m,r)=(2,2),(4,2),(6,2)$, one finds:
\begin{align*}
A_{2\Lambda,\eta}(D^\alpha G_{22}^{\Lambda\Lambda_0})\leq& K(\alpha)\Lambda_0^{2-|\alpha|}\quad\text{for }|\alpha|\leq1,\\
A_{2\Lambda,\eta}(D^\alpha G_{22}^{\Lambda\Lambda_0})\leq& K(\alpha)\log(\Lambda_0)\quad\text{for }|\alpha|=2,\\
A_{2\Lambda,\eta}(D^\alpha G_{22}^{\Lambda\Lambda_0})\leq& K(\alpha)\Lambda^{2-|\alpha|}\quad\text{for }|\alpha|\geq3,\\
A_{2\Lambda,\eta}(D^\alpha G_{42}^{\Lambda\Lambda_0})\leq&K(\alpha)\log(\Lambda_0)\quad\text{for }\alpha=0,\\
A_{2\Lambda,\eta}(D^\alpha G_{42}^{\Lambda\Lambda_0})\leq&K(\alpha)\Lambda^{-|\alpha|}\quad\text{for }|\alpha|\geq1,\\
A_{2\Lambda,\eta}(D^\alpha G_{62}^{\Lambda\Lambda_0})\leq&K(\alpha)\Lambda^{-2-|\alpha|}\quad\text{for }|\alpha|\geq0.
\end{align*}

We see that the terms to be renormalized are the 4-points and 2-points functions. Indeed, for $m=2,4$, $A_{2\Lambda,\eta}(G_{m2}^{\Lambda\Lambda_0})$ has a divergent bound for $\Lambda_0\to\infty$. We have computed $A_{2\Lambda,\eta}(D^\alpha G_{m2}^{\Lambda\Lambda_0})$ in order to see how many counterterms each Green function needs to be renormalized. A priori, the 2-points function needs 3 counterterms ($|\alpha|=0,1,2$) and the 4-points function just one ($\alpha=0$). But, since $G_{22}^{\Lambda\Lambda_0}(k)$ is even in $k$, it needs only 2 counterterms.
\medskip

As an intermediate step, we consider a (unrenormalized) {\it truncated flow} $\overline G_{mr}^{\Lambda\Lambda_0}$ obtained from $G_{mr}^{\Lambda\Lambda_0}$ by removing the divergences we have observed above, namely by removing terms with $m'=2$ or $m''=2$, and those with $(m'=4,r'>1)$ or $(m''=4,r''>1)$ in the RGE \eqref{eq-ren-rgecomp}. Of course, this is not the flow of an effective action because the RGE have been modified, but the study of this truncated flow by power counting gives some informations about the convergence of the Green functions. For this theory, it can be shown that the power counting of the truncated flow in $D=4$ is given by:
\begin{align}
&A_{2\Lambda,\eta}(D^\alpha \overline G_{mr}^{\Lambda\Lambda_0})\leq g_{mr|\alpha|}\Lambda^{4-m-|\alpha|},\quad\text{if }m+|\alpha|\geq5,\nonumber\\
&A_{2\Lambda,\eta}(D^\alpha \overline G_{mr}^{\Lambda\Lambda_0})\leq g_{mr|\alpha|}\log(\Lambda_0),\quad\text{if }m+|\alpha|=4,\nonumber\\
&A_{2\Lambda,\eta}(D^\alpha \overline G_{mr}^{\Lambda\Lambda_0})\leq g_{mr|\alpha|}\Lambda_0^{2-|\alpha|},\quad\text{if }m=2\text{ and }|\alpha|\leq1,\label{eq-ren-powerwilson}
\end{align}
where $g_{mr|\alpha|}$ are constants. Indeed, this power counting is obvious for $r=1$, and it can be proved inductively in $r$ by using Equation \eqref{eq-ren-majorgen}. We see that the truncated Green functions $\overline G_{mr}^{\Lambda\Lambda_0}$ are bounded in $\Lambda_0$ for $m\geq6$, but it is not the case for $\overline G_{2r}^{\Lambda\Lambda_0}$ and $\overline G_{4r}^{\Lambda\Lambda_0}$.

\subsubsection{Renormalization}

Let us now proceed to the renormalization of $G_{2r}^{\Lambda\Lambda_0}$ and $G_{4r}^{\Lambda\Lambda_0}$, so that the renormalized $G_{2r}^{\Lambda\Lambda_0}$ and $G_{4r}^{\Lambda\Lambda_0}$ become bounded in $\Lambda_0$. Then, the general renormalized Green functions $G_{mr}^{\Lambda\Lambda_0}$ are just the sum of the truncated $\overline G_{mr}^{\Lambda\Lambda_0}$ and the missing terms involving the (renormalized) $G_{2r}^{\Lambda\Lambda_0}$ and $G_{4r}^{\Lambda\Lambda_0}$ of the RGE.

As we have already announced before, the renormalization procedure consists to change the initial condition \eqref{eq-ren-initcondcomp} for the problematic terms, with a certain number of counterterms (here, 3 counterterms: $a_r^{\Lambda_0}$, $b_r^{\Lambda_0}$, $c_r^{\Lambda_0}$), functions of $\Lambda_0$:
\begin{align}
&G_{mr}^{\Lambda_0\Lambda_0}=V_{mr}^{(\Lambda_0)}=0\quad\text{if }m\geq6\text{ or }m\text{ is odd},\nonumber\\
&G_{4r}^{\Lambda_0\Lambda_0}(k_1,k_2,k_3)=V_{4r}^{(\Lambda_0)}(k_1,k_2,k_3)=-1\quad\text{if }r=1,\label{eq-ren-condren0}\\
&G_{4r}^{\Lambda_0\Lambda_0}(k_1,k_2,k_3)=V_{4r}^{(\Lambda_0)}(k_1,k_2,k_3)=c_r^{\Lambda_0}\quad\text{if }r\geq2,\nonumber\\
&G_{2r}^{\Lambda_0\Lambda_0}(k)=V_{2r}^{(\Lambda_0)}(k)=a_r^{\Lambda_0}+b_r^{\Lambda_0}k^2\quad\text{if }r\geq2.\nonumber
\end{align}

It means that the renormalized potential takes the form:
\begin{equation}
V^{(\Lambda_0)}(\phi)=\Omega_{C_{0\Lambda_0}}\left(\int_\Gamma\dd x\ -\frac{Z-1}{2}(\partial_\mu\phi(x))^2-\frac{\delta m_\phi^2}{2}\phi(x)^2-(\lambda+\delta\lambda)\phi(x)^4\right), \label{eq-ren-potren}
\end{equation}
with
\begin{equation*}
-\frac{Z-1}{2}=\sum_{r=2}^\infty b_r^{\Lambda_0}\lambda^r,\qquad -\frac{\delta m_\phi^2}{2}=\sum_{r=2}^\infty a_r^{\Lambda_0}\lambda^r,\qquad -\delta\lambda=\sum_{r=2}^\infty c_r^{\Lambda_0}\lambda^r.
\end{equation*}

Note that just a finite number (three) of parameters have been modified. Moreover, the formal series for $\delta m_\phi^2$ starts from $r=2$ because the term $r=1$ is removed by the Wick ordering. To fix the choice of the counterterms $(a_r^{\Lambda_0},b_r^{\Lambda_0},c_r^{\Lambda_0})$, we impose some {\it renormalization conditions} at a certain scale $\Lambda=\Lambda_1$:
\begin{align}
&G_{4r}^{\Lambda_1\Lambda_0}(0,0,0)=-1\quad\text{if }r=1,\nonumber\\
&G_{4r}^{\Lambda_1\Lambda_0}(0,0,0)=c_r^R\quad\text{if }r\geq2,\label{eq-ren-condren1}\\
&G_{2r}^{\Lambda_1\Lambda_0}(0)=a_r^R\quad\text{if }r\geq2,\nonumber\\
&\left(\frac{\partial^2}{\partial k_\mu\partial k_\nu}G_{2r}^{\Lambda_1\Lambda_0}\right)(0)=2\delta_{\mu\nu}b_r^R\quad\text{if }r\geq2.\nonumber
\end{align}
These conditions can be thought to be given by experimental measures.
\begin{proposition}
For any sequence $(a_r^R,b_r^R,c_r^R)$ in $\gR^3$, there exists a unique sequence $(a_r^{\Lambda_0},b_r^{\Lambda_0},c_r^{\Lambda_0})$ such that the Green functions $G_{mr}^{\Lambda\Lambda_0}$ ($\forall m,r$) obtained from the condition \eqref{eq-ren-condren0} and the RGE, satisfy the condition \eqref{eq-ren-condren1} at scale $\Lambda_1$.
\end{proposition}

The relation between $(a_r^R,b_r^R,c_r^R)$ and $(a_r^{\Lambda_0},b_r^{\Lambda_0},c_r^{\Lambda_0})$ can be made explicite by computing the renormalized Green functions $G_{mr}^{\Lambda\Lambda_0}$, with Feynman graphs, for the interaction given by the renormalized potential \eqref{eq-ren-potren} (in particular involving 2-legged vertices), in terms of $(a_r^{\Lambda_0},b_r^{\Lambda_0},c_r^{\Lambda_0})$ and $\Lambda_0$. Then, the renormalization conditions \eqref{eq-ren-condren1} are imposed, and one can inverse the resulting relation to obtain $(a_r^{\Lambda_0},b_r^{\Lambda_0},c_r^{\Lambda_0})$ in terms of $(a_r^R,b_r^R,c_r^R)$ and $\Lambda_0$. By this way, the counterterms $Z$, $\delta m_\phi^2$ and $\delta\lambda$ can be reexpressed in terms of $\Lambda_0$, $(a_r^R,b_r^R,c_r^R)$ and $\lambda$.

Setting now $c_r^R=0$ ($\forall r\geq2$) means that $\lambda$ is the renormalized coupling constant at scale $\Lambda_1$. The bare constant is:
\begin{equation*}
\lambda_b=\lambda+\delta\lambda=\lambda-\sum_{r=2}^\infty c_r^{\Lambda_0}\lambda^r,
\end{equation*}
where $(c_r^{\Lambda_0})$ has been computed in terms of $\Lambda_0$, $(a_r^R,b_r^R)$. $\lambda_b$ is divergent when $\Lambda_0\to\infty$.

\subsubsection{Convergence of the renormalized Green functions}

By adding the (unrenormalized) truncated Green functions and the terms involving the renormalized $G_{2r}^{\Lambda\Lambda_0}$ and $G_{4r}^{\Lambda\Lambda_0}$ in the RGE, one obtains the renormalized Green functions, still denoted by $G_{mr}^{\Lambda\Lambda_0}$, and satisfying (\cite{Salmhofer:1999}): $\forall\alpha$ multi-index, $\forall\eta\geq1$, $\forall m,r\geq1$, there exist $g_{mr|\alpha|}>0$ and $E_{mr}\in\gN$ such that $\forall\Lambda\geq\Lambda_1\geq3$,
\begin{equation*}
A_{2\Lambda,\eta}(D^\alpha G_{mr}^{\Lambda\Lambda_0})\leq g_{mr|\alpha|}\Lambda^{4-m-|\alpha|}(\log(\Lambda))^{E_{mr}}.
\end{equation*}

By applying $\partial_{\Lambda_0}$ to the Taylor expansion \eqref{eq-ren-taylor} and using the RGE, it can also be proved that:
\begin{equation*}
A_{2\Lambda,\eta}(\partial_{\Lambda_0}G_{mr}^{\Lambda\Lambda_0})\leq h_{mr}\frac{(\log(\Lambda_0))^{E_0}}{\Lambda_0^2}.
\end{equation*}

Since the renormalized Green functions $G_{mr}^{\Lambda\Lambda_0}$ are bounded and $\partial_{\Lambda_0}G_{mr}^{\Lambda\Lambda_0}$ converge to 0 for $\Lambda_0\to\infty$, one concludes that the $G_{mr}^{\Lambda\Lambda_0}$ converge for $\Lambda_0\to\infty$. Let us denote the limit by $G_{mr}^{\Lambda}=\lim_{\Lambda_0\to\infty}G_{mr}^{\Lambda\Lambda_0}$. These functions are independent from the regularization function K of \eqref{eq-ren-defpropreg}.

Since $\partial_\Lambda C_{\Lambda\Lambda_0}(p_0)$ is independent of $\Lambda_0$, it is possible to take the limit\\ $Q_{mr}^{\Lambda}=\lim_{\Lambda_0\to\infty}Q_{mr}^{\Lambda\Lambda_0}$ in \eqref{eq-ren-defQ}. Then, the Green functions $G_{mr}^{\Lambda}$ satisfy:
\begin{equation*}
\partial_\Lambda G_{mr}^{\Lambda}=\frac12\gS_mQ_{mr}^{\Lambda},
\end{equation*}
and to the conditions:
\begin{align}
&G_{4r}^{\Lambda_1}(0,0,0)=-1\quad\text{if }r=1,\nonumber\\
&G_{4r}^{\Lambda_1}(0,0,0)=c_r^R\quad\text{if }r\geq2,\nonumber\\
&G_{2r}^{\Lambda_1}(0)=a_r^R\quad\text{if }r\geq2,\label{eq-ren-condren2}\\
&\left(\frac{\partial^2}{\partial k_\mu\partial k_\nu}G_{2r}^{\Lambda_1}\right)(0)=2\delta_{\mu\nu}b_r^R \quad\text{if }r\geq2,\nonumber\\
&G_{mr}^\Lambda=0\quad\forall\Lambda\text{ and if }m\text{ is odd},\nonumber\\
&\lim_{\Lambda\to\infty}G_{mr}^\Lambda=0\quad\text{if }m\geq6,\nonumber
\end{align}
at a certain scale $\Lambda_1$.

\section{Relation with BPHZ renormalization}
\label{sec-ren-bphz}

For more details about BPHZ renormalization and the proofs, see \cite{Piguet:1995,VignesTourneret:2006xa}. We take the same action as above \eqref{eq-ren-defaction} (without Wick ordering for the potential), namely:
\begin{equation}
S(\phi)=\int \dd^Dx\left(\frac 12(\partial_\mu\phi)^2+\frac{m_\phi^2}{2}\phi^2+\lambda\phi^4\right).\label{eq-ren-defactionbphz}
\end{equation}
Between the two approaches, wilsonian and BPHZ, there are some differences. In the present (BPHZ) framework, one expands the functional in the number $l$ of loops rather than the number $r$ of vertices of the graphs. If $m$ is the number of external legs of a connected graph, one can find the following relation between $l$ and $r$:
\begin{equation*}
l=r+1-\frac m2,
\end{equation*}
in any dimension $D$, by using the Euler characteristic of the graph. The wilsonian approach deals with the effective action $\caG$ (see Equation \eqref{eq-ren-defeffective}), decomposed in the amputed connected Green functions between two cut-offs $\Lambda$ and $\Lambda_0$. Here, one introduces, by a Legendre transformation, the {\it vertex functional} (also called effective action in the literature):
\begin{equation*}
\Gamma(\phi)=-\log(Z_\lambda(J))-(J,\phi),
\end{equation*}
where $J(x)$ is replaced by the solution $J(\phi)(x)$ of the equation
\begin{equation*}
\phi(x)=-\frac{\delta\log(Z_\lambda(J))}{\delta J(x)}.
\end{equation*}
Then, with the decomposition:
\begin{equation*}
\Gamma(\phi)=\sum_{m=0}^\infty\int_{(\Gamma^\ast)^m}\prod_{k=1}^m\dd p_k\delta(p_1+\dots+p_m)\Gamma_m(p_1,p_2,\dots,p_m) \hat\phi(p_1)\dots\hat\phi(p_m)
\end{equation*}
(without Wick ordering), it can be proved that the Green functions $\Gamma_m(p_1,p_2,\dots,p_m)$ are the sum of the amplitudes of all 1PI (connected graph, and again connected after cutting any internal line) and amputated Feynman graphs with $m$ external lines (and without cut-offs). We denote by $\Gamma_{ml}(p_1,p_2,\dots,p_m)$ the contribution of $\Gamma_m(p_1,p_2,\dots,p_m)$ for a fixed number of loops $l$. However, these quantities are often divergent, so that they have to be regularized (for example by dimensional regularization).

\subsection{Power-counting}
\label{subsec-ren-powcount}

Let $G$ be a 1PI graph with $r$ vertices, $l$ loops and $m$ external legs, and let $\caA_G$ be its amplitude. Let us define the superficial degree of divergence of the graph $G$. The amplitude takes the following form:
\begin{equation}
\caA_G(p_2,\dots,p_m)=\int_{(\Gamma^\ast)^l}\prod_{i=1}^l\dd k_i\ I_G(p_2,\dots,p_m,k_1,\dots,k_l).\label{eq-ren-amplintegr}
\end{equation}
Then, by the scale transformation $p_i\mapsto\rho p_i$ and $k_i\mapsto\rho k_i$, the amplitude behaves as:
\begin{equation*}
\caA_G^\rho\sim \rho^{d(G)}
\end{equation*}
when $\rho\to\infty$ and up to a certain coefficient. $d(G)$ is called the {\it superficial UV degree of divergence} of the graph $G$ and can be expressed as:
\begin{equation}
d(G)=D+(D-4)r+(2-D)\frac m2.\label{eq-ren-powcount}
\end{equation}

\begin{theorem}
\label{thm-ren-powcount}
(\cite{Weinberg:1960}, see also \cite{Piguet:1995})\\
For the massive $\phi^4$ theory, the integral \eqref{eq-ren-amplintegr} associated to the graph $G$ is absolutely convergent if and only if the superficial degree of divergence of $G$ and each of its 1PI subgraphs is strictly negative.
\end{theorem}

One can see with the above theorem that if $D>4$, for any $m$, there exists a $r$ such that the superficial degree of divergence is positive. Therefore all the Green functions $\Gamma_m$ are divergent, and the theory is non renormalizable. If $D<4$, only a finite number of graphs are divergent, and the theory is called super-renormalizable. Finally, if $D=4$, $d(G)=4-m$, so that an infinite number of 1PI graphs are divergent, but they belong all to $\Gamma_2$ or $\Gamma_4$. We have seen (and we will see in the following) that it is possible to remove these divergences in a finite number of constants of the Lagrangian. The theory is then said to be power-counting renormalizable. Note that the superficial degree of divergence appears in the power of $\Lambda$ or $\Lambda_0$ in Equation \eqref{eq-ren-powerwilson} (for $D=4$). One says that a graph $G$ is {\it primitively divergent} (or prim. div.) if its superficial degree of divergence satisfies $d(G)\geq 0$.

\subsection{BPHZ subtraction scheme}
\label{subsec-ren-bphzsub}

The other ingredient of renormalization is the form of the counterterms. If the theory is local, as it is the case in the usual $\phi^4$ theory (but not in the Moyal case), it is then possible to prove that the divergence of a primitively divergent graph is contained in the Taylor expansion of its amplitude, which is of the form of the terms involved in the Lagrangian.

\begin{figure}[htb]
  \centering
  \includegraphics{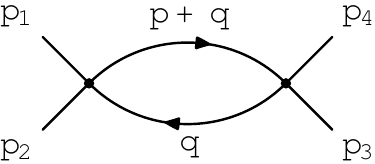}
  \caption{The bubble graph}
  \label{fig-ren-bubble}
\end{figure}

\subsubsection{Example of a primitively divergent graph}

Let us apply this on a simple example: the bubble graph $G$. If $p=p_1+p_2$ is the sum of the left external impulsions on Figure \ref{fig-ren-bubble}, the amplitude of $G$ is given by:
\begin{equation*}
\caA_G(p)=-\frac{36\lambda^2}{(2\pi)^{D}} \int \dd^Dk\frac{1}{((p+k)^2+m_\phi^2)(k^2+m_\phi^2)}.
\end{equation*}
Let us proceed to the dimensional regularization. We consider the integrals as functions of the dimension $D$. Then, we make an analytic continuation of these expressions and expand them in terms of $\eps=4-D\to 0$. After computation, it gives:
\begin{equation*}
\caA_G(p)=-\frac{36\lambda^2}{(2\pi)^{4-\eps}}\int_0^1\dd t\frac{\Gamma(\frac{\eps}{2})}{(4\pi)^{2-\frac{\eps}{2}}}(m_\phi^2+t(1-t)p^2)^{-\frac{\eps}{2}}.
\end{equation*}
Using the expansion of the $\Gamma$ function: $\Gamma(\frac{\eps}{2})=\frac{2}{\eps}-\gamma+O(\eps)$, where $\gamma$ is the Euler constant, one finds:
\begin{equation*}
\caA_G(p)=-\frac{9\lambda^2}{16\pi^6}\Big(\frac{2}{\eps}-\gamma+\log(16\pi^3)-\int_0^1\dd t\log(m_\phi^2+t(1-t)p^2)\Big)+O(\eps).
\end{equation*}
If we denote by $\tau$ the Taylor operator, evaluating the amplitude at vanishing external momenta, then:
\begin{equation}
\tau\caA_G=-\frac{36\lambda^2}{(2\pi)^{D}} \int \dd^Dk\frac{1}{(k^2+m_\phi^2)^2}=-\frac{9\lambda^2}{16\pi^6}\Big(\frac{2}{\eps}-\gamma+\log(\frac{16\pi^3}{m_\phi^2})\Big)+O(\eps)\label{eq-ren-countbubble}
\end{equation}
is divergent, local and of the type of the 4-legs vertex, while
\begin{equation*}
(1-\tau)\caA_G(p)=\frac{9\lambda^2}{16\pi^6}\int_0^1\dd t\log(1+\frac{t(1-t)p^2}{m_\phi^2}) +O(\eps)
\end{equation*}
is finite for $\eps\to 0$. The renormalized amplitude is therefore defined as
\begin{equation*}
\caA^R_G=\caA_G+c_G,
\end{equation*}
where the counterterm is $c_G=-\tau\caA_G$.

\subsubsection{Primitively divergent graphs (without prim. div. subgraphs)}

If $G$ is such a prim. div. graph with an even number $m$ of external legs, we will add the following counterterm to its amplitude:
\begin{equation}
-\tau\caA_G(p_1,\dots,p_m)=\sum_{j=0}^{d(G)}\frac{1}{j!}\frac{\dd^j}{\dd t^j}\caA_G(tp_1,\dots,tp_{m-1},p_m)|_{t=0}.\label{eq-ren-taylorexp}
\end{equation}
One can see that if $m=4$, $d(G)=0$ (like in the example of the bubble graph), and if $m=2$, $d(G)=2$ (the first order of the Taylor expansion $\tau$ vanishes by parity).

\begin{proposition}
The counterterm $c_G=-\tau\caA_G$ given by Equation \eqref{eq-ren-taylorexp} is local, of the type of the terms involved in the classical action \eqref{eq-ren-defactionbphz}, and contains the whole divergent part of the amplitude $\caA_G$.

We then write the renormalized amplitude: $\caA_G^R=\caA_G+c_G$.
\end{proposition}

\subsubsection{Graphs involving only one prim. div. subgraph}

Let $G$ be a graph with only one proper prim. div. subgraph $S\subset G$. One has to take care of the divergence coming from $S$, so that the renormalized amplitude will be defined recursively:
\begin{equation*}
\caA_G^R=\caA_G+c_S \caA_{G/S}^R+c_G,
\end{equation*}
with $G/S$: the graph $G$ where $S$ is reduced to a point. 

\subsubsection{Graphs involving disjoint families of prim. div. subgraphs}

Let $G$ be a graph with disjoint families $\caS$ of ordered (for the inclusion) prim. div. subgraphs $S\in \caS$. For each family $\caS$, the reduced graph $G/\caS$ is well-defined, and the renormalized amplitude takes this form:
\begin{equation*}
\caA_G^R=\sum_{\caS}\left(\caA_{G/\caS}\prod_{S\in\caS}c_S\right)+c_G,
\end{equation*}
where the sum is over all the families of prim. div. subgraphs (including $\caS=\emptyset$). Note that $\caA_G$ will appear for $\caS=\emptyset$.

\subsubsection{General method}

If $G$ contains non-disjoint families of ordered prim. div. subgraphs, one cannot apply the previous procedure, because some prim. div. subgraphs will be taken into account more than once. One then uses the Zimmermann's forest formula \cite{Zimmermann:1969}.

\begin{definition}
A {\it forest} $\caF$ of a graph $G$  is a set of subgraphs of $G$ such that $\forall S_1,S_2\in\caF$,
\begin{equation*}
S_1\subset S_2\text{ or } S_2\subset S_1\text{ or } S_1\cap S_2=\emptyset.
\end{equation*}
\end{definition}

With this notion, one can write the Zimmermann's formula which defines the renormalized amplitude of a graph $G$ in the general case:
\begin{equation}
\caA_G^R=\left(\sum_{\caF}\prod_{S\in\caF}(-\tau_S)\right)\caA_G,\label{eq-ren-zff}
\end{equation}
where the sum is over all forests of 1PI divergent subgraphs of $G$ (including $\caF=\emptyset$), and we have used the notation
\begin{equation*}
\tau_S\caA_G=\int_{(\Gamma^\ast)^l}\prod_{i=1}^l\dd k_i\ \tau_S I_G(p_2,\dots,p_m,k_1,\dots,k_l),
\end{equation*}
(see \eqref{eq-ren-amplintegr} for the definition of $I_G$). Indeed, $\tau_S$ can act on the $p_i$, but also on the $k_i$, depending on the subgraph $S$.

\begin{theorem}
The renormalized amplitudes $\caA_G^R$ are the graph-coefficients of the 1PI correlation functions expanded in terms of $\lambda_R$, and these correlation functions converge.
\end{theorem}
The full above procedure is called the {\it BPHZ subtraction scheme}.

\subsubsection{Renormalization conditions}

Note that terms of type $\int(\partial_\mu\phi)^2$ will also have to be absorbed in a redefinition of the bare constants, but there is no corresponding freedom for this term in the action \eqref{eq-ren-defactionbphz}. Consequently, one introduces the so-called renormalized field $\phi_R$ defined by
\begin{equation*}
\phi=Z^{\frac 12}\phi_R,
\end{equation*}
where $Z$ is a free parameter, called ``wave function normalization''. The action \eqref{eq-ren-defactionbphz} writes now:
\begin{equation*}
S(\phi)=\int \dd^Dx\left(\frac{Z}{2}(\partial_\mu\phi_R)^2+\frac{m_\phi^2Z}{2}\phi_R^2+\lambda Z^2\phi_R^4\right).
\end{equation*}
The divergences involved in terms of type $\int(\partial_\mu\phi)^2$ can now be absorbed in $Z$. Then, the vertex functional $\Gamma(\phi)$ associated to the action \eqref{eq-ren-defactionbphz} can be expressed in terms of the renormalized field $\phi_R$, and this defines the renormalized vertex functional:
\begin{equation*}
\Gamma^{(R)}(\phi_R)=\Gamma(\phi),
\end{equation*}
which is finite after BPHZ subtraction scheme. The corresponding relation for vertex functions is:
\begin{equation*}
\Gamma_m^{(R)}=Z^{\frac m2}\Gamma_m.
\end{equation*}
The BPHZ scheme permits to find the renormalized parameters $Z_R$, $m_R^2$ and $\lambda_R$ in function of the bare ones $Z$, $m_\phi^2 Z$ and $\lambda Z^2$, satisfying:
\begin{align}
&\Gamma^{(R)}_4(0,0,0,0)=\lambda_R\nonumber\\
&\Gamma^{(R)}_2(0,0)=\frac{m^2_R}{2}\label{eq-ren-condren3}\\
&\frac{\dd}{\dd p^2}\Gamma^{(R)}_2(p,-p)|_{p=0}=\frac{Z_R}{2}=\frac 12\nonumber
\end{align}
which look like very similar to \eqref{eq-ren-condren1}. To summarize the above discussion, one has:
\begin{equation*}
\Gamma(Z^{\frac 12}\phi_R,m_\phi,\lambda)=S(\phi_R,m_R,\lambda_R),
\end{equation*}
where the expression of $S(\phi,m_\phi,\lambda)$ is given by \eqref{eq-ren-defactionbphz}.
\medskip

The BPHZ renormalization scheme is not canonical in the mathematical sense, it is not unique. For example, we have subtracted $\tau\caA_G$ (see \eqref{eq-ren-countbubble}) to the bubble graph. In the dimensional renormalization scheme, one would have subtracted only the divergent part in $\eps$, that is $-\frac{9\lambda^2}{8\pi^6\eps}$, and it would also have provided a renormalization of the theory.

Moreover, the BPHZ renormalization leads to a Hopf algebra structure of the graphs \cite{Connes:1999yr,Connes:2000fe}. Indeed, by considering the algebra generated by 1PI graphs of the theory, one introduces the following coproduct:
\begin{equation*}
\Delta G=G\otimes\gone+\gone\otimes G+\sum_{S}S\otimes G/S,
\end{equation*}
for any 1PI graph $G$, and where the sum is performed over all the 1PI prim. div. subgraphs $S$ of $G$. This coproduct encodes the BPHZ procedure.

One can also see the correspondence with wilsonian renormalization because divergences of the Green functions $G_{mr}^{\Lambda\Lambda_0}(p_1,\dots,p_m)$ and $G_{mr}^{\Lambda\Lambda_0}(0,\dots,0)$ are related by Taylor expansion and the counterterms determined by $(a_r^{\Lambda_0},b_r^{\Lambda_0},c_r^{\Lambda_0})$ have to be local and of the form of the Lagrangian (see Equation \eqref{eq-ren-potren}). Note that the multi-scale analysis \cite{Rivasseau:1991} uses the tools of the BPHZ renormalization, and in particular the Zimmermann's forest formula \eqref{eq-ren-zff}, combined with ideas coming from wilsonian renormalization.

\subsection{Beta functions}
\label{subsec-ren-betafun}

For an introduction to beta functions, see for example \cite{Zinn-Justin:1996,Itzykson:1980}. Notice that it is possible to introduce a ``normalization'' mass $\mu$ in the renormalization conditions. In particular, it is a necessary task for a massless theory (see section \ref{sec-ren-gauge}):
\begin{align}
&\Gamma^{(R)}_4(p_1,p_2,p_3,p_4)|_{p_i^2=\mu^2}=\lambda_R\nonumber\\
&\Gamma^{(R)}_2(p,-p)|_{p^2=0}=0\label{eq-ren-condren4}\\
&\frac{\dd}{\dd p^2}\Gamma^{(R)}_2(p,-p)|_{p^2=\mu^2}=\frac 12\nonumber
\end{align}
where the impulsions $p_i$ satisfy also: $(p_i+p_j)^2=\frac 43\mu^2$, for $i\neq j$. This additional mass $\mu$ represents the energy scale of the measurements, like the IR cut-off $\Lambda$ in the wilsonian point of view.
\medskip

We have seen (see subsection \ref{subsec-ren-bphzsub}) that the BPHZ scheme permits one to obtain the renormalized constant $\lambda_R(\lambda,\mu)$, but also the wave function normalization $Z(\lambda,\mu)$ (after inversion). Note that they depend now from the scale $\mu$ by the conditions \eqref{eq-ren-condren4}. Then, we define the beta function of $\lambda$ as describing the response to a change of the scale $\mu$ in the renormalization conditions. The {\it beta function} and the $\eta$ function of $Z$ are given by:
\begin{equation}
\beta(\lambda_R)=\mu\frac{\partial}{\partial\mu}\lambda_R|_{\lambda},\qquad \eta(\lambda_R)=\mu\frac{\partial}{\partial\mu}\log(Z)|_{\lambda}.\label{eq-ren-betafunscal}
\end{equation}
Since the unrenormalized vertex functions do not depend from $\mu$, one derives the Renormalization Group Equation:
\begin{equation*}
\left(\mu\frac{\partial}{\partial\mu}+\beta(\lambda_R)\frac{\partial}{\partial\lambda_R}-\frac m2\eta(\lambda_R)\right)\Gamma_m^{(R)}=0,
\end{equation*}
for $m$ an even number. This equation corresponds to \eqref{eq-ren-rgecomp} in the wilsonian point of view.

For the $\phi^4$ theory, an approximation of the beta function gives: $\beta(\lambda_R)=c\lambda_R^2+\dots$, where $c>0$. This means that $\lambda_R$ diverges in a finite time in the ultraviolet sector ($\mu\to\infty$). This phenomenon is called the Landau ghost. On the contrary, if $\lambda_R\to_{\mu\to\infty}0$, one speaks about asymptotic freedom, like in the case of non-abelian gauge theories.

\section{Renormalization of gauge theories}
\label{sec-ren-gauge}

In this section, we will give an introduction to the renormalization of euclidean Yang-Mills theories, following \cite{Piguet:1995}. We also refer the reader to this reference for a more complete presentation on this subject. Yang-Mills theories are indeed important in physics since they appear among other in the standard model of particles. We will study in particular the renormalization of Yang-Mills theory in the formalism developed by Becchi, Rouet and Stora (BRS formalism), which has permitted to prove the unitarity of scattering matrix \cite{Becchi:1974,Becchi:1975}.

One considers here a pure Yang-Mills theory associated to a trivial principal fiber bundle on the basis manifold $\gR^4$ and with a simple compact Lie group $G$ as structure group. For a general Yang-Mills theory, with non-trivial fiber bundle, and topological effects, see \cite{Birmingham:1991} for example.

\subsection{Classical theory and BRS formalism}
\label{subsec-ren-brs}

\subsubsection{Classical Yang-Mills theory}

Let $(\tau_a)$ be a basis of the Lie algebra $\kg$ of $G$ satisfying:
\begin{equation*}
[\tau_a,\tau_b]=if_{abc}\tau_c,\qquad \tr(\tau_a\tau_b)=\delta_{ab},
\end{equation*}
where $f_{abc}$ are the constants structures of $\kg$. Due to the Jacobi identity, these constants satisfy:
\begin{equation*}
f_{bcd}f_{ade}+f_{ca}f_{bde}+f_{abd}f_{cde}=0.
\end{equation*}
Then, the gauge potential $A_\mu(x)$, which belongs to $\kg$, can be decomposed in $A_\mu(x)=A_\mu^a(x)\tau_a$. The gauge transformations, acting on $A_\mu(x)$, are given by:
\begin{equation}
{A_\mu}^g(x)=g(x)A_\mu(x)g^{-1}(x)+ig(x)\partial_\mu g^{-1}(x),\label{eq-ren-gaugetr}
\end{equation}
where $g:\gR^4\to G$, since the fiber bundle of the Yang-Mills theory is trivial. In an infinitesimal version, with $g(x)=e^{i\lambda(x)}$, one has:
\begin{equation*}
\delta_\lambda A_\mu(x)=\partial_\mu \lambda(x)-i[A_\mu(x),\lambda(x)]=D_\mu \lambda(x),
\end{equation*}
where $\lambda:\gR^4\to\kg$, and $D_\mu$ is the covariant derivative associated to the gauge potential $A_\mu$. The curvature of this potential $A_\mu$ writes:
\begin{equation*}
F_{\mu\nu}(x)=\partial_\mu A_\nu(x)-\partial_\nu A_\mu(x)-i[A_\mu(x),A_\nu(x)],
\end{equation*}
and transforms covariantly under a gauge transformation:
\begin{equation*}
{F_{\mu\nu}}^g(x)=g(x)F_{\mu\nu}(x)g^{-1}(x).
\end{equation*}

The most general gauge invariant and power-counting renormalizable (see section \ref{sec-ren-bphz}) corresponding to the above Yang-Mills theory is:
\begin{equation*}
S_{\text{inv}}=\int \dd^4x\tr\Big(\frac{1}{4g^2}F_{\mu\nu}F_{\mu\nu}\Big).
\end{equation*}
Notice that the euclidean metric has been implicitely used to construct this action, in the Hodge product. The power-counting of this action is the same as in the $\phi^4$ theory (see Equation \eqref{eq-ren-powcount}). Furthermore, topological terms do not appear here, since the fiber bundle considered is trivial.

\subsubsection{Ghosts and antighosts}

Because of the gauge symmetry of this action, some degrees of freedom of $A_\mu$ are unphysical, so that one has to gauge fix this action. We follow the Faddeev-Popov approach \cite{Faddeev:1967}. The functional integral writes:
\begin{equation*}
Z(0)=\int \dd A_\mu(x) e^{-S_{\text{inv}}(A)},
\end{equation*}
but one has to quotient the space of gauge potentials by the gauge group. One defines $\Delta(A)$ such that:
\begin{equation*}
\Delta(A)\int \dd g(x)\delta(\partial_\mu {A_\mu}^g(x))=1,
\end{equation*}
where $g:\gR^4\to G$ are gauge transformations. If one introduces this identity in the functional integral, one finds:
\begin{equation*}
Z(0)=\int \dd A_\mu(x) e^{-S_{\text{inv}}(A)}\int \dd g(x)\delta(\partial_\mu {A_\mu}^g(x))\Delta(A).
\end{equation*}
Since the measure $\dd A_\mu(x)$, the action $S_{\text{inv}}$ and $\Delta(A)$ are gauge invariant, one obtains after a change of variables:
\begin{equation*}
Z(0)=\int \dd g(x)\int \dd A_\mu(x) e^{-S_{\text{inv}}(A)}\delta(\partial_\mu {A_\mu}(x))\Delta(A).
\end{equation*}
The factor $\int \dd g(x)$ is exactly the volume of the gauge group, so it can be put away. For a gauge transformation $g(x)=e^{i\lambda(x)}$, one has:
\begin{equation*}
\partial_\mu {A_\mu}^g(x)=\partial_\mu A_\mu(x)+\partial^2\lambda(x)-i\partial_\mu[A_\mu(x),\lambda(x)],
\end{equation*}
at the first order in the infinitesimal transformation $\lambda(x)$. Therefore, $\Delta(A)$ writes like a determinant:
\begin{equation*}
\Delta(A)=\int \dd c(x)\dd\overline c(x) e^{\int \dd^4x \tr(\overline c(x)\partial_\mu(\partial_\mu c(x)-i[A_\mu,c(x)]))},
\end{equation*}
where $c(x)$ and $\overline c(x)$ are the $\kg$-valued anticommuting Faddeev-Popov {\it ghost} and {\it antighost}. For the reality of the action, $c(x)$ is real and $\overline c(x)$ imaginary. Then, let us modify the gauge condition in $\partial_\mu A_\mu(x)-\alpha B(x)$ in order to make appear the Lautrup-Nakanishi field $B(x)$. This field is $\kg$-valued, real and commuting. If one takes moreover a gaussian mean on the field $B(x)$, one obtains:
\begin{align*}
Z(0)=&\int \dd A_\mu(x)\int \dd B(x) e^{-\int \dd^4x\tr(\frac{\alpha}{2}B^2(x))} e^{-S_{\text{inv}}(A)}\delta(\partial_\mu A_\mu(x)-\alpha B(x))\Delta(A)\\
=&\int \dd A_\mu(x) e^{-S_{\text{inv}}(A)}e^{-\frac{1}{2\alpha}\int \dd^4x \tr((\partial_\mu A_\mu(x))^2)}\Delta(A)\\
=&\int \dd A_\mu(x)\dd c(x)\dd\overline c(x)\dd B(x) e^{-S_{\text{inv}}(A)-S_{\text{gf}}(A)},
\end{align*}
where $\alpha$ is the gauge parameter, and the gauge fixing action is given by:
\begin{equation*}
S_{\text{gf}}=\int \dd^4x\tr\Big(B\partial_\mu A_\mu-\frac{\alpha}{2}B^2-\overline c\partial_\mu(\partial_\mu c-i[A_\mu,c])\Big).
\end{equation*}

The {\it gauge condition} reads
\begin{equation}
\frac{\delta S}{\delta B}=\partial_\mu A_\mu-\alpha B.\label{eq-ren-gfcond}
\end{equation}

\subsubsection{BRS formalism and Slavnov-Taylor identity}

The BRS approach \cite{Becchi:1974,Becchi:1975} consists to introduce a new grading, the {\it ghost number} $gh$, and a corresponding differential denoted by $s$. It can be implemented by:
\begin{align}
&gh(A_\mu)=0,\qquad sA_\mu=\partial_\mu c-i[A_\mu,c]=D_\mu c, \qquad (\Rightarrow sF_{\mu\nu}=-i[F_{\mu\nu},c])\nonumber\\
&gh(c)=1,\qquad sc^a=-\frac 12 f_{bca}c^bc^c,\nonumber\\
&gh(\overline c)=-1,\qquad s\overline c=B,\nonumber\\
&gh(B)=0,\qquad sB=0.\label{eq-ren-brstransf}
\end{align}
$s$ is indeed a derivation of ghost number $1$, and nilpotent: $s^2=0$. Then, the gauge invariance of $S_{\text{inv}}$ is equivalent to the identity $sS_{\text{inv}}=0$ ($S_{\text{inv}}$ is $s$-closed). Moreover, the gauge fixing part can be expressed as a $s$-exact term:
\begin{equation*}
S_{\text{gf}}=s\int \dd^4x\tr\Big(\overline c(\partial_\mu A_\mu-\frac{\alpha}{2}B)\Big),
\end{equation*}
so that the action $S_{\text{inv}}+S_{\text{gf}}$ is BRS invariant.
\medskip

To derive functional identities, one also introduces the $\kg$-valued external fields $\rho_\mu(x)$ and $\sigma(x)$, where $\rho_\mu$ anticommutes, is imaginary and $gh(\rho_\mu)=-1$, while $\sigma$ commutes, is imaginary and $gh(\sigma)=-2$. The external part of the action can be taken as:
\begin{equation*}
S_{\text{ext}}=\int \dd^4x\tr\Big(\rho_\mu(x)sA_\mu(x)+\sigma(x)sc(x)\Big).
\end{equation*}
$\rho_\mu$ and $\sigma$ are the Batalin-Vilkovisky antifields relative to $A_\mu$ and $c$ \cite{Batalin:1981jr,Teitelboim:1992}. By denoting $S=S_{\text{inv}}+S_{\text{gf}}+S_{\text{ext}}$, the BRS invariance of the action can be reexpressed in a functional identity, called the Slavnov-Taylor identity:
\begin{equation}
\caS(S)=0,\label{eq-ren-brsinv}
\end{equation}
where the {\it Slavnov operator} $\caS$ is defined for any functional $\caF$ by:
\begin{equation*}
\caS(\caF)=\int \dd^4x\tr\Big(\frac{\delta\caF}{\delta \rho_\mu(x)}\frac{\delta\caF}{\delta A_\mu(x)}+\frac{\delta\caF}{\delta\sigma(x)}\frac{\delta\caF}{\delta c(x)}+B(x)\frac{\delta\caF}{\delta\overline c(x)}\Big).
\end{equation*}
Indeed, one has
\begin{equation*}
\caS(S)=\int \dd^4x\tr\Big((sA_\mu(x))\frac{\delta S}{\delta A_\mu(x)}+(sc(x))\frac{\delta S}{\delta c(x)}+(s\overline c(x))\frac{\delta S}{\delta\overline c(x)}\Big)=sS=0.
\end{equation*}
The action $S$ is then completely characterized by its power-counting renormalizability, its gauge fixing condition \eqref{eq-ren-gfcond}, and the Slavnov-Taylor identity \eqref{eq-ren-brsinv}. Indeed, the fields have the following mass dimensions:
\begin{align*}
&\dim(A_\mu)=1,\qquad \dim(c)=0,\qquad \dim(\overline c)=2,\\
&\dim(B)=2,\qquad \dim(\rho_\mu)=3,\qquad \dim(\sigma)=4, \qquad (\dim(\partial_\mu)=1),
\end{align*}
so that every term in the action $S$ has mass dimension $4$.

\subsection{Algebraic renormalization}
\label{subsec-ren-algren}

We suppose here that the full process of section \ref{sec-ren-bphz} has been done for the local power-counting renormalizable theory $S=S_{\text{inv}}+S_{\text{gf}}+S_{\text{ext}}$. The quantum field theory is then described by the ``renormalized'' vertex functional $\Gamma(A,c,\overline c,B,\rho,\sigma)$, in the sense that the divergent parts have been subtracted in $\Gamma$. But the subtraction procedure may have changed the symmetries initially satisfied by $S$. One has to check if $\Gamma$ satisfies also the gauge condition and the Slavnov-Taylor identity (eventually by finite subtraction of additional terms). Note that the dimensional regularization is well-adapted to the gauge theory, because this first process does not break gauge invariance in this regularization.

\begin{theorem}
Without anomalies (see further), it is possible to modify the vertex functional $\Gamma$ (order by order) such that it satisfies the gauge condition and the Slavnov-Taylor identity:
\begin{align}
\frac{\delta\Gamma}{\delta B}&=\partial_\mu A_\mu-\alpha B,\label{eq-ren-gfcondren}\\
\caS(\Gamma)&=0.\label{eq-ren-slavidren}
\end{align}
\end{theorem}
The rest of this subsection is devoted to the proof of this theorem following the algebraic renormalization approach \cite{Piguet:1995}.

\subsubsection{Gauge condition}

One defines $\Gamma^{(n)}$, the contribution of $\Gamma$ below the order $n$ of loops, and we prove recursively the gauge condition. Assuming that \eqref{eq-ren-gfcondren} holds at order $n-1$, it can be shown (by the quantum action principle, see \cite{Lowenstein:1971a,Lowenstein:1971b}) that the most general term at order $n$ is:
\begin{equation*}
\frac{\delta\Gamma^{(n)}}{\delta B^a(x)}=\partial_\mu A^a_\mu(x)-\alpha B^a(x)+\Delta^a(x),
\end{equation*}
where $\Delta^a$ is a local polynomial in the fields of dimension equal to $4-\dim(B)=2$, so that it reads:
\begin{equation*}
\Delta^a=F^a(A,c,\overline c)+\omega_{ab}B^b,
\end{equation*}
with $\dim(F^a)=2$ and $\omega_{ab}$ a parameter. Then,
\begin{equation*}
0=\frac{\delta^2\Gamma^{(n)}}{\delta B^a(x)\delta B^b(y)}-\frac{\delta^2\Gamma^{(n)}}{\delta B^b(y)\delta B^a(x)}=(\omega_{ba}-\omega_{ab})\delta(x-y).
\end{equation*}
As a consequence, $\omega_{ab}=\omega_{ba}$, and $\Delta^a(x)=\frac{\delta}{\delta B^a(x)}\widetilde\Delta$, where
\begin{equation*}
\widetilde\Delta=\int \dd^4y(B^bF^b(A,c,\overline c)+\frac 12\omega_{bc}B^bB^c).
\end{equation*}
Taking $S-\widetilde\Delta$ as the new action (note that $\widetilde\Delta$ is finite), the gauge condition writes at order $n$:
\begin{equation*}
\frac{\delta\Gamma^{(n)}}{\delta B}=\partial_\mu A_\mu-\alpha B.
\end{equation*}

\subsubsection{Ghost equation}

By the following change of variables:
\begin{equation*}
\widehat\rho_\mu=\rho_\mu+\partial_\mu\overline c,\qquad \widehat{\overline c}=\overline c,
\end{equation*}
where the derivatives are given by:
\begin{equation*}
\frac{\delta}{\delta\widehat\rho_\mu}=\frac{\delta}{\delta\rho_\mu},\qquad \frac{\delta}{\delta\widehat{\overline c}}=\frac{\delta}{\delta\overline c}+\partial_\mu\frac{\delta}{\delta\rho_\mu},
\end{equation*}
one obtains the ghost equation:
\begin{equation*}
\frac{\delta S}{\delta\widehat{\overline c}}=0,
\end{equation*}
which is a consequence of the functional identity:
\begin{equation*}
\frac{\delta \caF}{\delta\widehat{\overline c}}=\frac{\delta}{\delta B}\caS(\caF)-\caS_\caF\Big(\frac{\delta\caF}{\delta B}-\partial_\mu A_\mu\Big),
\end{equation*}
for any functional $\caF$, and where $\caS_\caF$ is the linearized Slavnov-Taylor operator:
\begin{equation*}
\caS_\caF=\int\dd^4x\tr\Big(\frac{\delta\caF}{\delta \rho_\mu(x)}\frac{\delta}{\delta A_\mu(x)}+\frac{\delta\caF}{\delta A_\mu(x)}\frac{\delta}{\delta \rho_\mu(x)}+\frac{\delta\caF}{\delta\sigma(x)}\frac{\delta}{\delta c(x)}+\frac{\delta\caF}{\delta c(x)}\frac{\delta}{\delta \sigma(x)}+B(x)\frac{\delta\caF}{\delta\overline c(x)}\Big).
\end{equation*}
For convenience, one forgets the hat over $\overline c$ (but not over $\rho_\mu$). Let us show recursively that the ghost equation writes also for $\Gamma$ as the gauge condition. Suppose that
\begin{equation*}
\frac{\delta\Gamma^{(n-1)}}{\delta\overline c}=0,
\end{equation*}
for a certain $n$, then:
\begin{equation*}
\frac{\delta\Gamma^{(n)}}{\delta\overline c^a(x)}=\Delta^a(x),
\end{equation*}
where $\Delta^a$ is a local polynomial in the fields of dimension equal to $4-\dim(\overline c)=2$, so that it reads:
\begin{equation*}
\Delta^a=F^a(A,c)+\omega_{ab}\overline c^b,
\end{equation*}
with $\dim(F^a)=2$ and $\omega_{ab}$ a parameter. Indeed, from \eqref{eq-ren-gfcondren}, it can be shown that $\Delta^a$ is independent from $B$. As above, one sees that $\Delta^a(x)=\frac{\delta\widetilde\Delta}{\delta \overline c^a(x)}$, with
\begin{equation*}
\widetilde\Delta=\int \dd^4y(\overline c^bF^b(A,c)+\frac 12\omega_{bc}\overline c^b\overline c^c),
\end{equation*}
and, by subtracting $\widetilde\Delta$ from the action $S$, one obtains: $\frac{\delta\Gamma^{(n)}}{\delta\overline c}=0$.

\subsubsection{Slavnov-Taylor identity}

Finally, the Slavnov-Taylor identity has to be satisfied by $\Gamma$. We will use the following functional relations:
\begin{align}
&\forall\caF,\quad\caS_\caF\caS(\caF)=0,\label{eq-ren-idfun1}\\
&\text{if }\caS(\caF)=0,\quad \caS_\caF\caS_\caF=0.\label{eq-ren-idfun2} 
\end{align}
Let us define $b=\caS_S$. Equation \eqref{eq-ren-idfun2} implies $b^2=0$. The action of $b$ on the fields is given by:
\begin{align*}
bA_\mu^a&=sA_\mu^a\\
bc^a&=sc^a\\
b\overline c^a&=s\overline c^a\\
bB^a&=sB^a\\
b\widehat\rho_\mu^a&=\frac{\delta S}{\delta A_\mu^a}=\frac{1}{g^2}D_\nu F_{\mu\nu}+f_{abc}\widehat\rho_\mu^c c^b\\
b\sigma^a&=\frac{\delta S}{\delta c^a}=\partial_\mu\widehat\rho_\mu^a+f_{abc}(\widehat\rho_\mu^c-\sigma^cc^b)
\end{align*}
Suppose that $\caS(\Gamma)=0$ is satisfied until the order $n-1$. Then, at order $n$, one has \cite{Piguet:1995}:
\begin{equation*}
\caS(\Gamma)=\Delta,
\end{equation*}
where $gh(\Delta)=1$ and $\dim(\Delta)=4$. By using the gauge condition and the ghost equation (satisfied by $\Gamma$), one has:
\begin{equation*}
\frac{\delta\Delta}{\delta B}=\frac{\delta\Delta}{\delta\overline c}=0.
\end{equation*}
Since $\caS_\Gamma=b$ at order $0$, and by using \eqref{eq-ren-idfun1}, one obtains $0=\caS_\Gamma\caS(\Gamma)=b\Delta$ at order $n$. This equation:
\begin{equation}
b\Delta=0\label{eq-ren-conscond}
\end{equation}
is called the consistency condition. Suppose that there is no possible anomaly, namely the cohomology class is $H^1(b)=\algzero$, then the solution of \eqref{eq-ren-conscond} is: $\Delta=b\widetilde\Delta$, where $gh(\widetilde\Delta)=0$ and $\dim(\widetilde\Delta)=4$. It is also possible to choose $\widetilde \Delta$ satisfying \cite{Piguet:1995}:
\begin{equation*}
\frac{\delta\widetilde\Delta}{\delta B}=\frac{\delta\widetilde\Delta}{\delta\overline c}=0.
\end{equation*}
Then, by changing the action $S$ into $S-\widetilde \Delta$, the new vertex functional obeys to the Slavnov-Taylor identity at the order $n$:
\begin{equation*}
\caS(\Gamma)=0.
\end{equation*}

\begin{proposition}
The freedom in the choice of $\widetilde\Delta$ is given by a term $L$ ($\widetilde\Delta\mapsto\widetilde\Delta+L$), such that $bL=0$, or $L\in H^0(b)$. The general expression of $L$ is:
\begin{equation*}
L=\int\dd^4x\tr\Big(\beta_1 F_{\mu\nu}F_{\mu\nu}+\beta_2\widehat\rho_\mu A_\mu+\beta_3\sigma c\Big),
\end{equation*}
where $\beta_i$ are free coefficients.
\end{proposition}
Like in the section \ref{sec-ren-bphz} (see Equation \eqref{eq-ren-condren4}), the coefficients $\beta_i$ can be fixed by renormalization conditions:
\begin{align*}
\frac{\dd}{\dd p^2}\Gamma^T(p^2)|_{p^2=\mu^2}&=\frac{1}{g^2}\\
\frac{\dd}{\dd p^2}\Gamma_{c\overline c}(p^2)|_{p^2=\mu^2}&=1\\
\Gamma_{\sigma^ac^bc^c}(p_1,p_2,p_3,p_4)|_{p_i^2=\mu^2}&=f_{abc},
\end{align*}
where one uses the decomposition:
\begin{equation*}
\Gamma_{A_\mu A_\nu}(p,-p)=\Big(\delta_{\mu\nu}-\frac{p_\mu p_\nu}{p^2}\Big)\Gamma^T(p^2)+\frac{p_\mu p_\nu}{p^2}\Gamma^L(p^2).
\end{equation*}

\numberwithin{equation}{section}
\chapter[QFT on Moyal space]{Quantum Field Theory on the Moyal space}
\label{cha-moy}

In this chapter, we will introduce the euclidean Moyal space, and study the most usual scalar (and fermionic) QFT defined on it.

\section{Presentation of the Moyal space}
\label{sec-moy-presmoy}

\subsection{Deformation quantization}
\label{subsec-moy-defqu}

Deformation quantization finds its origin in the quantum mechanics, where the algebra of quantum observables can be seen as a deformation of the algebra of classical observables. Indeed, the Weyl correspondence \cite{Weyl:1931,Wigner:1932} permits to associate an operator to a classical function of the standard phase space. Then, the product of such operators corresponds to a noncommutative product introduced by Moyal \cite{Moyal:1949} and Groenewold \cite{Groenewold:1946} on the functions of the phase space. This construction has been generalized to the formal deformations of the functions on a symplectic manifold \cite{Flato:1974,Vey:1975,Bayen:1978}, and this will be briefly exposed here. See \cite{Hazewinkel:1988} for more details, and for example \cite{Bordemann:1998a,Bordemann:1998b} for further results on formal deformations.

Note that this construction deals with formal series in a deformation parameter, while strict deformation quantization, introduced by Rieffel \cite{Rieffel:1989}, permits to obtain convergent deformations of $C^\ast$-algebras (see also \cite{Bieliavsky:2002}). Strict quantization is closely related to noncommutative geometry \cite{Connes:2000tj} (see chapter \ref{cha-gnc}).
\medskip

Let us recall some basic properties of formal power series. The commutative $\gC$-algebra of formal power series of $\gC$ is defined by $\gC[[\theta]]=\prod_{i=0}^\infty\gC$, where each element can be written in terms of the formal parameter $\theta$:
\begin{equation*}
z=\sum_{i=0}^\infty z_i \theta^i,
\end{equation*}
and the product is given by:
\begin{equation*}
z\fois z'=\sum_{i=0}^\infty\sum_{j=0}^i (z_j z'_{i-j})\theta^i.
\end{equation*}
This definition can be generalized to the case of a $\gC$-algebra $\algA$. The set $\algA[[\theta]]$ of formal power series of $\algA$ is naturally a $\gC$-algebra and a $\gC[[\theta]]$-module, which contains $\algA$.
\medskip

We can now give the definition of a formal deformation of a manifold.
\begin{definition}
\label{def-moy-fordef}
Let $M$ be a smooth manifold and $C^\infty(M)$ the $\gC$-algebra of its smooth functions. A {\it formal deformation quantization} of $M$ is a $\gC[[\theta]]$-bilinear associative product $\star$ on $C^\infty(M)[[\theta]]$, which can be written as: $\forall f,g\in C^\infty(M)$,
\begin{equation*}
f\star g=\sum_{i=0}^\infty c_i(f,g)\theta^i,
\end{equation*}
where $c_0(f,g)=f\fois g$ is the ordinary product, and the function $1$ is a unit of $\star$.
\end{definition}
Note that the product is uniquely defined by its values on $C^\infty(M)$ by $\gC[[\theta]]$-linearity. On the smooth manifold, let us define the notion of a bidifferential operator by a bilinear map $D:C^\infty(M)\times C^\infty(M)\to C^\infty(M)$, locally of the form: $\forall f,g\in C^\infty(M)$, $\forall x\in M$,
\begin{equation*}
D(f,g)(x)=\sum_{|\alpha|,|\beta|=0}^N d_{\alpha\beta}(x)\frac{\partial^{|\alpha|}f}{\partial x^\alpha}(x)\frac{\partial^{|\beta|}g}{\partial x^\beta}(x),
\end{equation*}
where $\alpha$ and $\beta$ are multi-indices, $N$ is a fixed integer boundary, and $d_{\alpha\beta}$ are smooth functions. In the following of this subsection, we will assume that the maps $c_i:C^\infty(M)\times C^\infty(M)\to C^\infty(M)$ of Definition \ref{def-moy-fordef} are bidifferential operators. Such deformations are called differentiable, and the product $\star$ occurring in Definition \ref{def-moy-fordef} is called the star product of the differentiable deformation.

The map $c_1$ can be interpreted as related to a Poisson bracket.
\begin{proposition}
Let $\star$ be a star product on a smooth manifold $M$. Then, in the notations of Definition \ref{def-moy-fordef}, $\forall f,g\in C^\infty(M)$,
\begin{equation*}
\{f,g\}=c_1(f,g)-c_1(g,f)=\left(\frac{f\star g-g\star f}{\theta}\right)_{|\theta=0}
\end{equation*}
defines a Poisson bracket on $M$.
\end{proposition}
One says that the deformation is in the direction of the Poisson bracket $\{,\}$.
\medskip

Let $T=\sum_{i=0}^\infty T_i\theta^i$ be a formal power series of differentiable operators on $C^\infty(M)$. Then, it acts on $f\in C^\infty(M)[[\theta]]$ by $\gC[[\theta]]$-linearity:
\begin{equation*}
T(f)=\sum_{i=0}^\infty\sum_{j=0}^i T_j(f_{i-j})\theta^i.
\end{equation*}
One can introduce a notion of equivalence on the star products as follows.
\begin{definition}
Two star products $\star$ and $\ast$ on the same smooth manifold $M$ are said to be equivalent if there exists a formal power series $T$ of differential operators $T_i:C^\infty(M)\to C^\infty(M)$ such that $T_0=\text{Id}$ and $\forall f,g\in C^\infty(M)[[\theta]]$,
\begin{equation*}
T(f\ast g)=T(f)\star T(g).
\end{equation*}
\end{definition}
Note that this relation is symmetric, since $T$ is invertible as a formal power series.

\begin{proposition}
Two equivalent star products on a smooth manifold have the same associated Poisson brackets.
\end{proposition}
Finally, the classification of star products on symplectic manifolds in the direction of their canonical Poisson bracket has been performed, and is related to the de Rham cohomology of the manifold.
\begin{theorem}
Let $M$ a smooth manifold with symplectic form $\omega$. Then, the formal deformations of $M$ in the direction of $\omega$ are classified by an element of
\begin{equation*}
\frac{1}{\theta}\omega+H^2(M,\gC[[\theta]]),
\end{equation*}
modulo the action by symplectomorphisms.
\end{theorem}

\subsection{The Moyal product on Schwartz functions}
\label{subsec-moy-schwartz}

In this subsection, we collect the definition and the main properties of the Moyal space, following in particular \cite{GraciaBondia:1987kw,Varilly:1988jk}. We recall that the Schwartz functions of the standard vector space $\gR^D$, namely the functions rapidly decreasing at infinity, are defined by:
\begin{equation*}
\caS(\gR^D)=\{f:\gR^D\to \gC\ C^\infty,\quad \forall\alpha,\beta,\ x^\alpha\partial^\beta f\text{ is bounded}\},
\end{equation*}
where $\alpha$ and $\beta$ are multi-indices. One also defines a family of seminorms: $\forall k\in\gN$, $\forall f\in\caS(\gR^D)$,
\begin{equation}
p_k(f)=\sup_{|\alpha|,|\beta|\leq k, x\in\gR^D}|x^\alpha\partial^\beta f|.\label{eq-moy-seminorm}
\end{equation}

Let us consider on $\gR^D$, where the dimension $D$ is even, the canonical euclidean structure $\delta_{\mu\nu}$ and the standard symplectic structure $\Sigma^{\text{st}}_{\mu\nu}$, given by:
\begin{equation*}
\Sigma^{\text{st}}=\begin{pmatrix} 0 & -1 & 0 & 0 & \\ 1 & 0 & 0 & 0 & \\ 0 & 0 & 0 & -1 & \ddots \\ 0 & 0 & 1 & 0 &  \\  &  & \ddots &  & \end{pmatrix} .
 \end{equation*}
By defining $\Theta=\theta\Sigma^{\text{st}}$, where $\theta$ is a positive deformation parameter of mass dimension $-2$, and $x\wedge y=2x_\mu\Theta^{-1}_{\mu\nu}y_\nu$ for $x,y\in\gR^D$, one can introduce on $\caS(\gR^D)$ the following product:
\begin{equation}
(f\star g) (x)=\frac{1}{\pi^D\theta^D}\int \dd^Dy\,\dd^Dz\ f(y)\,g(z)e^{-ix\wedge y-iy\wedge z-iz\wedge x},\label{eq-moy-moyprod}
\end{equation}
called the {\it Moyal product}.

Let us give the major properties of $\caS(\gR^D)$ with the Moyal product \cite{GraciaBondia:1987kw,Moyal:1949}.
\begin{proposition}
Endowed with the seminorms \eqref{eq-moy-seminorm}, $\caS(\gR^D)$ is a Fr\'echet space. With the Moyal product \eqref{eq-moy-moyprod} and the usual complex conjugation, $\caS(\gR^D)$ is an involutive associative topological algebra, for the topology of the seminorms \eqref{eq-moy-seminorm}. The usual derivations $\partial_\mu$ are continuous derivations for this algebra. Moreover, it can be extended to a formal deformation of $\gR^D$ \cite{Moyal:1949}, since $\forall f,g\in\caS(\gR^D)$, $\forall x\in\gR^D$,
\begin{equation*}
(f\star g)(x)=e^{\frac i2\Theta_{\mu\nu}\partial_\mu^y\partial_\nu^z}f(y)g(z)_{|y=z=x}.
\end{equation*}
\end{proposition}
The Moyal product is then a star-product on $\gR^D$ in the sense of subsection \ref{subsec-moy-defqu}, and in the direction of the Poisson bracket associated to the standard symplectic structure $\Sigma^{\text{st}}$. Note that the seminorms \eqref{eq-moy-seminorm} do not satisfy: $p_k(f\star g)\leq p_k(f)p_k(g)$, and they are not algebra seminorms.

By using the following identity
\begin{equation*}
\int \dd^Dx\, e^{i\alpha x\wedge y}=\left(\frac{\pi\theta}{\alpha}\right)^D\delta(y),
\end{equation*}
one shows the important tracial property: $\forall f,g\in\caS(\gR^D)$,
\begin{equation}
\int \dd^Dx\ (f\star g)(x)=\int \dd^Dx f(x)\fois g(x)=\int \dd^Dx (g\star f)(x),\label{eq-moy-trace}
\end{equation}
and the following formula: $\forall f,g,h\in\caS(\gR^D)$
\begin{equation*}
(f\star g\star h) (x)=\frac{1}{\pi^D\theta^D}\int \dd^Dy\,\dd^Dz\ f(y)\,g(z)\,h(x-y+z)e^{-ix\wedge y-iy\wedge z-iz\wedge x}.
\end{equation*}

By denoting $\langle f,g\rangle=\int \dd^Dx f(x)\fois g(x)=\langle g,f\rangle$, Equation \eqref{eq-moy-trace} implies: $\forall f,g,h\in\caS(\gR^D)$,
\begin{equation}
\langle f\star g,h\rangle=\langle f,g\star h\rangle=\langle g,h\star f\rangle,\label{eq-moy-cycl}
\end{equation}
which will allow to extend $\caS(\gR^D)$ by duality into a distribution algebra (see subsection \ref{subsec-moy-moyal}).

\subsection{The matrix basis}
\label{subsec-moy-matrix}

It has been shown \cite{GraciaBondia:1987kw} (see also \cite{Grosse:2003nw}) that the eigenfunctions $(b_{mn}^{(D)}(x))$ of the harmonic oscillator hamiltonian $H=\frac{x^2}{2}$ are a hilbertian basis of $\caS(\gR^D)$, called ``matrix basis''. They are defined by:
\begin{equation*}
H\star b_{mn}^{(D)}=\theta(|m|+\frac{1}{2})b_{mn}^{(D)} \qquad b_{mn}^{(D)}\star H=\theta(|n|+\frac{1}{2})b_{mn}^{(D)},
\end{equation*}
where $m,n\in\mathbb N^{\frac D2}$ and $|m|=\sum_{i=1}^{\frac D2}m_i$. In two dimensions, the expression of the elements $(b_{mn}^{(2)})=(f_{mn})$ of the matrix basis in polar coordinates
\begin{equation*}
x_1=r\cos(\varphi),\qquad x_2=r\sin(\varphi),
\end{equation*}
is given by
\begin{align}
f_{mn}(x)&=\frac{2e^{-\frac{x^2}{\theta}}}{\sqrt{m!n!}}\sum_{k=0}^{\min(m,n)}(-1)^k {\begin{pmatrix} m \\ k \end{pmatrix}} {\begin{pmatrix} n \\ k \end{pmatrix}} k!\left(\sqrt{\frac{2}{\theta}}\right)^{m+n-2k}r^{m+n-2k}e^{i\varphi(n-m)}\nonumber\\
f_{mn}(x)&=2(-1)^m\sqrt{\frac{m!}{n!}}e^{i(n-m)\varphi}\left(\frac{2r^2}{\theta}\right)^{\frac{n-m}{2}}L_m^{n-m}\left(\frac{2r^2}{\theta}\right) e^{-\frac{r^2}{\theta}},\label{eq-moy-laguerre}
\end{align}
where the $L_n^k(x)$ are the associated Laguerre polynomials. The extension in even $D$ dimensions is straightforward. Namely, one has for $m=(m_1,\dots,m_{\frac D2})$ and $n=(n_1,\dots,n_{\frac D2})$,
\begin{equation}
b_{mn}^{(D)}(x)=f_{m_1,n_1}(x_1,x_2)\dots f_{m_{\frac D2},n_{\frac D2}}(x_{D-1},x_D).\label{eq-moy-basematD}
\end{equation}

The matrix basis satisfies the following properties:
\begin{subequations}
\label{eq-moy-matrprop}
\begin{align}
&(b_{mn}^{(D)}\star b_{kl}^{(D)})(x)=\delta_{nk} b_{ml}^{(D)}(x),\\
&\int \dd^Dx\ b_{mn}^{(D)}(x)=(2\pi\theta)^{\frac D2}\delta_{mn},\\
&(b_{mn}^{(D)})^\dag(x)=b_{nm}^{(D)}(x).
\end{align}
\end{subequations}

Let $\Matr_{\infty,D}^S$ be the Fr\'echet space of rapidly decreasing double sequences $(a_{mn})$, namely such that $\forall k\in\gN$,
\begin{equation*}
r_k(a)=\left(\sum_{m,n\in\gN^{\frac D2}}(m+\frac 12)^{2k}(n+\frac 12)^{2k}|a_{mn}|^2\right)^{\frac 12}<\infty.
\end{equation*}
This space can be endowed with the product of (infinite) matrices, and it becomes an involutive (for the usual matrix conjugation) topological algebra. Then, the matrix basis provides an isomorphism of involutive topological algebras between $\caS(\gR^D)$ and $\Matr_{\infty,D}^S$ \cite{GraciaBondia:1987kw}:
\begin{equation*}
(g_{mn})\mapsto g(x)=\sum_{m,n\in\gN^{\frac D2}}g_{mn}b_{mn}^{(D)}(x),
\end{equation*}
and the inverse isomorphism is given by:
\begin{equation}
g_{mn}=\frac{1}{(2\pi\theta)^{\frac D2}}\int d^Dx\ g(x)b_{nm}^{(D)}(x).\label{eq-moy-coeffmatrix}
\end{equation}

\subsection{The Moyal algebra}
\label{subsec-moy-moyal}

In this subsection, we extend by duality the Moyal product \eqref{eq-moy-moyprod} to a certain class of distributions. The tempered distributions on $\gR^D$ are defined to be the continuous linear form on $\caS(\gR^D)$, namely:
\begin{equation*}
\caS'(\gR^D)=\{T\in\caL(\caS(\gR^D),\gC),\ \exists k\in\gN,\,C\geq0,\ \forall f\in\caS(\gR^D),\ |T(f)|\leq C\fois p_k(f)\}.
\end{equation*}
It is a topological vector space, endowed with the weak topology, and it contains $\caS(\gR^D)$. Let us denote the duality bracket: $\langle T,f\rangle=T(f)$ and define the usual involution: $\langle T^\dag,f\rangle=\overline{\langle T,f^\dag\rangle}$, for $T\in\caS'(\gR^D)$ and $f\in\caS(\gR^D)$. These notations are compatible with those of subsection \ref{subsec-moy-schwartz} if $T\in\caS(\gR^D)$.
\begin{definition}
For $T\in\caS'(\gR^D)$ and $f,g\in\caS(\gR^D)$, one defines the product $T\star f$ and $f\star T$ by:
\begin{align*}
&\langle T\star f,g\rangle=\langle T,f\star g\rangle,\\
&\langle f\star T,g\rangle=\langle T,g\star f\rangle.
\end{align*}
\end{definition}
Because of Equation \eqref{eq-moy-cycl}, these definitions are compatible when $T\in\caS(\gR^D)$, and by using the continuity of the Moyal product on $\caS(\gR^D)$, one can show that $T\star f$ and $f\star T$ are continuous, and belong to $\caS'(\gR^D)$. Note that the function $\gone:x\mapsto 1$ belongs to $\caS'(\gR^D)$ (but not to $\caS(\gR^D)$), and is the unit of the Moyal product: $\forall f\in\caS(\gR^D)$, $\gone\star f=f=f\star\gone$.

\begin{proposition}
The diagonal partial sum of the matrix basis:
\begin{equation*}
\forall k\in\gN^{\frac D2},\quad \forall x\in\gR^D,\quad e_k(x)=\sum_{m\leq k}b_{mm}^{(D)}(x),
\end{equation*}
is an approximate unit in $\caS(\gR^D)$, that is $\forall f\in\caS(\gR^D)$,
\begin{equation*}
\lim_{k\to\infty}(e_k\star f)=f=\lim_{k\to\infty}(f\star e_k).
\end{equation*}
\end{proposition}
\begin{proof}
Indeed, for $g(x)=\sum_{m,n}g_{mn}b_{mn}^{(D)}(x)$,
\begin{align*}
&(e_k\star g)(x)=\sum_{m\leq k}\sum_{n\in\gN^{\frac D2}}g_{mn}b_{mn}^{(D)}(x)\\
&(g\star e_k)(x)=\sum_{m\in\gN^{\frac D2}}\sum_{n\leq k}g_{mn}b_{mn}^{(D)}(x),
\end{align*}
and these sequences (in $k$) converge to $g(x)$.
\end{proof}

We can now define the {\it Moyal algebra} $\caM$ as the multiplier algebra of $\caS(\gR^D)$ (see \cite{Blackadar:2006} for multiplier algebras in the case of $C^\ast$-algebras).
\begin{definition}
\label{def-moy-alg}
\begin{align*}
\caM_L=&\{S\in\caS'(\gR^D),\quad \forall f\in\caS(\gR^D),\ S\star f\in\caS(\gR^D)\}\\
\caM_R=&\{R\in\caS'(\gR^D),\quad \forall f\in\caS(\gR^D),\ f\star R\in\caS(\gR^D)\}\\
\caM=&\caM_L\cap\caM_R.
\end{align*}
\end{definition}
Let us define the product of a tempered distribution by an element of $\caM_L$ and $\caM_R$.
\begin{lemma}
\label{lem-moy-compat}
For $S\in\caM_L$, $R\in\caM_R$ and $f\in\caS(\gR^D)$, one has:
\begin{equation*}
\langle R,S\star f\rangle=\langle S,f\star R\rangle.
\end{equation*}
\end{lemma}
\begin{proof}
Due to the associativity of the Moyal product on $\caS(\gR^D)$, it can be shown that $\forall T\in\caS'(\gR^D)$, $\forall f,g\in\caS(\gR^D)$, $(T\star f)\star g=T\star(f\star g)$ and $f\star(g\star T)=(f\star g)\star T$. Then, $\forall R\in\caM_R$, $\forall S\in\caM_L$, $\forall f\in\caS(\gR^D)$, $\forall k\in\gN^{\frac D2}$,
\begin{equation*}
\langle R,S\star e_k\star f\rangle=\langle f\star R,S\star e_k\rangle=\langle S\star e_k,f\star R\rangle=\langle S,e_k\star f\star R\rangle.
\end{equation*}
By taking the limit on $k$, since $e_k\star f\to f$, one obtains $\langle R,S\star f\rangle=\langle S,f\star R\rangle$.
\end{proof}

\begin{definition}
\label{def-moy-prodalg}
For $T\in\caS'(\gR^D)$, $S\in\caM_L$, $R\in\caM_R$ and $f\in\caS(\gR^D)$, one defines the product:
\begin{align*}
\langle T\star S,f\rangle=&\langle T,S\star f\rangle\\
\langle R\star T,f\rangle=&\langle T,f\star R\rangle.
\end{align*}
\end{definition}
It is important to notice that these definitions are compatible when $T\in\caM_R$ and $T\in\caM_L$ because of Lemma \ref{lem-moy-compat}.

The properties of $\caM$ are summarized here \cite{GraciaBondia:1987kw}.
\begin{proposition}
$\caM$, introduced in Definition \ref{def-moy-alg}, endowed with the product of Definition \ref{def-moy-prodalg}, is a complete unital involutive associative topological algebra.
\end{proposition}

In the picture of noncommutative geometry (see chapter \ref{cha-gnc}), the Moyal space can be defined as the ``noncommutative space'' associated to the noncommutative algebra $\caM$. Note that in \cite{Gayral:2003dm}, a certain (non-unital) algebra containing $\caS(\gR^D)$, but strictly contained in $\caM$, is introduced as a part of a spectral triple.

$\caM$ contains $\caS(\gR^D)$, the unit function $\gone$, polynomial functions, the Dirac distribution $\delta$ and its derivatives. Moreover, by defining the commutator $[a,b]_\star=a\star b-b\star a$, and the anticommutator $\{a,b\}_\star=a\star b+b\star a$, for $a,b\in\caM$, one has the following relations, true in $\caM$:
\begin{subequations}
\label{eq-moy-basident}
\begin{align*}
[x_\mu,x_\nu]_\star&=i\Theta_{\mu\nu}\\
[\widetilde x_\mu,a]_\star&=2i\partial_\mu a\\
\{\widetilde x_\mu,a\}_\star&=2\widetilde x_\mu\fois a,
\end{align*}
\end{subequations}
where $\widetilde x_\mu=2\Theta^{-1}_{\mu\nu}x_\nu$. Therefore, $\partial_\mu$ are inner derivations of the algebra $\caM$.

Note that the tempered distributions can also be written in terms of the matrix basis. For $T\in\caS'(\gR^D)$, define the coefficient
\begin{equation*}
t_{mn}=\frac{1}{(2\pi\theta)^{\frac D2}}\langle T,b_{nm}^{(D)}\rangle,
\end{equation*}
where $m,n\in\gN^{\frac D2}$. It can be proved that there exists $k\in\gN$ such that $r_{-k}(t)<\infty$ \cite{GraciaBondia:1987kw}. Finally, $\sum_{m,n}t_{mn}b_{mn}^{(D)}$ converges weakly to $T$.

\begin{example}
We give the matrix coefficients of the following distributions:
\begin{itemize}
\item For $T=\gone$, $t_{mn}=\delta_{mn}$.
\item For the Dirac distribution: $T=\delta(x)$,
\begin{equation*}
t_{mn}=\frac{(-1)^{|m|}}{(\pi\theta)^{\frac D2}}\delta_{mn}.
\end{equation*}
\item For $T=x_{2j}$ ($j\in\{1,\dots,\frac D2\}$),
\begin{equation*}
t_{mn}=\sqrt{\frac{\theta}{2}}\Big(\sqrt{m_j+1}\delta_{m_j+1,n_j}+\sqrt{n_j+1}\delta_{m_j,n_j+1}\Big)\left(\prod_{i\neq j}\delta_{m_i,n_i}\right).
\end{equation*}
\item For $T=x_{2j+1}$ ($j\in\{1,\dots,\frac D2\}$),
\begin{equation*}
t_{mn}=-i\sqrt{\frac{\theta}{2}}\Big(\sqrt{m_j+1}\delta_{m_j+1,n_j}-\sqrt{n_j+1}\delta_{m_j,n_j+1}\Big)\left(\prod_{i\neq j}\delta_{m_i,n_i}\right).
\end{equation*}
\item For $T=x^2$, $t_{mn}=\theta\big(2|m|+\frac D2\big)\delta_{mn}$.
\item For $T=e^{-\frac{x^2}{\theta}}$, $t_{mn}=\frac 12\delta_{m0}\delta_{n0}$.
\end{itemize}
\end{example}

\subsection{The symplectic Fourier transformation}
\label{subsec-moy-sympl}

Let us define the symplectic Fourier transformation.
\begin{definition}
For $f\in\caS(\gR^D)$ and $k\in\gR^D$, the {\it symplectic Fourier transform} of $f$ is given by:
\begin{equation*}
\hat f(k)=\caF(f)(k)=\frac{1}{(\pi\theta)^{\frac D2}}\int \dd^Dx\ f(x)e^{-ik\wedge x},
\end{equation*}
where we recall that $k\wedge x=2k_\mu\Theta^{-1}_{\mu\nu}x_\nu$.
\end{definition}

\begin{proposition}
The symplectic Fourier transformation is a surjective isometry on $\caS(\gR^D)$. Its inverse is given by:
\begin{equation*}
f(x)=\caF^{-1}(\hat f)(x)=\frac{1}{(\pi\theta)^{\frac D2}}\int \dd^Dk\hat f(k) e^{ik\wedge x}.
\end{equation*}
It can also be extended on $\caS'(\gR^D)$ by: $\forall T\in\caS'(\gR^D)$, $\forall f\in\caS(\gR^D)$,
\begin{equation*}
\langle \caF(T),f\rangle=\langle \hat T,f\rangle=\langle T,\hat f\rangle,
\end{equation*}
and it is an automorphism of $\caS'(\gR^D)$.
\end{proposition}

Recall that $\widetilde x_\mu=2\Theta^{-1}_{\mu\nu}x_\nu$. Then,
\begin{subequations}
\label{eq-moy-symplder}
\begin{align*}
&\caF(\partial_\mu^x f)(k)=-i\widetilde k_\mu\fois\caF(f)(k)\\
&\caF(\widetilde x_\mu\fois f)(k)=i\partial_\mu^k\caF(f)(k).
\end{align*}
\end{subequations}

\begin{example}
We have seen that $\gone$ is in $\caS'(\gR^D)$, and it is known that the gaussian function $e^{-\frac{x^2}{\theta}}$ belongs to $\caS(\gR^D)$. Their Fourier transform are computed here:
\begin{align*}
&\caF(\gone)(k)=(\pi\theta)^{\frac D2}\delta(k)\\
&\caF(e^{-\frac{x^2}{\theta}})(k)=e^{-\frac{k^2}{\theta}}.
\end{align*}
\end{example}

Let us define the symplectic convolution by: $\forall f,g\in\caS(\gR^D)$, $\forall x\in\gR^D$,
\begin{equation*}
(f\lozenge g)(x)=\frac{1}{(\pi\theta)^{\frac D2}}\int\dd^Dy\ f(y)g(x-y)e^{-ix\wedge y}=\frac{1}{(\pi\theta)^{\frac D2}}\int\dd^Dy\ f(x-y)g(y)e^{ix\wedge y}
\end{equation*}
This associative product, whose unit is the Dirac distribution, can also be extended on $\caS'(\gR^D)$. Then, the Moyal product and this symplectic convolution are related in the following way: $\forall f,g\in\caS(\gR^D)$, $\forall k\in\gR^D$,
\begin{multline*}
\caF(f\star g)(k)=\frac{1}{(\pi\theta)^{\frac D2}}\int \dd^Dx\ f(x+k)g(x)e^{-ik\wedge x}=\\
\frac{1}{(\pi\theta)^{\frac D2}}\int\dd^Dx\ \hat f(x)\hat g(k-x)e^{-ik\wedge x}=(\hat f\lozenge \hat g)(k).
\end{multline*}

Under the complex conjugation, symplectic Fourier transformation becomes:
\begin{equation*}
\caF(f^\dag)(k)=(\caF(f))^\dag(-k).
\end{equation*}
The usual Parseval-Plancherel equality is also valid in this framework of symplectic Fourier transformation:
\begin{equation}
\int \dd^Dx\ f^\dag(x)\fois g(x)=\int\dd^Dk\ (\hat f)^\dag(k)\fois \hat g(k).\label{eq-moy-parseval}
\end{equation}
Moreover, one has the following identity:
\begin{equation}
\int \dd^Dx\ f_1^\dag\star f_2\star f_3^\dag\star f_4(x)=\int\dd^Dk\ (\hat f_1)^\dag\star \hat f_2\star(\hat f_3)^\dag\star\hat f_4(k).\label{eq-moy-parsint}
\end{equation}
Note also that the symplectic Fourier transformation is deeply related with the Dirac distribution: $\forall f\in\caS(\gR^D)$ and $\forall x\in\gR^D$,
\begin{equation*}
(f\star\delta)(x)=\frac{1}{(\pi\theta)^{\frac D2}}\caF(f)(x),\qquad (\delta\star f)(x)=\frac{1}{(\pi\theta)^{\frac D2}}\caF^{-1}(f)(x)
\end{equation*}

\section{UV/IR mixing on the Moyal space}
\label{sec-moy-uvir}

The Moyal space could be a possibility to deal with quantum space-time \cite{Doplicher:1994zv,Doplicher:1994tu}, so that QFT defined on it are interesting in this way. However, we restrict our study here to the euclidean Moyal space. In the commutative theory (see chapter \ref{cha-ren}), scalar fields are simply functions on the configuration space, or sections of a (trivial) vector fiber bundle on this space. In the noncommutative point of view, the analog of scalar fields then belong to the noncommutative algebra, as ``functions'' on the ``noncommutative space''. The generalization of the usual scalar field theory to the Moyal space is straightforward: one simply replaces the usual commutative product by the Moyal star product. For $\phi\in\caS(\gR^D)$, the new action writes down:
\begin{equation*}
S(\phi)=\int \dd^Dx\big(\frac 12\partial_\mu\phi\star\partial_\mu\phi+\frac{\mu^2}{2}\phi\star\phi+\lambda\,\phi\star\phi\star\phi\star\phi\big),
\end{equation*}
where the euclidean metric has been used. With the tracial identity \eqref{eq-moy-trace}, it becomes
\begin{equation}
S(\phi)=\int \dd^Dx\big(\frac 12\partial_\mu\phi\fois\partial_\mu\phi +\frac{\mu^2}{2}\phi\fois\phi+\lambda\,\phi\star\phi\star\phi\star\phi\big),\label{eq-moy-actscal}
\end{equation}
where $\lambda$ is the coupling constant, and $\mu$ the mass associated to the field $\phi$.
\medskip

\begin{figure}[htbp]
  \centering
  \subfloat[Planar]{{\label{fig-moy-ribbon1}}\includegraphics[scale=1]{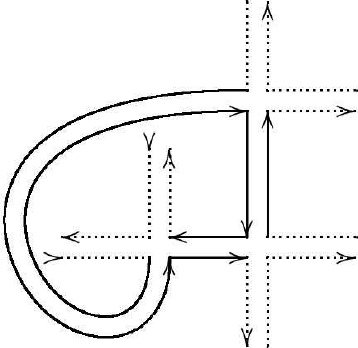}}\qquad
  \subfloat[Non-planar]{\label{fig-moy-ribbon2}\includegraphics[scale=1]{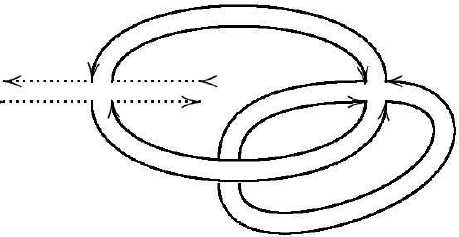}}
  \caption{Ribbon graphs}
  \label{fig-moy-ribbon}
\end{figure}

We will expose now the different steps toward renormalizability for scalar theories on the Moyal space (see \cite{Douglas:2001ba,Wulkenhaar:2006si,Rivasseau:2007rz} for a review). The Feynman rules corresponding to the above theory have been computed by Filk \cite{Filk:1996dm}. In the Fourier conventions \eqref{eq-ren-fourierconv} (do not confuse with symplectic Fourier transformation of subsection \ref{subsec-moy-sympl}), these are given by:
\begin{itemize}
\item a propagator: $\frac{(2\pi)^D}{p^2+\mu^2}$, where $p$ is the impulsion of the line.
\item a vertex:
\begin{equation*}
\frac{\lambda}{(2\pi)^{3D}\theta^{2D}}\delta(p_1+p_2+p_3+p_4)e^{\frac{i\theta^2}{4}(p_1\wedge p_2+p_1\wedge p_3+p_2\wedge p_3)},
\end{equation*}
where $p_i$ are the (incoming) impulsions.
\end{itemize}
Note that, because of the non-locality of the vertex and the noncommutativity of the product, the Feynman graphs can be suitably represented as ribbon diagrams drawn on an oriented Riemann surface with boundary, like on Figure \ref{fig-moy-ribbon}.

Let us define some topological notions associated to ribbon graphs.
\begin{definition}
For a given ribbon graph $G$, one defines:
\begin{itemize}
\item a {\it face} by a closed path drawn on the boundaries of the ribbons and possibly interrupted by external legs.
\item a {\it broken face} by a face which is interrupted by at least one external leg.
\item the {\it genus} $g$ of the graph $G$ by the identity:
\begin{equation*}
\chi=2-2g=n-I+F,
\end{equation*}
where $\chi$ is the Euler characteristic of the surface, $V$ the number of vertices of $G$, $I$ the number of internal (double) lines and $F$ the number of faces.
\item $G$ is called {\it planar} if its genus $g=0$. Otherwise, it is called {\it non-planar}.
\item $G$ is called {\it regular} if it has only one broken face.
\end{itemize}
\end{definition}
For the $\phi^4$-theory, one has the further identity: $I=2n-\frac N2$, where $N$ is the number of external legs, so that the Euler characteristic reads:
\begin{equation*}
\chi=2-2g=\frac N2-n+F.
\end{equation*}
Filk proved in \cite{Filk:1996dm} that the amplitudes of the planar regular graphs of the theory \eqref{eq-moy-actscal} are identical to those of the commutative theory, while planar graphs with at least two broken faces have the same amplitudes as their commutative counterparts but with oscillations which couple internal lines and external legs. Moreover, non-planar graphs can be constructed from planar ones. Then, Chepelev and Roiban established a power-counting of this theory related to the topology of the graphs \cite{Chepelev:1999tt}.

\begin{figure}[htb]
  \centering
  \includegraphics{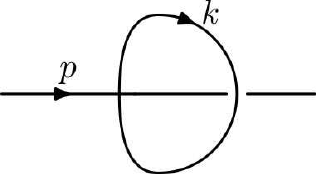}
  \caption{Non-planar tadpole}
  \label{fig-moy-nonplanartadpole}
\end{figure}

However, Minwalla {\it et al.} found a new type of infra-red (IR) divergences in this theory, which renders the renormalizability quite unlikely \cite{Minwalla:1999px}. Indeed, if one computes the amplitude of the so-called ``non-planar tadpole'' (which is planar but with two broken faces), represented on Figure \ref{fig-moy-nonplanartadpole}, one finds:
\begin{equation*}
\caA(p)=\frac{\lambda}{(2\pi\theta)^{2D}}\int\dd^Dk\frac{e^{\frac{i\theta^2}{2}k\wedge p}}{k^2+\mu^2}.
\end{equation*}
By using the identity:
\begin{equation*}
\int\dd^Dk\frac{e^{-ik\fois x}}{k^2+\mu^2}=(2\pi)^{\frac D2}\left(\frac{\mu^2}{x^2}\right)^{\frac{D-2}{4}}K_{\frac D2-1}\left(\mu\sqrt{x^2}\right),
\end{equation*}
where $K_n$ is the n${}^{\text{th}}$ modified Bessel function of the second kind, one obtains:
\begin{equation}
\caA(p)=\frac{\lambda(2\pi)^{\frac D2}}{(2\pi\theta)^{2D}}\left(\frac{\mu^2}{\theta^2p^2}\right)^{\frac{D-2}{4}}K_{\frac D2-1}\left(\mu\theta\sqrt{p^2}\right)\propto_{p\to0} \frac{1}{p^2}.\label{eq-moy-uvirtadpole}
\end{equation}
For $p\neq0$, this amplitude is finite, and does not need to be regularized and renormalized. But if $p$ goes to $0$, it diverges. It is an IR divergence (for small $p$) coming from the ultra-violet (UV) part of the integration over $k$ (it corresponds to large $k$), which is a typical feature of the so-called UV/IR mixing. It is not dangerous at the one-loop order, but if this tadpole diagram is inserted many times into higher loop order graphs, it can produce an IR divergence in the integrals over internal impulsions. This divergence is non-local and cannot be reabsorbed in a redefinition of the parameters of the action \eqref{eq-moy-actscal}.
\medskip

Note that the UV/IR mixing problem is present in other noncommutative spaces: it has been shown \cite{Gayral:2004cs} that all isospectral deformations contain it, so that the study of UV/IR mixing on the Moyal space can appear as generic to a certain class of noncommutative spaces. In the next section, we will see that by modifying the action with an harmonic term, this UV/IR mixing problem is solved. Numerical evidence for the effects induced by the UV/IR mixing can be found in \cite{Panero:2006cs,Panero:2006bx} for example.

\section{Renormalizable QFT on Moyal space}
\label{sec-moy-ren}

\subsection{Renormalization of the theory with harmonic term}
\label{subsec-moy-harm}

To avoid the problem of UV/IR mixing (see section \ref{sec-moy-uvir}), Grosse and Wulkenhaar introduced a harmonic term in the action \eqref{eq-moy-actscal}, so that the new action reads:
\begin{equation}
S(\phi)=\int \dd^Dx\big(\frac 12\partial_\mu\phi\fois\partial_\mu\phi +\frac{\Omega^2}{2}\widetilde x^2\phi^2 +\frac{\mu^2}{2}\phi^2+\lambda\,\phi\star\phi\star\phi\star\phi\big),\label{eq-moy-actharm}
\end{equation}
where $\Omega$ is a positive dimensionless parameter, and we recall that $\widetilde x_\mu=2\Theta^{-1}_{\mu\nu}x_\nu$. The model is then renormalizable up to all orders in perturbation in $D=4$ dimensions \cite{Grosse:2004yu}. In $D=2$, it is super-renormalizable \cite{Grosse:2003nw}.

\subsubsection{In the matrix basis}

The original proof of the renormalizability of \eqref{eq-moy-actharm} remains in the matrix basis (see subsection \ref{subsec-moy-matrix}) \cite{Grosse:2003nw,Grosse:2004yu}. See also \cite{Wulkenhaar:2006si,VignesTourneret:2006xa} for a review. Indeed, by decomposing the field $\phi$ in this basis,
\begin{equation*}
\phi(x)=\sum_{m,n\in\gN^{\frac D2}}\phi_{mn}b_{mn}^{(D)}(x)
\end{equation*}
and by using the identities \eqref{eq-moy-matrprop}, one can reexpress the action into:
\begin{equation}
S(\phi)=(2\pi\theta)^{\frac D2}\sum_{m,n,k,l\in\gN^{\frac D2}}\left(\frac 12\phi_{mn}\Delta_{mn,kl}\phi_{kl}+\lambda\,\phi_{mn}\phi_{nk}\phi_{kl}\phi_{lm}\right),\label{eq-moy-actmatrix}
\end{equation}
with
\begin{multline}
\Delta_{mn,kl}=\frac{4}{\theta}\left(|m|+|n|+\frac D2(1+\frac{\mu^2\theta}{4})\right)\delta_{ml}\delta_{nk}-\frac{2(1-\Omega^2)}{\theta}\\
\times\sum_{j=1}^{D/2}\left(\sqrt{(m_j+1)(n_j+1)}\delta_{m_j+1,l_j}\delta_{n_j+1,k_j} +\sqrt{m_jn_j}\delta_{m_j-1,l_j}\delta_{n_j-1,k_j}\right)\left(\prod_{i\neq j}\delta_{m_il_i}\delta_{n_ik_i}\right).\label{eq-moy-deltamatr}
\end{multline}

\begin{figure}[htb]
  \centering
  \includegraphics{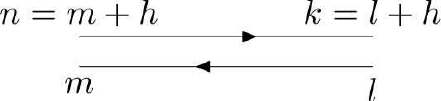}
  \caption{Matrix propagator}
  \label{fig-moy-matrixprop}
\end{figure}

To determine the propagator (in the matrix basis) of the theory, one has to inverse the matrix $\Delta$ with four indices, that is to find $C$ satisfying: $\sum_{p.q}C_{mn,pq}\Delta_{pq,kl}=\delta_{ml}\delta_{nk}=\sum_{p,q}\Delta_{mn,pq}C_{pq,kl}$, with $m,n,k,l\in\gN^{\frac D2}$ multi-indices. It has been shown that \cite{Grosse:2004yu,Rivasseau:2005bh}
\begin{multline*}
C_{mn,kl}=\frac{\theta}{8\Omega}\int_0^1\dd\alpha\frac{(1-\alpha)^{\frac{\mu^2\theta}{8\Omega}+\frac D4-1}}{(1+\frac{(1-\Omega)^2\alpha}{4\Omega})^{\frac D2}}\delta_{m+k,n+l}\prod_{j=1}^{D/2}\Big(\left(\frac{\sqrt{1-\alpha}}{1+\frac{(1-\Omega)^2\alpha}{4\Omega}}\right)^{n_j+l_j}\\ \sum_{i=\max(0,m_j-n_j)}^{\min(m_j,l_j)}\caA(m_j,l_j,n_j-m_j,i)\left(\frac{(1-\Omega^2)\alpha}{4\Omega\sqrt{1-\alpha}}\right)^{m_j+l_j-2i} \Big),
\end{multline*}
where $\caA(m_j,l_j,n_j-m_j,i)=\sqrt{{\begin{pmatrix} m_j \\ m_j-i \end{pmatrix}} {\begin{pmatrix} n_j \\ m_j-i \end{pmatrix}} {\begin{pmatrix} l_j \\ l_j-i \end{pmatrix}} {\begin{pmatrix} k_j \\ l_j-i \end{pmatrix}}}$. This is a very complicated expression, but note that for $\Omega=1$, it simplifies into:
\begin{equation*}
C_{mn,kl}=\frac{\theta}{4}\frac{1}{|m|+|n|+\frac{\mu^2\theta}{4}+\frac D2}\delta_{ml}\delta_{nk}.
\end{equation*}
We can see that $\Delta$ and $C$ are not trivial in the indices, namely the coefficient before $\delta_{ml}\delta_{nk}$ depends on $m$ and $n$, and the model \eqref{eq-moy-actmatrix} is then called a dynamical matrix model. In this picture, Feynman graphs are also represented by ribbon graphs (see section \ref{sec-moy-uvir}), and each boundary of a ribbon carries an index $m\in\gN^{\frac D2}$, like on Figure \ref{fig-moy-matrixprop}.

Grosse and Wulkenhaar have performed a smooth regularization directly on the multi-indices, by comparing $|m|$ and $\Lambda^2\theta$, where $\Lambda$ is a cut-off. Then, the Polchinski method has been used, in the wilsonian point of view of renormalization (see section \ref{sec-ren-wilson}). Indeed, the Polchinski equation $\Lambda\partial_\Lambda Z(J,\Lambda)=0$ (where $Z(J,\Lambda)$ is the partition function of the theory), which is another formulation of the renormalization group equation \eqref{eq-ren-rgefunc}, can produce boundaries for the Green functions, and by using these boundaries, ones shows that the theory is renormalizable.

Let $G$ be a graph of the theory \eqref{eq-moy-actmatrix}. We denote by $g$ its genus, $n$ the number of its vertices, $N$ the number of its external legs, and $B$ the number of its broken faces. The superficial degree of divergence of the graph $G$ in this theory is given by \cite{Grosse:2003aj,Grosse:2003nw,Grosse:2004yu}:
\begin{equation*}
d^{\text{nc}}(G)=d^{\text{c}}(G)-D(2g+B-1),
\end{equation*}
where $d^{\text{c}}(G)$ is the superficial degree of divergence of $G$ \eqref{eq-ren-powcount} corresponding to the commutative $\phi^4$ theory:
\begin{equation*}
d^{\text{c}}(G)=D+(D-4)n+(2-D)\frac N2.
\end{equation*}
Then, the analog of Theorem \ref{thm-ren-powcount} is valid:
\begin{theorem}
\label{thm-moy-powcountmatr}
For the theory given by \eqref{eq-moy-actmatrix}, the amplitude of a graph $G$ is absolutely convergent if and only if the degree $d^{\text{nc}}$ of $G$ and of its 1PI subgraphs are strictly negative.
\end{theorem}
One can notice that the planar ($g=0$) graphs with $1$ broken face have the same superficial degree of convergence as the corresponding graphs in the commutative theory. This result was already proven by Filk \cite{Filk:1996dm} in the theory \eqref{eq-moy-actscal} on the Moyal space. In $D=4$ dimensions, we deduce from Theorem \ref{thm-moy-powcountmatr} that the subgraphs which need to be renormalized in this theory are the 2- and 4-points planar subgraphs with one broken face ($g=0$, $B=1$, $N=2$ or $4$). Finally, the divergences can be absorbed by a change in the initial condition of the Polchinski equation (see subsection \ref{subsec-ren-renphi4} for the commutative case).

\begin{theorem}
\label{thm-moy-renphi4}
The scalar quantum field theory defined by the action \eqref{eq-moy-actharm}, or equivalently \eqref{eq-moy-actmatrix}, is renormalizable to all orders of perturbation, for $D=4$.
\end{theorem}

Another proof of this theorem, shorter than \cite{Grosse:2004yu}, has been given in \cite{Rivasseau:2005bh} by using the multiscale analysis (see \cite{Rivasseau:1991} and the end of section \ref{sec-ren-bphz}).

\subsubsection{In the configuration space}

When the propagator of the action \eqref{eq-moy-actharm} has been exhibited \cite{Gurau:2005qm} in the configuration space:
\begin{align}
C(x,y)&=\frac{\theta}{4\Omega}\left(\frac{\Omega}{\pi\theta}\right)^{\frac D2}\int \frac{\dd\alpha}{\sinh^{\frac D2}(\alpha)}e^{-\frac{\mu^2\alpha}{2\tilde\Omega}}C(x,y,\alpha),\nonumber\\
C(x,y,\alpha)&= \exp\left(-\frac{\tilde\Omega}{4}\coth(\frac{\alpha}{2})(x-y)^2-\frac{\tilde\Omega}{4}\tanh(\frac{\alpha}{2})(x+y)^2\right),\label{eq-moy-propagx}
\end{align}
where $\tilde\Omega=\frac{2\Omega}{\theta}$, a proof of the renormalizability of the theory in this space was possible. It has been performed in the multiscale analysis setting \cite{Gurau:2005gd}.
\begin{figure}[htb]
  \centering
  \includegraphics{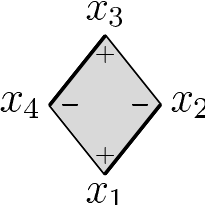}
  \caption{Vertex in the Moyal space}
  \label{fig-moy-vertexconf}
\end{figure}

The interaction of \eqref{eq-moy-actharm} can be reexpressed as:
\begin{multline}
\int\dd^Dx(\phi\star\phi\star\phi\star\phi)(x)=\\
\frac{1}{(\pi\theta)^{D}}\int\left(\prod_{a=1}^4\dd^Dx_a\,\phi(x_a)\right) \delta(x_1-x_2+x_3-x_4)e^{-i\sum_{a<b}(-1)^{a+b+1}x_a\wedge x_b},\label{eq-moy-inter}
\end{multline}
and represented by Figure \ref{fig-moy-vertexconf}, where the signs of the corners correspond to the signs in the Dirac distribution of \eqref{eq-moy-inter}. Since the interaction is invariant under a cyclic permutation on the indices $a$ of the positions $x_a$, one sign on the four can be chosen, the others are imposed. Such a choice is called an orientation of the vertex.

\begin{figure}[htbp]
  \centering
  \subfloat[Planar orientable]{{\label{fig-moy-orientgraph}}\includegraphics[scale=1]{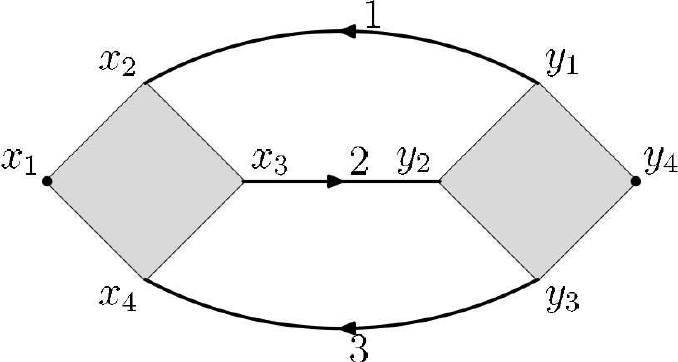}}\qquad
  \subfloat[Non-planar orientable]{\label{fig-moy-orientnonplanar}\includegraphics[scale=1]{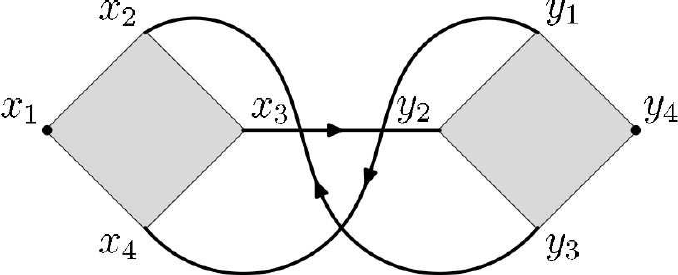}}
  \caption{Graphs in the Moyal space}
  \label{fig-moy-orientable}
\end{figure}

Let us introduce a topological notion for the graphs, which is used in \cite{Gurau:2005gd}.
\begin{definition}
\label{def-moy-orientable}
Let $G$ be a graph of the theory \eqref{eq-moy-actharm}.
\begin{itemize}
\item An internal line of $G$ is called {\it orientable} if it joins two corners (of vertices) of opposite signs.
\item The graph $G$ is said to be {\it orientable} if it is possible to choose the orientations of the vertices of $G$ such that all the internal lines are orientable. Otherwise, $G$ is called {\it non-orientable}.
\end{itemize}
\end{definition}
Figure \ref{fig-moy-orientable} depicts examples of orientable (planar and non-planar) graphs. The orientability of graphs is an important notion in the framework of Moyal space, because it can be shown that only non-orientable graphs suffer from UV/IR mixing (see section \ref{sec-moy-uvir}). Note also that all planar graphs with only one broken face are orientable, while the contrary is false.

Then, in $D=4$ dimensions, the new superficial degree of divergence found in \cite{Gurau:2005gd} satisfies:
\begin{itemize}
\item if $G$ is orientable, $d(G)\leq 4-N$.
\item if $G$ is orientable with $g=0$ and $B\geq2$, $d(G)\leq -N$.
\item if $G$ is orientable with $g\geq 1$, $d(G)\leq -4-N$.
\item if $G$ is non-orientable, $d(G)\leq -N$.
\end{itemize}
From the power-counting involving the above superficial degree of divergence $d(G)$, one can deduce, like in the matrix basis analysis, that only the 2- and 4-points planar subgraphs with one broken face have to be renormalized.

But how is solved the problem of UV/IR mixing (see section \ref{sec-moy-uvir})? The new theory breaks the translation invariance since $C(x,y)\neq C(x-y)$. And this lack of translation invariance is responsible of the removing of the UV/IR mixing. Indeed, the amplitude of the non-planar tadpole (Figure \ref{fig-moy-nonplanartadpole}) has the same IR-behavior as before, but the propagators joining these non-planar tadpoles in a higher loop order graph are now of the type \eqref{eq-moy-propagx} and remove the IR-divergence thanks to this translational symmetry breaking. Thus, both planar irregular and non-planar graphs are finite.

However, the locality of all the terms in the action, a key ingredient of BPHZ renormalization is no longer true on the Moyal space, because of the 4-points interaction. Nevertheless, it has been proven \cite{Gurau:2005gd} that a Taylor expansion at the first order for the 4-points planar regular subgraphs (with $g=0$ and $B=1$), and at the third order for the 2-points planar regular subgraphs (with $g=0$ and $B=1$) provides divergent terms of the form of the Lagrangian \eqref{eq-moy-actharm}, and that the rest of this Taylor expansion is convergent, so that BPHZ renormalization can be performed on the planar regular sector (see section \ref{subsec-ren-bphzsub}). The planar irregular and the non-planar sectors are already convergent, which means that renormalization is not needed.

\subsection{Principal properties}
\label{subsec-moy-prop}

In this subsection, we will study some properties of the noncommutative scalar field theory with harmonic term \eqref{eq-moy-actharm}. Let us start by a special symmetry of the theory, called the Langmann-Szabo duality.

\subsubsection{Langmann-Szabo duality}

A special symmetry of the interaction of the initial theory \eqref{eq-moy-actscal} has been pointed out by Langmann and Szabo in \cite{Langmann:2002cc}. Let us reformulate here this duality in terms of the symplectic Fourier transformation (see \ref{subsec-moy-sympl}). For a real scalar field $\phi\in\caS(\gR^D)$, from \eqref{eq-moy-parsint}, one finds
\begin{equation*}
\int\dd^Dx\ (\phi\star\phi\star\phi\star\phi)(x)=\int\dd^Dk\ ((\hat\phi)^\dag\star\hat\phi\star(\hat\phi)^\dag\star\hat\phi)(k).
\end{equation*}
Recall that $(\hat\phi)^\dag(k)=\hat\phi(-k)$ since $\phi$ is real. By using a modified symplectic Fourier transformation:
\begin{equation}
\tilde \phi(k_a)=\frac{1}{(\pi\theta)^{\frac D2}}\int \dd^Dx\ \phi(x)e^{-i(-1)^ak_a\wedge x},\label{eq-moy-cyclsymplFourier}
\end{equation}
where $a\in\{1,\dots,4\}$ refers to the indices in the Dirac distribution of \eqref{eq-moy-inter}, one obtains the Langmann-Szabo duality of the interaction:
\begin{equation}
\lambda\int\dd^Dx\ (\phi\star\phi\star\phi\star\phi)(x)=\lambda\int\dd^Dk\ (\tilde\phi\star\tilde\phi\star\tilde\phi\star\tilde\phi)(k).\label{eq-moy-lsint}
\end{equation}
This new convention satisfies:
\begin{align*}
&\tilde\phi(k_{2b})=\hat\phi(k_{2b}),\\
&\tilde\phi(k_{2b+1})=(\hat\phi)^\dag(k_{2b+1})=\hat\phi(-k_{2b+1}).
\end{align*}

However, the quadratic part of the initial action \eqref{eq-moy-actscal} does not possess this duality, and the theory exhibits UV/IR mixing, which spoils its renormalizability. The renormalizable theory with harmonic term involves this duality for the interaction term, which is the same as the initial theory, but also at the level of its quadratic terms. Indeed, by using the identities \eqref{eq-moy-symplder} together with the (symplectic) Parseval identity \eqref{eq-moy-parseval}, we find:
\begin{equation*}
\int\dd^Dx\Big(\frac 12(\partial_\mu\phi)^2 +\frac{\Omega^2}{2}\widetilde x^2\phi^2 +\frac{\mu^2}{2}\phi^2\Big)= \int\dd^Dk\Big(\frac 12\widetilde k^2\tilde\phi^2 +\frac{\Omega^2}{2}(\partial_\mu\tilde\phi)^2 +\frac{\mu^2}{2}\tilde\phi^2\Big)
\end{equation*}
Then, the action \eqref{eq-moy-actharm} satisfies:
\begin{equation*}
S[\phi;\mu,\lambda,\Omega]=\Omega^2S\Big[\tilde\phi;\frac{\mu}{\Omega},\frac{\lambda}{\Omega^2},\frac{1}{\Omega}\Big],
\end{equation*}
and is covariant under the Langmann-Szabo duality, or the (modified \eqref{eq-moy-cyclsymplFourier}) symplectic Fourier transformation. In the special case $\Omega=1$, the action is invariant under this duality. This symmetry seems therefore to play a crucial role in the renormalizability of quantum field theories to avoid the problem of the dangerous UV/IR mixing.

Note that the Langmann-Szabo duality is simpler reexpressed in terms of the symplectic Fourier transformation, and that this symplectic Fourier transformation is more natural on the Moyal space. This duality can also be described in terms of the metaplectic representation of the symplectic group \cite{Bieliavsky:2008qy}, and it is then related \cite{deGoursac:2010zb} to the superalgebraic interpretation presented in section \ref{sec-gauge-interp}.

\subsubsection{Beta functions}

Concerning the renormalization flows, it has been shown up to three loops, that the constant $\Omega$ was running towards a fixed point $\Omega=1$ \cite{Disertori:2006uy}, which is also a special value for the Langmann-Szabo duality. Moreover, for this value $\Omega=1$, the flow of the coupling constant has been shown to vanish up to irrelevant term to all orders in perturbation \cite{Disertori:2006nq} (see \cite{Grosse:2004by} for the one-loop order, \cite{Disertori:2006uy} for the two and three loops order, and also \cite{Gurau:2009ni}). Then, this theory does not involve a Landau ghost, contrary to the commutative $\phi^4$ theory (see subsection \ref{subsec-ren-betafun}).

\subsubsection{Parametric representation}

A parametric representation as in the commutative theory (see \cite{Bergere:1976}) has been constructed for this noncommutative model in $D=4$ dimensions \cite{Gurau:2006yc} (see also \cite{Rivasseau:2007qx,Gurau:2007az}, and \cite{Tanasa:2007ai,deGoursac:2007qi} for an introduction). The propagator is given by Equation \eqref{eq-moy-propagx} for $D=4$. Thanks to the following identity
\begin{equation*}
\delta(x_1-x_2+x_3-x_4)=\int\frac{\dd^4p}{\pi^4\theta^4}e^{-ip\wedge(x_1-x_2+x_3-x_4)},
\end{equation*}
the vertex \eqref{eq-moy-inter} becomes
\begin{align*}
\int \dd^4x(\phi^\dag\star\phi\star\phi^\dag\star\phi) &=\frac{1}{\pi^8\theta^8}\int\prod_{i=1}^4\dd^4x_i\,\dd^4p\,\phi^\dag(x_1)\phi(x_2)\phi^\dag(x_3)\phi(x_4) V(x_1,x_2,x_3,x_4,p),\\
V(x_1,x_2,x_3,x_4,p)&=\exp\left(-i\sum_{i<j}(-1)^{i+j+1}x_i\wedge x_j-i\sum_{i}(-1)^{i+1}p\wedge x_i\right).
\end{align*}

Consider now a graph $G$ for the theory \eqref{eq-moy-actharm} with a set $V$ of $n$ internal vertices, $N$ external legs and a set $L$ of $(2n-\frac N 2)$ internal lines or propagators. In this theory, there are four positions (called ``corners'') associated to each vertex $v\in V$, and each corner is bearing either a half internal line or an external field. These corners are denoted by $x^v_i$, where $i\in\{1,..,4\}$ is given by the cyclic order of the Moyal product. The set $I\subset V\times\{1,..,4\}$ of internal corners (hooked to some internal line) has $4n-N$ elements whereas the set $E=V\times\{1,..,4\}\setminus I$ of external corners has $N$ elements. Each vertex $v\in V$ carries also a hypermomentum, which is noted $p_v$. A line $l\in L$ of the graph $G$ joins two corners in $I$, and we note their positions by $x^{l,1}$ and $x^{l,2}$. We will also note the external corners $x_e$. Be careful that each corner has two notations for its position. In these notations, we can express the amplitude $\mathcal A_G$ of such a graph $G$
\begin{multline*}
\mathcal A_G(\{x_e\})=\left(\frac{\Omega}{4\pi^2\theta}\right)^{2n-\frac N 2}\left(\frac{1}{\pi^8\theta^8}\right)^n \int_0^\infty \prod_{l\in L}\frac{\dd\alpha_l}{\sinh^2(\alpha_l)} \int \prod_{(v,i)\in I}\dd^4x^v_i\prod_{v\in V}\dd^4p_v\\
\times\prod_{l\in L}C(x^{l,1},x^{l,2},\alpha_l)\prod_{v\in V} V(x^v_1,x^v_2,x^v_3,x^v_4,p_v).
\end{multline*}

We take here the mass $\mu=0$ to simplify calculations. The article \cite{Gurau:2006yc} dealing with the parametric representation tells us that by performing Gaussian integrations, we get a Gaussian function of the external variables, divided by a determinant. If we note $t_l=\tanh(\frac{\alpha_l}{2})$, using $\sinh(\alpha_l)=\frac{2t_l}{1-t_l^2}$, then the expression of the amplitude is given by
\begin{equation}
\mathcal A_G(\{x_e\})=\left(\frac{\theta}{8\Omega}\right)^{2n-\frac N 2}\left(\frac{1}{\pi^4\theta^4}\right)^n \int_0^\infty \prod_{l\in L}(\dd\alpha_l(1-t_l^2)^2) HU_G(t)^{-2}e^{-\frac{HV_G(x_e,t)}{HU_G(t)}},\label{eq-moy-polydef}
\end{equation}
and that stands for a definition of the polynom $HU_G(t)$ in the $t$ variables and the quadratic form $HV_G(x_e,t)$ in the external variables $x_e$, which is also polynomial in the $t$ variables. Let us now recall the computation of $HU$ and $HV$.
\begin{multline*}
\mathcal A_G(\{x_e\})=\left(\frac{\Omega}{16\pi^2\theta}\right)^{2n-\frac N 2}\left(\frac{1}{\pi^8\theta^8}\right)^n \int_0^\infty \prod_{l\in L}\frac{\dd\alpha_l(1-t_l^2)^2}{t_l^2} \int \prod_{(v,i)\in I}\dd^4x^v_i\prod_{v\in V}\dd^4p_v\\
\times\prod_{l\in L} e^{-\frac{\widetilde \Omega}{4t_l}(x^{l,1}-x^{l,2})^2 -\frac{\widetilde \Omega t_l}{4}(x^{l,1}+x^{l,2})^2} \prod_{v\in V} V(x^v_1,x^v_2,x^v_3,x^v_4,p_v).
\end{multline*}

At this point, we introduce the short variables $u$ and the long ones $v$
\begin{align*}
u_l=&\frac{x^{l,1}-x^{l,2}}{\sqrt 2}\\
v_l=&\frac{x^{l,1}+x^{l,2}}{\sqrt 2}
\end{align*}
associated to each line $l\in L$. Notice that the Jacobian of the transformation is 1. Each internal line $l\in L$ of the graph $G$ joins two vertices (or one two times). This fact will be expressed by the $((2n-\frac N 2)\times 4)$-dimensional incidence matrix $\epsilon^v$ \cite{Filk:1996dm} associated to each vertex $v\in V$. We define $\epsilon^v_{li}=(-1)^{i+1}$ if the line $l\in L$ hooks the vertex $v$ at corner $i\in\{1,..,4\}$, $\epsilon^v_{li}=0$ if not, and $\eta^v_{li}=|\epsilon^v_{li}|$. Then the short and long variables are given by
\begin{align*}
u_l&=\frac{1}{\sqrt 2}\sum_{v\in V}\sum_{i=1}^4\epsilon^v_{li}x^v_i\\
v_l&=\frac{1}{\sqrt 2}\sum_{v\in V}\sum_{i=1}^4\eta^v_{li}x^v_i,
\end{align*}
and for $(v,i)\in I$ (and not for $(v,i)\in E$),
\begin{equation*}
x^v_i=\frac{1}{\sqrt 2}\sum_{l\in L}(\epsilon^v_{li}u_l+\eta^v_{li}v_l).
\end{equation*}
The vector $\chi_i^v=1$ if $(v,i)\in E$ and $\chi_i^v=0$ if not, shows how the external legs are hooked on vertices. We define also $\omega_{ij}=1$ if $i<j$, $\omega_{ij}=-1$ if $j>i$ and $\omega_{ii}=0$. The amplitude is then given by
\begin{multline}
\mathcal A_G(\{x_e\})=\left(\frac{\Omega}{16\pi^2\theta}\right)^{2n-\frac N 2}\left(\frac{1}{\pi^8\theta^8}\right)^n \int_0^\infty \prod_{l\in L}\frac{\dd\alpha_l(1-t_l^2)^2}{t_l^2} \int \prod_{l\in L}\dd^4u_l\dd^4v_l\prod_{v\in V}\dd^4p_v \\
\prod_{l\in L} e^{-\frac{\widetilde \Omega}{2t_l}u_l^2 -\frac{\widetilde \Omega t_l}{2}v_l^2} \prod_{v\in V} \exp\Big(-i\sum_{i<j}\frac{(-1)^{i+j+1}}{2}\sum_{l,l'\in L} (\epsilon^v_{li}u_l+\eta^v_{li}v_l)\wedge (\epsilon^v_{l'j}u_{l'}+\eta^v_{l'j}v_{l'})\\
-i\sum_{i}\frac{(-1)^{i+1}}{\sqrt 2}\sum_{l\in L} p_v\wedge (\epsilon^v_{li}u_l+\eta^v_{li}v_l) -i\sum_{i\neq j}\frac{(-1)^{i+j+1}}{\sqrt 2}\sum_{l\in L} \chi^v_i\omega_{ij}x^v_i\wedge (\epsilon^v_{lj}u_l+\eta^v_{lj}v_l)\\
-i\sum_{i}(-1)^{i+1}\chi^v_i p_v\wedge x^v_i -i\sum_{i<j}(-1)^{i+j+1} \chi^v_i\chi_j^v x^v_i\wedge x^v_j\Big).\label{eq-moy-bigexp}
\end{multline}
The sums over $i$ and $j$ are between 1 and 4. We note
\begin{equation*}
X=\begin{pmatrix} x_e \\ u \\ v \\ p \end{pmatrix}\quad\text{and}\quad G=\begin{pmatrix} M & P \\ P^T & Q \end{pmatrix},
\end{equation*}
where $M$ is a $4N\times 4N$ matrix (quadratic form for external variables), $Q$ is a $[8(2n-\frac N 2)+4n]\times [8(2n-\frac N 2)+4n]$ matrix (quadratic form for short and long variables and hypermomenta), and $P$ is the coupling. The following expression together with \eqref{eq-moy-bigexp} define $M$, $P$ and $Q$,
\begin{equation*}
\mathcal A_G(\{x_e\})=\left(\frac{\Omega}{16\pi^2\theta}\right)^{2n-\frac N 2}\left(\frac{1}{\pi^8\theta^8}\right)^n \int_0^\infty \frac{\dd\alpha(1-t^2)^2}{t^2} \int \dd u\dd v\dd p\, e^{-\frac 1 2 X^TGX}.
\end{equation*}
By performing Gaussian integrations, we obtain
\begin{equation*}
\mathcal A_G(\{x_e\})=\left(\frac{\pi^2\Omega}{\theta}\right)^{2n-\frac N 2}\left(\frac{4}{\pi^6\theta^8}\right)^n \int_0^\infty \dd\alpha(1-t^2)^2\, \frac{e^{-\frac 1 2 (x_e)^T(M-PQ^{-1}P^T)(x_e)}}{\sqrt{\det Q}}.
\end{equation*}
We deduce from the latter equation the expression of polynoms $HU_G(t)$ and $HV_G(x_e,t)$ defined in \eqref{eq-moy-polydef}. The noncommutative analogs of Symanzik polynomials, $HU_G$ and $HV_G$, are topological characterizations of the graph $G$ of the theory \eqref{eq-moy-actharm}. Note that a dimensional renormalization of \eqref{eq-moy-actharm} has been performed \cite{Gurau:2007fy} by using this parametric representation. Moreover, a Hopf algebra structure ``\`a la Connes-Kreimer'' of the graphs has been exhibited for this theory \cite{Tanasa:2007xa}.

\subsection{Vacuum configurations}
\label{subsec-moy-vacuum}

In this subsection, we will study the equation of motion of the action \eqref{eq-moy-actharm} in the special case $\Omega=1$, and determine the vacuum configurations, that is the solutions of the equation of motion which are minima of the action \eqref{eq-moy-actharm} \cite{deGoursac:2007uv}. Note that $\Omega=1$ is stable under the renormalization group, so that it is consistent to consider the theory at this point $\Omega=1$. We will also restrict our study to the two-dimensional case to simplify the notations, but it will be extended to higher dimensions. These vacuum configurations are very important for quantum field theory since one has to expand around a vacuum in the perturbation theory to compute Feynman graphs. Of course, the configuration $\phi=0$ is a solution of the equation of motion, and it is the only vacuum if the mass parameter $-\mu^2\geq0$. In the case of negative mass term, we will see that other vacuum configurations appear, and this phenomenon is directly related to spontaneous symmetry breaking in the case of commutative theory.

\subsubsection{Equation of motion}

Let us reformulate the action \eqref{eq-moy-actharm}:
\begin{equation}
S(\phi)=\int \dd^Dx\big(\frac 12\partial_\mu\phi\fois\partial_\mu\phi +\frac{\Omega^2}{2}\widetilde x^2\phi^2 -\frac{\mu^2}{2}\phi^2+\lambda\,\phi\star\phi\star\phi\star\phi\big),\label{eq-moy-actvac}
\end{equation}
where the sign of the mass term has been changed and we take $\mu^2\geq0$. Then, the equation of motion derived from \eqref{eq-moy-actvac} writes  
\begin{equation*}
-\partial^2\phi+\Omega^2\widetilde{x}^2\phi-\mu^2\phi+4\lambda\phi\star\phi\star\phi=0,
\end{equation*}
which for $\Omega=1$ can be rewritten as
\begin{equation}
\frac 12(\wx^2\star\phi +\phi\star\wx^2)-\mu^2\phi+4\lambda\,\phi\star\phi\star\phi=0.\label{eq-moy-mvt}
\end{equation}
With the properties of subsection \ref{subsec-moy-matrix}, the equation of motion \eqref{eq-moy-mvt} can be reexpressed in the matrix basis into (for $D=2$):
\begin{equation}
\frac{4}{\theta}(m+n+1)\phi_{mn}-\mu^2\phi_{mn}+4\lambda\,\phi_{mk}\phi_{kl}\phi_{ln}=0.\label{eq-moy-mvtmatrix}
\end{equation}

\subsubsection{Reduction of solutions by symmetry}

In the general case, one has to solve a cubic matrix equation \eqref{eq-moy-mvtmatrix}, which is a difficult task. Let us take into account the symmetries of the theory in order to simplify the resolution of \eqref{eq-moy-mvtmatrix}. The symmetry group of the euclidean Moyal algebra $\caM$ in $D$ dimensions is {\it a priori}:
\begin{equation*}
G_D=SO(D)\cap Sp(D),
\end{equation*}
where $SO(D)$ is the group of rotations and $Sp(D)$ is the group of symplectic isomorphisms. In fact, it has been shown \cite{deGoursac:2009fm} that the whole orthogonal group $O(D)$ is a symmetry of the model \eqref{eq-moy-actharm} at both the classical and the quantum level, if it acts also on the symplectic structure $\Sigma$ used in the definition of the Moyal product (see subsection \ref{subsec-moy-schwartz}). But here, it is sufficient to consider only $G_D$, which acts only on the scalar fields $\phi$ by:
\begin{equation*}
\forall \Lambda\in G_D,\quad \phi^\Lambda(x)=\phi(\Lambda^{-1}x).
\end{equation*}
The action $S(\phi)$ \eqref{eq-moy-actharm} is invariant under these transformations. The requirement that the new action obtained from the expansion of $S(\phi)$ around a non-trivial vacuum, is also invariant under $G_D$, permits to obtain a condition on all these vacua $\phi$:
\begin{equation}
\forall\Lambda\in G_D,\quad \phi^\Lambda=\phi,\label{eq-moy-condvacsymm}
\end{equation}
namely, they are invariant under $G_D$. From the theory of invariants of groups \cite{Weyl:1946}, one can infer that $\phi$ depends only on $x^2$. Furthermore, by Formula \eqref{eq-moy-coeffmatrix}, one deduces that the matrix components of $\phi$ satisfy:
\begin{equation}
\phi_{mn}=a_m\delta_{mn},\label{eq-moy-vacformmatrix}
\end{equation}
where $a_m$ are arbitrary real numbers. Then, one has to solve the equation of motion \eqref{eq-moy-mvt} for fields $\phi$ of the form:
\begin{equation}
\phi(x)= \sum_{m=0}^\infty a_{m} f_{mm}(x),\label{eq-moy-vacform}
\end{equation}
where $(a_m)\in\gR^{\gN}$.

\subsubsection{Solutions of the equation of motion}

By insertion of \eqref{eq-moy-vacformmatrix} in \eqref{eq-moy-mvtmatrix}, one finds: $\forall m\in\gN$,
\begin{equation*}
a_{m}\Big(\frac{4}{\theta}(2m+1)-\mu^2+4\lambda\,a_{m}^2\Big)=0.
\end{equation*}
It then leads to the following condition on the coefficients $a_m$:
\begin{equation*}
a_m=0\quad\text{or}\quad a_{m}^2=\frac{1}{\lambda\theta}\left(\frac{\mu^2\theta}{4}-2m-1\right).
\end{equation*}
If the coefficient $a_m\neq0$, the mass has to satisfy:
\begin{equation}
\mu^2>\frac{4}{\theta}(2m+1).\label{eq-moy-condvac}
\end{equation}

Now, upon setting
\begin{equation*}
p=\lfloor \frac{\mu^2\theta}{8}-\frac 12 \rfloor,
\end{equation*}
where $\lfloor.\rfloor$ denotes the integer part, the discussion proceeds along the sign of $p$. Namely,
\begin{itemize}
\item if $p$ is negative, no index $m$ can satisfy the condition \eqref{eq-moy-condvac}.
\item if $p$ is positive, there are $(p+1)$ possible indices $m$, satisfying the constraint \eqref{eq-moy-condvac}.
\end{itemize}

Owing to the analysis given above, one readily infers that the sum involved in \eqref{eq-moy-vacform} cannot run to infinity simply because one must have $m\leq p$.\\
Therefore, $v(x)=\sum_{m=0}^p a_{m} f_{mm}(x)$ are the general solutions of the equation of motion \eqref{eq-moy-mvt} symmetric under rotations.

One has for the coefficients of $v\star v\star\phi$ and $v\star\phi\star v$:
\begin{align*}
(v\star v\star\phi)_{mn}&=\big(\sum_{k=0}^p a_k^2\delta_{mk}\big)\phi_{mn},\\
(v\star\phi\star v)_{mn}&=\big(\sum_{k,l=0}^p a_ka_l\delta_{mk}\delta_{nl}\big)\phi_{mn},
\end{align*}
where $v$ is a solution and $\phi$ a general field.

\subsubsection{Minima of the action}

Let us now check if these solutions $v(x)$ are minima for the action. The quadratic part of the action \eqref{eq-moy-actvac} writes
\begin{equation}  
S_{quadr}=2\pi\theta\Big(\frac{2}{\theta}(m+n+1)-\frac{\mu^2}{2}+2\lambda\sum_{k=0}^p a_k^2(\delta_{mk}+\delta_{nk})+ 2\lambda\sum_{k,l=0}^p a_ka_l\delta_{mk}\delta_{nl}\Big)\phi_{nm}\phi_{mn}.\label{eq-moy-actmatreexpr}
\end{equation}
Due to the form of the action, the propagator $C_{mn,kl}$ is diagonal $C_{mn,kl}=C_{mn}\delta_{m l}\delta_{nk}$. In order to have a minimum, one thus needs $C_{mn}$ to be positive, for all $m,n\in\gN$. From \eqref{eq-moy-actmatreexpr}, one has
\begin{equation*}
C_{mn}^{-1}=\alpha_{mn}+4\lambda\pi\theta\sum_{k=0}^p a_k^2(\delta_{mk}+\delta_{nk})+ 4\lambda\pi\theta\sum_{k,l=0}^p a_ka_l\delta_{mk} \delta_{nl},
\end{equation*}
where
\begin{equation*}
\alpha_{mn}=4\pi(m+n+1)-\mu^2\pi\theta,\mbox{ and } a_k^2=0\quad\text{or}\quad a_k^2=-\frac{\alpha_{kk}}{4\lambda\pi\theta}.
\end{equation*}

For $p\geq 0$, one has to distinguish between the following cases.
\begin{itemize}
\item $m>p$ and $n>p$:
\begin{equation*}
C^{-1}_{mn}=\alpha_{mn}=4\pi(m+n-2(\frac{\mu^2\theta}{8}-\frac 12))>0.
\end{equation*}
\item $m\leq p$ and $n>p$:
if $a_m^2=0$ then
\begin{equation}
C^{-1}_{mn}=\alpha_{mn}=4\pi(m+n-2(\frac{\mu^2\theta}{8}-\frac 12)).\label{eq-moy-condpropag}
\end{equation}
In order not to have for certain value of $n$ ($p< n\leq 2p-m$) $C_{mn}^{-1}<0$, one needs $a_m^2=-\frac{\alpha_{mm}}{4\lambda\pi\theta}$. In this case, we have
\begin{equation*}
C^{-1}_{mn}=\alpha_{mn}-\alpha_{mm}=4\pi(n-m)>0.
\end{equation*}
\item $m>p$ and $n\leq p$: is treated along the same lines as above. $C^{-1}_{mn}=4\pi(m-n)>0$ because we assumed that $\forall k\in \{0,..,p\}$, $a_k^2=-\frac{\alpha_{kk}}{4\lambda\pi\theta}$.
\item $m\leq p$ and $n\leq p$:
\begin{equation*}
C^{-1}_{mn}=\alpha_{mn}-\alpha_{mm}-\alpha_{nn}+\sqrt{\alpha_{mm}\alpha_{nn}}=-\alpha_{mn}+\sqrt{\alpha_{mm}\alpha_{nn}}\geq 0,
\end{equation*}
and $C^{-1}_{mn}=0$ holds if and only if $m=n=\frac{\mu^2\theta}{8}-\frac 12\in\mathbb N$.
\end{itemize}

From the above analysis, one can conclude that one has a positive defined propagator just for a single class of solutions. These solutions, which are minima of the action \eqref{eq-moy-actvac}, correspond to
\begin{equation}
v(x)=\sum_{k=0}^p a_k f_{kk}(x)\label{eq-moy-zevacuum}
\end{equation}
where $a_{k}^2=\frac{1}{\lambda\theta}\left(\frac{\mu^2\theta}{4}-2k-1\right)$.

For $\phi(x)=\sum_{k,l=0}^\infty\phi_{mn}f_{mn}(x)$ a solution of equation of motion \eqref{eq-moy-mvt}, we compute the value of the action \eqref{eq-moy-actvac} in the matrix basis
\begin{equation*}
S[\phi]=2\pi\sum_{k,l=0}^\infty\left( m+n-2\big(\frac{\mu^2\theta}{8}-\frac 12\big)\right)|\phi_{mn}|^2.
\end{equation*}
So for $p<0$,
\begin{equation}
S[\phi]\geq S[0]=0,\label{eq-moy-actmin}
\end{equation}
and for $p\geq 0$ and $v(x)$ the vacuum \eqref{eq-moy-zevacuum},
\begin{equation*}
S[v]=-\frac{8\pi}{\lambda \theta} \sum_{k=0}^p (\frac{\mu^2\theta}{8}-\frac12-k)^2<0.
\end{equation*}

At this point, a remark is to be done. In commutative QFT or in non-commutative QFT without the harmonic term, one has a phenomenon of spontaneously symmetry breaking as soon as the mass parameter is taken to be negative. For the model considered here this is not the case anymore. Indeed, if the mass parameter does not go beyond a certain limit $\mu^2=\frac{4}{\theta}$, then from \eqref{eq-moy-actmin}, we see that $\phi(x)=0$ is the global minimum of the action \eqref{eq-moy-actvac}, since $p$ is negative. So the harmonic term will prevent the phenomenon of spontaneously symmetry breaking from happening. 

When the mass parameter exceeds this critical value (i.e. $p\geq 0$), one can see from \eqref{eq-moy-condpropag} that $\phi(x)=0$ is no longer a local minimum of the action. Therefore one has to consider the non-trivial vacuum $v(x)$, given by \eqref{eq-moy-zevacuum}. Note that \eqref{eq-moy-zevacuum} corresponds to a different solution when changing the value of the limit 
parameter $p$. Let us now stress on some of the features of these vacuum configurations. Owing again to the properties of the matrix basis (see subsection \ref{subsec-moy-matrix}), it can be realized that \eqref{eq-moy-zevacuum} does not vanish for $x=0$ while it decays at infinity (as a {\it finite} linear combination of the Schwartz functions $f_{mn}(x)$).

\subsubsection{Extension in higher dimensions}

Let us further indicate that these results of \cite{deGoursac:2007uv} can be extended in arbitrary even $D$ dimensions. Indeed, the equation of motion in the matrix basis is given by
\begin{equation}
\frac{4}{\theta}\left(|m|+|n|+\frac D2\right)\phi_{mn}-\mu^2\phi_{mn}+4\lambda\phi_{mk}\phi_{kl}\phi_{ln}=0,\label{eq-moy-mvtD}
\end{equation}
where $m,n,k,l\in\gN^{\frac D2}$ are multi-indices, and $k$ and $l$ are summed over. The above analysis permits to obtain the following theorem (see \cite{deGoursac:2007uv}).
\begin{theorem}
\label{thm-moy-vac}
The set of all minima (vacuum configurations) of the action \eqref{eq-moy-actharm}, satisfying the condition \eqref{eq-moy-condvacsymm}, is given by:
\begin{equation}
v(x)=\sum_{k\in\gN^{\frac D2},\,|k|\leq p}a_k\, b_{kk}^{(D)}(x),\qquad a^2_k=\frac{1}{\lambda\theta}\left(\frac{\mu^2\theta}{4}-\frac D2-2|k|\right),\label{eq-moy-zevacuumD}
\end{equation}
where $p=\lfloor \frac{\mu^2\theta}{4}-\frac{D}{2}\rfloor\in\gZ$.
\end{theorem}

In four dimensions, it is well known that when computing radiative corrections, the mass parameter of a scalar field becomes huge (because of the quadratic divergence). In order to get a low value for the renormalized mass, one may thus consider a non-commutative scalar field theory with harmonic term, a negative mass term and this non-trivial vacuum. Note also that these vacua have been used in the study at the one-loop level of scalar field theory in $2+1$ dimensions \cite{Fosco:2007ww}.

\subsection{Possible spontaneous symmetry breaking?}
\label{subsec-moy-ssb}

We now consider a linear sigma model built from the renormalizable scalar action \eqref{eq-moy-actharm}, assuming again $\Omega=1$ (see \cite{deGoursac:2007uv}). The action involves $N$ real-valued fields $\phi_i$ and is given by
\begin{equation}
S_\sigma=\int \dd^Dx\Big( \frac 12(\partial_\mu\phi_i)(\partial_\mu\phi_i)+\frac 12\widetilde{x}^2\phi_i \phi_i-\frac{\mu^2}{2}\phi_i \phi_i +\lambda\phi_i \star\phi_i\star\phi_j \star\phi_j\Big) \label{eq-moy-actsigma}
\end{equation}
The above action is invariant under the action of the orthogonal group $O(N)$ (as it is also the case  in the absence of the harmonic  term, see \cite{Campbell:2001ek,RuizRuiz:2002hh,Liao:2002rp}). Let
\begin{equation*}
\langle\Phi\rangle=(0, \ldots, 0, v(x)),
\end{equation*}
be a non-zero vacuum expectation value, where $v(x)$ is a minimum \eqref{eq-moy-zevacuumD} obtained from the equation of motion \eqref{eq-moy-mvtD}, as  analyzed in subsection \ref{subsec-moy-vacuum}. Then, shifting $\Phi=(\phi_1,\dots,\phi_N)$ to $\langle\Phi\rangle+\delta\Phi$ with
\begin{equation*}
\delta \Phi=(\pi_1,\ldots, \pi_{N-1},\sigma (x)),
\end{equation*}
one obtains from \eqref{eq-moy-actsigma}:
\begin{align}
S_\sigma=\int \dd^Dx &\Big( \frac 12(\partial_\mu\pi_i) (\partial_\mu\pi_i)+ \frac 12\widetilde{x}^2\pi_i \pi_i-\frac{\mu^2}{2}\pi_i \pi_i+2\lambda v \star v\star\pi_i \star\pi_i+\lambda\pi_i \star\pi_i\star\pi_j \star\pi_j \nonumber\\  
& +\frac 12(\partial_\mu\sigma)(\partial_\mu\sigma)+ \frac 12\widetilde{x}^2\sigma \sigma-\frac{\mu^2}{2}\sigma \sigma+4\lambda v \star v\star\sigma \star\sigma +2\lambda v\star \sigma \star\ v\star\sigma \nonumber\\  
&  +2\lambda \sigma \star v\star\pi_i \star\pi_i+2\lambda v \star \sigma\star\pi_i \star\pi_i +2\lambda \sigma \star \sigma\star\pi_i \star\pi_i\nonumber\\  
&+4\lambda v\star\sigma\star\sigma\star\sigma+\lambda\sigma\star\sigma\star\sigma\star\sigma\Big).\label{eq-moy-actdev}
\end{align}

Consider closely the quadratic part of the action \eqref{eq-moy-actdev} in the fields $\pi$:
\begin{equation}
S_{\text{quadr}}(\pi)=\int \dd^D x \left(\frac 12(\partial_\mu\pi_i)^2+\frac 12\widetilde{x}^2\pi_i\pi_i-\frac{\mu^2}{2}\pi_i\pi_i+2\lambda v\star v\star\pi_i\star\pi_i\right).\label{eq-moy-actquadvac}
\end{equation}
In the absence of the harmonic term, the linear sigma model supports a constant non-zero vacuum configuration leading to the appearance of $N$ massless fields $\pi$. Indeed, the third and the fourth terms in \eqref{eq-moy-actquadvac} balance each other. This is an obvious analog of the Goldstone theorem at the classical level which has been further verified to the one and resp. two loop order in \cite{Campbell:2001ek,RuizRuiz:2002hh} and resp. \cite{Liao:2002rp}.

When the harmonic term is included in \eqref{eq-moy-actquadvac} the situation changes substantially. Indeed, in view of the discussion for the scalar field theory presented above, constant non-zero vacuum configurations are no longer supported by the action, since the invariance under translations is broken. Thus, the cancellation of the $\pi$ mass term does not occur automatically and a careful analysis has to be performed.

In the matrix basis, \eqref{eq-moy-actquadvac} takes the form
\begin{equation*}
S_{\text{quadr}}(\pi)=(2\pi\theta)^{\frac D2}\sum_{m,n\in\gN^{\frac D2}}\left(\frac{2}{\theta}(|m|+|n|+\frac D2-\frac{\mu^2\theta}{4})+2\lambda a_m^2\right)\phi_{mn}\phi_{nm}.
\end{equation*}
By using the explicit expression of $a_m^2$ related to the sign of $p=\lfloor \frac{\mu^2\theta}{4}-\frac{D}{2}\rfloor$, one gets:
\begin{equation*}
S_{\text{quadr}}(\pi)=(2\pi\theta)^{\frac D2}\sum_{m,n\in\gN^{\frac D2},\, |m|>p}\frac{2}{\theta}\left(|m|+|n|+\frac D2-\frac{\mu^2\theta}{4}\right)\phi_{mn}\phi_{nm},
\end{equation*}
which does not correspond to the standard action with harmonic term in the configuration space, because of the condition $|m|>p$ in the sum over $m$. The theory considered here is not stable when expanding around the vacuum $v(x)$.

Therefore, in the case of the noncommutative scalar field theory with harmonic term (and $\Omega=1$), the breaking of translations invariance does not allow a noncommutative analog of the Goldstone theorem, even at the classical level, like in the theory without harmonic term \cite{Campbell:2001ek,RuizRuiz:2002hh,Liao:2002rp}. As stated in the beginning of this subsection, all these calculations have been done in the case of a set of real fields $\phi$ and for the particular value $\Omega=1$. If one allows other values of the parameter $\Omega$, the situation is more intricate.

\subsection{Other renormalizable QFT on Moyal space}
\label{subsec-moy-other}

A degenerate version of \eqref{eq-moy-actharm} in $D=4$ has been studied in \cite{Grosse:2008df}, where the matrix $\Theta$ is not invertible, so that only two directions of the space $\gR^4$ are ``noncommutative''.

\subsubsection{Complex orientable $\phi^4$ theories}

The simplest complex scalar field theory on Moyal space is given by:
\begin{equation}
S(\phi)=\int\dd^Dx \left((\partial_\mu\phi^\dag)(\partial_\mu\phi)+\mu^2\phi^\dag\phi+\lambda_1\,\phi^\dag\star\phi\star\phi^\dag\star\phi+\lambda_2\,\phi^\dag\star\phi^\dag\star\phi\star\phi\right),\label{eq-moy-cplx}
\end{equation}
where $\mu$ is the mass of the field, and $\lambda_1$, $\lambda_2$ are coupling constants. It can be shown by a topological analysis that the theory \eqref{eq-moy-cplx} with $\lambda_2=0$ (and $\lambda_1\neq0$) gives rise only to orientable graphs (see Definition \ref{def-moy-orientable}), and is called the ``orientable'' complex theory. It has been proved to be renormalizable \cite{Aref'eva:2000hq,Chepelev:2000hm} without need of harmonic term or Langmann-Szabo duality, because of this orientability (see subsection \ref{subsec-moy-harm}).
\medskip

However, the orientable complex version of the Grosse-Wulkenhaar model (with harmonic term) is also renormalizable to all orders. The proof \cite{Gurau:2005gd} has been done for a generalized model with covariant derivatives (LSZ-type model) and with harmonic term:
\begin{multline}
S(\phi)=\int \dd^Dx\Big((\partial_\mu\phi-i\alpha\wx_\mu\phi)^\dag\star(\partial_\mu\phi-i\alpha\wx_\mu\phi) +\Omega^2(\wx_\mu\phi)^\dag\star(\wx_\mu\phi)\\
+\mu^2\phi^\dag\star\phi +\lambda\phi^\dag\star\phi\star\phi^\dag\star\phi\Big).\label{eq-moy-lszgen}
\end{multline}
For $\alpha=0$, we find the orientable complex version of the Grosse-Wulkenhaar model. For $\Omega=0$ and $\alpha\neq0$, we find the so-called Langmann-Szabo-Zarembo model, which is exactly solvable for $\alpha=1$ \cite{Langmann:2003cg,Langmann:2003if}.  Note that the LSZ model involves covariant derivatives, like in a magnetic field background $B_\mu=\alpha\wx_\mu$.

By introducing $\omega^2=\alpha^2+\Omega^2$, we can reformulate the action \eqref{eq-moy-lszgen} into:
\begin{equation*}
S[\phi;\mu,\lambda,\alpha,\omega]=\int \dd^Dx\left(|\partial_\mu\phi|^2-2i\alpha\wx_\mu\phi^\dag\partial_\mu\phi+\omega^2\wx^2|\phi|^2+\mu^2|\phi|^2+\lambda\phi^\dag\star\phi\star\phi^\dag\star\phi\right).
\end{equation*}
Then, subsection \ref{subsec-moy-sympl} permits to find a symmetry of the theory, the Langmann-Szabo duality \cite{Langmann:2002cc}, expressed here in terms of the symplectic Fourier transformation:
\begin{equation*}
S[\phi;\mu,\lambda,\alpha,\omega]=\omega^2 S\Big[\hat\phi,\frac{\mu}{\omega},\frac{\lambda}{\omega^2},\frac{\alpha}{\omega^2},\frac{1}{\omega}\Big].
\end{equation*}
We see that the theory is covariant under this Langmann-Szabo duality, and invariant for $\omega=1$. Note that for the complex theory, the Langmann-Szabo duality is given by the symplectic Fourier transformation without need of a cyclic Fourier convention like in the real case (see subsection \ref{subsec-moy-prop}). The propagator of this theory takes the form:
\begin{align*}
C(x,y)=&\left(\frac{\omega}{\pi\theta}\right)^{\frac D2}\int_0^\infty \frac{\dd t}{\sinh^{\frac D2}(\frac{4\omega t}{\theta})}e^{-\mu^2 t} \exp\Big(-\frac{\omega}{2\theta}\,\frac{\cosh(\frac{4\omega t}{\theta})+\cosh(\frac{4\alpha t}{\theta})}{\sinh(\frac{4\omega t}{\theta})}(x-y)^2\\
&-\frac{\omega}{2\theta}\,\frac{\cosh(\frac{4\omega t}{\theta})-\cosh(\frac{4\alpha t}{\theta})}{\sinh(\frac{4\omega t}{\theta})}(x+y)^2-i\omega\frac{\sinh(\frac{4\alpha t}{\theta})}{\sinh(\frac{4\omega t}{\theta})}x\wedge y\Big)
\end{align*}
Note that a study of the non-orientable case ($\lambda_2\neq0$ in \eqref{eq-moy-cplx}) remains to be done.

\subsubsection{$\phi^3$ theories}

The real $\phi^3$ theory on the Moyal space has also been considered:
\begin{equation*}
S(\phi)=\int\dd^Dx\left(\frac 12(\partial_\mu\phi)^2+\frac{\Omega^2}{2}\wx^2\phi^2+\frac{\mu^2}{2}\phi^2+\lambda\,\phi\star\phi\star\phi\right).
\end{equation*}
For $\Omega=1$ (the self-dual case), this model has been studied in the matrix basis and mapped to a Kontsevich model (see \cite{Kontsevich:1992ti}) by adding an additional counterterm interpreted as a divergent shift of the field $\phi$. By this procedure, the model is renormalizable in $D=2,4,6$ dimensions \cite{Grosse:2005ig,Grosse:2006qv,Grosse:2006tc}. It is also solvable genus by genus, and the renormalization depends only on the genus $0$ sector, like in the $\phi^4$ theory with harmonic term (see subsection \ref{subsec-moy-harm}).

\subsubsection{Fermionic models}

On the ``commutative'' $\gR^4$, the Gross-Neveu model is a pure fermionic theory in two dimensions, renormalizable, asymptotically free, and exhibiting dynamical mass generation \cite{Mitter:1973,Gross:1974,Kopper:1995}. The orientable noncommutative version of this theory has been considered in \cite{VignesTourneret:2006nb}. Its action is given by:
\begin{equation*}
S(\psi)=\int\dd^2x\left(\overline\psi(-i\gamma_\mu\partial_\mu+\Omega\gamma_\mu\wx_\mu+m+\mu\gamma_5)\psi\right) +V_{\text{o}}(\psi),
\end{equation*}
where $\psi$ is a spinor and $\{\gamma_\mu\}$ a Clifford algebra (in two dimensions) satisfying:
\begin{equation*}
\{\gamma_\mu,\gamma_\nu\}=-2\delta_{\mu\nu},\quad \gamma_\mu^\dag=-\gamma_\mu,\quad \gamma_5=i\gamma_0\gamma_1.
\end{equation*}
There are three possible orientable interactions (see Definition \ref{def-moy-orientable}):
\begin{align*}
V_{\text{o}}(\psi)=&\int\dd^2x\Big(\lambda_1\,\overline\psi\star\psi\star\overline\psi\star\psi+\lambda_2\,\overline\psi\star\gamma_\mu\psi\star\overline\psi\star\gamma_\mu\psi\\
&+\lambda_3\overline\psi\star\gamma_5\psi\star\overline\psi\star\gamma_5\psi\Big).
\end{align*}
The propagator is related to the one of the LSZ model, and has been showed \cite{Gurau:2005qm} to be (for $\mu=0$):
\begin{multline*}
C(x,y)=-\frac{\Omega}{\pi\theta}\int_0^\infty\dd t\frac{e^{-tm^2}}{\sinh(2\tilde\Omega t)}e^{-\frac{\tilde\Omega}{2}\coth(2\tilde\Omega t)(x-y)^2+i\Omega x\wedge y}\\
\Big(i\tilde\Omega\coth(2\tilde\Omega t)\gamma_\mu(x_\mu-y_\mu)+\Omega\gamma_\mu(\wx_\mu-\widetilde y_\mu)-m\Big)e^{-2i\Omega t\gamma\Theta^{-1}\gamma}.
\end{multline*}
This orientable two-dimensional theory is just renormalizable to all orders for $\Omega\in[0,1[$ \cite{VignesTourneret:2006nb}, and its one-loop beta function has been exhibited in \cite{Lakhoua:2007ra}.

\subsubsection{Models with translation invariance}

We have seen that the models with harmonic terms break the translation invariance. Another scalar field theory has been proposed in \cite{Gurau:2008vd} where the translation invariance is preserved (see \cite{Tanasa:2008gg} for a review). Indeed, the counterterm responsible of the UV/IR mixing (see section \ref{sec-moy-uvir}) has been added in the Lagrangian. In the Fourier conventions of section \ref{sec-moy-uvir}, the action takes the form:
\begin{equation}
S(\phi)=\frac{1}{(2\pi)^D}\int\dd^Dp\left(\frac 12p^2+\frac{\mu^2}{2}+\frac{a}{2\theta^2p^2}\right)\phi(-p)\phi(p)+V(\phi),\label{eq-moy-gmrt}
\end{equation}
where $V(\phi)$ is the quartic potential involved in the action \eqref{eq-moy-actharm}, and $a$ is some constant parameter. Note that the quadratic part of this action is no longer local in the $x$ variables, contrary to \eqref{eq-moy-actharm}, and that the dispersion relation is modified with respect to the commutative model. The model \eqref{eq-moy-gmrt} has been proven to be renormalizable to all orders \cite{Gurau:2008vd} in the multiscale analysis framework. Indeed, the IR-divergent terms coming from the UV/IR mixing (like in \eqref{eq-moy-uvirtadpole}) can now be reabsorbed in a (finite) renormalization of the parameter $a$ of \eqref{eq-moy-gmrt}.

Then, the parametric representation of this model has been performed \cite{Tanasa:2008bt}, and the beta functions have been proven at one-loop to be proportional of the beta functions of the commutative model \cite{Geloun:2008hr}. Of course, the beta function for the parameter $a$ vanishes since the counterterms of this type are always convergent. Moreover, the commutative limit of this model has been studied \cite{Magnen:2008pd}.

Such a scalar model has been adapted to gauge theories in \cite{Blaschke:2008yj} and some results have been obtained at the one-loop level in \cite{Blaschke:2008jh,Blaschke:2009gm}.

\numberwithin{equation}{section}
\chapter[Gauge theory on Moyal space]{Gauge Theory on the Moyal space}
\label{cha-gauge}

\section{Definition of gauge theory}
\label{sec-gauge-defgauge}

\subsection{Gauge theory associated to standard differential calculus}
\label{subsec-gauge-stdiffcalc}

In this subsection, we want to define a gauge theory appropriate to the Moyal space (see subsection \ref{subsec-moy-moyal}). Since the Moyal space is a deformation of the standard space $\gR^D$, and the standard derivations $\partial_\mu$ are also derivations for the deformed algebra, the standard differential calculus of the Moyal algebra can be defined as a noncommutative differential calculus based on these derivations $\partial_\mu$.

\subsubsection{Differential calculus of the Moyal space}

We have seen in the section \ref{sec-moy-presmoy} that the applications $\partial_\mu$ are derivations of the involutive unital algebra $\caM$, and that
\begin{equation}
\tr(\phi)=\int\dd^Dx\ \phi(x)\label{eq-gauge-deftrace}
\end{equation}
is a trace on a subalgebra of $\caM$ containing $\caS(\gR^D)$. The center of $\caM$ is 
\begin{equation*}
\caZ(\caM)=\gC\fois\gone.
\end{equation*}
The commutation relations of the applications $\partial_\mu$ can be computed:
\begin{equation*}
[\partial_\mu,\partial_\nu]=0,
\end{equation*}
due to Schwarz Lemma. Then, the vector space generated by these derivations $\kg_0=\langle\partial_\mu\rangle$ is a Lie algebra and a $\caZ(\caM)$-bimodule.
\medskip

By subsection \ref{subsec-gnc-diffcalc}, one can construct the differential calculus for the algebra $\caM$ restricted to the Lie algebra of derivations $\kg_0$:
\begin{equation*}
\Omder^n(\caM|\kg_0)=\{\omega:(\kg_0)^n\to\caM,\ n\text{-linear antisymmetric}\},
\end{equation*}
with $\Omder^0(\caM|\kg_0)=\caM$. This complex is endowed by a product and the standard Koszul differential $\dd$, which turn it into a noncommutative graded differential algebra. It satisfies: $\forall\omega\in\Omder^m(\caM|\kg_0)$, $\forall\eta\in\Omder^n(\caM|\kg_0)$,
\begin{equation*}
\dd (\omega\fois\eta)=(\dd\omega)\fois\eta+(-1)^m\omega\fois(\dd\eta),\qquad \dd^2=0,
\end{equation*}
and at low orders: $\forall a\in\caM$, $\forall\omega,\eta\in\Omder^1(\caM|\kg_0)$,
\begin{align*}
&\dd a(\partial_\mu)=\partial_\mu a\\
&\dd \omega(\partial_\mu,\partial_\nu)=\partial_\mu\omega(\partial_\nu)-\partial_\nu\omega(\partial_\mu)\\
&(\omega\fois\eta)(\partial_\mu,\partial_\nu)=\omega(\partial_\mu)\star\eta(\partial_\nu)-\omega(\partial_\nu)\star\eta(\partial\mu).
\end{align*}
The action of the differential on the functions $x_\mu$ gives: $\dd x_\mu(\partial_\nu)=\delta_{\mu\nu}$, so that the $\dd x_\mu$ form a basis of $\Omder^1(\caM|\kg_0)$, dual to the basis $\partial_\mu$. Indeed, $\forall\omega\in\Omder^1(\caM|\kg_0)$,
\begin{equation*}
\omega=\omega_\mu\dd x_\mu,\quad \text{with }\omega_\mu=\omega(\partial_\mu)\in\caM.
\end{equation*}

With this differential calculus, we can now construct a theory of connections and gauge transformations (see subsection \ref{subsec-gnc-diffcalc}). Note that other (non-standard) differential calculi have been exhibited for the Moyal space, as based on other Lie algebras of derivations \cite{Marmo:2004re,Cagnache:2008tz}, and these calculi give rise to other gauge theories than the one we will construct in this subsection.

\subsubsection{Theory of connections}

To define a $U(N)$ gauge theory on the Moyal space, with $N\in\gN^\ast$, we consider the free right module $\caM^N$ on the algebra $\caM$, acting by $\forall\phi\in\caM^N$, $\forall a\in\caM$,
\begin{equation*}
(\phi_1,\dots,\phi_N)\,a=(\phi_1\star a,\dots,\phi_N\star a).
\end{equation*}
It is endowed with the hermitean structure: $\forall\phi,\psi\in\caM^N$,
\begin{equation}
\langle \phi,\psi\rangle=\sum_{i=1}^N \phi_i^\dag\star\psi_i,\label{eq-gauge-hermstr}
\end{equation}
where $\phi_i\in\caM$ are the components of $\phi$. Following the subsection \ref{subsec-gnc-diffcalc}, one defines a connection by linear maps $\nabla_\mu:\caM^N\to\caM^N$ satisfying: $\forall \phi\in\caM^N$, $\forall a\in\caM$,
\begin{equation*}
\nabla_\mu(\phi\, a)=\nabla_\mu(\phi)\,a+\phi\, \partial_\mu a.
\end{equation*}
Note that $\nabla_\mu$ corresponds to $\nabla_{\partial_\mu}$, in the notations of subsection \ref{subsec-gnc-diffcalc}. By defining $e_i=(0,\dots,\gone,\dots,0)$ the canonical basis of the module $\caM^N$, and $\nabla_\mu(e_i)=-ie_j A_\mu^{ji}$, the connection $\nabla$ can be reexpressed in $\forall \phi\in\caM^N$,
\begin{equation}
\nabla_\mu(\phi)=\partial_\mu\phi-iA_\mu\fois\phi,\label{eq-gauge-connpot}
\end{equation}
where the gauge potential $A_\mu\in\Matr_N(\caM)$ is a function valued in the $N\times N$ matrices. Moreover, $\nabla$ is compatible with the hermitean structure \eqref{eq-gauge-hermstr} if and only if $A_\mu$ is a hermitean matrix. The curvature $F=i\nabla^2$ takes the following form:
\begin{equation*}
F_{\mu\nu}=\partial_\mu A_\nu-\partial_\nu A_\mu-i[A_\mu,A_\nu].
\end{equation*}

The unitary gauge transformations are given by $g\in\Unit_N(\caM)$ and acting by:
\begin{align*}
&\phi^g=g\fois\phi,\\
&(\nabla_\mu\phi)^g=g\fois(\nabla_\mu\phi),\\
&(A_\mu)^g=g\fois A_\mu\fois g^\dag+ig\fois\partial_\mu g^\dag,\\
&(F_{\mu\nu})^g=g\fois F_{\mu\nu}\fois g^\dag.
\end{align*}
If $g(x)=e^{i\lambda(x)}$, where $\lambda\in\Matr_N(\caM)$ is an infinitesimal (hermitean) transformation, it acts on the gauge potential by:
\begin{equation}
\delta_\lambda A_\mu=\partial_\mu\lambda-i[A_\mu,\lambda].\label{eq-gauge-infgaugtr}
\end{equation}

\subsubsection{Standard gauge action}

By restricting the algebra to the one which admits the trace \eqref{eq-gauge-deftrace}, one can construct a gauge invariant action:
\begin{equation*}
S(A)=\frac{1}{4g^2}\int\dd^Dx\tr_N(F_{\mu\nu}\fois F_{\mu\nu}),
\end{equation*}
where $\tr_N$ is the usual trace on $\Matr_N(\gC)$ and $g$ is the coupling constant. This action is the standard one for a gauge theory on the Moyal space, and it has the same form as the standard Yang-Mills action on the space $\gR^D$, by changing the commutative product into the Moyal product. However, we will see in the subsection \ref{subsec-gauge-uvir} that this action exhibits UV/IR mixing, which renders its renormalizability quite unlikely.

\subsubsection{Gauge invariant connection}

Since all the derivations considered in $\kg_0$ are inner, it is possible to construct a gauge invariant connection. By setting $\xi_\mu=-\frac 12\wx_\mu$, it turns out that the gauge potential
\begin{equation}
A^{\text{inv}}_\mu=\xi_\mu\,\gone_N\label{eq-gauge-invpot}
\end{equation}
defines a connection invariant under gauge transformation. Note that the occurrence of gauge-invariant connection is not new in noncommutative geometry and has been already mentionned in matrix-valued field theories for example \cite{DuboisViolette:1989vq,DuboisViolette:1998su,Masson:1999ea,Masson:2005ic} (see Example \ref{ex-gnc-matrix}). Indeed, according to \eqref{eq-gauge-connpot}, the connection $\nabla^{\text{inv}}$ associated to $A^{\text{inv}}$ verifies: $\forall\phi\in\caM^N$,
\begin{equation*}
\nabla^{\text{inv}}_\mu\phi=\partial_\mu\phi-iA_\mu^{\text{inv}}\fois\phi=-i\phi\,\xi_\mu,
\end{equation*}
where the second equality stems from $\partial_\mu\phi=iA_\mu^{\text{inv}}\phi-i\phi\,\xi_\mu$. Then, it is easy to realize that
\begin{equation*}
(\nabla^{\text{inv}}_\mu)^g\phi=\partial_\mu\phi-i(A_\mu^{\text{inv}})^g\fois\phi=\nabla^{\text{inv}}_\mu\phi,
\end{equation*}
since $(A_\mu^{\text{inv}})^g=A_\mu^{\text{inv}}$.

Let us introduce
\begin{equation*}
\caA_\mu=A_\mu-A_\mu^{\text{inv}},
\end{equation*}
the covariant coordinate associated to the gauge potential $A_\mu$, which is covariant under gauge transformations:
\begin{equation*}
(\caA_\mu)^g=g\fois \caA_\mu\fois g^\dag,
\end{equation*}
as the difference of two gauge potentials. Then, the curvature can be simply reexpressed in terms of this covariant coordinate:
\begin{equation}
F_{\mu\nu}=\Theta^{-1}_{\mu\nu}\gone-i[\caA_\mu,\caA_\nu].\label{eq-gauge-FcaA}
\end{equation}
Note that the curvature of the gauge invariant connection is $F_{\mu\nu}^{\text{inv}}=\Theta^{-1}_{\mu\nu}\gone$.

\subsection{U(N) versus U(1) gauge theory}
\label{subsec-gauge-un}

We adopt here the conventions of the subsection \ref{subsec-ren-brs}. Let $(\tau_a)$ be a basis of $\ksu(N)$, with $a\in\{1,\dots,N^2-1\}$ and $\tau_0=\gone_N$. Then, we define the constant structures and normalize this basis such that:
\begin{equation*}
[\tau_a,\tau_b]=if_{abc}\tau_c,\qquad \{\tau_a,\tau_b\}=d_{abc}\tau_c,\qquad \tr_N(\tau_a\tau_b)=\delta_{ab},
\end{equation*}
where the indices $a,b,c$ belong to $\{0,\dots,N^2-1\}$. Note that $f_{abc}=0$ if $a$ or $b$ or $c$ is equal to $0$, and $d_{0ab}=d_{a0b}=2\delta_{ab}$. We decompose the gauge potential on this basis: $A_\mu=A_\mu^a\tau_a$.

Then, a simple computation gives:
\begin{align*}
[A_\mu,A_\nu]=&\frac 12[A_\mu^a,A_\nu^b]_\star\{\tau_a,\tau_b\}+\frac 12\{A_\mu^a,A_\nu^b\}_\star[\tau_a,\tau_b]\\
=&\left(\frac 12d_{abc}[A_\mu^a,A_\nu^b]_\star+\frac i2 f_{abc}\{A_\mu^a,A_\nu^b\}\right)\tau_c,
\end{align*}
and the curvature is given by
\begin{equation*}
F_{\mu\nu}^a=\partial_\mu A_\nu^a-\partial_\nu A_\mu^a+\frac 12 f_{bca}\{A_\mu^b,A_\nu^c\}_\star-\frac i2 d_{bca}[A_\mu^b,A_\nu^c]_\star.
\end{equation*}

Let us separate the $\Unit(1)$ part and the $SU(N)$ part. If the indices $a,b,c$ belong now only to $\{1,\dots,N^2-1\}$, the two parts of curvature can be expressed as
\begin{align*}
F_{\mu\nu}^0=&\partial_\mu A_\nu^0-\partial_\nu A_\mu^0-i[A_\mu^0,A_\nu^0]_\star-\frac i2 d_{bc0}[A_\mu^b,A_\nu^c]_\star,\\
F_{\mu\nu}^a=&\partial_\mu A_\nu^a-\partial_\nu A_\mu^a+\frac 12f_{bca} \{A_\mu^b,A_\nu^c\}_\star-\frac i2d_{bca}[A_\mu^b,A_\nu^c]_\star-i[A_\mu^0,A_\nu^a]_\star-i[A_\mu^a,A_\nu^0]_\star.
\end{align*}
In the same way, the infinitesimal gauge transformations \eqref{eq-gauge-infgaugtr} take the form:
\begin{align*}
\delta_\lambda A^0_\mu=&\partial_\mu\lambda^0-i[A_\mu^0,\lambda^0]_\star-\frac i2d_{bc0}[A_\mu^b,A_\nu^c]_\star\\
\delta_\lambda A^a_\mu=&\partial_\mu \lambda^a+\frac 12f_{bca}\{A_\mu^b,\lambda^c\}_\star-\frac i2d_{bca}[A_\mu^b,\lambda^c]_\star-i[A_\mu^0,\lambda^a]_\star-i[A_\mu^a,\lambda^0]_\star.
\end{align*}
with $\lambda\in\Matr_N(\caM)$ hermitean.

Therefore, the $\Unit(1)$ part is stable under this decomposition: if $A_\mu^a=\lambda^a=0$, then $\delta_\lambda A_\mu^0=\partial_\mu \lambda^0-i[A_\mu^0,\lambda^0]_\star$, $\delta_\lambda A_\mu^a=0$, and it corresponds to the Yang-Mills theory with $N=1$. But the $SU(N)$ part is not stable under this decomposition. Namely, for $A_\mu^0=\lambda^0=0$, $\delta_\lambda A_\mu^0=-\frac i2d_{bc0}[A_\mu^b,\lambda^c]_\star$, which does not vanish for some $b,c$.

Then, a $SU(N)$ theory would be ill defined in this framework. Moreover, it has been shown \cite{Minwalla:1999px} that one-loop planar $\Unit(N)$ graphs with two broken faces, which exhibit the UV/IR mixing (see below), contribute only to the $\Unit(1)$ part of the theory. In the following, we will restrict the study to the $\Unit(1)$ gauge theory, but the conclusions will be generic to $\Unit(N)$ theory.

\subsection{UV/IR mixing in gauge theory}
\label{subsec-gauge-uvir}

The gauge action proposed in subsection \ref{subsec-gauge-stdiffcalc}:
\begin{equation}
S(A)=\frac{1}{4g^2}\int\dd^Dx\tr_N(F_{\mu\nu}\fois F_{\mu\nu}),\label{eq-gauge-actgauge}
\end{equation}
arises also from Connes-Lott action functional \cite{Gayral:2003ye}, from the spectral action principle \cite{Gayral:2004ww} (see chapter \ref{cha-gnc}), and also as a limiting regime of string theory \cite{Seiberg:1999vs}. The $U(1)$ Yang-Mills theory defined by \eqref{eq-gauge-actgauge} with $N=1$ has been proven to be renormalizable at the one-loop order \cite{Martin:1999aq}.
\medskip

However, the same phenomenon as in the scalar theory on the Moyal space appears at the one-loop order. Some contributions (planar with two broken faces) are finite, but IR singular in their external momentum: this theory exhibits also UV/IR mixing \cite{Minwalla:1999px,Matusis:2000jf} (see subsection \ref{sec-moy-uvir}).

Let us do the computation of the one-loop polarization tensor of the (on-shell) four-dimensional $U(1)$ Yang-Mills theory. The $\Unit(1)$ gauge action, with ghost sector, writes down (see subsection \ref{subsec-ren-brs}):
\begin{align*}
S_{\text{inv}}=&\int\dd^4x\Big(-\frac 12A_\mu\partial^2 A_\mu-\frac 12(\partial_\mu A_\mu)^2-ig(\partial_\mu A_\nu)[A_\mu,A_\nu]_\star-g^2[A_\mu,A_\nu]_\star^2\Big),\\
S_{\text{gf}}=&\int \dd^4x\Big(-\overline c\partial^2 c+ig\overline c\partial_\mu[A_\mu,c]_\star+\frac{g^2}{2\alpha}(\partial_\mu A_\mu)^2\Big),
\end{align*}
after the rescaling $A_\mu\to gA_\mu$. We set $\alpha=g^2$.

Once the propagator has been regularized by an $\epsilon\to0$ (see subsection \ref{subsec-gauge-compeffact}), one can compute the divergent part of the one-loop polarization tensor (obtained by the planar contributions with one broken face), which is proportional to:
\begin{equation}
\Pi_{\mu\nu}^{\text{div}}\propto \frac{g^2\theta^2}{16\pi^2}\left(\frac{\delta_{\mu\nu}}{p^2}-\frac{p_\mu p_\nu}{p^4}\right)\ln(\epsilon)\label{eq-gauge-poltensdiv}
\end{equation}
in the Fourier space, and where $p$ is the incoming momentum. The IR singular part of the finite contribution is given by the planar graphs with two broken faces, and is proportional to:
\begin{equation}
\Pi_{\mu\nu}^{\text{UV/IR}}\propto \frac{g^2\theta^4}{16\pi^2}\frac{\widetilde p_\mu\widetilde p_\nu}{p^4}.\label{eq-gauge-poltensuvir}
\end{equation}
These two contributions \eqref{eq-gauge-poltensdiv} and \eqref{eq-gauge-poltensuvir} are transverse, namely: $p_\mu \Pi_{\mu\nu}=0$, so that these results are compatible with the Slavnov-Taylor identities.

If $p\neq0$, $\Pi_{\mu\nu}^{\text{UV/IR}}$ is finite, but it is singular in the infrared region $p\to0$. Then, if this diagram is inserted into higher order graphs, it can produce an infrared divergence of the global amplitude, which cannot be renormalized as a counterterm of the lagrangian, like in the scalar case (see subsection \ref{sec-moy-uvir}).

\section{The effective action}
\label{sec-gauge-effact}

Consequently to the above section, we want to modify the gauge action \eqref{eq-gauge-actgauge} with additional terms in order to remove the UV/IR mixing, like in the scalar case where a harmonic term has been added. But, in the case of gauge theory, one has also to preserve the gauge invariance, so that such a harmonic term cannot be added alone.

A reasonable way to produce a candidate of action can be achieved by computing, at least at the one-loop order, the effective gauge theory stemming from the renormalizable scalar complex field theory with harmonic term, coupled to an external gauge potential in a gauge-invariant way (see \cite{deGoursac:2007gq,deGoursac:2007qi}). We recall that the present study is for a $\Unit(1)$ gauge theory, such that the right $\caM$-module considered here is $\caM$ itself. See \cite{deGoursac:2007gq,Wallet:2007em} for a construction of $\Unit(1)$ gauge theory on the Moyal space. The gauge potentials are hermitean elements $A_\mu\in\caM$, and gauge transformations are unitary elements $g\in\caM$, acting by: $\forall\phi\in\caM$,
\begin{subequations}
\label{eq-gauge-trgauge}
\begin{align}
&\phi^g=g\star \phi,\\
&(A_\mu)^g=g\star A_\mu\star g^\dag+ig\star\partial_\mu g^\dag.
\end{align}
\end{subequations}
Note that to construct an action, the fields $\phi$ and $A_\mu$ have to be restricted to a certain subalgebra of $\caM$, so that the integral is defined.

\subsection{Minimal coupling}
\label{sec-gauge-mincoupl}

In this subsection, we give the minimal coupling prescription, so that external gauge potentials can be coupled in a gauge invariant way to a complex scalar field theory described by the action:
\begin{equation}
S(\phi)=\int\dd^Dx\Big(|\partial_\mu\phi|^2+\Omega^2\wx^2|\phi|^2+m^2|\phi|^2+\lambda\,\phi^\dag\star\phi\star\phi^\dag\star\phi\Big).\label{eq-gauge-actscalcplx}
\end{equation}
Indeed, owing to the special role played by the coordinate functions $x_\mu$ through the gauge-invariant potential 
\begin{equation*}
A^{\text{inv}}_\mu=\xi_\mu=-\frac 12\wx_\mu,
\end{equation*}
(see \eqref{eq-gauge-invpot}) involved in $\nabla^{\text{inv}}_\mu$ and the expression for the inner derivatives $\partial_\mu\phi=i[\xi_\mu,\phi]_\star$, it follows that a natural choice for the minimal coupling of the action \eqref{eq-gauge-actscalcplx} to an external real gauge field $A_\mu$ is obtained by performing the usual substitution
\begin{equation*}
\partial_\mu\phi\to\nabla_\mu^A\phi=\partial_\mu\phi-iA_\mu\star\phi,
\end{equation*}
on the action \eqref{eq-gauge-actscalcplx}, provided this latter is reexpressed in terms of $\partial_\mu$ and $\nabla_\mu^{\text{inv}}$, using in particular the following identity:
\begin{equation}
\wx_\mu\,\phi=\wx_\mu\star\phi-i\partial_\mu\phi=-i(\partial_\mu\phi-2i\xi_\mu\star\phi)= -2i\nabla_\mu^\xi\phi+i\partial_\mu\phi. \label{eq-gauge-identitycouplage}
\end{equation}
By using \eqref{eq-gauge-identitycouplage}, one easily infers that the minimal coupling prescription can be conveniently written as
\begin{subequations}
\label{eq-gauge-mincoupl}
\begin{align}
\partial_\mu\phi&\mapsto \nabla_\mu^A\phi=\partial_\mu\phi-iA_\mu\star\phi,\\
\wx_\mu\,\phi &\mapsto -2i\nabla_\mu^\xi\phi+i\nabla_\mu^A\phi=\widetilde{x}_\mu\phi+A_\mu\star\phi.
\end{align}
\end{subequations}
Note that gauge invariance of the resulting action functional is obviously obtained thanks to the relation
$(\nabla^{A,\text{inv}}_\mu(\phi))^g = g\star(\nabla^{A,\text{inv}}_\mu(\phi))$.

By applying the above minimal coupling prescription to \eqref{eq-gauge-actscalcplx}, we obtain the following gauge-invariant action
\begin{align}
S(\phi,A) =& \int\dd^Dx\Big(|\partial_\mu\phi-iA_\mu\star\phi|^2+\Omega^2|\wx_\mu+A_\mu\star\phi|^2+m^2|\phi|^2+\lambda\,\phi^\dag\star\phi\star\phi^\dag\star\phi\Big)\nonumber\\
=&S(\phi)+ \int \dd^Dx \ \Big((1+\Omega^2)\phi^\dag\star (\widetilde{x}_\mu A_\mu)\star\phi\nonumber\\ 
&-(1-\Omega^2)\phi^\dag\star A_\mu \star\phi\star \widetilde{x}_\mu+(1+\Omega^2)\phi^\dag\star A_\mu\star A_\mu\star \phi\Big),\label{eq-gauge-harmcoupled}
\end{align}
where $S(\phi)$ is given by \eqref{eq-gauge-actscalcplx}.

\subsection{Computation of the effective action}
\label{subsec-gauge-compeffact}

We will calculate the one-loop effective action for $D=4$ in this subsection, starting from the action $S(\phi,A)$ \eqref{eq-gauge-harmcoupled}. Recall that the effective action (or vertex functional) is formally obtained from (see section \ref{sec-ren-bphz}):
\begin{equation}
e^{-\Gamma(A)}= \int \dd\phi \dd\phi^\dag e^{-S(\phi,A)} =\int \dd\phi \dd\phi^\dag e^{-S(\phi)} e^{-S_{int}(\phi,A)}, \label{eq-gauge-defact}
\end{equation}
where $S(\phi)$ is given by \eqref{eq-gauge-actscalcplx} and $S_{int}(\phi,A)$ can be read off from \eqref{eq-gauge-harmcoupled} and \eqref{eq-gauge-actscalcplx}. At the one-loop order, \eqref{eq-gauge-defact} reduces to
\begin{equation*}
e^{-\Gamma_{1l}(A)}=\int \dd\phi \dd\phi^\dag e^{-S_{\text{free}}(\phi)} e^{-S_{int}(\phi,A)},
\end{equation*}
where $S_{\text{free}}(\phi)$ is simply the quadratic part of \eqref{eq-gauge-actscalcplx}. The corresponding diagrams are depicted on Figures \ref{fig-gauge-1-point}-\ref{fig-gauge-4-point}.

\begin{figure}[!htb]
  \centering
  \includegraphics[scale=1]{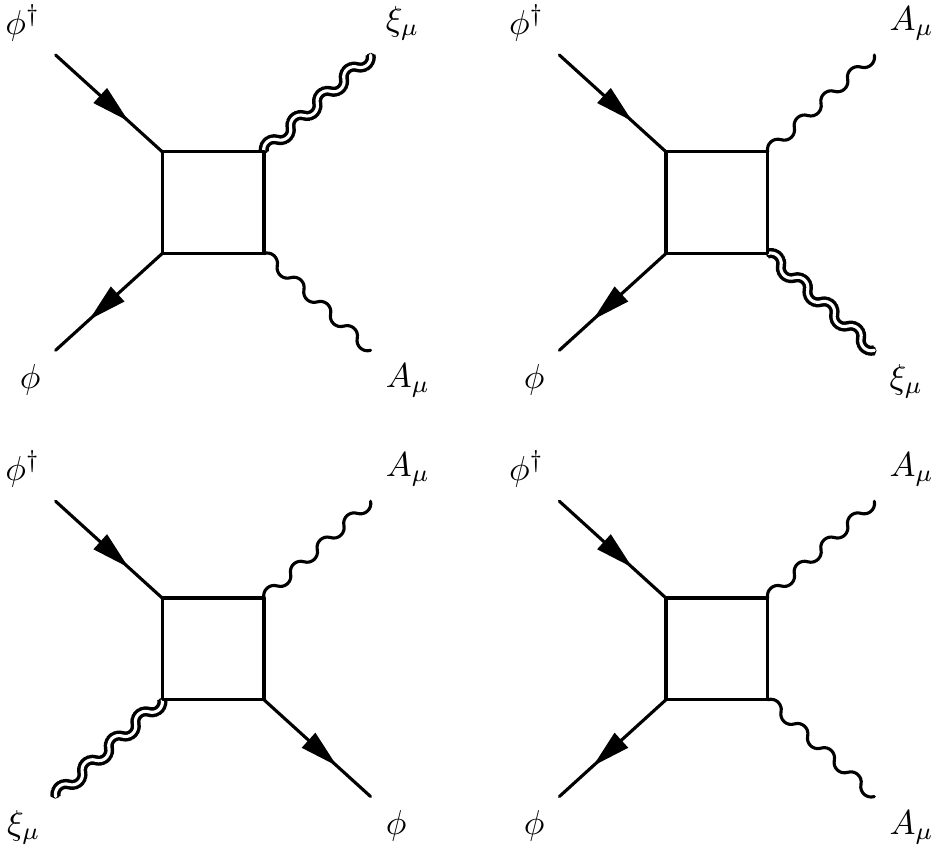}
  \caption[The vertices of the gauge theory]{\footnotesize{Graphical representation for the vertices carrying the external gauge potential $A_\mu$ involved in the action \eqref{eq-gauge-harmcoupled}. The overall factor affecting the two uppermost vertices is $(1 + \Omega^2)$.
      From left to right, the overall factors affecting the lower vertices are respectively equal to $- 2(1 - \Omega^2)$ and $- (1+ \Omega^2)$.}}
  \label{fig-gauge-vertices}
\end{figure}

The additional vertices involving $A_\mu$ and/or $\xi_\mu$ (or $\wx_\mu$) and generated by the minimal coupling can be obtained by combining \eqref{eq-gauge-harmcoupled} with the generic relation: $\forall f_i\in\caS(\gR^D)$,
\begin{align*}
\int \dd^4x(f_1\star f_2\star f_3\star f_4)(x)=&\frac{1}{\pi^4\theta^4}\int \prod_{i=1}^4\dd^4x_i\, f_1(x_1) f_2(x_2) f_3(x_3) f_4(x_4) \\
&\times\delta(x_1-x_2+x_3-x_4)e^{-i\sum_{i<j}(-1)^{i+j+1}x_i\wedge x_j }.
\end{align*}
These vertices are depicted on Figure \ref{fig-gauge-vertices}. Note that additional overall factors must be taken into account. These are indicated on Figure \ref{fig-gauge-vertices}.

\subsubsection{Computation of the tadpole}

\begin{figure}[!htb]
  \centering
  \includegraphics[scale=1]{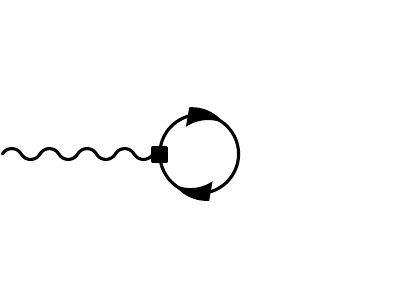}
  \caption[1-point]{\footnotesize{The non vanishing tadpole diagram. To simplify the figure, we do not explicitly draw all the diagrams that would be obtained from the vertices given on Figure \ref{fig-gauge-vertices} but indicate only the overall topology of the corresponding diagrams. Notice that the background lines are not explicitly depicted.}}
  \label{fig-gauge-1-point}
\end{figure}
Using the expression for the vertices and the minimal coupling, the amplitude corresponding to the tadpole on Figure \ref{fig-gauge-1-point} is
\begin{equation}
\mathcal{T}_1= \frac{1}{\pi^4\theta^4}\int \dd^4x\dd^4u\dd^4z\ A_\mu(u)\  e^{-i(u-x)\wedge z}\ C(x+z,x)\  ((1-\Omega^2)(2\widetilde{x}_\mu+\widetilde{z}_\mu)-2\widetilde{u}_\mu). \label{eq-gauge-tadpole1}
\end{equation}
Combining this with the explicit expression for the propagator \eqref{eq-moy-propagx}, \eqref{eq-gauge-tadpole1} can be reexpressed as
\begin{multline}
\mathcal{T}_1= \frac{\Omega^2}{4\pi^6\theta^6}\int \dd^4x\dd^4u\dd^4z \int_0^\infty  \frac{dt\ e^{-tm^2}}{ \sinh^2(\widetilde{\Omega}t)
\cosh^2(\widetilde{\Omega}t)}\  A_\mu(u)\  e^{-i(u-x)\wedge z}\\
\times e^{-\frac{\widetilde{\Omega}}{4} (\coth(\widetilde{\Omega}t)z^2+\tanh(\widetilde{\Omega}t)(2x+z)^2} ((1-\Omega^2)(2\widetilde{x}_\mu+ \widetilde{z}_\mu)-2\widetilde{u}_\mu). \label{eq-gauge-tadpole2}
\end{multline}
At this point, we find convenient to introduce the following 8-dimensional vectors $X$, $J$ and the $8\times 8$ matrix $K$ defined by
\begin{equation}
X=\begin{pmatrix} x\\ z \end{pmatrix}, \quad K=\begin{pmatrix} 4\tanh(\widetilde{\Omega}t) \mathbb{I} & 2\tanh(\widetilde{\Omega}t)\mathbb{I} -2i\Theta^{-1} \\  2\tanh(\widetilde{\Omega}t)\mathbb{I} +2i\Theta^{-1} &  (\tanh(\widetilde{\Omega}t)+ \coth(\widetilde{\Omega}t))\mathbb{I}
\end{pmatrix} ,\quad \ J=\begin{pmatrix} 0\\ i\tilde{u} \end{pmatrix}. \label{eq-gauge-tadpole2bis}
\end{equation}
This permits one to reexpress \eqref{eq-gauge-tadpole2} in a form such that some Gaussian integrals can be easily performed. Note that this latter procedure can be adapted to the calculation of the higher order Green functions (see below). The combination of \eqref{eq-gauge-tadpole2bis} with \eqref{eq-gauge-tadpole2} then yields
\begin{multline*}
\mathcal{T}_1= \frac{\Omega^2}{4\pi^6\theta^6} \int \dd^4x\dd^4u\dd^4z \int_0^\infty  \frac{\dd t\ e^{-tm^2}}{\sinh^2(\widetilde{\Omega}t)\cosh^2(\widetilde{\Omega}t)}\  A_\mu(u) \\
\times e^{-\frac{1}{2}X.K.X+J.X} ((1-\Omega^2)(2\widetilde{x}_\mu+ \widetilde{z}_\mu)-2\widetilde{u}_\mu).
\end{multline*}
By performing the Gaussian integrals on $X$, one finds:
\begin{equation}
\mathcal{T}_1=-\frac{\Omega^4}{\pi^2\theta^2(1+\Omega^2)^3}\int \dd^4u \int_0^\infty  \frac{\dd t\ e^{-tm^2}}{\sinh^2(\widetilde{\Omega}t)\cosh^2(\widetilde{\Omega}t)}\ A_\mu(u)\widetilde{u}_\mu\ e^{-\frac{2\Omega}{\theta(1+\Omega^2)}
\tanh(\widetilde{\Omega}t)u^2}. \label{eq-gauge-tadpole3}
\end{equation}
Then, inspection of the behavior of \eqref{eq-gauge-tadpole3} for $t \to0$ shows that this latter expression has a quadratic as well as a
logarithmic UV divergence. Indeed, by performing a Taylor expansion of \eqref{eq-gauge-tadpole3}, one obtains:
\begin{multline}
\mathcal{T}_1 = -\frac{\Omega^2}{4\pi^2(1+\Omega^2)^3} \left( \int \dd^4u\ \widetilde{u}_\mu A_\mu(u)\right)\ \frac{1}{\epsilon}\ -\frac{m^2\Omega^2}{4\pi^2(1+\Omega^2)^3}\left( \int \dd^4u\ \widetilde{u}_\mu A_\mu(u)\right)\ \ln(\epsilon) \\
-\frac{\Omega^4}{\pi^2\theta^2(1+\Omega^2)^4}\left( \int \dd^4u\ u^2\widetilde{u}_\mu A_\mu(u)\right)\  \ln(\epsilon)\ + \dots,\label{eq-gauge-1ptfun}
\end{multline}
where $\epsilon \to 0$ is a cut-off and the ellipses denote finite contributions. The fact that the tadpole is (a priori) non-vanishing is a rather unusual feature for a Yang-Mills type theory. This will be discussed more closely in section \ref{sec-gauge-prop}.

\subsubsection{Choice of a regularization scheme}

Let us explain and justify the special regularization scheme \cite{deGoursac:2007qi} used in the calculations (see $\mathcal T_2'$ below). In \cite{Gayral:2004cs}, the one-loop effective action can be expressed in terms of heat kernels:
\begin{equation}
\Gamma_{1l}(\phi,A)=\frac 12 \ln(\det(H(\phi,A)H(0,0)^{-1})),\label{eq-gauge-defact2}
\end{equation}
where $H(\phi,A)=\frac{\delta^2 S(\phi,A)}{\delta \phi \,\delta \phi^\dag}$ is the effective potential. In the Schwinger representation, one obtains:
\begin{align*}
\Gamma_{1l}(\phi,A)&=-\frac{1}{2} \int_0^\infty \frac{\dd t}{t} \,\tr\big( e^{-t  H(\phi,A)} - e^{-t H(0,0)}\big)\\
&= -\frac{1}{2} \lim_{s \to 0} \Gamma(s)\,\tr\big(H^{-s}(\phi,A)-H^{-s}(0,0)\big).
\end{align*}
By using $\Gamma(s+1)=s\Gamma(s)$ and expanding \cite{Connes:2006qj}
\begin{equation*}
H^{-s}(\phi,A) =\big(1+a_1(\phi,A) s + a_2(\phi,A) s^2 + \dots\big)  H^{-s}(0,0),
\end{equation*}
one obtains the following expression for the effective action
\begin{equation}
\Gamma_{1l}(\phi,A)= -\frac{1}{2} \lim_{s \to 0} \tr\Big(\big(\Gamma(s+1)a_1(\phi,A) + s \Gamma(s+1) a_2(\phi,A) +\dots \big) H^{-s}(0,0)\Big).\label{eq-gauge-hkact}
\end{equation}
The expansion $\Gamma(s+1)=1-s\gamma+\dots$ permits one to reexpress \eqref{eq-gauge-hkact} in the form:
\begin{align*}
\Gamma_{1l}(\phi,A)&= -\frac{1}{2} \lim_{s \to 0} \tr\big(a_1(\phi,A) H^{-s}(0,0)\big) \\
&-\frac{1}{2} \lim_{s \to 0} \tr \Big( s\big(a_2(\phi,A) -\gamma a_1(\phi,A)\big)  H^{-s}(0,0)\Big).
\end{align*}
We can reconcile both definitions of the effective action \eqref{eq-gauge-defact} and \eqref{eq-gauge-defact2} by $\Gamma(A)=\Gamma_{1l}(0,A)$, so that we get:
\begin{align*}
\Gamma(A)&= -\frac{1}{2} \lim_{s \to 0} \tr\big(a_1(A) H^{-s}(0,0)\big) \\
&-\frac{1}{2} \lim_{s \to 0} \tr\Big( s\big(a_2(A) -\gamma a_1(A)\big)  H^{-s}(0,0)\Big),
\end{align*}
where the ellipses denote finite contributions. The second part of the effective action is called the Wodzicki residue \cite{Wodzicki:1984} and corresponds to the logarithmically divergent part of $\Gamma(A)$. This residue is a trace and is gauge invariant. But the first part of $\Gamma(A)$, which corresponds to the quadratically divergent part, is not gauge invariant. That is why the naive $\epsilon$-regularization of the Schwinger integrals breaks the gauge invariance of the theory in the quadratically divergent part. One possible way to solve this problem could be the restoration of gauge invariance by using methods from algebraic renormalization \cite{Piguet:1995} (see subsection \ref{subsec-ren-algren}). Notice also that a dimensional regularization scheme could have been used \cite{Gurau:2007fy}, where the divergent contributions of the effective action were automatically gauge invariant.
\medskip

We give here another way, which works for this special type of theory and which has been used for the calculations (see below). To find a convenient regularization scheme, one can do the following transformation on the cut-off $\epsilon\to\lambda\epsilon$, where $\lambda\in\mathbb R_+^\ast$, and it can be different for each contribution. The logarithmically divergent part of the effective action is insensitive to a finite scaling of the cut-off, so these transformations will affect only the quadratically divergent part, which is not gauge invariant.

Let us consider the contribution of a graph $G_p$ with $p$ internal lines and suppose that it involves a quadratic divergence. If not, this contribution does not need any change for the restoration of the gauge invariance of the effective action. Using the polynom $HU_{G_p}$ (see subsection \ref{subsec-moy-prop}) and integrating on the appropriated variables like in \eqref{eq-gauge-reg1}, one obtains at the one-loop order that the quadratically divergent part is proportional (after Taylor expansion on $\epsilon\to 0$) to $\int_\epsilon^1\frac{\dd t_i}{(t_1+...+t_p)^{p+1}}$. Indeed the integration on the $t_i$ between 1 and $+\infty$ gives only finite contributions. The expression of the propagator \eqref{eq-moy-propagx} implies that a part of the logarithmic divergence is proportional to $-m^2\int_\epsilon^1\frac{\dd t_i}{(t_1+...+t_p)^{p}}$ with exactly the same factor as the quadratic divergence. Since
\begin{align*}
\int_\epsilon^1\frac{\dd t_i}{(t_1+...+t_p)^{p+1}}&=\frac{1}{p(p!)\epsilon}+\dots,\\
\int_\epsilon^1\frac{\dd t_i}{(t_1+...+t_p)^{p}}&=-\frac{\ln(\epsilon)}{(p-1)!}+\dots,
\end{align*}
the contribution of $G_p$ writes $\frac{1}{p(p!)\epsilon}K_p+\frac{m^2\ln(\epsilon)}{(p-1)!}K_p+\dots$, where the ellipses denote other logarithmic contributions and finite terms, and $K_p$ is a factor depending on the graph $G_p$. The logarithmically divergent term involving $m^2$ in the effective action is gauge invariant. That is why one has to choose a regularization scheme $\epsilon\to\lambda_p\epsilon$ so that $\frac{1}{p(p!)\lambda_p}=\frac{1}{(p-1)!}$, i.e. $\lambda_p=\frac{1}{p^2}$.

Therefore, with the regularization scheme $\epsilon\to\frac{\epsilon}{p^2}$ for a graph with $p$ internal lines, one can conclude that the gauge invariance is restored in the effective action. Indeed, we made the transformation $\epsilon\to\frac{\epsilon}{4}$ for the $\mathcal T_2'$ contribution (see \eqref{eq-gauge-2aptfun}). Finally, notice that this method is very specific to this theory and to the one-loop order. As noted above, the dimensional regularization or the algebraic renormalization provide more general results.

\subsubsection{Computation of the multi-point contributions}

The 2, 3 and 4-point functions can be computed in a way similar to the one used for the tadpole. The algebraic manipulations are standard
but cumbersome so that we only give below the final expressions for the various contributions.
\begin{figure}[!htb]
  \centering
  \includegraphics[scale=1]{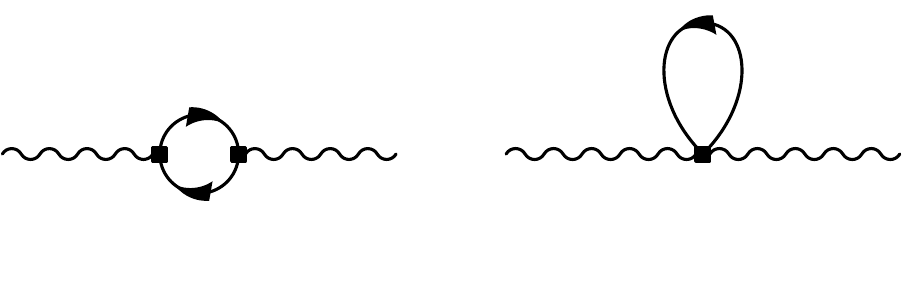}
  \caption[Two-point]{\footnotesize{Relevant one-loop diagrams contributing to the two-point function. To simplify the figure, we do not explicitly draw all the diagrams that would be obtained from the vertices given in Figure \ref{fig-gauge-vertices} but indicate only the overall topology of the corresponding diagrams. Notice that the background lines are not explicitly depicted. The leftmost (resp.\ rightmost) diagram corresponds to the contribution $\mathcal{T}_2'$ (resp.\ $\mathcal{T}_2''$).}}
  \label{fig-gauge-2-point}
\end{figure}
Let us start with the two-point function. We use the special regularization scheme described above. After some tedious calculations, we find the following final expressions for the diagrams on Figure \ref{fig-gauge-2-point} are:
\begin{subequations}
\label{eq-gauge-2ptfun}
\begin{align}
\mathcal{T}_2' &= \frac{(1{-}\Omega^2)^2}{16\pi^2(1{+}\Omega^2)^3} \left(\int \! \dd^4u\ A_\mu(u)A_\mu(u)\right) \frac{1}{\epsilon}
+ \frac{m^2(1{-}\Omega^2)^2}{16\pi^2(1{+}\Omega^2)^3} \left(\int \! \dd^4u\ A_\mu(u)A_\mu(u)\right) \ln(\epsilon) \nonumber\\
& +\frac{\Omega^2(1{-}\Omega^2)^2}{4\pi^2\theta^2(1{+}\Omega^2)^4} \left(\int \!\dd^4u\ u^2A_\mu(u)A_\mu(u)\right) \ln(\epsilon)\nonumber\\
&- \frac{\Omega^4}{2\pi^2(1{+}\Omega^2)^4}\left(\int \!\dd^4u\ (\widetilde{u}_\mu A_\mu(u))^2\right) \ln(\epsilon) \nonumber\\
& -\frac{(1{-}\Omega^2)^2(1{+}4\Omega^2{+}\Omega^4)}{96\pi^2(1{+}\Omega^2)^4}\left(\int \! \dd^4u\ A_\mu(u)\partial^2 A_\mu(u)\right) \ln(\epsilon) \nonumber\\
& -\frac{(1{-}\Omega^2)^4}{96\pi^2(1{+}\Omega^2)^4} \left(\int \! \dd^4u\ (\partial_\mu A_\mu(u))^2\right)\ln(\epsilon)+ \dots\label{eq-gauge-2aptfun}\\
\mathcal{T}_2'' &= -\frac{1}{16\pi^2(1{+}\Omega^2)}\left(\int \!\dd^4u\ A_\mu(u)A_\mu(u)\right) \frac{1}{\epsilon} -\frac{m^2}{16\pi^2(1{+}\Omega^2)}\left(\int\! \dd^4u\ A_\mu(u)A_\mu(u)\right) \ln(\epsilon) \nonumber\\
& -\frac{\Omega^2}{4\pi^2\theta^2(1+\Omega^2)^2}\left(\int\! \dd^4u\ u^2A_\mu(u)A_\mu(u)\right) \ln(\epsilon) \nonumber\\
& + \frac{\Omega^2}{16\pi^2(1+\Omega^2)^2}\left(\int \dd^4u\ A_\mu(u)\partial^2 A_\mu(u)\right) \ln(\epsilon) + \dots
\end{align}
\end{subequations}

The computation of the 3-point function contributions can be conveniently carried out by further using the following identity:
\begin{equation*}
\int \!\dd^4u\ \widetilde{u}_\mu A_\mu(u) (A_\nu\star A_\nu)(u) =\frac{1}{2}\int \! \dd^4u\ \left(\widetilde{u}_\mu A_\nu(u)\{A_\mu,A_\nu\}_\star(u) \, -i(\partial_\mu A_\nu(u))[A_\mu,A_\nu]_\star(u) \right).
\end{equation*}
\begin{figure}[!htb]
  \centering
  \includegraphics[scale=1]{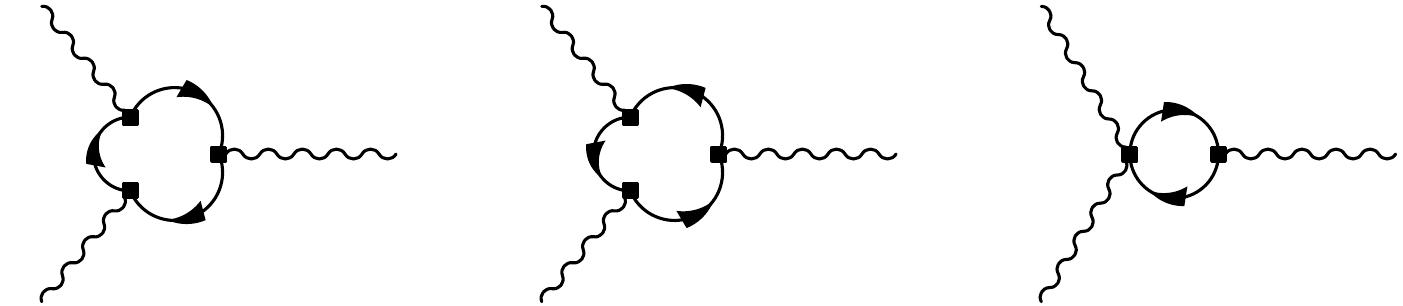}
  \caption[3-point]{\footnotesize{Relevant one-loop diagrams contributing to the 3-point function. Comments similar to those related to Figure \ref{fig-gauge-2-point} apply. The rightmost (resp.\ two leftmost) diagram(s) corresponds to the contribution $\mathcal{T}_3''$ (resp.\ $\mathcal{T}_3'$).}}
  \label{fig-gauge-3-point}
\end{figure}

The contributions corresponding to the diagrams of Figure \ref{fig-gauge-3-point} can then be expressed as:
\begin{subequations}
\label{eq-gauge-3ptfun}
\begin{align}
\mathcal{T}_3' &=
\frac{\Omega^2(1{-}\Omega^2)^2}{8\pi^2(1+\Omega^2)^4}\left( \int \! \dd^4u\ \widetilde{u}_\mu A_\nu(u)\{A_\mu,A_\nu\}_\star(u) \right)\ln(\epsilon) \nonumber\\
& +\frac{(1-\Omega^2)^2(1{+}4\Omega^2{+}\Omega^4)}{48\pi^2(1{+}\Omega^2)^4}\left(\int \!\dd^4u\ ((-i\partial_\mu A_\nu(u))[A_\mu,A_\nu]_\star(u) )\right)\ln(\epsilon)+ \dots \\
\mathcal{T}_3'' &= -\frac{\Omega^2}{8\pi^2(1+\Omega^2)^2}\left( \int \!\dd^4u\ (\widetilde{u}_\mu A_\nu(u)\{A_\mu,A_\nu\}_\star(u) )\right)\ln(\epsilon) \nonumber\\
& +\frac{i\Omega^2}{8\pi^2(1+\Omega^2)^2}\left(\int \!\dd^4u\ (\partial_\mu A_\nu(u))[A_\mu,A_\nu]_\star(u)\right)\ln(\epsilon)+ \dots
\end{align}
\end{subequations}
\begin{figure}[!htb]
  \centering
  \includegraphics[scale=1]{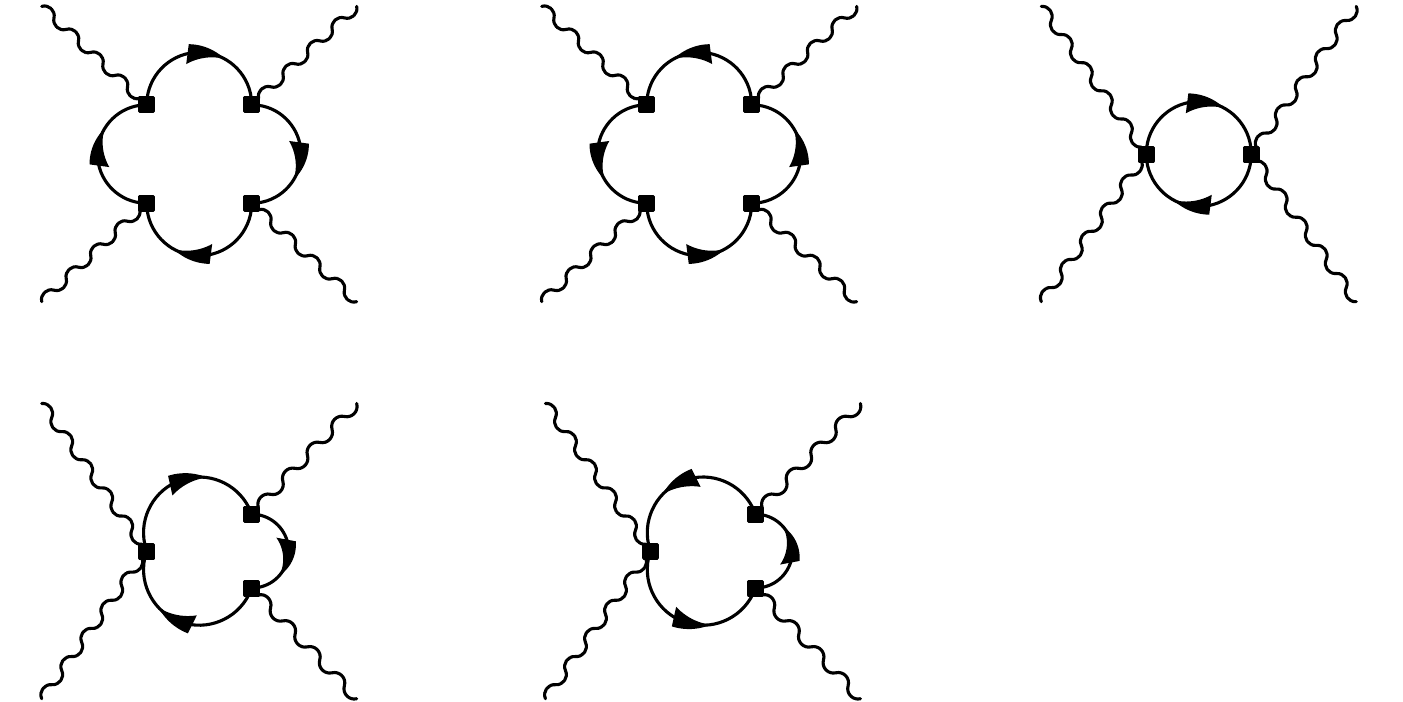}
  \caption[4-point]{\footnotesize{Relevant one-loop diagrams contributing to the 4-point function. Comments similar to those related to Figure \ref{fig-gauge-2-point} apply. Among the upper figures, the rightmost figure (resp.\ the two leftmost) diagram(s) corresponds to the contribution $\mathcal{T}_4'''$ (resp.\ $\mathcal{T}_4'$). The lower diagrams correspond to $\mathcal{T}_4''$. }}
  \label{fig-gauge-4-point}
\end{figure}

In the same way, the 4-point contributions depicted on Figure \ref{fig-gauge-4-point} are given by:
\begin{subequations}
\label{eq-gauge-4ptfun}
\begin{align}
\mathcal{T}_4' &= -\frac{(1-\Omega^2)^4}{96\pi^2(1+\Omega^2)^4} \left( \int \!\dd^4u\ \left( (A_\mu\star A_\nu (u))^2+2(A_\mu\star A_\nu(u))^2\right) \right) \ln(\epsilon)+ \dots\\
\mathcal{T}_4'' &= \frac{(1-\Omega^2)^2}{16\pi^2(1+\Omega^2)^2}\left(\int \!\dd^4u\ (A_\mu\star A_\mu(u))^2 \right) \ln(\epsilon)+ \dots\\
\mathcal{T}_4''' &= -\frac{1}{32\pi^2}\left(\int \! \dd^4u\ (A_\mu\star A_\mu(u))^2 \right) \ln(\epsilon)+ \dots\label{eq-gauge-4cptfun}
\end{align}
\end{subequations}

\subsection{Discussion on the effective action}
\label{subsec-gauge-disceffact}

By collecting the various contributions given in \eqref{eq-gauge-1ptfun}, \eqref{eq-gauge-2ptfun}, \eqref{eq-gauge-3ptfun} and \eqref{eq-gauge-4ptfun}, we find that the effective action $\Gamma(A)$ can be written as \cite{deGoursac:2007gq}:
\begin{multline}
\Gamma(A) = \frac{\Omega^2}{4\pi^2(1+\Omega^2)^3}\left(\int \!\dd^4u\ (\mathcal{A}_\mu\star\mathcal{A}_\mu -\frac{1}{4}\widetilde{u}^2) \right) \left(\frac{1}{\epsilon}+m^2\ln(\epsilon)\right)\\
 -\frac{(1-\Omega^2)^4}{192\pi^2(1+\Omega^2)^4} \left(\int \!\dd^4u\ F_{\mu\nu}\star F_{\mu\nu}\right)  \ln(\epsilon)\\
+\frac{\Omega^4}{8\pi^2(1+\Omega^2)^4}\left(\int \! \dd^4u\ (F_{\mu\nu}\star  F_{\mu\nu}+\{\mathcal{A}_\mu, \mathcal{A}_\nu\}_\star^2 -\frac{1}{4}(\widetilde{u}^2)^2)\right)\ln(\epsilon)+ \dots , \label{eq-gauge-zegamma}
\end{multline}
where $\mathcal{A}_\mu(u)= A_\mu(u)+\frac 12\widetilde{u}_\mu$ is the covariant coordinate (see subsection \ref{subsec-gauge-stdiffcalc}) and $F_{\mu\nu}=\partial_\mu A_\nu-\partial_\nu A_\mu-i[A_\mu,A_\nu]_\star$ is the curvature. To put the effective action into the form \eqref{eq-gauge-zegamma}, it is convenient to use the following formulae:
\begin{subequations}
\label{eq-gauge-reexpr}
\begin{align}
\int \!\dd^4x\ &\caA_\mu\star\caA_\mu =\int \!\dd^4x(\frac{1}{4}{\widetilde{x}}^2+{\widetilde{x}}_\mu A_\mu+A_\mu A_\mu),\\
\int \!\dd^4x\ &F_{\mu\nu}\star F_{\mu\nu} = \int \dd^4x\big(-2(A_\mu\partial^2A_\mu+(\partial_\mu A_\mu)^2)-4i\partial_\mu A_\nu[A_\mu,A_\nu]_\star-[A_\mu,A_\nu]_\star^2 \big),\\
\int \!\dd^4x\; &\{{\cal{A}}_\mu,{\cal{A}}_\nu\}^2_\star=\int \!\dd^4x\; \big(\frac{1}{4}({\widetilde{x}}^2)^2+2{\widetilde{x}}^2{\widetilde{x}}_\mu A_\mu
+4({\widetilde{x}}_\mu A_\mu)^2+2{\widetilde{x}}^2A_\mu A_\mu\nonumber\\
&+2(\partial_\mu A_\mu)^2+4{\widetilde{x}}_\mu A_\nu\{A_\mu,A_\nu\}_\star+\{A_\mu,A_\nu\}^2_\star\big)\\
\int \!\dd^4x\; &\widetilde{x}_\mu\widetilde{x}_\nu (A_\mu\star A_\nu)(x)=\int \!\dd^4x\; \big((\widetilde{x}_\mu A_\mu)^2+(\partial_\mu A_\mu)^2\big).
\end{align}
\end{subequations}
The effective action \eqref{eq-gauge-zegamma} is then a good candidate to solve the UV/IR mixing problem of the action \eqref{eq-gauge-actgauge}. Note that a one-loop computation of the effective action stemming from the noncommutative scalar field theory with harmonic term coupled to external field has also been done in the matrix basis \cite{Grosse:2007dm}, but with scalar fields transforming as
\begin{equation*}
\phi^g=g\star\phi \star g^\dag
\end{equation*}
under gauge transformations, instead of $\phi^g=g\star \phi$ in the present framework. The effective action computed in \cite{Grosse:2007dm} has the same expression as \eqref{eq-gauge-zegamma}.

\subsubsection{Relation with the parametric representation}

We will see here how the above calculations of subsection \ref{subsec-gauge-compeffact} could have been done in the parametric representation of the noncommutative scalar field theory with harmonic term. See subsection \ref{subsec-moy-prop} for a presentation of this parametric representation.
\medskip

\begin{figure}[!htb]
  \centering
  \includegraphics[scale=1]{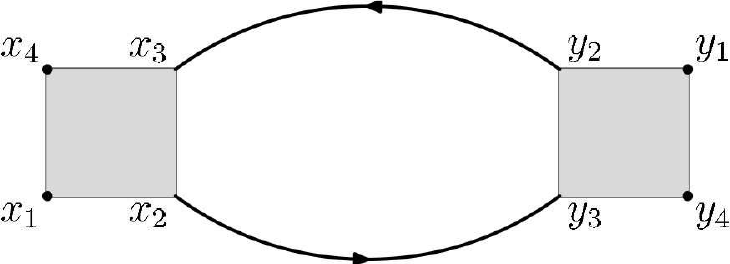}
  \caption[bubble]{\footnotesize{The bubble graph on the Moyal space}}
  \label{fig-gauge-bubble}
\end{figure}

It is possible to compute all the contributions to the effective action $\Gamma(A)$ we found in subsection \ref{subsec-gauge-compeffact}. We will nevertheless reduce our study to the contribution $\mathcal T_4'''$. One can notice on Figures \ref{fig-gauge-4-point} and \ref{fig-gauge-vertices} that this contribution involves only the ``bubble graph'' (see Figure \ref{fig-gauge-bubble}).
In \cite{Gurau:2006yc}, the expression for the polynom $HU_G(t)$ and the real part of $HV_G(x_e,t)$ of the ``bubble graph'' have been found. Notice that we have integrated over the hypermomentum of the rooted vertex (see \cite{Gurau:2006yc}). We have computed in \cite{deGoursac:2007qi} the imaginary part of $HV_G(x_e,t)$ and we have obtained:
\begin{align*}
HU_G(t)=&4(t_1+t_2)^2,\\
HV_G(x_e,t)=&\frac{2(1+\Omega^2)}{\Omega\theta}(t_1+t_2)((x_1-x_4+y_1-y_4)^2+t_1t_2(x_1-x_4-y_1+y_4)^2)\\
&-4i(t_1^2-t_2^2)(x_1-x_4)\wedge (y_1-y_4)-4i(t_1+t_2)^2(x_1\wedge x_4+y_1\wedge y_4).
\end{align*}
As a consequence, the amplitude $\mathcal T_4'''$ is given by
\begin{multline*}
\mathcal T_4'''=K \int_{\epsilon'}^\infty \frac{\dd\alpha_1 \dd\alpha_2(1-t_1^2)^2(1-t_2^2)^2}{16(t_1+t_2)^4} \int \dd^4x_1\dd^4x_4\dd^4y_1\dd^4y_4 A_\mu(x_1)A_\mu(x_4)A_\nu(y_1)A_\nu(y_4)\\ \exp\Big(-\frac{m^2\theta(\alpha_1+\alpha_2)}{4\Omega}-\frac{1+\Omega^2}{2\Omega\theta(t_1+t_2)}((x_1-x_4+y_1-y_4)^2
+t_1t_2(x_1-x_4-y_1+y_4)^2)\\
+i\frac{t_1-t_2}{t_1+t_2}(x_1-x_4)\wedge (y_1-y_4)+i(x_1\wedge x_4+y_1\wedge y_4) \Big),
\end{multline*}
where $K=\frac{2(1+\Omega^2)^2}{64\Omega^2\pi^8\theta^6}$ and $\epsilon'=2\widetilde \Omega \epsilon$ an UV cut-off. In the expression of $K$, $2(1+\Omega^2)^2$ comes from the combinatory of $\mathcal T_4'''$ and the overall factors affecting the vertices (see Figure \ref{fig-gauge-vertices}), while the other part of $K$ comes from the overall factor of the equation \eqref{eq-moy-polydef}. If we keep only the leading term in $\epsilon'$, we get
\begin{multline*}
\mathcal T_4'''= K \int_{\epsilon'}^\infty  \frac{\dd\alpha_1 \dd\alpha_2}{16(t_1+t_2)^4}\int \dd^4x_1\dd^4x_4\dd^4y_1\dd^4y_4 A_\mu(x_1)A_\mu(x_4)A_\nu(y_1)A_\nu(y_4)\\
\exp\Big(-\frac{1+\Omega^2}{2\Omega\theta(t_1+t_2)}(x_1-x_4+y_1-y_4)^2
+i\frac{t_1-t_2}{t_1+t_2}(x_1-x_4)\wedge (y_1-y_4)\\
+i(x_1\wedge x_4+y_1\wedge y_4) \Big).
\end{multline*}
Now we can make a change of variables $y_4=x_1-x_4+y_1+z$, and a Taylor expansion of $A_\nu(y_4)$:
\begin{multline}
\mathcal T_4'''=\\
K \int_{\epsilon'}^\infty \frac{\dd\alpha_1 \dd\alpha_2}{16(t_1+t_2)^4}\int \dd^4x_1\dd^4x_4\dd^4y_1\dd^4z A_\mu(x_1)A_\mu(x_4)A_\nu(y_1)(A_\nu(x_1-x_4+y_1)+O_{z\to0}(z)_\nu)\\
\exp\Big(-\frac{1+\Omega^2}{2\Omega\theta(t_1+t_2)}z^2 -i\frac{t_1-t_2}{t_1+t_2}(x_1-x_4)\wedge z+i(x_1\wedge x_4+y_1\wedge x_1-y_1\wedge x_4+y_1\wedge z) \Big).\label{eq-gauge-reg1}
\end{multline}
By performing a Gaussian integration over $z$ together with a Taylor expansion on the variable $\epsilon'\to0$ at the first order, the amplitude becomes:
\begin{multline*}
\mathcal T_4'''= K' \int_{\epsilon'}^\infty \frac{\dd\alpha_1 \dd\alpha_2}{(t_1+t_2)^2}\int \dd^4x_1\dd^4x_4\dd^4y_1 A_\mu(x_1)A_\mu(x_4)A_\nu(y_1)\\
(A_\nu(x_1-x_4+y_1)+O_{(t_1,t_2)\to(0,0)}(t_1,t_2)_\nu)e^{+i(x_1\wedge x_4+y_1\wedge x_1-y_1\wedge x_4)},
\end{multline*}
where $K'=\frac{\Omega^2\pi^2\theta^2}{4(1+\Omega^2)^2}K$. If we simplify this expression, we obtain:
\begin{equation*}
\mathcal T_4'''= -\frac{\ln(\epsilon)}{32\pi^2} \int \dd^4x (A_\mu\star A_\mu\star A_\nu\star A_\nu)(x)+O_{\epsilon\to0}(1),
\end{equation*}
and we recognize the equation \eqref{eq-gauge-4cptfun}. This was just an example to show the calculation power of the parametric representation in the above effective action's computation.

\section{Properties of the effective action}
\label{sec-gauge-prop}

From the analysis of section \ref{sec-gauge-effact}, we deduce that the following action:
\begin{equation}
S(A)=\int\dd^Dx\left(\frac 14 F_{\mu\nu}\star F_{\mu\nu}+\frac{\Omega^2}{4}\{\caA_\mu,\caA_\nu\}_\star^2+\kappa\caA_\mu\star\caA_\mu\right),\label{eq-gauge-actax}
\end{equation}
corresponding to the effective action \eqref{eq-gauge-zegamma}, is a good candidate to renormalizability. The constants are of mass dimension: $[\Omega]=0$, $[\kappa]=2$, and $\Omega$ do not correspond to the one of the scalar action \eqref{eq-moy-actharm}, even if it is related at the quantum level. We set the coupling constant $g=1$ to simplify the present analysis.
\medskip

We recall here that the covariant coordinate $\caA_\mu=A_\mu+\frac 12\wx_\mu$, and the curvature $F_{\mu\nu}=\partial_\mu A_\nu-\partial_\nu A_\mu-i[A_\mu,A_\nu]_\star$ transform covariantly under a unitary gauge transformation $g(x)$. Consequently, the action is invariant under gauge transformations, due to the tracial identity \eqref{eq-moy-trace}. Moreover, as a one-loop effective action, this action is closely related (at the quantum level) to the renormalizable scalar field theory with harmonic term \eqref{eq-moy-actharm}, and it is a good argument for its possible renormalizability.

By using Equation \eqref{eq-gauge-FcaA}, the action \eqref{eq-gauge-actax} can be reexpressed in terms of the covariant coordinate, up to a constant term:
\begin{equation}
S(A)=\int\dd^Dx\left(-\frac 14[\caA_\mu,\caA_\nu]_\star^2 +\frac{\Omega^2}{4}\{\caA_\mu,\caA_\nu\}_\star^2+\kappa\caA_\mu\star\caA_\mu\right).\label{eq-gauge-actax2}
\end{equation}
One can notice a symmetry $i[\caA_\mu,\caA_\nu]_\star\rightleftarrows\{\caA_\mu,\caA_\nu\}_\star$ in the first two terms of the action. This will be discussed more closely and interpreted in section \ref{sec-gauge-interp}.

Thanks to the equations \eqref{eq-gauge-reexpr}, the action can also be completely expanded in terms of the gauge potential:
\begin{multline*}
S(A)=\int\dd^Dx\Big(\frac{\Omega^2}{4}\wx^2\wx_\mu A_\mu+\kappa\wx_\mu A_\mu-\frac 12A_\mu\partial^2 A_\mu+\frac{\Omega^2}{2}\wx^2 A_\mu A_\mu+\kappa A_\mu A_\mu-\frac 12(1-\Omega^2)(\partial_\mu A_\mu)^2\\
+\Omega^2(\wx_\mu A_\mu)^2-i(\partial_\mu A_\nu)[A_\mu,A_\nu]_\star+\Omega^2\wx_\mu A_\nu\{A_\mu,A_\nu\}_\star-\frac 14[A_\mu,A_\nu]_\star^2+\frac{\Omega^2}{4}\{A_\mu,A_\nu\}_\star^2\Big),
\end{multline*}
up to a constant term. In this expression, one notes that the quadratic part involves the operator $(-\partial^2+\Omega^2\wx^2+2\kappa)$, whose inverse is the Mehler kernel \eqref{eq-moy-propagx}, responsible of the renormalizability in the scalar case. It turns out that there is a mass-type term in the action \eqref{eq-gauge-actax}, without spontaneous symmetry breaking and Higgs mechanism, and without loss of gauge invariance. This is a particular feature in this noncommutative field theory.

Moreover, one sees that this action is not covariant under the Langmann-Szabo duality (for example, the cubic part is not, see subsection \ref{subsec-moy-prop}). Finally, the first two terms are linear in the field $A_\mu$, so that the action \eqref{eq-gauge-actax} does not admit $A_\mu=0$ as a vacuum configuration. In the rest of this section, we will study the vacuum solutions for this action \cite{deGoursac:2008rb}.

\subsection{Symmetries of vacuum configurations}
\label{subsec-gauge-symvac}

Let us first take into account the symmetries of the action to reduce the study of the vacuum configurations, like in subsection \ref{subsec-moy-vacuum}. Indeed, we will also assume that the new action, obtained from \eqref{eq-gauge-actax} by expanding around a vacuum configuration, is invariant under the group $G_D=SO(D)\cap Sp(D)$. Since $G_D$ acts on the gauge potentials as:
\begin{equation*}
\forall\Lambda\in G_D,\quad A_\mu^\Lambda(x)=\Lambda_{\mu\nu}\, A_\nu(\Lambda^{-1}x),
\end{equation*}
one obtains the following condition \cite{deGoursac:2008rb}:
\begin{equation}
\forall\Lambda\in G_D,\quad A_\mu^\Lambda(x)=A_\mu(x),\label{eq-gauge-condvac}
\end{equation}
for any vacuum configuration $A_\mu$, which is equivalent to: $\caA_\mu^\Lambda=\caA_\mu$, for the covariant coordinate vacuum. Owing to the properties of the theory of groups invariants \cite{Weyl:1946}, one finds that the general expression for a vacuum configuration can be written as:
\begin{equation}
\caA_\mu(x)=\Phi_1(x^2)x_\mu+\Phi_2(x^2)\wx_\mu,\label{eq-gauge-condvac2}
\end{equation}
where $\Phi_i$ are arbitrary smooth functions on $x^2$.

\subsection{Equation of motion}
\label{subsec-gauge-motion}

The equation of motion $\frac{\delta S}{\delta \mathcal A_\mu(x)}=0$, of the action \eqref{eq-gauge-actax}, reexpressed in terms of $\caA_\mu$, is given by:
\begin{equation}
-2(1-\Omega^2)\mathcal{A}_\nu\star\mathcal{A}_\mu\star\mathcal{A}_\nu + (1+\Omega^2)\mathcal{A}_\mu\star\mathcal{A}_\nu\star\mathcal{A}_\nu + (1+\Omega^2)\mathcal{A}_\nu\star\mathcal{A}_\nu\star\mathcal{A}_\mu + 2\kappa\mathcal{A}_\mu=0.\label{eq-gauge-motion}
\end{equation}
Due to the very structure of the Moyal product, this is a complicated integro-differential equation for which no known algorithm to solve it does exist so far. Let us solve this equation in the case $D=2$, by using the matrix basis and assuming the condition \eqref{eq-gauge-condvac}.

In two dimensions, it is convenient to define
\begin{equation*}
Z(x)=\frac{\mathcal A_1(x)+i\mathcal A_2(x)}{\sqrt{2}}\qquad Z^\dag(x)=\frac{\mathcal A_1(x)-i\mathcal A_2(x)}{\sqrt{2}}.
\end{equation*}
Then, the action can be expressed as:
\begin{equation}
S=\int d^2x\Big((-1+3\Omega^2)Z\star Z\star Z^\dag\star Z^\dag+(1+\Omega^2)Z\star Z^\dag\star Z\star Z^\dag+2\kappa Z\star Z^\dag\Big),\label{eq-gauge-actmatrix}
\end{equation}
so that the equation of motion takes the form:
\begin{equation}
(3\Omega^2-1)(Z^\dag\star Z\star Z+Z\star Z\star Z^\dag)+2(1+\Omega^2)Z\star Z^\dag\star Z+2\kappa Z=0.\label{eq-gauge-motmatr}
\end{equation}
Expressing now $Z(x)$ in the matrix base, namely:
\begin{equation*}
Z(x)=\sum_{m,n=0}^\infty Z_{mn} f_{mn}(x),
\end{equation*}
\eqref{eq-gauge-motmatr} becomes a cubic infinite-dimensional matrix equation. In view of the discussion carried out in subsection \ref{subsec-gauge-symvac}, we now look for the symmetric solutions of the form given by \eqref{eq-gauge-condvac2}, namely:
\begin{equation*}
Z(x)=\Phi_1(x^2)\left(\frac{x_1+ix_2}{\sqrt 2}\right)+\Phi_2(x^2)\left(\frac{\widetilde x_1+i\widetilde x_2}{\sqrt 2}\right).
\end{equation*}
By using the properties of the matrix basis (see subsection \ref{subsec-moy-matrix}) in the polar coordinates, one obtains the general form of the matrix coefficients of $Z(x)$:
\begin{equation}
Z_{mn}=-ia_m\delta_{m+1,n},\label{eq-gauge-vacmatrix}
\end{equation}
with $a_m\in\gC$, for a vacuum configuration satisfying \eqref{eq-gauge-condvac}.

This, inserted into \eqref{eq-gauge-motmatr} yields: $\forall m\in\gN$,
\begin{equation}
a_m\left((3\Omega^2-1)(|a_{m-1}|^2+|a_{m+1}|^2)+2(1+\Omega^2)|a_m|^2+2\kappa\right)=0,\label{eq-gauge-mvtmatr}
\end{equation}
where it is understood that $a_{-1}=0$. Then, \eqref{eq-gauge-mvtmatr} implies: $\forall m\in\gN$,
\begin{equation}
(i)\ a_m=0\quad\text{or}\quad (ii)\ (3\Omega^2-1)(|a_{m-1}|^2+|a_{m+1}|^2) +2(1+\Omega^2)|a_m|^2+2\kappa=0.\label{eq-gauge-mvtcond}
\end{equation}

From now on, we will focus only on the second condition $(ii)$ (this will be more closely discussed in subsection \ref{subsec-gauge-minact}), that is, upon setting $u_{m+1}=|a_m|^2$: $\forall m\in\gN$,
\begin{equation}
(3\Omega^2-1)(u_m+u_{m+2})+2(1+\Omega^2)u_{m+1}+2\kappa=0,\label{eq-gauge-mvtprinc}
\end{equation}
which is a non homogenous linear iterative equation of second order with boundary condition $u_0=0$.

\subsection{Solutions of the equation of motion}
\label{subsec-gauge-solvac}

Let us solve the equation \eqref{eq-gauge-mvtprinc}, with $u_m\geq0$ and $u_0=0$ \cite{deGoursac:2008rb}. For $\Omega^2\neq \frac 13$, we define
\begin{equation*}
r=\frac{1+\Omega^2+\sqrt{8\Omega^2(1-\Omega^2)}}{1-3\Omega^2},
\end{equation*}
and one has $\frac 1r=\frac{1+\Omega^2-\sqrt{8\Omega^2(1-\Omega^2)}}{1-3\Omega^2}$. Then, it is easy to realize that \eqref{eq-gauge-mvtprinc} supports different types of solutions according to the range for the values taken by $\Omega$. Namely, one has:
\begin{itemize}
\item for $\Omega^2=0$, $\kappa=0$, $u_m=\alpha m$ and $\alpha\geq0$;
\item for $0<\Omega^2<\frac 13$, $u_m=\alpha(r^{m}-r^{-m})-\frac{\kappa}{4\Omega^2}(1-r^{-m})$, $\alpha\geq0$ and $r>1$;
\item for $\Omega^2=\frac 13$, $\kappa\leq0$ and $u_m=-\frac{3\kappa}{4}$;
\item for $\frac 13<\Omega^2<1$, $\kappa\leq0$, $u_m=-\frac{\kappa}{4\Omega^2}(1-r^{-m})$ and $r<-1$;
\item for $\Omega^2=1$, $\kappa\leq0$ and $u_m=-\frac{\kappa}{4}(1-(-1)^{-m})$.
\end{itemize}

Consider the asymptotic behavior of the vacuum in the configuration space for $x^2\to\infty$. If $0<\Omega^2<\frac 13$ and $\alpha\neq0$, then $\sqrt{u_m}\thicksim_{m\to\infty}r^{\frac m2}$ and $r>1$. As a consequence (see subsection \ref{subsec-moy-moyal}), the solution $\mathcal A_\mu(x)$ of the equation of motion does not belong to the Moyal algebra. So we require that $\alpha=0$. Then, for $\Omega\neq0$, $\kappa$ has to be negative and $u_m$ has a finite limit. This indicates that $\mathcal A_\mu(x)$ has a constant limit when $x^2\to\infty$.

Notice also that the solution for $\Omega=0$ corresponds to the commutative case: $u_m=\frac{m}{\theta}$ is equivalent to a vacuum $\mathcal A_\mu(x)=\frac 12\widetilde x_\mu$ or $A_\mu=0$. Then, it is possible to choose $\kappa$ depending on $\Omega$ so that the solution is continuous in $\Omega$ near 0. From the Taylor expansion
\begin{equation*}
1-r^{-m}=2\sqrt 2m\Omega+O(\Omega^2),
\end{equation*}
one deduces that $\kappa$ must have the same asymptotic behavior as $\Omega$ near 0. If one sets:
\begin{equation*}
 \kappa=-\frac{\Omega\sqrt 2}{\theta}+O(\Omega^2),
\end{equation*}
then $u_m=\frac{m}{\theta}+O(\Omega)$, so that the commutative limit for the vacuum is recovered: $A_\mu=0$, and moreover the gauge potential is massless ($\lim_{\Omega\to0}\kappa=0$).

Finally, there is a freedom in the choice of $a_m$ in function of $u_m$, described by a phase $\xi_m\in\gR$:
\begin{equation*}
a_m=e^{i\xi_m}\sqrt{u_m},
\end{equation*}
where the expression of $u_m$ has been found above. Then, by using the expression of the matrix basis in two dimensions \eqref{eq-moy-laguerre}, it can be shown \cite{deGoursac:2008rb} that the vacuum configurations satisfying \eqref{eq-gauge-condvac} are given by:
\begin{equation}
\caA_\mu(x)=2\sqrt{\theta}\frac{e^{\frac z2}}{\sqrt z}\int_0^\infty dt\ e^{-t}\sqrt tJ_1(2\sqrt{tz}) \sum_{m=0}^\infty \frac{(-1)^m\sqrt{u_{m+1}}}{m!\sqrt{m+1}}t^m\left(\widetilde x_\mu\cos(\xi_m)+\frac{2}{\theta}x_\mu\sin(\xi_m)\right)\label{eq-gauge-vacsol}
\end{equation}
where $J_k(x)$ is the $k$-th function of Bessel of the first kind, and $z=\frac{2x^2}{\theta}$.

Note that: $\forall m\in\gN$, $u_m=0$, or equivalently $\caA_\mu=0$ or $A_\mu=-\frac 12\wx_\mu$, is a trivial solution of the equation of motion (by using the condition (i) of \eqref{eq-gauge-mvtcond}). By expanding the action \eqref{eq-gauge-actax} around this solution, one finds:
\begin{equation}
S=\int\dd^Dx\Big(\kappa\caA_\mu\star\caA_\mu+\frac{1+\Omega^2}{2}\caA_\mu\star\caA_\mu\star\caA_\nu\star\caA_\nu -\frac{1-\Omega^2}{2}\caA_\mu\star\caA_\nu\star\caA_\mu\star\caA_\nu\Big),\label{eq-gauge-actexp}
\end{equation}
up to an inessential constant term. $\caA_\mu$ is then the quantum gauge field with vanishing expectation value. Note that for $\kappa\geq0$, the propagator $\frac 2 \kappa \delta_{\mu\nu}\delta(x-y)$, which is trivial, is positive, so $\mathcal A_\mu=0$ is a minimum of the action \eqref{eq-gauge-actax}. However, since the action \eqref{eq-gauge-actexp} involves only this trivial quadratic term and two quartic vertices, it gives rise to a non-dynamical matrix theory, where the fields are the four infinite-dimensional matrices associated to the $\caA_\mu$ variable.

\subsection{Minima of the action}
\label{subsec-gauge-minact}

In this subsection, we discuss about the condition of minimality of the solutions of \eqref{eq-gauge-mvtcond}. For $\kappa\geq0$, the action, reexpressed in terms of $\caA_\mu$ \eqref{eq-gauge-actax2}, is positive. Since this action vanishes for $\caA_\mu=0$, and since $\caA_\mu=0$ is a solution of the equation of motion (see above), one deduces that $\caA_\mu=0$ is the global minimum of the action for $\kappa\geq0$. Moreover, owing to the properties of the solutions of \eqref{eq-gauge-mvtprinc}, discussed in the latter subsection, one can say that it is the only local minimum of the action. Therefore, the action \eqref{eq-gauge-actax} has a single vacuum configuration for $\kappa\geq0$: $\caA_\mu=0$, or equivalently $A_\mu=-\frac 12\wx_\mu$. The situation is very similar to the scalar case (see subsection \ref{subsec-moy-vacuum}).

\medskip

If $\kappa<0$, the solutions of \eqref{eq-gauge-mvtcond} using only condition (ii) have been exhibited in the previous subsection, but one can deduce the general solutions (using (i) and (ii)) from this study. However, it is difficult to verify if these solutions are minima of the action or not, even in the matrix basis.

Note that in the scalar case, which looks very similar, the minima use maximally the condition (ii). If it is also the case here, the minima would then be given by the solutions \eqref{eq-gauge-vacsol}. Furthermore, in the special case $\Omega^2=\frac 13$ (and $\kappa<0$), it has been checked explicitely \cite{deGoursac:2008rb} that the solutions \eqref{eq-gauge-vacsol} correspond to degenerated minima of the action \eqref{eq-gauge-actax}.

\subsection{Extension in higher dimensions}
\label{subsec-gauge-extdim}

The previous results concerning the vacuum configurations of the action \eqref{eq-gauge-actax} in two dimensions have been generalized to the case of four dimensions \cite{deGoursac:2008rb}. Defining
\begin{equation*}
Z_1(x)=\frac{\mathcal A_1(x)+i\mathcal A_2(x)}{\sqrt{2}}\qquad Z_2(x)=\frac{\mathcal A_3(x)+i\mathcal A_4(x)}{\sqrt{2}},
\end{equation*}
one can show with the symmetry of the action that the vacuum has to be of the following form in the matrix basis:
\begin{align*}
(Z_1)_{m,n}&=-ia_{m_1+m_2}\sqrt{m_1+1}\delta_{m_1+1,n_1}\delta_{m_2,n_2},\\
(Z_2)_{m,n}&=-ia_{m_1+m_2}\sqrt{m_2+1}\delta_{m_1,n_1}\delta_{m_2+1,n_2},
\end{align*}
where $a_{m_1+m_2}\in\mathbb C$. Then, if one assumes that $a_m\neq0$, the equation of motion is equivalent to: $\forall m\in\gN$,
\begin{equation}
(3\Omega^2-1)(mv_m+(m+3)v_{m+2})+(1+\Omega^2)(2m+3)v_{m+1}+2\kappa=0,\label{eq-gauge-mvtprinc2}
\end{equation}
with $v_{m+1}=|a_m|^2$, and the boundary condition $v_0=0$.

\begin{theorem}
\label{thm-gauge-vac}
For $\kappa=0$, the solutions of \eqref{eq-gauge-mvtprinc2} are given by: $\forall m\geq0$,
\begin{multline}
v_{m+1}=\frac{(1+\Omega^2)^2v_1}{4\sqrt{\pi}\Omega^2(1-\Omega^2)} \frac{\Gamma(3/2)\Gamma(m+3/2)}{\Gamma(m/2+3/2)\Gamma(m/2+2)} \Big(\frac{1+\Omega^2}{1-3\Omega^2}\Big)^m\\
{}_2F_1\left(-\frac m2-\frac 12,-\frac m2-1;-m-\frac 12; \frac{(1-3\Omega^2)^2}{(1+\Omega^2)^2}\right),\label{eq-gauge-bigseq}
\end{multline}
where ${}_2F_1$ is a hypergeometric function. Moreover, they correspond in the configuration space to:
\begin{equation*}
\caA_\mu(x)=2\sqrt{2\theta}\frac{e^{\frac z2}}{z}\int_0^\infty dt\ e^{-t}J_2(2\sqrt{tz})\sum_{m=0}^\infty \frac{(-1)^m}{m!}\sqrt{v_{m+1}}t^{m+1}\Big(\widetilde x_\mu\cos(\xi_m)+\frac{2}{\theta}x_\mu\sin(\xi_m)\Big),
\end{equation*}
with $z=\frac{2x^2}{\theta}$, and where $\xi_m\in\gR$ are arbitrary phases.
\end{theorem}
Notice that for $\frac{(1-3\Omega^2)^2}{(1+\Omega^2)^2}$ small enough, this expression is positive if we choose $v_1\geq0$. Furthermore, ${}_2F_1\left(-\frac m2-\frac 12,-\frac m2-1;-m-\frac 12; z\right)$ is a polynom of degree $\lfloor \frac{m+2}{2}\rfloor$ in $z$. Some numerical studies have been done on these vacuum configurations by the group of R. Wulkenhaar in M\"unster.

\section{Interpretation of the effective action}
\label{sec-gauge-interp}

We have seen in subsection \ref{subsec-moy-prop} that the renormalizable scalar field theory with harmonic term \eqref{eq-moy-actharm} is symmetric under the Langmann-Szabo duality \cite{Langmann:2002cc}. Note that the action \eqref{eq-moy-actharm} can be reformulated into:
\begin{equation*}
S(\phi)=\int\dd^Dx\big(\frac 12(i[\xi_\mu,\phi]_\star)^2 +\frac{\Omega^2}{2}(\{\xi_\mu,\phi\}_\star)^2 +\frac{\mu^2}{2}\phi^2+\lambda\,\phi\star\phi\star\phi\star\phi\big),
\end{equation*}
due to the identities \eqref{eq-moy-basident}. Recall that $\xi_\mu=-\frac 12\wx_\mu$. Then, at the quadratic level of the action, the Langmann-Szabo duality can be seen as a symmetry between commutator and anticommutator: $i[\xi_\mu,\fois]_\star\rightleftarrows\{\xi_\mu,\fois\}_\star$.

The gauge action \eqref{eq-gauge-actax} can be reexpressed in terms of the covariant coordinate (see \eqref{eq-gauge-actax2}):
\begin{equation*}
S(A)=\int\dd^Dx\left(-\frac 14[\caA_\mu,\caA_\nu]_\star^2 +\frac{\Omega^2}{4}\{\caA_\mu,\caA_\nu\}_\star^2+\kappa\caA_\mu\star\caA_\mu\right).
\end{equation*}
We have also seen that this action is not covariant under Langmann-Szabo duality, and this duality seems to be not well adapted to gauge theory. However, one can notice in this gauge theory a symmetry between commutator and anticommutator: $i[\fois,\fois]_\star\rightleftarrows\{\fois,\fois\}_\star$, like in the scalar case. This is the sign that the Langmann-Szabo duality could be the scalar part of a more general symmetry, exchanging commutators and anticommutators. Of course, this more general symmetry is then very reminiscent to a grading symmetry.
\medskip

In this section, we will follow \cite{deGoursac:2008bd} and  give an explicit grading symmetry adapted to the above field theories. First, we find a superalgebra (algebra with a grading) constructed from the Moyal space. Then, we construct an appropriate framework to a gauge theory with this superalgebra (see section \ref{sec-eps-epsncg} for the general theory). Finally we compute the classical action $\tr(F^2)$, where $F$ is now the graded curvature of this theory.

\subsection{A superalgebra constructed from Moyal space}
\label{subsec-gauge-epssuper}

Let us define a superalgebra ($\gZ_2$-graded algebra) from the Moyal algebra \cite{deGoursac:2008bd}.

\begin{definition}
Let $\algrA$ be the $\gZ_2$-graded complex vector space defined by $\algA^0=\caM$ and $\algA^1=\caM$. Let us also introduce the following product on $\algrA=\algA^0\oplus\algA^1$: $\forall a,b,c,d\in\caM$,
\begin{equation}
(a,b)\fois (c,d)=(a\star c+\alpha\ b\star d,a\star d+b\star c),\label{defprodmoy}
\end{equation}
where $\alpha$ is a real parameter.
\end{definition}
\begin{proposition}
The vector space $\algrA$, endowed with the product \eqref{defprodmoy}, is a superalgebra. The bracket of its associated Lie superalgebra has the following expression: for $\phi=(\phi_0,\phi_1)\in\algrA$ and $\psi=(\psi_0,\psi_1)\in\algrA$,
\begin{equation}
[\phi,\psi]=([\phi_0,\psi_0]_\star+\alpha\{\phi_1,\psi_1\}_\star,[\phi_0,\psi_1]_\star+[\phi_1,\psi_0]_\star).\label{eq-gauge-epsbracket}
\end{equation}
Moreover, $(\phi_0,\phi_1)^\ast=(\phi_0^\dag,\phi_1^\dag)$ is an involution for $\algrA$ ($\alpha\in\gR$).
\end{proposition}
\begin{proof}
For example, let us check the associativity of the product: $\forall a,b,c,d,e,f\in\caM$,
\begin{align*}
((a,b)&\fois(c,d))\fois(e,f)\\
=&(a\star c\star e+\alpha(b\star d\star e+a\star d\star f+b\star c\star f),\\
& a\star c\star f+a\star d\star e+b\star c\star e+\alpha\ b\star d\star f)\\
=&(a,b)\fois((c,d)\fois(e,f)).
\end{align*}
In the same way, distributivity and the other axioms are verified. Notice that $\gone=(1,0)$ is the unit element of this algebra.
\end{proof}
Note that, like the other superalgebras, $\algrA$ is an $\eps$-graded algebra (see subsection \ref{subsec-eps-defalg}) for the commutation factor $\eps:\gZ_2\times\gZ_2\to\{-1,1\}$ defined by: $\eps(i,j)=(-1)^{ij}$, for $i,j\in\gZ_2$.

Since the center of the Moyal algebra is trivial ($\caZ(\caM)=\gC$), we find the graded center of $\algrA$:
\begin{equation*}
\caZ^\bullet(\algA)=\gC\gone=\gC\oplus\algzero.
\end{equation*}
Furthermore, a (non-graded) trace of this algebra is given by: $\forall\phi=(\phi_0,\phi_1)\in\algrA$,
\begin{equation*}
\tr(\phi)=\int d^Dx\ \phi_0(x).
\end{equation*}

Let us now define the following objects $\gamma=1$, $\xi_\mu=-\frac 12\wx_\mu$ and $\eta_{\mu\nu}=\frac12\wx_\mu\wx_\nu=2\xi_\mu\xi_\nu$, and give some calculation rules, which will be useful in the following: $\forall\phi\in\caM$,
\begin{align}
[i\gamma,\phi]_\star=0,\qquad  \{i\gamma,\phi\}_\star=2i\gamma\phi,\nonumber\\
[i\xi_\mu,\phi]_\star=\partial_\mu\phi, \qquad \{i\xi_\mu,\phi\}_\star=2i\xi_\mu.\phi,\qquad [i\eta_{\mu\nu},\phi]_\star=\frac 12\xi_\mu\partial_\nu\phi+\frac 12\xi_\nu\partial_\mu\phi.\label{eq-gauge-calcrules}
\end{align}
Recall the definition of the adjoint representation: for $\phi=(\phi_0,\phi_1)\in\algrA$ and $\psi=(\psi_0,\psi_1)\in\algrA$,
\begin{equation*}
\ad_{\phi}\psi=[\phi,\psi],
\end{equation*}
where the bracket is defined in \eqref{eq-gauge-epsbracket}.
\begin{proposition}
Then, $\ad_{(0,i\gamma)}$, $\ad_{(i\xi_\mu,0)}$, $\ad_{(0,i\xi_\mu)}$ and $\ad_{(i\eta_{\mu\nu},0)}$ are real graded derivations of $\algrA$ of respective degrees 1, 0, 1 and 0. Moreover, the vector space $\kg^\bullet$ generated by these graded derivations is a graded Lie subalgebra of $\Der^\bullet(\algA)$ and a right $\caZ^\bullet(\algA)$-module.
\end{proposition}
\begin{proof}
The following relations, computed from \eqref{eq-gauge-calcrules},
\begin{subequations}
\begin{align*}
[(0,i\gamma),(0,i\gamma)]&=(-2\alpha,0)\\
[(i\xi_\mu,0),(0,i\gamma)]&=(0,0)\\
[(0,i\xi_\mu),(0,i\gamma)]&=(-2\alpha\xi_\mu,0)\\
[(i\eta_{\mu\nu}),(0,i\gamma)]&=(0,0)\\
[(i\xi_\mu,0),(i\xi_\nu,0)]&=(i\Theta^{-1}_{\mu\nu},0)\\
[(i\xi_\mu,0),(0,i\xi_\nu)]&=(0,i\Theta^{-1}_{\mu\nu}\gamma)\\
[(0,i\xi_\mu),(0,i\xi_\nu)]&=(-\alpha\eta_{\mu\nu},0)\\
[(i\eta_{\mu\nu},0),(i\xi_\rho,0)]&=(\frac i2\xi_\mu\Theta^{-1}_{\nu\rho}+\frac i2\xi_\nu\Theta^{-1}_{\mu\rho},0)\\
[(i\eta_{\mu\nu},0),(0,i\xi_\rho)]&=(0,\frac i2\xi_\mu\Theta^{-1}_{\nu\rho}+\frac i2\xi_\nu\Theta^{-1}_{\mu\rho})\\
[(i\eta_{\mu\nu},0),(i\eta_{\rho\sigma},0)]&=(\frac i2\eta_{\mu\rho}\Theta^{-1}_{\nu\sigma} +\frac i2\eta_{\mu\sigma}\Theta^{-1}_{\nu\rho} +\frac i2\eta_{\nu\rho}\Theta^{-1}_{\mu\sigma} +\frac i2\eta_{\nu\sigma}\Theta^{-1}_{\mu\rho},0),
\end{align*}
\end{subequations}
combined with $\forall a,b\in\algrA$, $[\ad_a,\ad_b]=\ad_{[a,b]}$, show that $\kg^\bullet$ is an $\eps$-Lie algebra.
\end{proof}
Notice that the vector space generated only by $\ad_{(i\xi_\mu,0)}$ and $\ad_{(0,i\xi_\mu)}$ is not a graded Lie subalgebra. Indeed, $\kg^\bullet$ is the smallest subalgebra of $\Der^\bullet(\algA)$ involving $\ad_{(i\xi_\mu,0)}$ and $\ad_{(0,i\xi_\mu)}$.

\subsection{Differential calculus and scalar theory}
\label{subsec-gauge-epscalcdiff}

Within the formalism developed in subsection \ref{subsec-eps-diffcalc}, it is possible to construct a differential calculus for the superalgebra $\algrA$ \cite{deGoursac:2008bd}. Indeed, since $\kg^\bullet$ is a graded Lie subalgebra of $\Der^\bullet(\algA)$ and a $\caZ^\bullet(\algA)$-module, one can consider the graded derivation-based differential calculus $\Omega^{\bullet,\bullet}(\algA|\kg)$ for $\algrA$ restricted to the graded derivations of $\kg^\bullet$. Recall that $\Omega^{n,\bullet}(\algA|\kg)$ is defined as the space of $n$-linear maps $\omega:(\kg^\bullet)^n\to\algrA$ such that $\forall\kX_j\in\kg^\bullet$,
\begin{equation*}
\omega(\kX_1,\dots,\kX_i,\kX_{i+1},\dots,\kX_n)=-(-1)^{|\kX_i||\kX_{i+1}|}\omega(\kX_1,\dots,\kX_{i+1},\kX_i,\dots,\kX_n),
\end{equation*}
if $\kX_i$ and $\kX_{i+1}$ are homogeneous. Note that the conditions involving elements of the center of $\algrA$ disappear since $\caZ^\bullet(\algA)$ is trivial.

$\Omega^{\bullet,\bullet}(\algA|\kg)$ is endowed with the standard product and the standard Koszul differential $\dd$ (see subsection \ref{subsec-eps-diffcalc}), so that it is a ($\gN\times\gZ_2$) bigraded differential algebra. In degree $0$, the differential is given by: $\forall\phi=(\phi_0,\phi_1)\in\algrA$,
\begin{subequations}
\label{eq-gauge-diffeps}
\begin{align}
&\dd(\phi_0,\phi_1)(\ad_{(0,i\gamma)})=(-2i\alpha\phi_1,0),\\
&\dd(\phi_0,\phi_1)(\ad_{(i\xi_\mu,0)})=(\partial_\mu\phi_0,\partial_\mu\phi_1),\\
&\dd(\phi_0,\phi_1)(\ad_{(0,i\xi_\mu)})=(i\alpha\wx_\mu\phi_1,\partial_\mu\phi_0),\\
&\dd(\phi_0,\phi_1)(\ad_{(i\eta_{\mu\nu},0)})=-\frac 14(\wx_\mu\partial_\nu\phi_0+\wx_\nu\partial_\mu\phi_0, \wx_\mu\partial_\nu\phi_1+\wx_\nu\partial_\mu\phi_1),
\end{align}
\end{subequations}

Then, by identifying $\phi_0=\phi_1=\phi\in\caM$, one obtains a part of an action for the scalar field $\phi$:
\begin{equation*}
\tr\Big(|\dd(\phi,\phi)(\ad_a)|^2\Big)\\
=\int d^Dx\Big((1+2\alpha)(\partial_\mu\phi)^2+\alpha^2(\wx_\mu\phi)^2+\frac{4\alpha^2}{\theta}\phi^2\Big),
\end{equation*}
where $|b|^2=b^\ast\fois b\in\algrA$ and if $a$ is summed over $\{(0,\frac{i}{\sqrt{\theta}}\gamma),(i\xi_\mu,0),(0,i\xi_\mu)\}$ . Setting $\Omega^2=\frac{\alpha^2}{1+2\alpha}$ and $\mu^2=\frac{4\alpha^2}{\theta(1+2\alpha)}$, we find that the latter expression is the quadratic part of the Grosse-Wulkenhaar action with harmonic term \eqref{eq-moy-actharm}. This stems to the fact that this graduation of Moyal mimics the Langmann-Szabo duality. Indeed, the Langmann-Szabo duality ($\partial_\mu\rightleftarrows\wx_\mu$) can be related to the symmetry $(i\xi_\mu,0)\rightleftarrows(0,i\xi_\mu)$ thanks to the relations \eqref{eq-gauge-diffeps}, and further assuming $\phi_0=\phi_1=\phi$. In fact, an exact correspondence between Langmann-Szabo duality and this grading exchange has been proven in \cite{deGoursac:2010zb}.

\subsection{Graded connections and gauge theory}
\label{subsec-gauge-epsconn}

Let us consider the module $\modrM=\algrA$, with the hermitian structure $\langle a,b\rangle=a^\ast\fois b$, $\forall a,b\in\algrA$. We describe here the graded connections (see subsection \ref{subsec-eps-conn}) of degree $0$ and their curvatures in terms of the gauge potentials and the covariant coordinates \cite{deGoursac:2008bd}.

\begin{proposition}
Let $\nabla$ be a hermitean graded connection on $\algrA$ of degree $0$. For $\kX\in\Der^\bullet(\algA)$, we define the gauge potential associated to $\nabla$: $-iA_\kX=\nabla(\gone)(\kX)$. The map $A:\kX\mapsto A_\kX$ is in $\Omega^{1,0}(\algA|\kg)$ and if $\kX$ is real, $A_\kX$ is real too.

Then, the graded connection $\nabla$ takes the form: $\forall\kX\in\Der^\bullet(\algA)$, $\forall a\in\algrA$,
\begin{equation*}
\nabla_\kX a=\kX(a)-iA_\kX\fois a,
\end{equation*}
where $\nabla_\kX a=(-1)^{|\kX||a|}\nabla(a)(\kX)$. Defining $F_{\kX,\kY}\fois a=i(-1)^{(|\kX|+|\kY|)|a|}R(a)(\kX,\kY)$, we obtain the curvature: $\forall\kX,\kY\in\Der^\bullet(\algA)$,
\begin{equation*}
F_{\kX,\kY}=\kX(A_\kY)-(-1)^{|\kX||\kY|}\kY(A_\kX)-i[A_\kX,A_\kY]-A_{[\kX,\kY]}.
\end{equation*}
\end{proposition}
\begin{proof}
Indeed, for $\kX\in\Der^\bullet(\algA)$ and $a\in\algrA$,
\begin{equation*}
\nabla(a)(\kX)=\nabla(\gone\fois a)(\kX)=\nabla(\gone)(\kX)+\gone\fois \dd a(\kX).
\end{equation*}
The curvature simplifies as:
\begin{align*}
R(a)&(\kX,\kY)=(-1)^{|\kX||\kY|}\nabla(\nabla(a)(\kY))(\kX)-\nabla(\nabla(a)(\kX))(\kY)-\nabla(a)([\kX,\kY])\\
=&(-1)^{|\kX||\kY|}\dd(\dd a(\kY))(\kX)-\dd(\dd a(\kX))(\kY)-\dd a([\kX,\kY])\\
&-i(-1)^{(|\kX|+|a|)|\kY|}\dd(A_\kY\fois a)(\kX)-i(-1)^{|a||\kX|}A_\kX\fois\dd a(\kY)-(-1)^{|a|(|\kX|+|\kY|)}A_\kX\fois A_\kY\fois a\\
&+i(-1)^{|a||\kX|}\dd(A_\kX\fois a)(\kY)+i(-1)^{(|\kX|+|a|)|\kY|}A_\kY\fois\dd a(\kX)\\
&+(-1)^{|\kX||\kY|+|a|(|\kX|+|\kY|)}A_\kY\fois A_\kX\fois a+i(-1)^{|a|(|\kX|+|\kY|)}A_{[\kX,\kY]}\fois a.
\end{align*}
By using
\begin{equation}
0=\dd^2a(\kX,\kY)=(-1)^{|\kX||\kY|}\dd(\dd a(\kY))(\kX)-\dd(\dd a(\kX))(\kY)-\dd a([\kX,\kY]),\label{eq-gauge-d2ident}
\end{equation}
we find:
\begin{equation*}
R(a)(\kX,\kY)=(-1)^{|a|(|\kX|+|\kY|)}(-i\kX(A_\kY)+i(-1)^{|\kX||\kY|}\kY(A_\kX)-[A_\kX,A_\kY]+iA_{[\kX,\kY]})\fois a.
\end{equation*}
\end{proof}

Therefore, we set:
\begin{subequations}
\label{eq-gauge-potmoy}
\begin{align}
\nabla(\gone)(\ad_{(0,i\gamma)})&=(0,-i\varphi),\\
\nabla(\gone)(\ad_{(i\xi_\mu,0)})&=(-iA_\mu^0,0),\\
\nabla(\gone)(\ad_{(0,i\xi_\mu)})&=(0,-iA_\mu^1),\\
\nabla(\gone)(\ad_{(i\eta_{\mu\nu},0)})&=(-iG_{\mu\nu},0).
\end{align}
\end{subequations}
The associated curvature can then be expressed as:
\begin{subequations}
\begin{align*}
F_{(0,i\gamma),(0,i\gamma)}&=(2i\alpha\varphi-2i\alpha\varphi\star\varphi,0)\\
F_{(i\xi_\mu,0),(0,i\gamma)}&=(0,\partial_\mu\varphi-i[A_\mu^0,\varphi]_\star)\\
F_{(0,i\xi_\mu),(0,i\gamma)}&=(-i\alpha\wx_\mu\varphi-i\alpha\{A_\mu^1,\varphi\}_\star+2i\alpha(A_\mu^1-A_\mu^0),0)\\
F_{(i\eta_{\mu\nu},0),(0,i\gamma)}&=(0,-\frac 14\wx_\mu\partial_\nu\varphi-\frac 14\wx_\nu\partial_\mu\varphi-i[G_{\mu\nu},\varphi]_\star)\\
F_{(i\xi_\mu,0),(i\xi_\nu,0)}&=(\partial_\mu A_\nu^0-\partial_\nu A_\mu^0-i[A_\mu^0,A_\nu^0]_\star,0)\\
F_{(i\xi_\mu,0),(0,i\xi_\nu)}&=(0,\partial_\mu A_\nu^1-\partial_\nu A_\mu^0-i[A_\mu^0,A_\nu^1]_\star-\Theta_{\mu\nu}^{-1}\varphi)\\
F_{(0,i\xi_\mu),(0,i\xi_\nu)}&=(-i\alpha\wx_\mu A_\nu^1-i\alpha\wx_\nu A_\mu^1-i\alpha\{A_\mu^1,A_\nu^1\}_\star-i\alpha G_{\mu\nu},0)\\
F_{(i\xi_\mu,0),(i\eta_{\nu\rho},0)}&=(\partial_\mu G_{\nu\rho}+\wx_\nu\partial_\rho A_\mu^0+\wx_\rho\partial_\nu A_\mu^0-i[A_\mu^0,G_{\nu\rho}]_\star+2\Theta^{-1}_{\nu\mu}A_\rho^0+2\Theta^{-1}_{\rho\mu}A_\nu^0,0)\\
F_{(0,i\xi_\mu),(i\eta_{\nu\rho},0)}&=(0,\partial_\mu G_{\nu\rho}+\wx_\nu\partial_\rho A_\mu^1+\wx_\rho\partial_\nu A_\mu^1-i[A_\mu^1,G_{\nu\rho}]_\star+2\Theta^{-1}_{\nu\mu}A_\rho^1+2\Theta^{-1}_{\rho\mu}A_\nu^1)\\
F_{(i\eta_{\mu\nu},0),(i\eta_{\rho\sigma},0)}&=(-\wx_\mu\partial_\nu G_{\rho\sigma}-\wx_\nu\partial_\mu G_{\rho\sigma}+\wx_\rho\partial_\sigma G_{\mu\nu}+\wx_\sigma\partial_\rho G_{\mu\nu}-i[G_{\mu\nu},G_{\rho\sigma}]_\star\\
&-2\Theta^{-1}_{\mu\rho}G_{\nu\sigma}-2\Theta^{-1}_{\nu\rho}G_{\mu\sigma}-2\Theta^{-1}_{\mu\sigma}G_{\nu\rho} -2\Theta^{-1}_{\nu\sigma}G_{\mu\rho},0),
\end{align*}
\end{subequations}
where we have omitted the $ad$ in the indices of $F$ to simplify the notations. Note that for dimensional consistency, one has to consider $i\sqrt{\theta}\eta_{\mu\nu}$ rather than $i\eta_{\mu\nu}$ in the action, $\frac{i}{\sqrt{\theta}}\gamma$ rather than $i\gamma$, and to rescale conveniently the fields $\varphi$ and $G_{\mu\nu}$.

\begin{proposition}
The unitary gauge transformations $\Phi$ of $\algrA$ of degree $0$ are completely determined by $g=\Phi(\gone)$: $\forall a\in\algrA$, $\Phi(a)=g\fois a$, and $g$ is a unitary element of $\algrA$ of degree 0: $g^\ast\fois g=\gone$. Then, this gauge transformation acts on the gauge potential and the curvature as: $\forall\kX,\kY\in\Der^\bullet(\algA)$,
\begin{align}
A_\kX^g &=g\fois A_\kX\fois g^\ast+ig\fois \kX(g^\ast),\label{eq-gauge-gtpot}\\
F_{\kX,\kY}^g &=g\fois F_{\kX,\kY}\fois g^\ast.\label{eq-gauge-gtcurv}
\end{align}
\end{proposition}
\begin{proof}
$\forall \kX\in\Der^\bullet(\algA)$ and $\forall a\in\algrA$,
\begin{align*}
\nabla_\kX^g a=&g\fois (\kX(g^\ast\fois a)-iA_\kX\fois g^\ast\fois a)\\
=& \kX(a)+g\fois\kX(g^\ast)\fois a-ig\fois A_\kX\fois g^\ast\fois a.
\end{align*}
Since $\nabla^g_\kX a=\kX(a)-iA_\kX^g\fois a$, we obtain the result \eqref{eq-gauge-gtpot}. The proof is similar for \eqref{eq-gauge-gtcurv}.
\end{proof}

In the following, we will denote the gauge transformations (of degree 0) by $(g,0)$, where $g\in\caM$ and $g^\dag\star g=1$. Then, the gauge potentials \eqref{eq-gauge-potmoy} transform as:
\begin{subequations}
\begin{align}
(0,\varphi)^g&=(0,g\star\varphi\star g^\dag),\label{eq-gauge-gtpot0}\\
(A_\mu^0,0)^g&=(g\star A_\mu^0\star g^\dag+ig\star\partial_\mu g^\dag,0),\label{eq-gauge-gtpot1}\\
(0,A_\mu^1)^g&=(0,g\star A_\mu^1\star g^\dag+ig\star\partial_\mu g^\dag),\label{eq-gauge-gtpot2}\\
(G_{\mu\nu},0)^g&=(g\star G_{\mu\nu}\star g^\dag-\frac i4g\star(\wx_\mu\partial_\nu g^\dag)-\frac i4g\star(\wx_\nu\partial_\mu g^\dag),0).\label{eq-gauge-gtpot3}
\end{align}
\end{subequations}

\begin{proposition}
\label{prop-gauge-conninv}
In this (graded) Moyal case, there exists a form $A^\invar\in\Omega^{1,0}(\algA|\kg)$ such that: $\forall a\in\algrA$, $\forall \kX\in\Der^\bullet(\algA)$,
\begin{equation*}
\dd a(\kX)=[iA^\invar,a](\kX).
\end{equation*}
Moreover, the graded connection $\nabla^\invar$ canonically associated to $A^\invar$ is gauge invariant: $(A^\invar_\kX)^g=A^\invar_\kX$, and the covariant coordinate associated to any graded connection: $\caA_\kX=A_\kX-A^\invar_\kX$ is gauge covariant.
\end{proposition}
\begin{proof}
Let us define
\begin{align*}
iA^\invar(\ad_{(0,i\gamma)})&=(0,i\gamma),\quad& iA^\invar(\ad_{(i\xi_\mu,0)})&=(i\xi_\mu,0),\\ iA^\invar(\ad_{(0,i\xi_\mu)})&=(0,i\xi_\mu), \quad& iA^\invar(\ad_{(i\eta_{\mu\nu},0)})&=(i\eta_{\mu\nu},0).
\end{align*}
We see that for $a\in\{(0,i\gamma),(i\xi_\mu,0),(0,i\xi_\mu),(i\eta_{\mu\nu},0)\}$, $iA^\invar_{\ad_a}=a$. Then,
\begin{equation*}
(A^\invar_{\ad_a})^g=g\fois A^\invar_{\ad_a}\fois g^\ast+ig\fois[a,g^\ast]_\eps=-ia=A^\invar_{\ad_a}.
\end{equation*}
\end{proof}

In function of the covariant coordinates
\begin{subequations}
\begin{align*}
\Phi&=\varphi-1,\\
\caA_\mu^0&=A_\mu^0+\frac 12\wx_\mu,\\
\caA_\mu^1&=A_\mu^1+\frac 12\wx_\mu,\\
\caG_{\mu\nu}&=G_{\mu\nu}-\frac 12\wx_\mu\wx_\nu,
\end{align*}
\end{subequations}
the curvature takes the new form:
\begin{subequations}
\begin{align*}
F_{(0,i\gamma),(0,i\gamma)}&=(2i\alpha-2i\alpha\Phi\star\Phi,0)\\
F_{(i\xi_\mu,0),(0,i\gamma)}&=(0,-i[\caA_\mu^0,\Phi]_\star)\\
F_{(0,i\xi_\mu),(0,i\gamma)}&=(-i\alpha\{\caA_\mu^1,\Phi\}_\star-2i\alpha\caA_\mu^0,0)\\
F_{(i\eta_{\mu\nu},0),(0,i\gamma)}&=(0,-i[\caG_{\mu\nu},\Phi]_\star)\\
F_{(i\xi_\mu,0),(i\xi_\nu,0)}&=(\Theta^{-1}_{\mu\nu}-i[\caA_\mu^0,\caA_\nu^0]_\star,0)\\
F_{(i\xi_\mu,0),(0,i\xi_\nu)}&=(0,-i[\caA_\mu^0,\caA_\nu^1]_\star-\Theta^{-1}_{\mu\nu}\Phi)\\
F_{(0,i\xi_\mu),(0,i\xi_\nu)}&=(-i\alpha\{\caA_\mu^1,\caA_\nu^1\}_\star-i\alpha\caG_{\mu\nu},0)\\
F_{(i\xi_\mu,0),(i\eta_{\nu\rho},0)}&=(-i[\caA_\mu^0,\caG_{\nu\rho}]_\star+2\Theta^{-1}_{\nu\mu}\caA_\rho^0 +2\Theta^{-1}_{\rho\mu}\caA_\nu^0,0)\\
F_{(0,i\xi_\mu),(i\eta_{\nu\rho},0)}&=(0,-i[\caA_\mu^1,\caG_{\nu\rho}]_\star+2\Theta^{-1}_{\nu\mu}\caA_\rho^1 +2\Theta^{-1}_{\rho\mu}\caA_\nu^1)\\
F_{(i\eta_{\mu\nu},0),(i\eta_{\rho\sigma},0)}&=(-i[\caG_{\mu\nu},\caG_{\rho\sigma}]_\star-2\Theta^{-1}_{\mu\rho}\caG_{\nu\sigma} -2\Theta^{-1}_{\nu\rho}\caG_{\mu\sigma}-2\Theta^{-1}_{\mu\sigma}\caG_{\nu\rho}-2\Theta^{-1}_{\nu\sigma}\caG_{\mu\rho},0).
\end{align*}
\end{subequations}

\subsection{Discussion and interpretation}
\label{subsec-gauge-epsdisc}

Like in the scalar case (see subsection \ref{subsec-gauge-epscalcdiff}), we identify $A_\mu^0=A_\mu^1=A_\mu$ for the computation of the action. This can be done because they have the same gauge transformations \eqref{eq-gauge-gtpot1} and \eqref{eq-gauge-gtpot2}. We then obtain for the Yang-Mills action\footnote{there is an implicit summation on $a,b\in\{(0,\frac{i}{\sqrt{\theta}}\gamma),(i\xi_\mu,0),(0,i\xi_\mu),(i\sqrt{\theta}\eta_{\mu\nu},0)\}$.} $\tr\Big(\sum_{a,b}|F_{\ad_a,\ad_b}|^2\Big)$:
\begin{multline}
S(A,\varphi,G)=\int d^Dx\Big((1+2\alpha)F_{\mu\nu}\star F_{\mu\nu}+\alpha^2\{\caA_\mu,\caA_\nu\}_\star^2 +\frac{8}{\theta}(2(D+1)(1+\alpha)+\alpha^2)\caA_\mu\star\caA_\mu\\
+2\alpha(\partial_\mu\varphi-i[A_\mu,\varphi]_\star)^2+2\alpha^2(\wx_\mu\varphi+\{A_\mu,\varphi\}_\star)^2-4\alpha\sqrt{\theta}\varphi\Theta_{\mu\nu}^{-1}F_{\mu\nu}
+\frac{2\alpha(D+2\alpha)}{\theta}\varphi^2\\
-\frac{8\alpha^2}{\sqrt{\theta}}\varphi\star\varphi\star\varphi +4\alpha^2\varphi\star\varphi\star\varphi\star\varphi -\alpha[\caG_{\mu\nu},\varphi]_\star^2 +2\alpha^2\{\caA_\mu,\caA_\nu\}_\star\star\caG_{\mu\nu}\\
-8i(1+\alpha)\Theta^{-1}_{\nu\mu}[\caA_\rho,\caA_\mu]_\star\star\caG_{\nu\rho} +8i\caG_{\mu\nu}\star\caG_{\rho\sigma}\star(\Theta^{-1}_{\mu\rho}\caG_{\nu\sigma} +\Theta^{-1}_{\nu\sigma}\caG_{\mu\rho} +\Theta^{-1}_{\nu\rho}\caG_{\mu\sigma} +\Theta^{-1}_{\mu\sigma}\caG_{\nu\rho})\\
+(\frac{16}{\theta^2}(D+2)\alpha^2)\caG_{\mu\nu}\star\caG_{\mu\nu} +2\caG_{\mu\nu}\star\caG_{\mu\nu}\star\caG_{\rho\sigma}\star\caG_{\rho\sigma} -2\caG_{\mu\nu}\star\caG_{\rho\sigma}\star\caG_{\mu\nu}\star\caG_{\rho\sigma}\Big),\label{eq-gauge-actymh}
\end{multline}

Taking for the moment $\caG_{\mu\nu}=0$ and $\Phi=0$, which again can be done because these fields transform covariantly, \eqref{eq-gauge-actymh} can be rewritten, up to constant terms, as:
\begin{equation}
S(A)=
\int d^Dx\Big((1+2\alpha)F_{\mu\nu}\star F_{\mu\nu}+\alpha^2\{\caA_\mu,\caA_\nu\}_\star^2+\frac{8}{\theta}(2(D+1)(1+\alpha)+\alpha^2)\caA_\mu\star\caA_\mu\Big),\label{eq-gauge-actgaugeresult}
\end{equation}
where $F_{\mu\nu}=\partial_\mu A_\nu-\partial_\nu A_\mu-i[A_\mu,A_\nu]_\star$. By setting $\Omega^2=\frac{\alpha^2}{1+2\alpha}$ (like in the scalar case) and $\kappa=\frac{8(\alpha^2+2(D+1)(\alpha+1))}{\theta(1+2\alpha)}$, we easily recognize the action \eqref{eq-gauge-actax}, proposed as a candidate for a renormalizable gauge action on Moyal space. When taking $\Phi\neq0$ and $\caG_{\mu\nu}=0$, we obtain an additional real scalar field action coupled to gauge fields:
\begin{multline}
S(A,\varphi)=\int d^Dx\Big(2\alpha|\partial_\mu\varphi-i[A_\mu,\varphi]_\star|^2 +2\alpha^2|\wx_\mu\varphi+\{A_\mu,\varphi\}_\star|^2-4\alpha\sqrt{\theta}\varphi\Theta_{\mu\nu}^{-1}F_{\mu\nu}\\
+\frac{2\alpha(D+2\alpha)}{\theta}\varphi^2-\frac{8\alpha^2}{\sqrt{\theta}}\varphi\star\varphi\star\varphi+4\alpha^2\varphi\star\varphi\star\varphi\star\varphi \Big). \label{eq-gauge-acthiggs}
\end{multline}
Note that the scalar field $\varphi$ is in the adjoint representation of the gauge group \eqref{eq-gauge-gtpot0}. A part of this action can be interpreted as a Higgs action coupled to gauge fields, with harmonic term ``\`a la Grosse-Wulkenhaar'' \cite{Grosse:2004yu} and a positive potential term built from the $\star$-polynomial part involving $\varphi$. Notice that the gauge fields $A_\mu$ are already massive (see the last term in \eqref{eq-gauge-actgaugeresult}). Furthermore, the action \eqref{eq-gauge-acthiggs} involves a BF-like term $\int \varphi\Theta^{-1}_{\mu\nu}F_{\mu\nu}$ similar to the one introduced by Slavnov \cite{Slavnov:2003ae} in the simplest noncommutative extension of the Yang-Mills theory on Moyal space. For more details on BF models, see e.g \cite{Wallet:1989wr,Birmingham:1991} and references therein.

It is not surprising that we obtain a Higgs field in this theory as a part of a connection. This was already the case e.g. for spectral triples approach of the Standard Model \cite{Chamseddine:2006ep} and for the models stemming from the derivation-based differential calculus of \cite{DuboisViolette:1989vq} (see subsection \ref{subsec-gnc-diffcalc}). Notice that a somewhat similar interpretation of covariant coordinates as Higgs fields in the context of gauge theory models on Moyal algebras has also been given in \cite{Cagnache:2008tz}. Note that an additional BF-like term in \eqref{eq-gauge-acthiggs} appears in the present situation. Such a Slavnov term has been shown \cite{Slavnov:2003ae,Blaschke:2005dv}  to improve the IR (and UV) behavior of the following action:
\begin{equation}
\int d^Dx\Big((1+2\alpha)F_{\mu\nu}\star F_{\mu\nu}-4\alpha\sqrt{\theta}\varphi\Theta_{\mu\nu}^{-1}F_{\mu\nu}\Big), \label{eq-gauge-actBF}
\end{equation}
when the field $\varphi$ is not dynamical \cite{Slavnov:2003ae}. The corresponding impact on the UV/IR mixing of the full action in the present situation remains to be analyzed.
\medskip

In view of the discussion given in subsection \ref{subsec-gauge-epscalcdiff}, the grading of the $\eps$-associative algebra $\algrA$ mimics the Langmann-Szabo duality in the scalar case. As far as the gauge theory built from the square of the curvature is concerned, this is reflected in particular into the action \eqref{eq-gauge-actgaugeresult} (The $\caG_{\mu\nu}=\Phi=0$ part). As a consequence, one can view this grading as a generalization of the Langmann-Szabo duality to the scalar and gauge case. Observe that in \eqref{eq-gauge-actgaugeresult}, the part $(\{\caA_\mu,\caA_\nu\}_\star-\frac14\{\wx_\mu,\wx_\nu\}_\star)$ can be viewed as the symmetric counterpart of the usual antisymmetric curvature $F_{\mu\nu}=\frac i4[\wx_\mu,\wx_\nu]_\star-i[\caA_\mu,\caA_\nu]_\star$. Then, the general action \eqref{eq-gauge-actymh} involves the terms
\begin{equation}
\int d^Dx\Big(\alpha^2(\{\caA_\mu,\caA_\nu\}_\star-\frac 12\wx_\mu\wx_\nu)^2+ 2\alpha^2G_{\mu\nu}(\{\caA_\mu,\caA_\nu\}_\star-\frac 12\wx_\mu\wx_\nu)\Big),\label{eq-gauge-actBFsym}
\end{equation}
which can also be seen as the symmetric counterpart of the BF action \eqref{eq-gauge-actBF}, if we interpret $G_{\mu\nu}$ as the ``symmetric'' analog of the BF multiplier $\Theta_{\mu\nu}^{-1}\varphi$.
\medskip

Observe further that the gauge action built from the square of the curvature has a trivial vacuum (and in particular $A_\mu=0$) which avoid therefore the difficult problem \cite{deGoursac:2007qi} to deal with the non-trivial vacuum configurations of \eqref{eq-gauge-actgaugeresult} that have been determined in \cite{deGoursac:2008rb} (see section \ref{sec-gauge-prop}). Notice finally that the gauge action involves as a natural part the Grosse-Wulkenhaar model. This relationship linking the Grosse-Wulkenhaar model with a Langmann-Szabo duality symmetry and a gauge theory model deserves further study that may well reveal so far unexpected classical properties related to mathematical structures underlying these models as well as interesting quantum properties of the gauge action defined on $\algrA$.

\numberwithin{equation}{section}
\chapter*{Conclusion}
\chaptermark{Conclusion}

In this thesis, we have essentially studied a new gauge theory on the noncommutative Moyal space, which is candidate to renormalizability. Indeed, in the introduction to noncommutative geometry of Chapter \ref{cha-gnc}, we have recalled the essence of noncommutative geometry, namely the duality between spaces and algebras, and we have presented in particular a noncommutative analog of the de Rham differential calculus and the theory of connections, that is the differential calculus based on the derivations and its noncommutative connections. This differential calculus permitted us to define a gauge theory on the Moyal space in section \ref{sec-gauge-defgauge}.

We have also constructed the Moyal space by its noncommutative algebra, which is a deformation quantization (see section \ref{sec-moy-presmoy}), exposed the problem of UV/IR mixing, a new type of divergence (see section \ref{sec-moy-uvir}) which does not appear in the renormalization of usual scalar theories, presented in Chapter \ref{cha-ren}. The solution to this problem, discovered by Harald Grosse and Raimar Wulkenhaar, has then been exposed in section \ref{sec-moy-ren}, and an important property of this new scalar model, the vacuum configurations, has been studied. The equation of motion has indeed been solved, by assuming some natural symmetry, and all the minima have been found among these solutions (see Theorem \ref{thm-moy-vac}), in view of a possible spontaneous symmetry breaking. A brief presentation of other scalar and fermionic theories on the Moyal space has also been given.
\medskip

In section \ref{sec-gauge-effact}, we have computed the one loop effective action from the scalar model of Grosse-Wulkenhaar coupled to external gauge fields, and we have obtained the following new gauge action:
\begin{equation*}
S=\int\dd^Dx\left(\frac 14 F_{\mu\nu}\star F_{\mu\nu}+\frac{\Omega^2}{4}\{\caA_\mu,\caA_\nu\}_\star^2+\kappa\caA_\mu\star\caA_\mu\right).
\end{equation*}
This action, related to the renormalizable Grosse-Wulkenhaar model at the quantum level, is then a strong candidate to renormalizability. Moreover, by looking at the diagonal quadratic part of this action (in the $A_\mu$ variables), one sees the same quadratic terms as the one of the Grosse-Wulkenhaar model. It means that after a suitable gauge fixing, the gauge part of this model could avoid the problem of UV/IR mixing.

However, the equation of motion studied in section \ref{sec-gauge-prop} shows that $A_\mu=0$ is not a solution. All the solutions, assuming a natural symmetry, like in the scalar case, have then been computed in Theorem \ref{thm-gauge-vac}. In particular, $\caA_\mu=0$, or equivalently $A_\mu=-\frac 12\wx_\mu$, is a solution of the equation of motion. It leads to non-dynamical matrix model (with trivial propagator), which could be studied in a future work. For the other solutions, it gives rise to complicated computations, so that the properties of these vacua can be first investigated by numerical simulations.
\bigskip

In the Chapter \ref{cha-eps}, we generalized the notion of $\gZ$-graded associative algebras to the one of $\eps$-graded algebras, inspired by the work of Scheunert on commutation factors $\eps$, and $\eps$-Lie algebras (also called color algebras). Two important and compatible structures are then associated to these $\eps$-graded algebras: the $\eps$-bracket, and the $\eps$-derivations of the algebra. We have thus introduced the differential calculus based on these $\eps$-derivations, the notion of $\eps$-connections, and shown that they satisfy globally the same properties as in the non-graded case. This provides a (purely algebraic) step in the study of what could be called ``noncommutative graded spaces''. It has been also illustrated on two opposite types of examples: the $\eps$-commutative algebras, coming from a manifold and a supermanifold, and the matrix algebras, with various elementary and fine gradings. By the way, an interesting continuity of this work would be to study other examples of non-trivial $\eps$-graded algebras, and more generally, to apply noncommutative geometry to graded algebras.
\medskip

Finally, we have introduced in section \ref{sec-gauge-interp} a superalgebra constructed from two copies of the Moyal space, its differential calculus and its theory of connections, by restricting the above notions to the $\gZ_2$-graded case. It has permitted us to interpret the Grosse-Wulkenhaar model, where the Langmann-Szabo duality can be seen a grading exchange, and the new gauge theory as a part of a general action built from a graded structure in the above framework. This more general action involves in fact two additional fields: one can be interpreted as a scalar Higgs field, or as a BF field, depending on the terms of the action. The other field, a symmetric tensor of order two, could take the signification of a ``symmetric analog'' of a BF field. A complementary study of this graded model remains to be done, especially to see which gauge-invariant terms of the action would be necessary to the renormalizability of the theory. Moreover, the vacuum of this action becomes trivial: $A_\mu=0$, thanks to this additional field, and it solves the complicated problem of expanding around a non-trivial vacuum configuration.
\medskip

Thus, the section \ref{sec-ren-gauge} shows us the future steps which remain to be done before obtaining the renormalization of this theory. First, one has to find a suitably chosen gauge fixing. It can be done in the BRS formalism, but it is not sure that this formalism does not have to be adapted to the noncommutative case of Moyal. A general study of the BRS formalism in the noncommutative framework would be an interesting future direction of research, and could also produce results for other noncommutative spaces.
Finally, after the gauge fixing and the investigation of the ghost sector, if we suppose that all UV/IR mixing problems disappear, one must study the Slavnov-Taylor identities to all orders, the cohomology of the possibly new BRS complex adapted to the noncommutative case, and the possible anomalies for this theory. That would be the end of the renormalization proof of this noncommutative gauge theory. Then, it would be interesting to compute the flows of this model to see wether they vanish like in the scalar case, or not.
\bigskip

If one succeeds in the renormalization program described above, one could then couple this gauge theory to fermionic fields, study the renormalizability of a noncommutative quantum electrodynamics, and particularly the flows of its coupling. The question of a suitable commutative limit would be crucial, and one could compare the behavior described by this model at this commutative limit with the predictions of the usual quantum electrodynamics. Furthermore, one can also whish to use this work in order to find renormalizable gauge theories on other noncommutative spaces, like for example on the deformation quantization of symplectic symmetric spaces. In a further perspective, it would be interesting to try to use noncommutative geometry for better understanding quantum properties of a physical theory, without needing to quantisize it.

\bibliographystyle{utcaps}
\bibliography{biblio-these,biblio-perso,biblio-recents}

\end{document}